\documentclass[11pt, a4paper, twoside]{mythesis}
\pdfoutput=1
\usepackage[centering]{geometry}
\def\sls#1{\rlap{\kern .10em /}#1}
\usepackage[scaled]{helvet}
\usepackage{amsmath,amsthm,amssymb}
\usepackage{mathrsfs, enumerate}
\usepackage{ascmac, anyfontsize}
\usepackage{epic, eepic}
\usepackage{graphicx, subfigure, comment}
\usepackage{float, cite}
\usepackage{bm,braket, fancyhdr, url}
\usepackage{makeidx}
\usepackage[pdftex]{xcolor}
\newcommand{\ms}[1]{\mathscr{#1}}
\newcommand{\mc}[1]{\mathcal{#1}}

\definecolor{murasaki}{rgb}{0.44 ,0.33, 0.49}
\definecolor{midori}{rgb}{0, 0.6, 0.3}
\definecolor{ao}{rgb}{0.2,0, 0.9}

 \renewcommand{\headrulewidth}{1.1 pt}
\makeatletter
\def\headrule{{\color{murasaki}\if@fancyplain\let\headrulewidth\plainheadrulewidth\fi
    \hrule\@height\headrulewidth\@width\headwidth
\vskip-\headrulewidth}}
\def\footrule{{\color{murasaki}\if@fancyplain\let\footrulewidth\plainfootrulewidth\fi
    \hrule\@height\footrulewidth\@width\headwidth
\vskip-\footrulewidth}}
\makeatother

\newcommand{\vev}[1]{\left \langle #1 \right \rangle}

\usepackage{hyperref}
\hypersetup{bookmarks=true, bookmarksnumbered=true, linkcolor=ao,
citecolor=magenta,
urlcolor=midori,
colorlinks=true
}

\begin{document}

\thispagestyle{empty}

\begin{center}
\leavevmode \\
\begin{figure}[!h]
\begin{center}
\vspace{-10mm}
\includegraphics[width=310pt,clip]{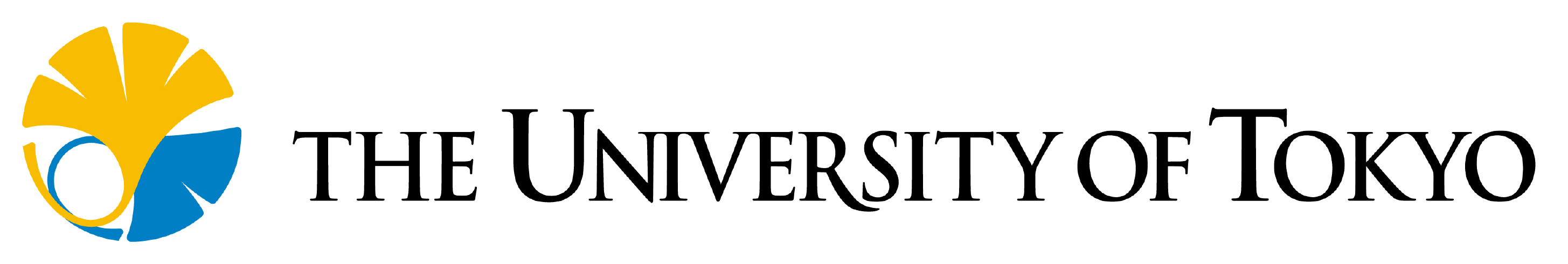}
\end{center}
\end{figure}
\vspace{5mm}
{\fontsize{16pt}{20.74pt} \underline{\phantom{physphysphysphysphysph}{ Ph.D. Thesis}\phantom{physphysphysphyphysphy}} }\\
\vspace{9pt}
\textsf{
{\fontsize{74pt}{0pt} \selectfont Inflation in \\}
{\fontsize{60pt}{50pt} \selectfont Supergravity \\}
{\fontsize{31.5pt}{42pt} \selectfont with a Single Superfield \\}
}
\vspace{-9pt}
{\fontsize{16pt}{20.74pt} \underline{\phantom{physphysphysphysphysph aPh.D. Thesis physphysphysphyphysphy}} }\\

\vspace{2cm}{\LARGE Takahiro Terada
}\\
\vspace{3mm}
{ {\it Advanced Leading Graduate Course for Photon Science, \\
and Department of Physics, Graduate School of Science, The University of Tokyo\\}}
\vspace{3mm}
{ {\it Research Fellow of Japan Society for the Promotion of Science }}

\vspace{60mm}

\begin{table}[!hbt]
\begin{center}
\begin{tabular}{l l}
Submitted:& \phantom{.} December 16, 2014
 \\
Defended:& \phantom{.} January 20, 2015 \\
Revised for Library:& \phantom{.} February 17, 2015  \\
Revised for arXiv: & \phantom{.} August 20, 2015 \\
 \end{tabular}
 \end{center}
 \end{table}

\end{center}
\pagenumbering{roman}

\newpage
\thispagestyle{empty}

\begin{figure}[!htbp]
\begin{center}
\vspace{55pt}
\includegraphics[width=350pt,clip]{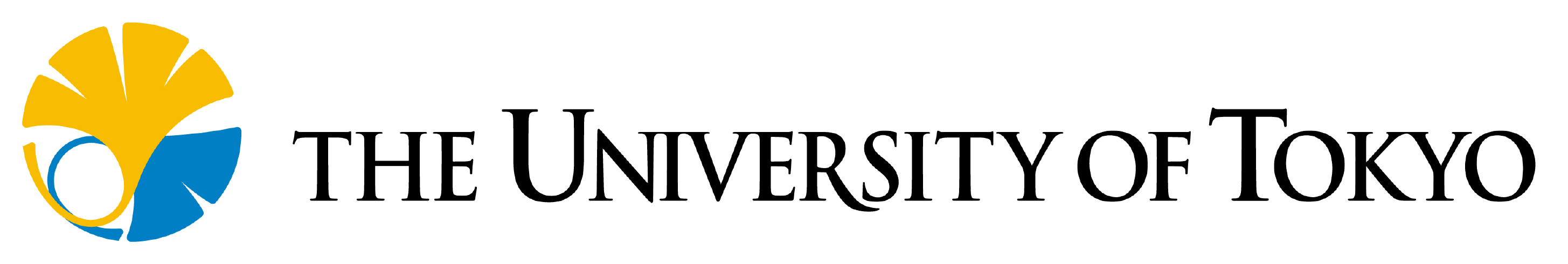}
\end{center}
\end{figure}
\vspace{30pt}
\begin{figure}[!htbp]
\begin{center}
\includegraphics[width=260pt,clip]{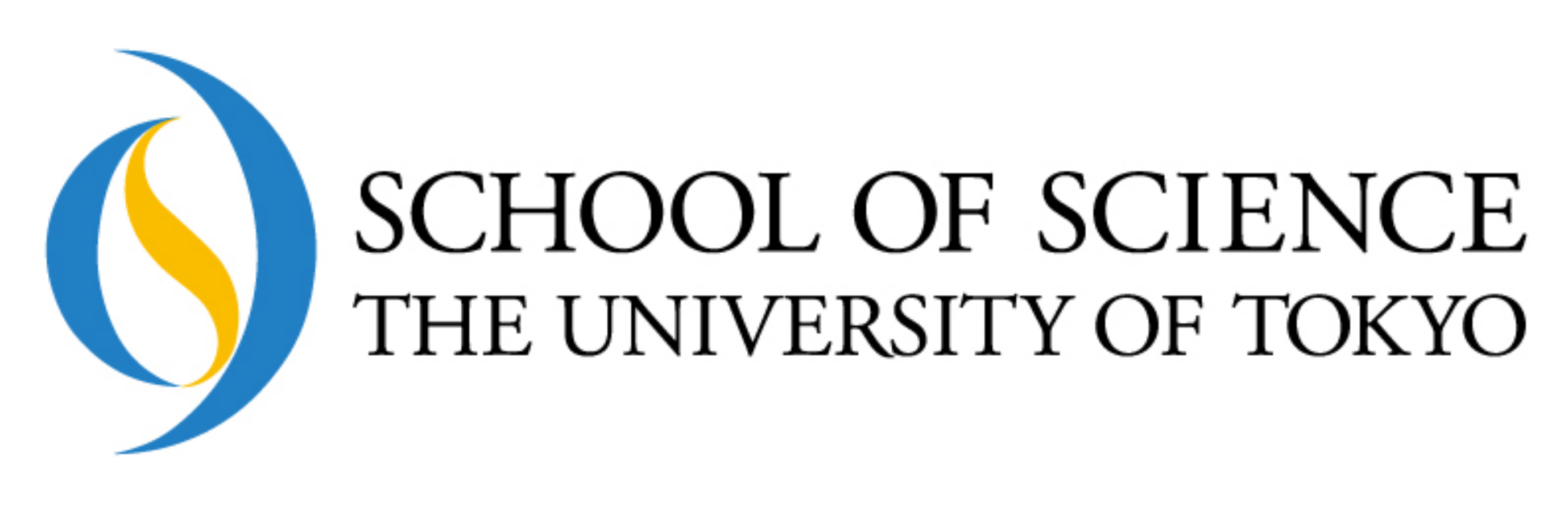}
\end{center}
\end{figure}
\vspace{30pt}
\begin{figure}[!htbp]
\begin{center}
\includegraphics[width=180pt,clip]{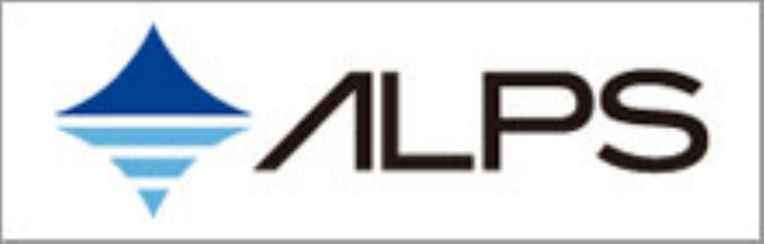}
\end{center}
\end{figure}

\vspace{40pt}

\begin{center}

\large{Thesis Advisor:  Prof.  Koichi Hamaguchi,}\\
\vspace{5mm}

{\large Chief Examiner:  Prof.  Hitoshi Murayama,}
\vspace{0pt}
\end{center}
\begin{table}[!hbtp]
\begin{center}
\begin{tabular}{l l l}
\normalsize Co-Examiner: Prof. Masaki Ando,
& &  Co-Examiner: Prof. Shigeki Matsumoto, \\
& &\\
 Co-Examiner: Prof. Yuji Tachikawa,
& &   Co-Examiner: Prof. Taizan Watari.
\end{tabular}
\end{center}
\end{table}

\large
\newpage

\begin{abstract}

Supergravity is a well-motivated theory beyond the standard model of particle physics, and a suitable arena to study high-energy physics at the early universe including inflation, whose observational evidences are growing more and more.  Inflation in supergravity, however, can not be trivially described because of restrictions from supersymmetry.   The scalar potential has an exponential factor and a large negative term whereas a flat and positive potential is needed to realize inflation.  The standard method to obtain a suitable inflationary scalar potential requires an additional superfield to the one containing inflaton.  In this thesis, we propose and develop an alternative method which does not require the additional superfield and thus reduces the necessary degrees of freedom by half.
  That is, we study inflation in supergravity with only a single chiral superfield which contains inflaton.
  We accomplish it by introducing a higher dimensional term in the inflaton K\"{a}hler potential, which plays an important dual role: fixing the value of the scalar superpartner of the inflaton resulting in effective single field models, and ensuring the positivity of the inflaton potential at the large field region.
  Our proposal is not just particular models but rather a new framework to realize various inflationary models in supergravity.
  In particular, large field inflation in supergravity using one superfield without tuning has become possible for the first time.
  In our generic models, supersymmetry breaking at the inflationary scale by inflaton is not completely restored after inflation, so null results for supersymmetry search at the LHC are predicted for the simplest cases.
  Remarkably, however, it is possible with tuning to embed arbitrary positive semidefinite scalar potentials into supergravity preserving supersymmetry at the vacuum.
 Our discovery opens up an entirely new branch of model building of inflation in supergravity.
\end{abstract}

\newgeometry{top=0pt, left=60pt, right=75pt}

\leavevmode \\
\vspace{50pt}

\section*{Preface for the arXiv version}

\thispagestyle{empty}

This is the arXiv version of the author's Ph.D.~thesis.
Only a few minor changes were made for this version.
The original version of the thesis will be available at \href{http://repository.dl.itc.u-tokyo.ac.jp/index_e.html}{UTokyo Repository}.
Also, the contents of Section~\ref{sec:generalization} were summarized in a proceeding of the HPNP 2015 workshop~\cite{Terada2015a} though some parts have been updated in this version of the thesis.

The achievement of the works~\cite{Ketov2014, Ketov2014a} that this thesis is based on is a breakthrough in the field of inflation in supergravity, and it triggered subsequent interesting developments~\cite{Linde2014c, Linde2015a, Gao2015, Roest2015, Linde2015, Carrasco2015, Nastase2015, Scalisi2015}.
In particular, another approach to inflation in supergravity with a single chiral superfield was proposed by Roest and Scalisi in Ref.~\cite{Roest2015} based on generalization of no-scale supergravity, which was further developed in Ref.~\cite{Scalisi2015}.
A similar model was proposed by Linde~\cite{Linde2015} as a generalization of Goncharov-Linde model~\cite{Goncharov1984d, Goncharov1984, Linde2015a, Kallosh2015b}.
These works appeared after the completion of this thesis, so these are not covered in the review part of the thesis.  Instead, we here briefly comment on some relations between their and our approaches.

The models in Refs.~\cite{Roest2015, Linde2015} are unstable for some parameter choice, and they requires stabilization~\cite{Carrasco2015} (see also Ref.~\cite{Scalisi2015} for the generalization which does not require stabilization in the K\"{a}hler potential).
When we take $c_2=c_4=0$ in eq.~(23) in Ref.~\cite{Carrasco2015}, it reduces to the model in Ref.~\cite{Roest2015}.
When we take $c_4=0$ and $c_2=(\alpha-1)/2$ in the equation, it reduces to the model in Ref.~\cite{Linde2015}.
Now, when we take $c_2=0$ and $\alpha=1$ in the same equation, it is essentially the same structure as eq.~(39) in Ref.~\cite{Ketov2014} (or eq.~\eqref{K-imaginary-fixed}).
(Note also the similarities of superpotentials between eqs.~(14) and (17) in Ref.~\cite{Roest2015} and eq.~(32) in Ref.~\cite{Ketov2014} or its extension, eq.~(B.2) in Ref.~\cite{Ketov2014a}, or equivalently eq.~\eqref{W-n03}.)
Thus, the apparently ad-hoc choice of K\"{a}hler potential in the equation can be understood in terms of the dilatation symmetry of the real part~\cite{Carrasco2015}.
In other examples of K\"{a}hler potential in Refs.~\cite{Ketov2014, Ketov2014a}, we exploit the shift symmetry of the imaginary part.  The two kinds of symmetry transformations form M\"{o}bius group~\cite{Carrasco2015}, so these models are related in the perspective of K\"{a}her geometry.

For this opportunity, it is useful to answer the question the author is sometimes asked: whether our models are the low-energy effective descriptions of the double superfield models~\cite{Kawasaki2000a, Kawasaki2001, Kallosh2010, Kallosh2011} after integrating out the stabilizer superfield.
We think the answer is no for the following reason.
In the latter case, inflation is driven by the supersymmetry (SUSY) breaking of the stabilizer superfield and inflaton does not break SUSY during inflation.  On the other hand, in our case, it is driven by the SUSY breaking of the inflaton superfield.  Thus, SUSY breaking property during inflation is qualitatively different.
Because of this argument, the single superfield models~\cite{Ketov2014, Ketov2014a} are distinguished from the double superfield models.

Since the mechanism of single-superfield inflation has just been discovered, there is room to be explored.
For example, we will show that potentials with very rich structures and SUSY preserving vacuum can be made for a simple polynomial K\"{a}hler potential in an upcoming publication.
It is our pleasure that the readers of the thesis further develop this exciting field.

Some parts of the preface and revision are motivated or benefited from useful discussions with I.~Ben-Dayan, E.~Dudas, S.~Shirai, F.~Takahashi, and A.~Westphal.  \hyperlink{acknowledgement}{Acknowledgement} is attached to the end of the thesis.

\newpage
\leavevmode \\
\vspace{100pt}
\section*{Preface}

This is a Ph.D.~thesis on cosmological inflation in supergravity theory submitted to the University of Tokyo.
The thesis is based on the author's works~[e,~f] in collaboration with S.~V.~Ketov.
It is newly written, and presentations and explanations have been reorganized and integrated.
There are some new contents beyond Refs.~[e,~f] such as Sections~\ref{sec:OtherTerms} and \ref{sec:generalization}.

\section*{List of the author's publication in Ph.D. course}
\begin{enumerate}[{[}a{]}]
\setlength{\itemsep}{5pt}

\item Motoi Endo, Koichi Hamaguchi, and Takahiro Terada, 
``Scalar decay into gravitinos in the presence of $D$-term supersymmetry breaking'', 
 \href{http://dx.doi.org/10.1103/PhysRevD.86.083543}{\textit{Phys.~Rev.}~D 86, 083543 (2012)}, \href{http://arxiv.org/abs/1208.4432}{\texttt{arXiv:1208.4432 [hep-ph]}}.

\item Sergei V. Ketov and Takahiro Terada,
``New Actions for Modified Gravity and Supergravity'', 
\href{http://dx.doi.org/10.1007/JHEP07(2013)127}{\textit{JHEP} 1307 (2013) 127}, \href{http://arxiv.org/abs/1304.4319}{\texttt{arXiv:1304.4319 [hep-th]}}.

\item Sergei V. Ketov and Takahiro Terada,
``Old-minimal supergravity models of inflation'', 
\href{http://dx.doi.org/10.1007/JHEP12(2013)040}{\textit{JHEP} 1312 (2013) 040}, \href{http://arxiv.org/abs/1309.7494}{\texttt{arXiv:1309.7494 [hep-th]}}.

\item Koichi Hamaguchi, Takeo Moroi, and Takahiro Terada, 
``Complexified Starobinsky Inflation in Supergravity in the Light of
  Recent BICEP2 Result'', 
\href{http://dx.doi.org/10.1016/j.physletb.2014.05.006}{\textit{Phys.~Lett.}~B 733 (2014), 305}, \href{http://arxiv.org/abs/1403.7521}{\texttt{arXiv:1403.7521 [hep-ph]}}.

\item Sergei V. Ketov and Takahiro Terada,
``Inflation in Supergravity with a Single Chiral Superfield'',
\href{http://dx.doi.org/10.1016/j.physletb.2014.07.036}{\textit{Phys.~Lett.}~B 736 (2014) 272}, \href{http://arxiv.org/abs/1406.0252}{\texttt{arXiv:1406.0252 [hep-th]}}.

\item Sergei V. Ketov and Takahiro Terada,
``Generic Scalar Potentials for Inflation in Supergravity with a Single Chiral Superfield'', \href{http://dx.doi.org/10.1007/JHEP12(2014)062}{\textit{JHEP} 1412 (2014) 062}, 
 \href{http://arxiv.org/abs/1408.6524}{\texttt{arXiv:1408.6524 [hep-th]}}.

\item Takahiro Terada, Yuki Watanabe, Yusuke Yamada, and Jun'ichi Yokoyama,
``Reheating processes after Starobinsky inflation in old-minimal supergravity'',  \href{http://dx.doi.org/10.1007/JHEP02(2015)105}{\textit{JHEP} 1502 (2015) 105}, 
\href{http://arxiv.org/abs/1411.6746}{\texttt{arXiv:1411.6746 [hep-ph]}}.

\end{enumerate}

\restoregeometry

\newgeometry{top=1.2cm}
\tableofcontents
\restoregeometry
\large
\newpage
\pagenumbering{arabic}

\pagestyle{fancy}
  \fancyhf{}
  \cfoot{\thepage}
    \lhead[]{\leftmark}
    \rhead[\rightmark]{}
    
\chapter{Introduction}
\label{ch:intro}

``What are the fundamental laws of Nature?''
``What happened at the beginning of the universe?''
These deep and fundamental questions have been fascinating people since the ancient time.
Scientifically, these questions are addressed in particle physics and cosmology.
The aim of particle physics or high energy physics is to understand the fundamental ingredients of the world --- particles, strings, or some unknown elements --- and interactions among them as well as to understand the stage these actors play --- spacetime.
The aim of cosmology is to understand the history and destiny of the universe.
Going back in time, the expansion of our universe implies that it began with an extremely hot and dense state, the so-called Big-Bang cosmology.
At very early stage of the universe, everything is decomposed into elementary particles due to high energy interactions.
This is the point where particle physics and cosmology meet.
We try to describe the universe in a way motivated by particle physics.

As of 21st century, our understanding of elementary particle physics is summarized as the standard model of particle physics.
It is based on quantum field theory (QFT) with the gauge group $\text{SU}(3)_c\times \text{SU}(2)_L \times \text{U}(1)_Y$.  The recently discovered Higgs boson~\cite{TheATLASCollaboration2012, TheCMSCollaboration2012}, the final piece of the Standard Model, plays a crucial role in the spontaneous symmetry breaking of the group into $\text{SU}(3)_c \times \text{U}(1)_{\text{EM}}$, and gives masses to quarks and leptons in the gauge invariant manner.
As for gravity, it is usually neglected in low-energy description of particle physics because of its tiny interaction strength at low-energy.
But it becomes important in the cosmological context.
The standard gravitational theory is General Relativity, the theory of gravity as distortion of spacetime.
These theories succeed to explain almost all experiments and observations except for a few anomalies of minor significance, \textit{e.g.}~mismatch of theoretical and experimental values of the muon anomalous magnetic moment.  At least as low-energy effective theories, viability of these theories are intact.
It is remarkable that although the Higgs sector is not as restrictive as the gauge sector, the Higgs observables are consistent with the minimal standard model at the level of the current experimental results obtained at the Large Hadron Collider (LHC) (see \textit{e.g.} Refs.~\cite{ATLASCollaboration2013, ATLASCollaboration2013a, CMSCollaboration2014, CMSCollaboration2015}).

On the other hand, the concordance cosmological model is called $\Lambda$CDM model with $\Lambda$ and CDM referring to the cosmological constant (dark energy) and cold dark matter respectively.
From recent precision observational cosmology~\cite{Hinshaw2013, Ade2015a, Ade2015}, we have been establishing this concordance model with the initial condition supposedly realized by cosmological inflation~\cite{Starobinsky1980, Sato1981, Guth1981a}.
Inflation solves the horizon, flatness, and monopole problems of the ``standard'' Big-Bang model. (It should be safe nowadays to say that the standard cosmology is now inflationary one.)
Moreover, it generates the seeds of densify fluctuation which eventually grows to the Large Scale Structure of the universe.
In the vanilla inflationary model, the potential energy of a scalar field, inflaton, drives exponential expansion of the universe.
The old models of inflation~\cite{Sato1981, Guth1981a} utilize the energy of false vacuum, and it ends with phase transition.
But it does not lead to a graceful exit and the phase transition to the true vacuum does not efficiently happen.
The new models of inflation~\cite{Linde1982, Albrecht1982} and subsequent models are based on the inflaton slowly rolling down its potential.  They are called slow-roll inflation.
Its Gaussian and almost scale invariant quantum fluctuation is the source of the primordial density perturbation.
The inflationary $\Lambda$CDM cosmology is consistent with precise observations such as Planck~\cite{Ade2015a, Ade2015}.

Although agreement of these theories with experiments and observations are spectacular, they are phenomenological. 
In particular, we do not know the nature of dark matter and dark energy despite the fact that it dominates the current energy density of the universe.
We do not know the nature of inflaton neither.
We want to understand them in a unified framework with particle physics.
Dark matter, dark energy, and inflation as well as neutrino oscillation suggests extension of the standard model of particle physics.
In addition to these motivations, the standard model of particle physics has its own reasons to be considered as not the fundamental theory and to be extended.
They are related to the theoretical problem of naturalness.
First, quantum gravity is perturbatively non-renormalizable.
Although the standard model is a renormalizable gauge theory, the whole theory of the standard model coupled to perturbative quantum gravity on fixed background is non-renormalizable.
Then, the theory is effective field theory, and we can easily understand the fact that we can neglect higher dimensional operators in the standard model because they are irrelevant in the renormalization sense.
They are suppressed by some high scale such as the Planck scale.
However, we can not understand the magnitude of coefficients of some of relevant and marginal operators:  the cosmological constant problem~\cite{Weinberg1989}, the (gauge) hierarchy problem of Higgs self-energy~\cite{Gildener1976, Weinberg1979, Susskind1979}, and the strong CP problem~\cite{Kim2010}.

One of the resolution to the hierarchy problems is supersymmetry (SUSY)~\cite{Volkov1972,WessZumino1974b}.
It is a fermionic symmetry which extends the Poincar\'{e} symmetry of spacetime leading to a concept of superspace.
When particle goes around a quantum loop in spacetime, its superpartner also goes around in superspace.  These contributions cancel not to produce large Higgs self-energy and cosmological constant.
Advantages of (low-energy) SUSY include the possibility of gauge coupling unification implying the Grand Unified Theory (GUT) and presence of candidates of dark matter as lightest supersymmetric particles (LSPs).
If Nature is supersymmetric, gravitational sector should be supersymmetric too.
It is the gauged (local) version of SUSY, and called supergravity (SUGRA)~\cite{FreedmanNieuwenhuizenFerrara1976, DeserZumino1976a} since it automatically includes gravity.
It is worth noting that SUSY is required to describe fermions in Superstring or M-Theory~\cite{GreenSchwarzWitten, GreenSchwarzWittena, Polchinskia, Polchinskim}, a candidate of the Theory of Everything.
Low energy landscape of Superstring Theory includes supergravity.

The topic of this thesis is inflation in supergravity theories.
In the context of inflation, SUSY ensures absence of the potential hierarchy problem of inflaton self-energy and of the Higgs in the standard model when they are coupled.
Once adopting SUSY, it should be upgraded to supergravity partly because inflaton traverse field range compatible with or larger than the Planck scale in some models, and partly because inflation, expansion of spacetime, is a phenomenon of gravitation, which is now part of supergravity.
On the other hand, effects of unknown ultraviolet (UV) effects presumably at the Planck scale can be neglected because the energy scale of inflation is typically $V^{1/4}\simeq 10^{-2.5}M_{\text{G}}$ and smaller than the Planck scale.  Here, $V$ is the value of the scalar potential during inflation, and $M_{\text{G}}$ is the reduced Planck mass (the subscript G denotes Gravitational).
Since it is not far below the Planck scale, it may be better to describe inflation in a fundamental theory like superstring.
The advantage of describing inflation in supergravity to such frameworks is that supergravity is better understood and we are not restricted to specific known situations by our limited knowledge of the UV theory.

Describing or even realizing inflation in supergravity is, however, highly non-trivial and not easy~\cite{Yamaguchi2011}.
The primary reason is because the scalar potential of supergravity involves an exponential factor of fields.
Without tuning or symmetrical reasons, any scalar field receives the mass contribution of order the Hubble variable (the Hubble induced mass).
It defies the inflaton to slowly roll down its potential.
Instead, it rapidly rolls down and settles to the minimum of the potential in a few Hubble times.
This is the so-called $\eta$ problem in supergravity~\cite{CopelandLiddleLythStewartWands1994, Yamaguchi2011}.
The details of these arguments will be presented in Section~\ref{ch:review}.
This difficulty made people to tune the K\"{a}hler potential~\cite{Murayama1994, Gelmini1985} and the superpotential~\cite{Nanopoulos1983, Ovrut1983,Goncharov1984, Holman1984}, which defines a model in supergravity, to obtain a suitable scalar potential for inflation.
There are no legitimate reasons to justify their choices of intricate expressions.
These features are shared by various inflationary models including small field and large field models, but the latter type is particularly difficult to be described in supergravity.

A true innovation was made in 2000 by Kawasaki, Yamaguchi, and Yanagida~\cite{Kawasaki2000a}.
They noticed a simple but important fact: a shift symmetry ensures that the exponential factor does not depend on the inflaton.
This means that the inflaton potential no longer has the exponential dependence on the inflaton, and the $\eta$ problem is naturally solved.
Since an exact shift symmetry means the exact flat scalar potential $V=\text{(constant)}$, we have to softly break the shift symmetry after all.
The shift symmetry in their model is broken by the superpotential, and the scale of the superpotential is much smaller than the reduced Planck scale.
So the soft breaking of the shift symmetry is controlled by the small scale of the inflaton superpotential.
In this sense, the shift symmetry breaking is natural~\cite{Hooft1980}.
They successfully embedded the chaotic inflationary model with a quadratic potential into supergravity.
To do that, the shift symmetry is not enough, and they were led to introduce another superfield in addition to the inflaton superfield.
Without the additional superfield, the inflaton potential becomes negative and even unbounded below.
The instability is fixed by introducing the additional superfield whose expectation value vanishes, so let us call it the stabilizer superfield.

Ten years later, Kallosh and Linde generalized the method to more generic superpotentials~\cite{Kallosh2010}.
The superpotential is taken to be the product of the stabilizer superfield and an arbitrary real-coefficient holomorphic function.
It can be used to approximately embed arbitrary inflationary models with positive semidefinite scalar potentials.
The conditions that other scalars than the inflaton are frozen at their origins and do not affect inflation are obtained in their subsequent paper with Rube~\cite{Kallosh2011}.
According to these works, it has become much easier to describe inflation in supergravity.

An alternative framework for inflation utilizing vector or linear supermultiplet was proposed by Farakos, Kehagias, and Riotto in the context of Starobinsky inflation in the new-minimal formulation of supergravity~\cite{Farakos2013}.
The idea to use these supermultiplets was generalized for generic supergravity models by Ferrara, Kallosh, Linde, and Porrati~\cite{Ferrara2013}.
Their theories are also capable of embedding approximately arbitrary positive semidefinite scalar potentials.
In this approach, the inflaton is the scalar component of a massive vector or massive linear supermultiplet.
Prior to gauge fixing, the theory contains a chiral or linear superfield containing the inflaton and a real superfield containing a vector field.  Thus, two superfields are introduced also in this framework.
In fact, the bosonic part of the resultant Lagrangian contains a massive vector or two-form tensor in addition to a real scalar inflaton as well as gravity.
An advantage of this approach is there is no other scalar in the inflaton sector, and we need not to worry about instability of the scalar potential in directions of other scalar fields.

All of the above models or frameworks introduce an additional superfield other than the inflaton superfield.
It is not inconsistent at all to use two superfields to describe inflation in supergravity by itself, but it is not favorable from the perspective of minimality.
We would like to ask the following questions.
\emph{``Are these frameworks the simplest ways of realizing inflation in supergravity?''
``Are they the only methods to embed arbitrary positive scalar potentials into supergravity?''
``Are there any other simple ways of describing inflation in supergravity?''}
Since the vector or tensor supermultiplet plays an essential role in the second approach~\cite{Farakos2013, Ferrara2013}, we focus on the first approach~\cite{Kawasaki2000a, Kallosh2010, Kallosh2011} based solely on chiral superfields.
More concretely, our aim is to examine the possibility of removing the stabilizer superfield.

However, this na\"{i}vely looks impossible.
Actually, the authors of the pioneering work~\cite{Kawasaki2000a} found that inflaton potential becomes unbounded below in the large field region once we impose the shift symmetry.
That is why they introduced the stabilizer superfield.
Because the shift symmetry is crucial for the flatness of the inflaton potential, we also impose it on the theory.
For this setup with a single chiral superfield and a shift symmetry, it was widely believed that large field inflation does not occur.  We show some of them in quotation.
\begin{quote}
	\textit{
	``Large field sgoldstino inflation does not work, at least not for potentials that grow at most polynomial.''
	}~\cite{Achucarro2012}
	\end{quote}
	\begin{quote}
	\textit{
	``We conclude that large field sgoldstino inflation in a sugra model does not work as it is plagued by an instability in the scalar potential.''
	}~\cite{Achucarro2012}
	\end{quote}
	\begin{quote}
	\textit{
	``We note that it is certainly not impossible to have large field inflation in sugra, only that it does not work with a single chiral superfield.''
	}~\cite{Achucarro2012}
	\end{quote}
	See also Section 2 of Ref.~\cite{Roest2013} for the discussion against large filed or chaotic inflation with a single superfield setup.

We have found in Refs.~\cite{Ketov2014, Ketov2014a} that it is actually possible to removing the stabilizer superfield in realizing large field inflation in supergravity.
The previous claims and arguments against this possibility shows how our new findings are groundbreaking and revolutionary.
Our proposal is \emph{a framework} for inflation in supergravity rather than some specific models.
The framework uses only a single chiral superfield which contains the inflaton, aside from the supergravity chiral supermultiplet consisting of graviton, gravitino, and auxiliary fields.
Regarding the inflaton sector, \emph{we have reduced the necessary number of degrees of freedom by half} of the above standard frameworks~\cite{Kallosh2010, Kallosh2011, Ferrara2013} from two superfields to one superfield.

Moreover, we proposed a class of single superfield models in supergravity that can accommodate almost \emph{arbitrary positive semidefinite scalar potentials}~\cite{Ketov2014a}.
Thus, our theory is as powerful as other frameworks~\cite{Kallosh2010, Kallosh2011, Ferrara2013} in spite of the fact that the former has only the half degrees of freedom of the latter.
Our discoveries open up novel possibilities of embedding vast kinds of inflationary models in supergravity with a single chiral superfield.

The purpose of this thesis is to extensively study various aspects of the proposed mechanism of inflation in supergravity with a single chiral superfield, and establish its viability.
Instead of introducing the stabilizer superfield, we introduce a higher dimensional operator in the K\"{a}hler potential.
This is the key ingredient in our method and we extensively study the effects of the stabilization term.
Other topics in the thesis include modification of the K\"{a}hler potential, coupling to other fields in the theory, and their effects on inflationary dynamics.

The organization of the thesis is as follows.
In Chapter~\ref{ch:review}, we review inflation in supergravity.
In Chapter~\ref{ch:main}, we present our new, simple, and powerful framework for inflation in supergravity without the need of the stabilizer superfield.
We introduce a higher order term (stabilization term) in the K\"{a}hler potential instead of the stabilizer superfield.
Its effect is extensively studied both analytically and numerically.
To address a naturalness or tuning issue, we study effects of other terms in the K\"{a}hler potential.
At the end of the Chapter, we generalize the method to charged superfields, thereby embed large-field Higgs inflation into the minimal supersymmetric standard model (MSSM) in SUGRA (without stringy effects) for the first time.
The summary, discussion, and conclusions are presented in Chapter~\ref{ch:conclusion}.
We give mini reviews on supergravity and inflation in Appendices~\ref{ch:SUGRA} and \ref{ch:inflation} respectively.

We take the reduced Planck unit where $c=\hbar=M_{\text{G}}=1$ where $M_{\text{G}}=M_{\text{Pl}}/\sqrt{8\pi}=1/\sqrt{8\pi G}$ is the reduced Planck mass.  When we clarify or emphasize dimensionality, we indicates the reduced Planck mass. 
The complex conjugate or Hermitian adjoint of a quantity $A$ is denoted by $\bar{A}$.
The base notation and convention are those of Ref.~\cite{WB}.
For example, the sign convention of the spacetime metric is $(-, +, +, +)$.
The sign of the coefficient of the Einstein-Hilbert action of General Relativity is negative.
For superconformal discussion, we follow Ref.~\cite{Freedman2012} with interpreted normalization augmented.
Appendix~\ref{ch:SUGRA} is not tailored for self-contained detailed introduction for supergravity.
It is an overview and summary of various formulations of supergravity, and help readers compare different  formulations appearing in the thesis.
It may serve as a conceptual introduction for beginners and a concise summary for professionals.  For detailed reviews of supergravity, see \textit{e.g.}~Refs.~\cite{WB, Buchbinder1998d, Gates1983, Weinberg3, Freedman2012}.
Similarly, Appendix~\ref{ch:inflation} is to provide minimal requisite knowledge for inflation to read the thesis.  There are many books or reviews on inflation. See \textit{e.g.}~Refs.~\cite{Linde1990, Mukhanov2005, Liddle2000, Lyth2009, ModernCosmology} and references therein.  The Planck papers~\cite{Ade2014, Ade2015} are also helpful.

\chapter{Review of Inflation in Supergravity}
\label{ch:review}

In this Chapter, we review inflation in supergravity.
There is already a review of inflation in sueprgravity by Yamaguchi~\cite{Yamaguchi2011}.  In the review, emphases are put on various inflationary models such as new, chaotic, hybrid, topological, and Higgs inflation models based on supergravity.  In contrast, we are interested more in general embedding mechanism itself of scalar potential into supergravity and some special classes of models with theoretical interests.
Also we discuss known results for inflation with a single chiral superfield (the so-called sGoldstino inflation), which is to be extended in the next Chapter.
For complimentary information on inflation in non-SUSY, supergravity, or superstring frameworks, see also some recent articles~\cite{Linde2014, Kallosh2014a, Kallosh2014c, Baumann2014a, Westphal2014} and references therein.

First we see non-trivial points in trying to embed inflationary models, especially large-field inflationary models, in supergravity in Section~\ref{sec:diff}.
The main theme of this thesis is the inflationary framework in supergravity with a single chiral superfield, in which inflaton field coincides with sGoldstino field which breaks SUSY during inflation.
We review earlier works on such a framework, sGoldstino inflation, and its limitation in the original form in Section~\ref{sec:sGoldstino}.
Then, we proceed to the standard approach based on a shift symmetry in Section~\ref{sec:SSSS} where another superfield is introduced to stabilize the inflationary potential.
In Section~\ref{sec:vector}, we review an alternative framework with a chiral supermultiplet and a vector or tensor supermultiplet. 
Finally, in relation to minimalistic approaches in Section~\ref{sec:sGoldstino} and in Chapter~\ref{ch:main}, we discuss other ``minimal'' directions to inflation in supergravity based on purely (super)gravitational theories in Section~\ref{sec:pureSUGRA}.

\section{Difficulties of inflation in supergravity}
\label{sec:diff}
The kinetic terms of the scalar sector in the flat space limit of the four dimensional $\mathcal{N}=1$ supergravity is
\begin{align}
\mc{L}_{\text{kin}}=-K_{i\bar{j}}\partial_{\mu}\bar{\phi}^{\bar{j}}\partial^{\mu}\phi^{i}
\end{align}
where a bar denotes complex conjugation, $K_{i\bar{j}}=\partial^{2}K/\partial \phi^{i}\partial\bar{\phi}^{\bar{j}}$ is the K\"{a}hler metric, and $K$ is the K\"{a}hler potential.

The scalar potential  is given by
\begin{align}
V=e^{K}\left( K^{\bar{j}i}D_{i}WD_{\bar{j}}\bar{W}-3\left|W\right|^{2}  \right) +\frac{g^{2}}{2}H_{R}^{AB}D_{A}D_{B} , \label{Vgeneral}
\end{align}
where $W$, $H_{R}^{AB}$, $D_{A}$, and $g$ are the superpotential, the real part of the gauge kinetic function, the Killing potential, and the gauge coupling constant, respectively.
$i, j, \dots$ are field indices while $A, B,\dots$ are gauge indices. The K\"{a}hler covariant derivative is defined as
\begin{align}
D_{i}W= W_{i}+K_{i}W , 
\end{align}
where subscripts denote differentiation, \textit{e.g.} $W_{i}=\partial W / \partial \phi^{i}$.
$K^{\bar{j}i}$ and $H_{R}^{AB}$ are the inverse matrices of the K\"{a}hler metric and the real part of the gauge kinetic function, respectively.

For the time being, let us consider $F$-term inflationary models and ignore $D$-term.
Taking the simplest, minimal K\"{a}hler potential,
\begin{align}
K=\delta_{i\bar{j}} \bar{\Phi}^{\bar{j}}\Phi^{i} ,
\end{align}
the exponential factor in the $F$-term scalar potential becomes
\begin{align}
e^{K}=e^{\sum_{i} | \Phi^{i} |^2}.
\end{align}
Thus, the scalar potential in supergravity typically has an exponential factor of field variables, and it is too steep in the large-field region $|\Phi|>1$ to have a flat potential suitable for inflation.
This obstacle is not limited to the large-field region.
Writing the $F$-term potential as $V=e^{K}\tilde{V}$, $\tilde{V}\simeq V \simeq 3H^2$ due to the Friedmann equation. Expanding the exponential factor, mass terms of order the Hubble parameter are induced,
\begin{align}
V=\tilde{V}+\tilde{V}|\Phi |^2 + \cdots .
\end{align}
This contributes to the second slow-roll parameter $\eta=V''/V$ as $\eta = \eta_0+1 +\cdots$ where $\eta_0$ is the $\eta$ parameter determined from the $\tilde{V}$ part.
To make it small, $\eta \ll 1$, tuning is required unless some symmetries control flatness.
Supersymmetry does not help because it must be broken during inflation to obtain positive vacuum energy.
This is the so-called $\eta$-problem.  See \textit{e.g.} Ref.~\cite{CopelandLiddleLythStewartWands1994}.

\subsection{Approaches to evade the problem} \label{subsec:approaches}
Because of this difficulty, tuning in the superpotential~\cite{Nanopoulos1983, Ovrut1983, Goncharov1984d,  Goncharov1984, Holman1984} or particular choice of the K\"{a}hler potential~\cite{Murayama1994, Gelmini1985} were required in the early attempts for inflation in supergravity.\footnote{
For example, the superpotential in Refs.~\cite{Goncharov1984d, Goncharov1984} is
\begin{align}
W(\Phi)=e^{-\Phi^2/4}\tanh \sqrt{\frac{3}{2}}\left(\Phi -\Phi_0 \right) \sinh \sqrt{\frac{3}{2}}\left(\Phi -\Phi_0 \right),
\end{align}
and the K\"{a}hler potentials studied in Refs.~\cite{Murayama1994} and \cite{Gelmini1985} are
\begin{align}
K=\frac{3}{8}\ln \left( Z+\bar{Z} + \bar{\Phi}\Phi \right) + \left( Z+\bar{Z} + \bar{\Phi}\Phi \right)^2, \label{MSYYKahler}
\end{align}
and
\begin{align}
K=-3\ln \left( Z+\bar{Z}-\frac{2}{a}\bar{\Phi}\Phi + \frac{\beta}{a}\left(\bar{\Phi}\Phi \right)^2 \right) -2 \bar{\Phi}\Phi +\frac{\beta}{2} \left( \bar{\Phi}\Phi \right)^2 , \label{GKNKahler}
\end{align}
respectively, where $Z$ is a no-scale modulus.
These complicated or particularly tailored forms imply how non-trivial it is to obtain a suitable inflationary potential in the supergravity framework.
}
The model in Refs.~\cite{Goncharov1984d, Goncharov1984} is the first chaotic inflation model in supergravity, which is consistent with observation, and it was recently revisited and slightly generalized by one of the authors~\cite{Linde2015a} (see also Ref.~\cite{Kallosh2015b} in which other forms of the model leading to the same inflaton potential are found).
The authors of Refs.~\cite{Murayama1994, Gelmini1985} utilized logarithmic K\"{a}hler potential to have global SUSY-like potentials for the inflaton.\footnote{
For example, the kinetic term and the scalar potential following from eq.~\eqref{MSYYKahler} is~\cite{Murayama1994}
\begin{align}
\mc{L}_{\text{kin}}=& \frac{16\eta^2-3}{32\eta^2} \left( \left(\partial_{\mu}\eta \right)^2+\left(I_{\mu}\right)^2 \right) + \frac{16\eta^2+3}{8\eta} \left| \partial_{\mu} \Phi \right|^2 ,\\
V=& \eta^{3/8} e^{\eta^2}\left( \frac{8\eta}{16\eta^2+3}|W_{\Phi}|^2 + \frac{\left(16\eta^2 -9\right)^2}{8\left(16\eta^2-3\right)}|W|^2  \right),
\end{align}
where $\eta=Z+\bar{Z}+\bar{\Phi}\Phi$, and $I_{\mu}=i\partial_{\mu}(Z-\bar{Z})+i(\Phi\partial_{\mu}\bar{\Phi}-\bar{\Phi}\partial_{\mu}\Phi)$ is a $U(1)$ current.
At the minimum of the $\eta=3/4$, the $|W|^2$ term vanishes and the first term proportional to $|W_{\Phi}|^2$ remains.
Thus, inflationary model building in global SUSY can seamlessly transformed into the supergravity framework.
} 
It was argued in Ref.~\cite{Gelmini1985} that to separate the inflation scale from the low-energy SUSY breaking scale, the no-scale model~\cite{Cremmer1983a, Ellis1984, Lahanas1987} is useful, but (without the stabilization term which is heavily used in this thesis,) the inflaton K\"{a}hler potential has to be generalized to a non-minimal form like eq.~\eqref{GKNKahler} to avoid the run-away behavior of the $Z$ field. 

Contrary to tuning, arguments based on symmetries are favorable.
Discrete $R$-symmetries \cite{Kumekawa1994, Stewart1995, Dine2012} and the Heisenberg symmetry~\cite{Gaillard1995} are useful to ensure flatness of the potential.
We briefly review the model with discrete $R$-symmetry in the next Section.
A simple method utilizing a shift symmetry was proposed in Ref.~\cite{Kawasaki2000a} and generalized in Refs.~\cite{Kallosh2010, Kallosh2011}, which will be discussed in Section~\ref{sec:SSSS} in more detail.

With the help of moduli fields whose values are driven to a quasi-fixed point where the inflaton mass term vanishes, a flat potential can be obtained without tuning, leading to inflection point inflation~\cite{Adams1997}.

The above problem originates from the exponential factor in the $F$-term scalar potential, so it can be avoided by using $D$-term as the driving source of inflation.
$D$-term inflation was proposed in Refs.~\cite{BinetruyDvali1996, Halyo1996} and revisited in Ref.~\cite{Binetruy2004}. 
The usual $D$-term inflation is one of hybrid inflationary models with charged and neutral chiral superfields as well as a real superfield utilizing the Fayet-Iliopoulos (FI) term.
It is also analyzed recently in the superconformal framework, named superconformal $D$-term inflation~\cite{Buchmuller2013z, Buchmuller2013}.
Recently, however, chaotic regime of $D$-term inflation in the hybrid-type model was discovered~\cite{Buchmuller2014}.
See also other simple $D$-term chaotic inflation models~\cite{Kadota2007, Kadota2008, Kawano2008}.
In this thesis, we are mainly interested in $F$-term inflation because inflation with minimal number of degrees of freedom is possible in the $F$-term case.

	Before moving to the discussion on small field inflationary models with a single chiral superfield, we briefly review a model with plural superfields, the SUSY hybrid inflation, because it is a famous, important, natural and typical inflationary model in SUSY.
	The hybrid inflation was proposed in Refs.~\cite{Linde1991, Linde1994b} and embedded in SUSY and supergravity in Refs.~\cite{CopelandLiddleLythStewartWands1994, Dvali1994, Linde1997, Panagiotakopoulos1997a,  Panagiotakopoulos1997}.
	This is an example of sGoldstino inflation, which is discussed in the next Section.
	The superpotential of the SUSY hybrid inflation is simple,
	\begin{align}
	W= \Phi \left( \kappa \hat{\Psi} \check{\Psi} - \mu^2 \right),
	\end{align}
	where $\Phi$ is an inflaton, and $\hat{\Psi}$ and $\check{\Psi}$ are complex conjugate representations of each other with respect to some symmetry (such as GUT gauge symmetry). 
	This form is ensured by an $R$-symmetry under which $\Phi$ is charged and $\hat{\Psi}\check{\Psi}$ is a singlet.
	The parameters can be taken to be real without loss of generality.
	As we will see below, $\Phi$ breaks SUSY during inflation, so it is the sGoldstino during inflation.
	
	For simplicity, we take a minimal K\"{a}hler potential.
	Possible higher order terms of the inflaton allowed by the $R$-symmetry are powers of $\bar{\Phi}\Phi$ in the K\"{a}hler potential.  Its quartic term produces a mass contribution, and it must be suppressed anyway.
	The scalar potential of the model is
	\begin{align}
	V=& e^{|\Phi|^2 + |\hat{\Psi}|^2+|\check{\Psi}|^2} \left( \left(1-|\Phi|^2+|\Phi|^4  \right) \left| \kappa \hat{\Psi}\check{\Psi}-\mu^2 \right|^2 +|\Phi|^2 \left( \left| \kappa \check \Psi +\bar{\hat{\Psi}} \left( \kappa \hat{\Psi} \check{\Psi} - \mu^2 \right) \right|^2  \right. \right. \nonumber \\
	& \qquad \qquad \qquad \left.\left. +\left| \kappa \hat{\Psi} +\bar{\check{\Psi}}\left( \kappa \hat{\Psi} \check{\Psi} - \mu^2 \right) \right|^2 \right) \right)+\frac{g^2}{2}\left( \bar{\hat{\Psi}} t_{A} \hat{\Psi} -\bar{\check{\Psi}}\bar{t_A} \check{\Psi} \right)^2, \label{VhybridSUGRA}
	\end{align}
	where $t_{A}$ is the generator of the gauge algebra with $A$ its index and implicitly summed.
	The essential feature is well described in the global SUSY version, in the decoupling limit of gravity $M_{\text{G}}\rightarrow \infty$ (with $W$ fixed) or the small field limit $|\Phi|, |\hat{\Psi}|, |\check{\Psi}| \ll M_{\text{G}}$. The $F$-term potential in the global SUSY is
	\begin{align}
	V_{F}^{\text{(global)}}= \left| \kappa \hat{\Psi}\check{\Psi}-\mu^2 \right|^2  + \kappa ^2 |\Phi|^2 \left( | \hat{\Psi} |^2 + | \check{\Psi} |^2  \right).
	\end{align}
	The vacuum is located at $\Phi=0, |\hat{\Psi}|=|\check{\Psi}|=\mu/\sqrt{\kappa}$ and supersymmetric.
	Here we have taken the $D$-flat direction as $|\hat{\Psi}|^2=|\check{\Psi}|^2$ as in the $U(1)$ case for simplicity.
	When $|\Phi|$ is larger than $\mu/\sqrt{2\kappa}$, it is energetically favorable for $\hat{\Psi}$ and $\check{\Psi}$ to vanish.  The tree-level masses of the particles along the line $\hat{\Psi}=\check{\Psi}=0$ at the global SUSY level are
	\begin{align}
	m^2_{\Phi}=&0 , \\
	m^2_{(\hat{\Psi}\mp \check{\Psi})/\sqrt{2}}=&\kappa^2 |\Phi |^2 \pm \kappa \mu^2  .
	\end{align}
	Note the vanishing mass of the inflaton at the leading order.  The potential is approximately a constant, $V=\mu^4$.
	This is a consequence of the $R$-symmetry, which forbids quadratic or cubic terms as well as a constant in the superpotential.
	The lighter nonzero mass becomes negative when the inflaton value becomes small.  The critical value of the inflaton is $|\Phi_{\text{c}}|=\mu / \sqrt{\kappa}$.
	As we will see shortly, the inflaton is driven to the smaller value by the quantum (and SUGRA) corrected potential.  The inflation ends with the phase transition of the waterfall field $(\hat{\Psi}+\check{\Psi})/\sqrt{2}$.
	
	Setting $\hat{\Psi}=\check{\Psi}=0$ (note that this actually kills the positive semidefinite SUSY breaking terms in the full supergravity potential~\eqref{VhybridSUGRA}), the full tree-level potential reduces to
	\begin{align}
	V|_{\hat{\Psi}=\check{\Psi}=0}=&e^{|\Phi|^2}\mu^4 \left( 1-|\Phi|^2+|\Phi|^4  \right) \nonumber \\
	=&\mu^4 \left( 1+ \frac{1}{8}\phi^4 +\mc{O}(\phi^6) \right), \label{hybridTree}
	\end{align}
	where we have identified the radial part $\phi$ of $\Phi=\phi e ^{i\theta}/\sqrt{2}$ as the inflaton.
	We see that quadratic terms has canceled each other in this potential, which preserves flatness obtained at the rigid SUSY.
	We again emphasize that this is a consequence of the $R$-symmetry.
	We also consider models with a discrete $R$-symmetry in the next Section.
	Note that the quartic term has arisen due to the supergravity effects.
	The coefficients may change if we include the quartic term $(\bar{\Phi}\Phi)^2$ in the K\"{a}hler potential.
	On top of this tree-level potential, the one-loop quantum correction, Coleman-Weinberg potential~\cite{ColemanWeinberg1973}, should be taken into account.  This is induced by splitting of masses in $\hat{\Psi}$ and $\check{\Psi}$ supermultiplets, and given by~\cite{Linde1997, Yamaguchi2011}
	\begin{align}
	V_{1\text{-loop}}=& \frac{\kappa^2 \mc{N}}{128\pi^2} \left( \left( \kappa \phi^2 -2\mu^2 \right)^2 \ln \frac{\kappa^2 \phi^2 /2 -\kappa \mu^2 }{\Lambda^2} + \left( \kappa \phi^2 + 2\mu^2 \right)^2 \ln \frac{\kappa^2 \phi^2 /2 +\kappa \mu^2 }{\Lambda^2} -2\phi^4 \ln \frac{\kappa^2\phi^2}{\Lambda^2} \right) \nonumber \\
	\simeq & \frac{\mu^4 \kappa^2 \mc{N}}{8\pi ^2} \ln \frac{\phi}{\phi_{\text{c}}} , \label{hybrid1loop}
	\end{align}
	where $\mc{N}$ is the dimension of the gauge group $\hat{\Psi}$ and $\check{\Psi}$ belong to, $\Lambda$ is the renormalization scale, and $\phi_{\text{c}}=\sqrt{2}|\Phi_{\text{c}}|$.
	We have assumed $\phi \gg \phi_{\text{c}}$ in the second equality.
	
	The inflationary observables obtained from the sum of the tree-level and one-loop potentials (eqs.~\eqref{hybridTree} and \eqref{hybrid1loop}) are summarized in Ref.~\cite{Yamaguchi2011}.
	When $\kappa \sqrt{\mc{N}}\gtrsim 0.1$, the inflaton experiences dynamics on the quartic potential, followed by dynamics on the logarithmic potential, during the last 50 or 60 $e$-foldings.
	In this case, the spectral index $n_{\text{s}}$ becomes larger than one, which is excluded. 
	On the other hand, when $\kappa \sqrt{\mc{N}}\lesssim 0.1$, the spectral index is obtained as $n_{\text{s}}\gtrsim 0.985$. 
	Observationally, the spectral index is measured as $n_{\text{s}}=0.9666+0.0062$ ($\Lambda$CDM+tensor; Planck TT+lowP)~\cite{Ade2015} (see Appendix~\ref{ch:inflation}).
	Thus, one of the simplest and motivated model, the SUSY hybrid inflation, has been excluded by recent precise observation at more than 2$\sigma$ level.
	This drives us to search for a more realistic and data-compatible models of inflation in supergravity.

\section{Small field inflation with a single chiral superfield}
\label{sec:sGoldstino}

In this thesis, we pursue mainly inflation in supergravity with a single chiral superfield.
The method introduced and studied in detail in the next Chapter is useful to embed various kinds of inflationary models.  For large field inflation, the formalism is particularly useful because corrections  stemming from finite strength of stabilization to the leading order potential are well suppressed and controlled.
Actually, it is virtually the unique method\footnote{
There is also a way to obtain a viable inflationary model in which the only scalar in the theory is the inflaton imposing the nilpotency condition to the stabilizer superfield~\cite{Antoniadis2014, Ferrara2014a}.  See Subsection~\ref{subsec:nilpotent}.
} to realize large field inflation in supergravity with a single chiral superfield without tuning, to the best of our knowledge.
On the other hand, there are studies on small field inflation using a single chiral superfield in supergravity.
We here review such models or scenarios.

\subsection{SGoldstino inflation}
\label{subsec:sGoldstino}
One such scenario is sGoldstino inflation, a minimal inflationary scenario~\cite{Alvarez-Gaume2010, Alvarez-Gaume2011, Achucarro2012}.
Inflation requires positive energy, and it in turn requires SUSY breaking in supersymmetric theories, global or local.
That is, the superfield driving inflation inevitably breaks SUSY at least during inflation.
The inflaton can thus be identified with sGoldstino field, the scalar partner of Goldstino, which is the analogue of Nambu-Goldstone boson in the case of the fermionic symmetry (supersymmetry).
Recall the supersymmetric conservation equation~\cite{Komargodski2009, Weinberg3} for the Ferrara-Zumino supermultiplet~\cite{Ferrara1975}.
\begin{align}
\bar{D}^{\dot{\alpha}}J_{\alpha\dot{\alpha}}=X, \label{SUSYconservation}
\end{align}
where $J_{\alpha\dot{\alpha}}$ the supercurrent, whose leading bosonic component contains the $U(1)_{R}$ current, the spinor component contains the current of supersymmetry, and the tensor part contains the energy-momentum tensor, the current of translational symmetries of spacetime, and $X$ is a chiral superfield representing breaking of the conformal symmetry.
The $X$ can be represented by superfields in the theory uniquely in the UV up to an ambiguity corresponding to the degrees of freedom of the K\"{a}hler transformation~\cite{Komargodski2009}.  The auxiliary component of $X$ is the order parameter of SUSY breaking, and the spinor component of $X$ is Goldstino in the IR~\cite{Komargodski2009}.  The important point noted in Ref.~\cite{Alvarez-Gaume2010} is that we can naturally identify $X$ as the inflaton without specifying the details of microscopic physics.

The $X$ in the right hand side of eq.~\eqref{SUSYconservation} flows into a constrained superfield $\frac{8f}{3}X_{\text{NL}}$ in the IR~\cite{Komargodski2009}, where $f$ is the order parameter of SUSY breaking, such that it satisfies the nilpotency condition,
\begin{align}
X_{\text{NL}}^2=0. \label{nilpotency}
\end{align}
This implies that the first scalar component  becomes not a dynamical independent scalar field, but a Goldstino bilinear,
\begin{align}
X_\text{NL}=\frac{GG}{2F}+\sqrt{2} \theta G + \theta \theta F, \label{X_NL}
\end{align}
where $G$ is the canonically normalized Goldstino.  The auxiliary field $F$ has an vacuum expectation value (VEV) whose leading term in the Goldstino expansion coincides with the order parameter of SUSY breaking $f$ up to a sign~\cite{Komargodski2009}. 
The proposal in Ref.~\cite{Alvarez-Gaume2010} is to identify the RG flow of $X$ into $X_{\text{NL}}$ as the inflaton trajectory.
Depending on the hierarchy of the inflaton mass and the gravitino mass, the inflaton transforms itself to a pair of the longitudinal component of gravitino (essentially, Goldstino)~\cite{Alvarez-Gaume2012}.

This scenario is interesting in the sense that it depends only on the universal structure of the SUSY theory and does not depend on the detail of the model.
Precisely because of this, however, it does not shed light on a concrete microphysical mechanism of inflation.
In fact, it is just a scenario, and it is argued in subsection 3.3.1 of Ref.~\cite{Achucarro2012} that the simple models by the original authors does not work without tuning.

If we completely identify $X$ as the inflaton, it remains to be the sGoldstino field after inflation, and the SUSY breaking scale is roughly same order as the inflation scale~\cite{Alvarez-Gaume2010, Alvarez-Gaume2011}.
But in general, SUSY breaking driving inflation and the low-energy SUSY breaking can be independent phenomena.
On the other hand, it is quite natural to identify the inflaton as the sGoldstino only during inflation in the case of $F$-term SUSY breaking.  In this broad sense, all inflationary models with a single chiral superfield are models of sGoldstino inflation, and it has nothing to do with whether it is large field, small field, or hybrid inflation.
In this sense, our new framework in the next Chapter is also regarded as sGoldstino inflation, but as we discussed here, the philosophy and explicit implementations are quite different than the original proposals of sGoldstino inflation.

Even if we deal with a single superfield inflation, sGoldstino inflation, there are realistically other sectors in the theory.
We should check whether these other fields affect the inflationary dynamics or not, or require they do not.
In general situations, we have to solve multi-field dynamics during inflation, which is not an easy task.
Various fields interact with each other at least the Planck-suppressed strength.
On the other hand, sGoldstino decouples from other scalar fields by requiring that other fields keep sitting at their minima and preserving SUSY during inflation, \textit{e.g.}~the only SUSY breaking field (sGoldstino) is the inflaton~\cite{Achucarro2012}.  Explicitly, we first change the field basis so that the only SUSY breaking field is $\Phi$ and other fields collectively denoted by $Y$ do not break SUSY: $G_{\Phi}\neq 0$ and $G_{Y}=0$ at a configuration point $Y=Y_0$ ($\bar{Y}=\bar{Y}_0$).  Here we consider the $F$-term SUSY breaking.  Second, we require that the latter condition is preserved along the inflaton trajectory,
\begin{align}
\partial_{\Phi} G_{Y}|_{Y=Y_0}=\partial_{\bar{\Phi}}G_{Y}|_{Y=Y_0}=0, \label{sGrequirement}
\end{align}
where $\partial_{X}$ is a shorthand notation of a derivative operation $\partial / \partial _{X}$.
This particularly means there is no kinetic mixing between $\Phi$ and $Y$.
It also implies block diagonalization of the mass matrix between $\Phi$ and $Y$~\cite{Achucarro2012}.
Under the above requirement~\eqref{sGrequirement}, the general total K\"{a}hler potential, $G=K+\ln |W|^2$, can be written as~\cite{Achucarro2012}
\begin{align}
G(\Phi, \bar{\Phi}, Y, \bar{Y})= &G_0 (\Phi, \bar{\Phi})\nonumber \\
&+ \frac{1}{2} \sum_{i \geq j} \left(  (Y-Y_0)(Y-Y_0) f(\Phi,\bar{\Phi},Y,\bar{Y})+(Y-Y_0)(\bar{Y}-\bar{Y}_0)h(\Phi,\bar{\Phi},Y,\bar{Y})  \right),
\end{align}
where $G_0$, $f$, and $h$ are real functions.
The general mass formula of $Y$ fields with this total K\"{a}hler potential was summarized in Ref.~\cite{Achucarro2012}.
We only consider the simplest case here: there is only one field in $Y$, and the total K\"{a}hler potential can be (additively) separable, \textit{i.e.} the functions $f$ and $h$ do not depend on the inflaton $\Phi$ and its conjugate.  The simplest case tells us basic features and essential things.
The mass eigenvalues are~\cite{Achucarro2012}
\begin{align}
m_{\pm}^2|_{Y_0}=e^{G_0} \left( \left( |x|^2 \pm \frac{1}{2} |b+2| \right)^2 -\frac{b^2}{4}  \right),
\end{align}
where $x=h^{-1}(f-f_{\bar{\Phi}}G_0^{\bar{\Phi}})$ and $b=G_{0\Phi}G_0^{\Phi}-3$.
$b$ is approximately the energy driving inflation in units of the gravitino mass, and $x$ very roughly corresponds to the mass parameter of $Y$.
During inflation, the value of $b$ changes, so do the masses of $Y$.
These should be positive, $|x|<1$ or $|x|>b+1$, (or greater than the Hubble scale) during inflation.
In the case of hybrid inflation, the sign change of these masses is an indication of the end of inflation.
In more general cases not described by the total K\"{a}hler potential~\eqref{sGrequirement}, a complete analysis involving complicated expressions, or model-specific studies are required.

\subsection{Model with discrete $R$-symmetry}
\label{subsec:discreteR}
To obtain a flat potential, let us consider the condition to eliminate the mass term at the origin of the field space for a simple setup.
Consider the minimal K\"{a}hler potential and a general superpotential at the renormalizable level (Wess-Zumino model~\cite{WessZumino1974b}),
\begin{align}
K=&\bar{\Phi}\Phi, \label{minimalK} \\
W=& c_0 + c_1 \Phi + \frac{c_2}{2}\Phi^2 + \frac{c_3}{3}\Phi^3.
\end{align}
There are no reasons to restrict to the minimal K\"{a}hler potential at this stage, but a consistent discussion are given below.
The potential up to the quadratic order is
\begin{align}
V_{\text{quad}}=\left( |c_2|^2 -3 |c_0|^2 \right)|\Phi|^2+ 2\text{Re}\left( \bar{c_1}c_3 \Phi^2 \right) -3 \text{Re} \left( \bar{c_0}c_2\Phi^2 \right).
\end{align}
Note that the terms proportional to $|c_1|^2 |\Phi|^2$ are cancelled each other: The coefficient of canonical kinetic term (1) from $e^{K}$, a binomial coefficient (2) in the $|W_{\Phi}+K_{\Phi}W|^2$ part, and the coefficient of $|W|^2$ term (-3) add up to vanish, $1+2-3=0$.
Barring tuning, the quadratic terms are eliminated by requiring $c_0$, $c_2$, and $c_1$ or $c_3$ vanish.
One can eliminate $c_0$, $c_2$, and $c_3$ with the assumption of $R$-symmetry under which $\Phi$ is charged.
Higher order terms in the superpotential are irrelevant for the mass term at the origin, and higher order terms in the K\"{a}hler potential are also irrelevant thanks to the $R$-symmetry.

The $R$-symmetry is not necessarily a continuous one, and we here take it as a discrete $Z_n$ symmetry.
  The continuous $U(1)$ $R$-symmetry may be realized as an accidental symmetry.  The discrete $Z_n$ $R$-symmetry $n>2$ is sufficient to forbid the terms of zeroth, second, and third power.  Instead, it brings $U(1)_{R}$ breaking terms like $\Phi^{n+1}$, $\Phi^{2n+1}, \dots$, into the superpotential.
The general K\"{a}hler potential and superpotential under the discrete $R$-symmetry are
\begin{align}
K=&\sum_{p, q}   a_{p, q}   \Phi^{qn}   \left( \bar{\Phi} \Phi \right)^p +\text{H.c.},\\
W=& \sum_{r} b_r  \Phi^{rn+1}.
\end{align}
The higher order terms proportional to $a_{p,0}$ have negligible effects on inflationary dynamics as long as they are not large, $|a_{2,0}|<3/8n(n+1)$ and $|a_{p,0}|\lesssim \mc{O}(1)$ for $p\geq 3$~\cite{Kumekawa1994}.
The terms proportional to $a_{p,q}$ with $q\geq 1$ are even higher terms and we neglect them.
We take the terms in the superpotential up to the next to leading order $\propto \Phi^{n+1}$ because we expect higher order terms are suppressed by higher powers of the parameter ($\propto b_1$) that breaks the $U(1)_R$ symmetry.  In summary, we consider the minimal K\"{a}hler potential~\eqref{minimalK} and the superpotential $W=b_0\Phi+b_1 \Phi^{n+1}$.  We can take these two coefficients as real without loss of generality.
Then, the scalar potential reads
\begin{align}
V=& e^{\bar{\Phi}\Phi}\left( b_0^2 \left( 1- |\Phi|^2+|\Phi|^4 \right) +b_1^2 |\Phi|^{2n} \left( (n+1)^2 +(2n-1) |\Phi|^2+|\Phi|^4  \right) \right. \nonumber \\
& \left. \qquad \qquad  +2b_0 b_1 \left( \text{Re}\Phi^n \right) \left( n+1+(n-1)|\Phi|^2+|\Phi|^4  \right)  \right). \label{VdiscreteR}
\end{align}
Assuming all the dimensionful parameters in the superpotential are scaled by a single dimensionful parameter $v (\ll M_{\text{G}})$ with order 1 coefficients (in contrast to the terms in K\"{a}hler potential whose dimensionality is compensated by the reduced Planck mass $M_{\text{G}}$), the potential can be approximated at $|\Phi | \lesssim v \ll M_{\text{G}}$ as
\begin{align}
V= b_0^2 +2b_0 b_1 (n+1)  \left( \text{Re}\Phi^n \right) . \label{VdiscreteRapprox}
\end{align}
The quartic term has been neglected assuming a suitable initial condition for $n\geq 5$~\cite{Kumekawa1994}.

The potential is very flat near the origin, and it can be used as a hilltop-type new inflation model.
Note that the hilltop model is consistent with the Planck constraint (see Figs.~12 and 54 of Ref.~\cite{Ade2015} for the prediction of the quartic ($n=4$) hilltop model).\footnote{
Some comments are in order.
For favored parameter region, inflation does not occur on the hilltop but at the edge of the hill or even on the cliff.  It is misleading that the prediction of the quartic hilltop model asymptotes to that of the linear potential since the presence of only the constant and quartic terms is assumed in drawing the figure.  To ensure the positivity of the potential, it must be deformed to create a vacuum, so the potential shape will look like quadratic near the minimum.
Also, for the case of small VEV $v$, the spectral index is smaller than the 95\% CL constraint.
Higher powers $n\geq 5$ are still consistent with the observation.
}
A complete analysis of the above model would require two-field analysis because the potential near the origin is so flat that both scalar degrees of freedom in $\Phi$ are light. Indeed, the mass squared eigenvalues for the real and imaginary parts, $\Phi=(\phi+i \chi)/\sqrt{2}$, are
\begin{align}
m^2_{\pm}=3\sqrt{2}(n+1)b_0 b_1 \left( \phi - \chi \pm \sqrt{5\phi^2 -2 \phi \chi +\chi^2} \right),
\end{align}
and these are suppressed by the small field values during inflation.
 We will not go into detail of the multi-field behavior, and we neglect the imaginary part assuming that it is located at the minimum as an initial condition along the lines of Ref.~\cite{Kumekawa1994}.
Balancing the $(\text{Re}\Phi^n)$ term in eq.~\eqref{VdiscreteRapprox} with the $b_1^2 |\Phi|^{2n}$ term in eq.~\eqref{VdiscreteR}, the location of the minimum (VEV), $v$, along the real axis is derived as
\begin{align}
v\simeq \left( - \frac{b_0}{(n+1)b_1} \right)^{1/n}.
\end{align}
Note that the $F$-term of the inflaton $\Phi$ vanishes at the vacuum.
Also, the value of the full potential~\eqref{VdiscreteR} is negative at the vacuum,
\begin{align}
V\simeq -3 \left( \frac{n}{n+1} \lambda v^3  \right)^2,
\end{align}
where we have defined $b_0=\lambda v^2$, so uplifting of the potential is necessary.
In Ref.~\cite{Kumekawa1994}, $D$-term symmetry breaking with FI term\footnote{
Some issues on the FI term in supergravity are discussed in Refs.~\cite{Komargodski2009b, DienesThomas2010, CatinoVilladoroZwirner2011} and references therein.
} \cite{FayetIliopoulos1974} was used to uplift the potential.  This does not affect the inflationary dynamics constructed by $F$-term.
On the other hand, uplifting by a constant term in the superpotential, which breaks the $R$-symmetry, was discussed in Refs.~\cite{Dine2012, Achucarro2012}. The constant to be added to the superpotential is $W_0=-\frac{n}{n+1}\lambda v^3$.
This induces a mass contribution $-2|W_0|^2|\Phi|^2$ and also the linear term $-2b_0 (\bar{W_0}\Phi+W_0\bar{\Phi})$.  In our case (as in Ref.~\cite{Kumekawa1994}), this just perturbs the potential, but in the case of Refs.~\cite{Dine2012, Achucarro2012} in which $b_1$ is assumed to be suppressed not by $v$ but by $M_{\text{G}}$,\footnote{
The reason of the Planck suppressed terms are because of the expectation that all global symmetries are broken by quantum gravity effects.  See \textit{e.g.} Ref.~\cite{Banks2011} and references therein.
} the effect is significant so that inflation no longer happens.
In such a case, the $D$-term uplifting is a solution to maintain inflation.

	\subsection{Inflection point inflation}\label{subsec:inflection}
	In small field inflation, one does not have to tune the flatness of the potential along a very long trajectory.
	One can utilize an (accidental) extremal point such as a maximum, saddle point, or inflection point to realize inflation around the point.
	At the inflection point, the second derivative of the potential vanishes by definition, leading to a vanishing $\eta$ parameter.
	If we tune parameters so that the first derivative also vanishes at the inflection point, the other slow-roll parameter $\epsilon$ also vanishes.
	
	Sets of parameters of the superpotential that lead to inflection point inflation was numerically obtained in Ref.~\cite{Achucarro2012} as an existence proof of working models of small field sGoldstino inflation.
	They used five parameters taken to be real, which are the coefficients of the zeroth to fifth power in the superpotential, to satisfy five constraints: (1, 2) The first and second derivatives of the potential with respect to the inflaton vanish at a point, which was taken to be the origin, (3) the height of the potential at the point is determined by the so-called COBE normalization, (4) There exists a minimum the inflaton settles in after the end of inflation, and (5) the cosmological constant vanishes there.
	In addition, is was checked that (6) the inflaton has a positive mass at the vacuum, (7) the non-inflaton field sits at its minimum and does not mix with the inflaton along the trajectory, and (8) its mass is larger than the Hubble scale, for the obtained parameter sets.
	
	One of the parameter set results in a SUSY preserving vacuum, but the other set leads to a SUSY breaking vacuum.
	For both of the parameter sets, the point where the $e$-folding is gained was found to be the inflection point.
	Inflection point inflation predicts the scalar spectral index too low ($n_{\text{s}}\lesssim 0.92$)~\cite{Achucarro2012} to reproduce the results of recent CMB observations (see the end of Section~\ref{sec:diff}).
	If one further tunes the parameter to suppress the cubic term at the inflection point, it can be modified to become a maximal point, and the upper bound of the spectral index is pushed to a larger value, $n_{\text{s}}\lesssim 0.95$~\cite{Achucarro2012}.
	This is allowed at the 2$\sigma$ level.
	The tensor-to-scalar ratio is much suppressed by the almost vanishing $\epsilon$ parameter.

\section{Shift symmetry and a stabilizer superfield}
\label{sec:SSSS}
We review the standard method to describe inflation in supergravity in this Section.
The approach based on a symmetry is introduced in Subsection~\ref{subsec:shift}.
An additional superfield to the inflaton superfield is introduced to avoid negative potential in Subsection~\ref{subsec:stabilizer}.
	\subsection{Shift symmetry}\label{subsec:shift}
	The main difficulty of embedding a positive and flat inflationary potential into supergravity comes from the exponential factor depending on the K\"{a}hler potential in the scalar potential~\eqref{Vgeneral}.
	The minimal K\"{a}hler potential leads to the exponential function and it is too steep.
	Making the coefficient of the exponent does not help because it is always one after canonical normalization.
	One can take a logarithmic K\"{a}hler potential to make the $e^K$ factor non-exponential, but the potential gets back to exponential form after canonical normalization. See the discussion around eq.~\eqref{logKexponentialV}.
	Even if one could find a good K\"{a}hler potential that does not have exponential dependence after canonical normalization, one still has to make sure that the potential is flat enough to avoid the Hubble-induced mass (the $\eta$ problem).
	Such a successful potential also receive radiative corrections and will be destroyed.
	The almost only way around this is to use symmetries to protect the structure of the K\"{a}hler potential.
	
	The clever way is to completely eliminate an inflaton candidate from the K\"{a}hler potential by virtue of a shift symmetry~\cite{Kawasaki2000a}.
	Let us consider the following shift symmetry,
	\begin{align}
	\Phi \rightarrow \Phi' = \Phi - i a,
	\end{align}
	where $a$ is a real transformation parameter.
	This is a matter of convention, and one can always change the direction of the shift in the complex field space.
	A function invariant under this symmetry must be a function only of the real part, $K(\Phi,\bar{\Phi})=K(\Phi+\bar{\Phi})$.  
	In particular, the inflaton candidate (the imaginary part) does not appear in the exponential factor.
	We take this symmetry as a global symmetry.
	Note that this axionic symmetry is also used in natural inflation~\cite{Freese1990,Adams1993}.
	It may have the origin like $U(1)$ symmetry at a higher energy scale, but we do not specify the origin of the shift symmetry.
	
	If the symmetry is exact, the superpotential must be a constant up to a factor absorbable by the K\"{a}hler potential, and the scalar potential becomes just a constant.
	Therefore, we must break the shift symmetry softly.
	It should be an approximate symmetry of the Planck scale perturbed at the inflationary scale.
	It is a good idea to break the shift symmetry by the superpotential since it is more useful for generating interesting scalar potentials than the K\"{a}hler potential is.
	Smallness of parameters breaking the shift symmetry are natural in 't~Hooft's sense~\cite{Hooft1980}, \textit{i.e.} the symmetry breaking is controlled by the small parameters such as inflationary scale in the superpotential, and symmetries are enhanced (the exact shift symmetry appears) in the vanishing limit of the parameters.
	We assume that inflation can be described supersymmetrically, \textit{i.e.} SUSY is not broken above the inflationary scale.\footnote{
	There are inflationary models with SUSY broken at a scale higher than the inflation scale. 
	In particular, multi-natural inflation~\cite{Czerny2014b} in supergravity~\cite{Czerny2014, Czerny2014a} by Czerny, Higaki, and Takahashi is a large-field model with a single chiral superfield.  See also Ref.~\cite{Buchmuller2015} in the context of moduli stabilization.}
	In other words, we assume that the constant term in the superpotential, which respects the shift symmetry, is at most as large as the inflationary scale.
	For typical models of chaotic inflation such as the quadratic model and Starobinsky model, the magnitude of the superpotential (the inflaton mass scale $m$) is fixed to match the amplitude of CMB anisotropy, and it is of order $m\sim \mc{O}(10^{-5})$.
	
	There are no reasons why shift-symmetry breaking is present only in the superpotential.
	For example, quantum corrections produce shift-symmetry breaking terms in the K\"{a}hler potential.
	The lowest order term of such kind is $\Delta K \sim m^2 \bar{\Phi}\Phi$.
	More generally, shift-symmetry breaking in the K\"{a}hler potential may have a different origin from the superpotential.
	For example, let us consider a shift-symmetry breaking term $\Delta K = -\frac{\mc{E}}{2}(\Phi -\bar{\Phi})^2$ following Ref.~\cite{Harigaya2014}.
	Suppose that the original theory without $\Delta K$ leads to an effective single field potential $V_0(\chi )$ where $\chi=\sqrt{2}\text{Im}\Phi$.
	The deformed potential is
	\begin{align}
	V\simeq e^{\mc{E}\chi^2}V_0 (\chi).
	\end{align}
	The scalar spectral index and the tensor-to-scalar ratio are given by
	\begin{align}
	n_{\text{s}}(\chi)=&n_{\text{s},0}(\chi)+4\mc{E}-\sqrt{2 r_0(\chi)}\mc{E}\chi -4 \mc{E}^2\chi^2 , \\
	r(\chi)=& r_0 (\chi)+8\sqrt{2r_0(\chi)}\mc{E}\chi+32\mc{E}^2\chi^2,
	\end{align}
	where the subscript 0 denotes the quantity in the absence of  $\Delta K$.
	Precisely speaking, the field value that corresponds to an $e$-folding number changes to induce same order corrections to the above formulae.
	We neglect this effect since we are now interested in the order of magnitude.
	Requiring that the corrections to the spectral index $n_{\text{s}}$ are much smaller than the current uncertainty of order $6\times 10^{-3}$ (see Appendix~\ref{ch:inflation}),  the constraint on the shift-symmetry breaking parameter $\mc{E}$ in the K\"{a}hler potential is 
	\begin{align}
	\mc{E} \ll 2\times 10^{-3},
	\end{align}
	 assuming $r_0 \lesssim 10^{-3}$.
	If we require that the corrections to the tensor-to-scalar ratio $r$ are much smaller than the current upper bound of order $10^{-1}$, the constraint is not as strong as above, $\mc{E}\ll 4 \times 10^{-3}$ (for $\chi=15$).
	But, if we require that the corrections to $r$ are much smaller than \textit{e.g.} $10^{-3}$, the constraint becomes severer, $\mc{E}\ll 4 \times 10^{-4}$ (for $\chi=15$) and $\mc{E}\ll 1 \times 10^{-3}$ (for $\chi=5$).
	
	These constrains not only parameters of inflaton potential but also coupling constants between inflaton and other fields, which inflaton may decay into to reheat the universe.
	This is because such couplings generate corrections to the inflaton potential in the effective Lagrangian after integrating out high momentum modes.
	For example, coupling like $W\sim c \Phi H_u H_d$ induces the effective K\"{a}hler potential $\Delta K =\mc{E}\Phi\bar{\Phi}$ with $\mc{E}\sim \left( \frac{c}{4 \pi }\right)^2 \ln \left( \frac{\Lambda}{\mu} \right)$ where $\Lambda$ is the cut-off scale such as the reduced Planck scale $M_{\text{G}}$, and $\mu$ is the energy scale of the effective Lagrangian.
	For concreteness, we take $\mc{E}\ll 10^{-3}$ and $\left( \frac{1}{4 \pi }\right)^2 \ln \left( \frac{\Lambda}{\mu} \right)\simeq 10^{-1}$. Then, the constraint on the coupling constant $c$ is
	\begin{align}
	c \ll 10^{-1}. \label{coupling_constraint}
	\end{align}
	For further discussion on the shift-symmetry breaking in the K\"{a}hler potential, see Refs.~\cite{Kawasaki2001, Li2014, Harigaya2014}.
	We conclude this topic here, and neglect the shift-symmetry breaking in the K\"{a}hler potential in the following.
		
	Before proceeding to Subsection~\ref{subsec:stabilizer}, we comment on variants of shift symmetry in the literature.
	We have imposed a shift symmetry on one real scalar field in $\Phi$ above, but what will happen if we impose two shift symmetries on the two real scalar fields in $\Phi$?
	Then, the potential on the entire complex plane of $\Phi$ becomes `flat' since the factor $e^{K}$ is just a constant number.
	The kinetic term for these fields are absent, and they are auxiliary fields.
	However, kinetic terms for $\Phi$ can be restored by introducing a higher-derivative term in the K\"{a}hler potential without breaking the shift symmetries.
	This is exactly the situation studied in Ref.~\cite{Aoki2014a}.
		
	Li \textit{et al.}~proposed helical phase inflation in Ref.~\cite{Li2014a}.
	They considered theories with a global $U(1)$ symmetry, and used the phase direction of a complex field in a chiral superfield as an inflaton.
	$U(1)$ symmetry can be regarded as a special type of shift symmetry.
	The transformation is not a shift in the real or imaginary direction but the phase direction.
	In fact, $U(1)$ transformation  becomes the usual shift symmetry  by field redefinition: 
	\begin{align}
	&\Phi= e^{\phi} , \label{U(1)shiftR} \\
	&\begin{cases}
	\Phi\rightarrow& \Phi e^{i\theta} \qquad \text{($U(1)$ transformation)}  \\
	\phi \rightarrow& \phi + i \theta \qquad \text{(shift transformation)}
	\end{cases}.
	\end{align}
	A $U(1)$ invariant combination becomes a shift invariant combination,
	 \begin{align}
	\bar{\Phi}\Phi=e^{\phi+\bar{\phi}}.
	\end{align}
	We will come back to these models after we introduce the stabilizer superfield.
	
	\subsection{Stabilizer superfield}\label{subsec:stabilizer}
	The story does not end with the shift symmetry.
	Once we impose the shift symmetry on the K\"{a}hler potential, the potential tends to become negative in the large field region of the imaginary part (inflaton).
	This is because the K\"{a}hler potential appears not only in the exponential factor $e^{K}$ but also in the form of the inverse of the K\"{a}hler metric $K^{\bar{\Phi}\Phi}$ and the first derivative $K_{\Phi}$.
	Let us look into it more in detail.
	Take the minimal K\"{a}hler symmetry with shift symmetry.
	\begin{align}
	K=\frac{1}{2}\left(\Phi+\bar{\Phi}\right)^2.
	\end{align}
	Note that this leads to the canonical kinetic term for both the real and imaginary parts of the inflaton $\Phi$.
	Now, the scalar potential is
	\begin{align}
	V=e^{\frac{1}{2}\left(\Phi+\bar{\Phi}\right)^2}\left( \left| W_{\Phi} + \left( \Phi+\bar{\Phi} \right) W \right|^2 -3 |W|^2  \right).
	\end{align}
	In this expression, $W_{\Phi}$ and $W$ depend on both the real and imaginary part of $\Phi$.
	The point is that $K_{\Phi}=(\Phi+\bar{\Phi})$, which is the coefficient of $W$ in the $F$-term SUSY breaking part, does not depend on the imaginary part (inflaton) due to the very shift symmetry.
	If it depended on the imaginary part, the strongest power of the imaginary part would be contained in the $K_{\Phi}W$ part.  Here and hereafter, we assume that the superpotential obeys a simple dimensional analysis $W_{\Phi} \sim W/\Phi$ in the large inflaton field $\Phi$ (or $\text{Im}\Phi$).
	Without the inflaton in the $K_{\Phi}$ factor, the strongest power of the inflaton comes from two $|W|^2$ terms.
	Neglecting the subdominant term $W_{\Phi}$, the potential possesses an approximate $Z_2$ symmetry, $(\Phi+\bar{\Phi})\rightarrow - (\Phi+\bar{\Phi})$. Thus, its expectation value vanishes.
	In conclusion, the potential becomes negative and unbounded below, $V\simeq -3 |W|^2$, in the large inflaton region, $|\Phi |\gg 1$.
	
	This problem seems generic for other choices of K\"{a}hler potential.
	As far as the strongest dependence of the real part comes from the exponential factor, it is a good approximation to minimize the exponential factor instead of the whole potential to obtain the effective single field potential of the inflaton.
	The stationary condition of the real part approximately minimize the K\"{a}hler potential, and $K$ also becomes stationary with respect to the real part, $K_{\Phi} \simeq 0$.  This is the same conclusion as above.  The effective potential for the imaginary part becomes negative.
	
	A solution to this problem is to eliminate the negative definite term $|W|^2$ by introduction of a field $S$ whose VEV vanishes, $\vev{S}=0$~\cite{Kawasaki2000a, Kallosh2010, Kallosh2011}. This field is sometimes called the stabilizer field or the sGoldstino field because it stabilize the inflationary potential to be positive, and its $F$-term breaks SUSY during inflation.  The superpotential is taken to be proportional to this field $S$.
	\begin{align}
	W(\Phi, S)= S f(\Phi), \label{W=Sf}
	\end{align}
	where $f(\Phi)$ is a holomorphic function of $\Phi$.
	Because of the vanishing value of $S$, the superpotential $W$ and its derivative $W_{\Phi}$ with respect to the inflaton $\Phi$ vanish too.
	The only nonzero part is what involves the derivative of the superpotential with respect to the stabilizer $S$.
	\begin{align}
	V=e^{K((\Phi+\bar{\Phi}),S,\bar{S})|_{S=\bar{S}=0}} \left. K^{\bar{S}S}\right|_{S=\bar{S}=0}|f(\Phi)|^2.
	\end{align}
	With this form, it is possible to describe almost arbitrary positive semidefinite scalar potentials in supergravity utilizing the functional freedom to choose $f(\Phi)$.
	
	This method was used first by Kawasaki, Yamaguchi, and Yanagida in Refs.~\cite{Kawasaki2000a} for a K\"{a}hler potential and a superpotential,
	\begin{align}
	K=&\frac{1}{2}(\Phi+\bar{\Phi})^2 + \bar{S}S, \\
	W=&mS\Phi,
	\end{align}
	and developed in Ref.~\cite{ Kawasaki2001}.  These potentials are obtained by imposing a $Z_2$ symmetry under which both $\Phi$ and $S$ are odd, and the $U(1)_{R}$ symmetry under which only the $S$ field is charged, in addition to the shift symmetry.  The possible $m^3 S \Phi^3$ has negligible effects~\cite{ Kawasaki2001}.  The real part has larger mass than the inflaton during large field inflation, and is rapidly damped.   On the other hand,  the real and imaginary parts of $S$ field have comparable masses to the inflaton, but their effects to observables are negligible compared to the inflaton (the imaginary part of $\Phi$)~\cite{Kawasaki2000a, Kawasaki2001}.
	
	There are many works on inflationary model building etc.~following Refs.~\cite{Kawasaki2000a, Kawasaki2001}.
	See \textit{e.g.}~Refs.~\cite{Yamaguchi2001,  Yamaguchi2001a, Kawasaki2002, Yamaguchi2003, Brax2005, Kallosh2008, Davis2008, Takahashi2010} and references therein.
	Among others, Kallosh and Linde generalized the method to general superpotentials and K\"{a}hler potentials in Refs.~\cite{Kallosh2010, Kallosh2011}, and studied the consequences of coupling the inflaton sector to the SUSY breaking sector (the KL model~\cite{Kallosh2004, Kallosh2007}) in Ref.~\cite{Kallosh2011a}.
	Here we review the discussion in Ref~\cite{Kallosh2011}.
	Let us impose the following three symmetries.
	\begin{align}
	S \rightarrow& -S  &   &(Z_2 \text{ for } S), \label{Z2forS} \\
	\Phi \rightarrow & -\bar{\Phi} &  &(Z_2 \text{ for } \text{Re}\Phi), \label{Z2forRePhi} \\
	 \Phi \rightarrow & \Phi+i a &  &(\text{shift symmetry for Im} \Phi) \label{ShiftforImPhi},
	\end{align}
	where $a$ is a real parameter.  The superpotential~\eqref{W=Sf} respects $Z_2$ for $S$ because the sign change of the superpotential does not affect the potential, but breaks the other symmetries.  
	Because of the $Z_2$ symmetry for $S$, the point $S=0$ is an extremum of the potential.  We will derive the condition that it is a minimum.  At $S=0$, there are no kinetic mixing between $\Phi$ and $S$, $K_{\Phi\bar{S}}=K_{S\bar{\Phi}}=0$, and the potential becomes
	\begin{align}
	V=e^{K((\Phi+\bar{\Phi})^2, S=0, \bar{S}=0)}K^{\bar{S}S}((\Phi+\bar{\Phi})^2, S=0, \bar{S}=0) |f(\Phi)|^2. \label{VformulaKLR}
	\end{align}
	The inflaton $\Phi$ enters in the K\"{a}hler potential in the combination $(\Phi+\bar{\Phi})^2$ because of the shift symmetry and the $Z_2$ for $\text{Re}\Phi$.
 	We assume the function $\tilde{f}(\Phi)=f(i\Phi/\sqrt{2})$ is a real holomorphic function, \text{i.e.} all the coefficients of its Taylor series expansion are real up to an overall phase.  Then, $|f(\Phi)|^2$ preserves the $Z_2$ symmetry for $\text{Re}\Phi$ too.
	The scalar potential at $S=0$ preserves $Z_2$ for Re$\Phi$, so Re$\Phi=0$ is an extremum of the potential.  We will derive the condition that it is a minimum of the potential.
	The potential reduces to
	\begin{align}
	V=&e^{K(0,0,0,0)}K^{\bar{S}S}(0,0,0,0)|f(i\text{Im}\Phi)|^2, 
	\end{align}
	where we have set $\Phi$ and $\bar{\Phi}$ to $0$ (implying $\text{Im}\Phi=0$) in the K\"{a}hler potential because it does not depend on the imaginary part, for simplicity of expression.
	It is always possible to normalize $K(0,0,0,0)$ to 1 by a K\"{a}hler transformation.  The factor is absorbed by the superpotential.  Also, $K^{\bar{S}S}(0,0,0,0)$ is normalized to 1 for canonically normalized fields.  Therefore, in terms of the canonically normalized field $\chi=\sqrt{2}\text{Im}\Phi$, the potential finally becomes
	\begin{align}
	V=\tilde{f}^2(\chi).
	\end{align}
	
	Let us compute the mass of the fields other than the inflaton at $\text{Re}\Phi=S=\bar{S}=0$.  
	The first derivatives with respect to $S$ vanish so the mass matrix factorizes to the $\Phi$ block and $S$ block.
	Moreover, because $\phi=\sqrt{2}\text{Re}\Phi$ has the $Z_2$ symmetry, the mixing between the real and imaginary part of $\Phi$ is absent.
	The mass of $\phi$ is 
	\begin{align}
	V_{\phi\phi}=2V\left(K_{\Phi\bar{\Phi}}-K_{S\bar{S}\Phi\bar{\Phi}}K^{\bar{S}S} +\epsilon - \frac{1}{2}\eta \right),
	\end{align}
	where the slow-roll parameters are given by
	\begin{align}
	\epsilon=& \frac{1}{2}\left( \frac{V_{\chi}}{V}  \right)^2=2\left(\frac{\tilde{f}'}{\tilde{f}}\right)^2, &  \eta=&\frac{V_{\chi\chi}}{V}=2\frac{\tilde{f}''}{\tilde{f}}+2\left(\frac{\tilde{f}'}{\tilde{f}}\right)^2. \label{slow-roll}
	\end{align}
	The second derivatives of the $S$ sector at $\text{Re}\Phi=S=\bar{S}=0$ are
	\begin{align}
	V_{SS}=& V(3K_{SS}-K_{SSS\bar{S}}K^{\bar{S}S}), \\
	V_{S\bar{S}}=& V \left( -K_{S\bar{S}S\bar{S}}K^{\bar{S}S}+\epsilon \right).
	\end{align}
	The mass eigenvalues are obtained by diagonalizing these.
	In the simplest case in which K\"{a}hler potential depends on $S$ and $\bar{S}$ only through $\bar{S}S$, we can neglect the off diagonal component $V_{SS}$ and $V_{\bar{S}\bar{S}}$.
	In such a case, the condition $V_{\phi\phi}, V_{S\bar{S}} \gtrsim H^2$ that other fields than the inflaton $\chi$ is rapidly damped to the minimum (origin) is recast in the following condition on the fourth derivative of the K\"{a}hler potential (curvature of the K\"{a}hler manifold).
	\begin{align}
	K_{S\bar{S}\Phi\bar{\Phi}}\lesssim & \frac{5}{6}, & K_{S\bar{S}S\bar{S}}\lesssim & -\frac{1}{3}.	
	\end{align}
	Here we have assumed canonical normalization $K_{\Phi\bar{\Phi}}=K^{\bar{S}S}=1$, neglected slow-roll parameters, and used the Friedmann equation, $V=3H^2$.
	
	In more general cases such that some of the symmetries in eqs.~\eqref{Z2forS}, \eqref{Z2forRePhi}, and \eqref{ShiftforImPhi} are not applicable, separate investigations are required model by model to confirm stability of other fields than the inflaton in the theory.
	
	Let us reconsider two variants of shift symmetry mentioned at the last of Subsection~\ref{subsec:shift}.
	The first is higher-superderivative theory obtained by imposing shift symmetries in both real and imaginary directions.
	The higher-derivative term introduced in Ref.~\cite{Aoki2014a} is (in the global SUSY notation)
	\begin{align}
	\int \text{d}\theta^4 c D^{\alpha}\Phi D_{\alpha}S \bar{D}_{\dot{\alpha}}\bar{\Phi}\bar{D}^{\dot{\alpha}}\bar{S},
	\end{align}
	where $c$ is a coupling constant, and $S$ is the stabilizer (sGoldstino). 
	The kinetic term of $\Phi$ is generated when $S$ breaks SUSY.  Its $F$-term SUSY breaking is the coefficient of the kinetic term of $\Phi$.
	 With the superpotential $W=S f (\Phi )$, they obtained the Lagrangian density of the following form after setting $S=0$ (actually they used Akulov-Volkov SUSY $S^2\equiv 0$, see Subsection~\ref{subsec:nilpotent}) and field redefinition $\varphi= 4 \sqrt{c}\int f(\Phi) \text{d}\Phi$,~\cite{Aoki2014a}
	\begin{align}
	\mc{L}=\frac{-\partial^{\mu}\bar{\varphi}\partial_{\mu}\varphi}{\left( 1- \frac{1}{|f(\Phi(\varphi))|^2 }\partial^{\mu}\bar{\varphi}\partial_{\mu}\varphi \right)^2} - \frac{|f(\Phi(\varphi) )|^2}{ 1- \frac{1}{|f(\Phi(\varphi))|^2 }\partial^{\mu}\bar{\varphi}\partial_{\mu}\varphi } \left( 2- \frac{1}{ 1- \frac{1}{|f(\Phi(\varphi))|^2 }\partial^{\mu}\bar{\varphi}\partial_{\mu}\varphi } \right).
	\end{align}
	This expression is not valid at $f(\Phi )=0$.
	Assuming slow-roll inflation and neglecting derivatives, the leading derivative term is the canonical form, $-\partial^{\mu}\bar{\varphi}\partial_{\mu}\varphi$, and the potential is the dominant term, $\mc{L}\simeq -|f(\Phi(\varphi))|^2$.
	This looks similar to the standard case~\eqref{VformulaKLR}, but the field redefinition may drastically change the shape of the potential.
	In fact, the superpotential of a monomial with an arbitrary power, $f(\Phi)=\lambda_n \Phi^n$, leads to a fractional power potential, $V=|\lambda_n \Phi^{n}|^2 = \tilde{\lambda}_n^2 \phi ^{2n/(n+1)}$, where $\phi=\sqrt{2}|\varphi|$ is the approximately canonically normalized inflaton, and $\tilde{\lambda}_n^2=(2^{-n}(16c)^{-n}(n+1)^{2n}\lambda_n^2 )^{1/(n+1)}$. The power of the potential is bounded as $0\leq 2n/(n+1) \leq 2$ for non-negative $n$.

	Next, we translate helical phase inflation~\cite{Li2014a} into shift symmetric inflation reviewed in this Section.
	They studied a model with its K\"{a}hler potential and superpotential given by
	\begin{align}
	K=& \bar{\Phi}\Phi+\bar{S}S - \zeta (\bar{S}S)^2, \label{helicalK}\\
	W=& a \frac{S}{\Phi} \ln \Phi, \label{helicalW}
	\end{align}
	where $a$ is a coupling constant, 
	  and the quartic term of $S$ is to stabilize the potential of $S$.
	The idea of the quartic term was present from early days~\cite{Yamaguchi2001, Kitano2006, Kallosh2007a}.  Its importance was recognized in inflationary context in Refs.~\cite{Lee2010, Ferrara2011}.
	This kind of the quartic term is often introduced in SUSY description of inflation~\cite{Kallosh2013a, Ellis2013}.
	Also, the quartic term is the key to realize large field inflation in supergravity without the field `$S$', which will be extensively discussed in Chapter~\ref{ch:main}. 
	Setting $S$=0, the scalar potential is
	\begin{align}
	V=a^2 e^{r^2} \frac{1}{r^2}\left( \left( \ln r \right) ^2 + \theta^2 \right),
	\end{align}
	where $\Phi = r e^{i\theta}$.  At the minimum of the radial part, $\vev{r}=1$, its mass is larger than the Hubble parameter $H$.
	Therefore, the potential of the phase part becomes
	\begin{align}
	V=a^2 e^2 \theta^2 .
	\end{align}
	Thus, a simple quadratic term has been obtained.
	Modification of the superpotential leads to other potentials.
	Now we redefine the inflaton as in eq.~\eqref{U(1)shiftR}, $\Phi=e^{\phi}$.
	In terms of $r$ and $\theta$, the new (super)field is $\phi=\ln r + i \theta$.
	The K\"{a}hler potential~\eqref{helicalK} and superpotential~\eqref{helicalW} are now
	\begin{align}
	K= & e^{\phi+\bar{\phi}} +\bar{S}S - \zeta (\bar{S}S)^2, \label{helicalK2} \\
	W=& a Se^{-\phi} \phi. \label{helicalW2}
	\end{align}
	The K\"{a}hler potential for $\phi$ is now shift-symmetric in the imaginary direction though the whole combination $\phi+\bar{\phi}$ is exponentiated.
	The exponential factor in the superpotential does not affect the potential for the imaginary part $\theta$ because the inflaton dependent part in the exponential is the unphysical overall phase of the superpotential.  Thus, the superpotential is essentially linear in the inflaton.
	This is why these $K$ and $W$ reproduces the quadratic potential.
	Results in Ref.~\cite{Li2014a} can be interpreted in terms of the shift symmetry.
	Both descriptions represented by eqs.~\eqref{helicalK} and \eqref{helicalW} or eqs.~\eqref{helicalK2} and \eqref{helicalW2} are equivalent.
	The question is which description is simpler or more natural.
	
		\subsection{Nilpotent stabilizer superfield}\label{subsec:nilpotent}
	The stabilizer field $S$ discussed in Subsection~\ref{subsec:stabilizer} is also the sGoldstino field, whose auxiliary $F$-term breaks SUSY.
	If SUSY is non-linearly realized (during and after inflation), the sGoldstino field obeys the nilpotency condition (\textit{cf.}~\eqref{nilpotency} in Subsection~\ref{subsec:sGoldstino}),
	\begin{align}
	S^2=0. \label{nilpotentS}
	\end{align}
	This removes the fundamental scalar from the superfield $S$, and the leading component is replaced by the Goldstino bilinear (see eq.~\eqref{X_NL}).
	Along with this nilpotent stabilizer, inflationary cosmology becomes simplified because the bosonic sector contains the inflaton field but not the stabilizing scalar $S$.
	Such consideration was recently made in Ref.~\cite{Antoniadis2014} in the context of the SUSY Starobinsky model, and it was promoted to more general cases in Ref.~\cite{Ferrara2014a}.
	The idea was applied to uplifting the Minkowski vacuum to de Sitter one in Refs.~\cite{Kallosh2014b, DallAgata2014, Kallosh2014d}.  Relations to superstring theory were discussed in Refs.~\cite{Ferrara2014a, Kallosh2014e}.
	
	In the standard linear SUSY case, \textit{i.e.}~the case in which $S$ is not nilpotent, its scalar component may obtain a finite value.
	It has to be checked that such a nonzero value of $S$ does not affect the inflationary dynamics along the course of inflationary trajectory.
	To stabilize it to the origin to simplify the scalar potential, the quartic term of $S$ is often introduced in the K\"{a}hler potential~\cite{Lee2010, Ellis2013}.
	This kind of care is not required if $S$ is nilpotent simply because there are no dynamical scalar degrees of freedom in $S$.
	Because its leading component becomes fermion bilinear, it is irrelevant in the bosonic action, which is the essential part to describe inflation.
	Thus, after the standard calculation of supergravity, one may simply set $S=0$ as if the scalar component of $S$ vanishes as far as the purely bosonic action is involved in the question~\cite{Ferrara2014a}.
	
	The fermionic part would be complicated because of the non-linearity.
	The nilpotency condition leads to the Volkov-Akulov non-linear Goldstino action~\cite{Volkov1972, Volkov1973} with higher derivative and higher fermion corrections~\cite{Komargodski2009}.
	On the relation between the nilpotency condition~\eqref{nilpotency} and the non-linear SUSY, see also Refs.~\cite{Rocek1978, Ivanov1978, Lindstrom1979, Casalbuoni1989c}.
	Though it is complicated, the non-linear Goldstino terms are absent in the unitary gauge, in which the Goldstino is absorbed by gravitino and becomes its ``longitudinal'' components provided that we ensure that SUSY breaking is triggered solely by the sGoldstino direction, $G_{S}\neq 0$ and $G_{I}=0$ ($I \neq S$).
	This is the super-Higgs effect in supergravity~\cite{DeserZumino1977, Cremmer1979}.
	
	An interesting feature of this class of models is that there are no higher terms in sGoldstino like $S^3$ or $S^2\bar{S}$ because of the nilpotency~\eqref{nilpotentS}.
	The holomorphic terms $S$ and $\bar{S}$ in the K\"{a}hler potential can be moved to the superpotential by the K\"{a}hler transformation.
	Therefore, the most general form of the super- and K\"{a}hler potentials are 
	\begin{align}
	K=& K_0(\Phi, \bar{\Phi} )+K_1 (\Phi, \bar{\Phi}) \bar{S}S , \\
	W=& W_0 (\Phi ) + W_1 (\Phi)S,
	\end{align}
	where $\Phi$ here denotes other superfields than $S$ collectively.
	For simplicity, we consider the case $\Phi$ denotes one chiral superfield.
	The scalar potential of these general models is
	\begin{align}
	V=e^{K_0} \left( K_{0\Phi\bar{\Phi}}^{-1}\left| W_{0\Phi}+K_{0\Phi}W_0 \right|^2 +K_1^{-1}|W_1 |^2 -3 |W_0 |^2  \right).
	\end{align}
	We require $D_{\Phi}W=W_{0\Phi}+K_{0\Phi}W_0=0$ and $D_{S}W=W_1\equiv e^{-K_0/2}K_1^{1/2} M\neq 0$ at the vacuum.
	Then, the cosmological constant is $V=V_0=M^2-3m_{3/2}^2$ and this should be a quite small number, $V_0 \simeq 10^{-120}$.
	It is obtained by fine-tuning.
	
	The nilpotency~\eqref{nilpotentS} can be implemented by the method of Lagrange multiplier by adding the following term to the superpotential, $W\rightarrow W+\Delta W$,
	\begin{align}
	\Delta W= \Lambda S^2 ,
	\end{align}
	where $\Lambda$ is a Lagrange multiplier chiral superfield.
	The equation of motion of this superfield reproduces eq.~\eqref{nilpotentS}.
	
	Following Refs.~\cite{Kallosh2011, Kallosh2014d}, let us simplify the analysis by assuming $K_0=K_0 ((\Phi+\bar{\Phi})^2)$ and $K_1=K_1((\Phi+\bar{\Phi})^2)$, and that $\widetilde{W}_0(\Phi)=W_0(i\Phi/\sqrt{2})$ and $\widetilde{W}_1=W_1(i\Phi/\sqrt{2})$ are real holomorphic functions.
	The assumption about $K_1$ could be weakened because it is not exponentiated in contrast to $K_0$, but it ensures the symmetry~\eqref{Z2forRePhi}.
	The point $\text{Re}\Phi=0$ is always an extremum.
	Let us first assume it is actually a minimum, and later check the condition.
	We can also take $K_0(0)=0$ (rescale of the superpotential), $K_1(0)=1$ (rescale of $S$), and $K_{0\Phi\bar{\Phi}}=1$ (rescale of $\Phi$), without loss of generality since these functions does not depend on the inflaton value.
	The potential at $\phi\equiv \text{Re}\Phi=S=\bar{S}=0$ is 
	\begin{align}
	V=&|W_{0\Phi}|^2+|W_1|^2-3|W_0|^2 , \\
	=& 2 \widetilde{W}_0 '^{2}(\chi)+ \widetilde{W}_1^2(\chi) -3 \widetilde{W}_0^2(\chi), \label{VGNSinflaton}
	\end{align}
	where $\chi=\sqrt{2}\text{Im}\Phi$ is the inflaton. The first term in each line vanishes at the vacuum because of the assumption $D_{\Phi}W=0$ at the vacuum and the $Z_2$ symmetry.
	The first derivative of the potential with respect to the inflaton at $\phi\equiv \text{Re}\Phi=S=\bar{S}=0$ is
	\begin{align}
	V_{\chi}=& 4\widetilde{W}_0' \widetilde{W}_0'' + 2\widetilde{W}_1 \widetilde{W}_1' -6 \widetilde{W}_0 \widetilde{W}_0'  . \label{NVchi}
	\end{align}
	This is small during inflation and has to vanish, $V_{\chi}=0$, at the vacuum after the end of inflation.
	It is satisfied if we take $\widetilde{W}_0'=\widetilde{W}_1'=0(=W_{0\Phi}=W_{1\Phi})$ at the vacuum.
	Because of the $Z_2$ symmetry, the real and imaginary parts of the inflaton do not mix with each other.
	The mass of the imaginary part (inflaton) $\chi=\text{Im}\Phi$ at $\phi\equiv \text{Re}\Phi=S=\bar{S}=0$ is
	\begin{align}
	V_{\chi\chi}=& 4\left( \widetilde{W}_0''^2+ \widetilde{W}_0' \widetilde{W}_0''' \right)+ 2\left( \widetilde{W}_1'^2 + \widetilde{W}_1 \widetilde{W}_1'' \right) -6 \left( \widetilde{W}_0'^2+\widetilde{W}_0 \widetilde{W}_0''\right).
	\end{align} 
	The slow-roll parameters can be calculated straightforwardly, but they are not as simple as eqs.~\eqref{slow-roll}.
	The mass of the real part (non-inflaton) $\phi=\text{Re}\Phi$ at $\phi\equiv \text{Re}\Phi=S=\bar{S}=0$ is 
	\begin{align}
	V_{\phi\phi}=& 2K_{0 \Phi\bar{\Phi}} V+e^{K_0} \left( \left( -2 K_{0\Phi\bar{\Phi}}^{-2}K_{0\Phi\bar{\Phi}\Phi\bar{\Phi}}-3\right)|W_{0\Phi}|^2  -\frac{3}{2}\left(W_{0\Phi\Phi}\bar{W}_0   + \text{H.c.}\right)  \right. \nonumber \\
	& + K_{0\Phi\bar{\Phi}}^{-1} \left( \left( K_{0\Phi\bar{\Phi}}^2+|K_{0\Phi\Phi}|^2 +K_{0\Phi\bar{\Phi}}\left(K_{0\Phi\Phi} +K_{0\bar{\Phi}\bar{\Phi}}\right) \right)|W_0|^2  +\frac{1}{2}\left( W_{0\Phi\Phi\Phi}\bar{W}_{0\bar{\Phi}} +\text{H.c.} \right)  \right. \nonumber \\
	& \left.  +\left(2K_{0\Phi\bar{\Phi}}+K_{0\Phi\Phi}+K_{0\bar{\Phi}\bar{\Phi}} \right)|W_{0\Phi}|^2 +|W_{0\Phi\Phi}|^2 + \left( \left( K_{0\Phi\bar{\Phi}}+K_{0\bar{\Phi}\bar{\Phi}} \right) \bar{W}_0W_{0\Phi\Phi} +\text{H.c.} \right)      \right) \nonumber \\
	& \left.-2K_1^{-2}K_{1\Phi\bar{\Phi}}|W_1|^2 + K_1^{-1}|W_{1\Phi}|^2  +\frac{1}{2}K_1^{-1}\left( W_{1\Phi\Phi}\bar{W}_1+\text{H.c.}\right) \right) \nonumber \\
	=& 2V - \left(4K_{0\Phi\bar{\Phi}\Phi\bar{\Phi}}-2\right) \widetilde{W}_0'^2 +4 \widetilde{W}_0^2 +4\widetilde{W}_0''^2 - 2\widetilde{W}_0''\widetilde{W}_0 -4 \widetilde{W}_0'\widetilde{W}_0''' \nonumber \\
	& -2K_{1\Phi\bar{\Phi}}\widetilde{W}_1^2+2\widetilde{W}_1'^2-2\widetilde{W}_1''\widetilde{W}_1 \nonumber \\
	\simeq & 4 \widetilde{W}_0^2 +4\widetilde{W}_0''^2 - 2\widetilde{W}_0''\widetilde{W}_0 -2K_{1\Phi\bar{\Phi}}\widetilde{W}_1^2-2\widetilde{W}_1''\widetilde{W}_1 ,
	\end{align}
	where we have used the fact $K_{0\Phi\Phi}=K_{0\Phi\bar{\Phi}}=K_{0\bar{\Phi}\bar{\Phi}}=1$ \textit{etc.} in the second equality, and the last equality is valid only at the vacuum since we have used $V\simeq 0$ and $\widetilde{W}_0'=\widetilde{W}_1'=0$ (see eq.~\eqref{NVchi} and the texts after it).
	To consistently neglect the real part, this mass should be larger than the Hubble scale during inflation.
	The mass squared has to be positive at the vacuum too.
	Let us turn to the fermion masses.
	The gravitino mass is
	\begin{align}
	e^{G/2}=e^{K_0/2}|W_0|=|W_0|, \label{GravitinoMassNS}
	\end{align}
	where we have followed the convention $K_0=0$ at the vacuum in the second equality.
	The inflatino mass (at the vacuum) is
	\begin{align}
	e^{G/2}\left|\nabla_{\Phi}G_{\Phi}+\frac{1}{3}G_{\Phi}G_{\Phi} \right|=& \left|-2 \widetilde{W}''_0+\widetilde{W}_0 K_{0 \Phi \Phi}\right|,
	\end{align}
	where we have applied assumptions that $\Phi$ does not break SUSY at all at the vacuum, and $Z_2$ symmetry of $\text{Re}\Phi$, \textit{i.e.}~$K_{0\Phi}=W_{0\Phi}=0$. 
	When we restrict ourselves to the case $K_0=\frac{1}{2}(\Phi+\bar{\Phi})^2$ and $K_1=1$, these formulae reduce to the results in Ref.~\cite{Kallosh2014d}.
	The roles of the real and imaginary parts are opposite between ours and the reference.
	
	As we discussed in Subsection~\ref{subsec:stabilizer}, originally the models with $W_0=0$ had been studied extensively in the literature with or without the nilpotency constraint~\eqref{nilpotentS}.
	A tiny SUSY breaking or cosmological constant parametrically much suppressed compared to the inflationary scale can be realized if we introduce another sector unrelated to the inflation sector provided the inflation sector does not break SUSY in the Minkowski vacuum after inflation.
	For the purpose of describing both the inflation and present acceleration of the cosmic expansion in one sector, a single function $\widetilde{W}_1$ is not flexible enough because separation of scales is not easy.
	For example, one may try to describe a simple potential, $V=V_0 +\frac{m^2}{2}\chi^2$, with a simple function $W_1=\sqrt{V_0}-\frac{m^2 }{2\sqrt{V_0}}\Phi^2$~\cite{Kallosh2014d}, but the coefficient of the second term is enormously large because of the enormously small parameter $V_0\simeq 10^{-120}$.  Perturbation breaks down.  Also, when the second term is dominant over the first one, the quartic term in the potential is more important than the quadratic one, $V=V_0+\frac{m^2}{2}\chi^2+\frac{m^4}{16V_0}\chi^4 \simeq \frac{m^4}{16V_0}\chi^4$.
	
	The situation is changed when we introduce a new function $W_0(\Phi)$ which is not multiplied by the nilpotent stabilizer $S$. 
	Take the same $W_1$ with another parametrization, $W_1=M-\frac{m^2}{2M}\Phi^2$, and a constant $W_0$.  Then, the potential is $V=(M^2 - 3W_0^2)+\frac{m^2}{2}\chi^2 + \frac{m^4}{16M^2}\chi^4$.
	If we take $M$ somewhat larger than $m$, $M \gtrsim 15m/2\sqrt{2}$, the quartic term is subdominant compared to the quadratic term during the last 50 or 60 $e$-foldings of accelerated expansion.
	The cosmological constant can be cancelled by tuning the difference between $M^2$ and $3W_0^2$. 
	In this argument, the nilpotency of $S$ is important because $S$ may develop a finite value when we arbitrarily introduce $W_0$ though it could be stabilized by the higher dimensional term in the K\"{a}hler potential.  In contrast, if $S$ is nilpotent, its scalar component does not have a nonzero value unless the gravitino bilinear condensates (Subsection~\ref{subsec:gravitino}).
		
	Similarly to our discussion above, the generic cases with a real holomorphic function $\widetilde{W}_0$ and $\widetilde{W}_1$ were studied in Ref.~\cite{Kallosh2014d} where examples $W_0(\Phi)=\text{const.}$~\cite{Kallosh2014b}, $W_0(\Phi)=\text{const.}\times W_1(\Phi)$~\cite{DallAgata2014}, and $W_0(\Phi)=\text{const.}+\text{const.}\times W_1(\Phi)$ were presented.
	The above example with $W_0(\Phi)=W_0$ (const.) and $W_1(\Phi)=M-\frac{m^2}{2M}\Phi^2$ falls into the first class.
	Let us comment on the second class, $W_1(\Phi)=b W_0(\Phi)$ with $b$ a (real) constant.
	The scalar potential~\eqref{VGNSinflaton} becomes
	\begin{align}
	V=2 \widetilde{W}_0'^2 +(b^2 -3) \widetilde{W}_0^2,
	\end{align}
	so if we take $b=\sqrt{3}$, the second term vanishes, 
	\begin{align}
	V=2\widetilde{W}_0'^2.
	\end{align}
	Interestingly, this potential discussed in Ref.~\cite{DallAgata2014} is completely same as the one in our work~\cite{Ketov2014a}, which will extensively discussed in Section~\ref{sec:arbitrary} of this thesis, up to notational difference although the mechanisms to obtain the potential are different.
	As we saw above, the SUSY breaking scale is generically within an order of magnitude from the inflaton mass scale.
	This is because the gravitino mass, which is related to the magnitude of SUSY breaking at the vacuum by the condition of vanishing of the cosmological constant, is determined by the scale in $W_0$ (see eq.~\eqref{GravitinoMassNS}).  To separate the scale of inflation and SUSY breaking, separation of the scales between $W_1$, which would drive inflation, and $W_0$, which would trigger SUSY breaking, is implied.
	However, it is non-trivial since these scales coincides at the vacuum, $|W_1|^2\simeq 3|W_0|^2$ (see eq.~\eqref{VGNSinflaton}).
	In the case of the models in Ref.~\cite{DallAgata2014} and Ref.~\cite{Ketov2014a}, the potential depends only on the derivative of a function, so it is easy to adjust the SUSY breaking at the vacuum  to any desirable value by adding a constant in the function, which does not change the potential itself.

	\subsection{Example: Higgs inflation in (SM, MSSM, and) NMSSM} \label{subsec:HiggsReview}
	
	As an example of the mechanism employing shift symmetry and the stabilizer superfield, and for preparation for an application in Section~\ref{sec:generalization} of our mechanism presented in Chapter~\ref{ch:main}, we review Higgs inflation in the next-to-minimal supersymmetric standard model (NMSSM).
	
	There are many other examples and applications in the literature, and we do not cover them all.
	But it is worth mentioning in passing that there are particularly interesting class of models called ``cosmological attractors''~\cite{Kallosh2013, Ferrara2013, Kallosh2013b, Kallosh2013c}.
	These do not refer to attractor solutions of a theory but to classes of theories whose predictions on inflationary observables (like $n_{\text{s}}$ and $r$) converges to some values in some limits of parameters in the theories.  There are ``attractor models'' in theory space.
	We refer the interested readers to a recent paper~\cite{Galante2014} on a unified understanding of attractor theories and references therein.
	The Higgs inflation we will discuss now is also at the attractor point.
	
	We start discussion on Higgs inflation in the standard model.
	The potential of the Higgs is controlled by SU(2)${}_{L}\times$U(1)${}_{Y}$, and its quartic term, which is dominant in the large field region, is not suitable for inflation consistent with observation.
This would not be improved by simply adding higher order gauge invariant terms like $(\bar{H}H)^3$ or $(\bar{H}H)^4$ since these are steeper than the quartic term.
Instead, the potential is flattened by non-minimal coupling to gravity like $\xi \bar{H}HR$~\cite{Bezrukov2008}, or by modification of the kinetic term of Higgs~\cite{Nakayama2011, Nakayama2014} (running kinetic inflation~\cite{Takahashi2010, Nakayama2010}). 
We would like to emphasize that these two approaches are related.
For example, in Higgs inflation with non-minimal coupling to gravity, the kinetic term of the Higgs becomes non-canonical after transition from the Jordan frame to the Einstein frame.
In supergravity, these non-minimal coupling to gravity or non-minimal kinetic term are described by non-minimal K\"{a}hler potential.  This fact was also clarified in Ref.~\cite{Nakayama2010a}.

	We begin with the Higgs inflation of the non-minimal coupling type~\cite{Bezrukov2008}.
	\begin{align}
	\mc{L}=& -\frac{1}{2} \left( 1 +2 \xi \bar{H}H \right) R -\partial^{\mu}\bar{H}\partial_{\mu}H - \lambda (\bar{H}H- v^2/2 )^2 \nonumber \\
	=& -\frac{1}{2} \left( 1+ \xi h^2 \right) R -\frac{1}{2} \partial^{\mu}h \partial_{\mu}h - \frac{\lambda}{4}(h^2-v^2)^2, \label{LHiggsInflation}
	\end{align}
	where $\xi$ is the non-minimal coupling constant between the Higgs $H$ and Ricci scalar $R$, $\lambda$ is the quartic coupling constant, and $v$ is the VEV of the Higgs.
	Since three would-be Nambu-Goldstone bosons are eaten by the gauge fields, we have rewritten it in terms of the physical Higgs, $H=h/\sqrt{2}$.
	Under the Weyl transformation, field combinations transform as
	\begin{align}
	g_{\mu\nu}\rightarrow & g_{\mu\nu} e^{2\sigma} , \label{WeylT_metric} \\
	(e=)\sqrt{-g} \rightarrow & \sqrt{-g} e^{4\sigma}, \label{WeylT_det} \\
	eR \rightarrow & eR e^{2\sigma}+ \frac{3}{2}e e^{6\sigma}\partial_{\mu}e^{-2\sigma}\partial^{\mu}e^{-2\sigma}-3e^{2\sigma}\partial_{\mu} \left( e g^{\mu\nu} e^{2\sigma}\partial_{\nu}e^{-2\sigma} \right). \label{WeylT_R}
	\end{align}
	Choosing $e^{-2\sigma}=1+\xi h^2$, eq.~\eqref{LHiggsInflation} becomes
	\begin{align}
	\mc{L}=& -\frac{1}{2}R-\frac{1+\xi h^2 + 6\xi^2 h^2}{2(1+\xi h^2 )^2}\partial^{\mu}h\partial_{\mu}h - \frac{\lambda (h^2 -v^2 )^2}{4 (1+\xi h^2)^2} \nonumber  \\
	=& -\frac{1}{2}R - \frac{1}{2}\partial^{\mu}\widetilde{h}\partial_{\mu}\widetilde{h} - V\left(\widetilde{h}\right),
	\end{align}
	where we defined the canonical normalized inflaton $\widetilde{h}=\sqrt{6+\frac{1}{\xi}}\sinh^{-1}(\sqrt{\xi (1+ 6\xi )} h) - \sqrt{6} \tanh^{-1} (\sqrt{6} \xi h/ \sqrt{1+\xi (1+6\xi) h^2})$ and its potential $V\left(\widetilde{h}\right)$.
	In the large field $h\gg \sqrt{\xi}$ and large $\sqrt{\xi}$, this behaves as $\widetilde{h}\simeq \sqrt{3/2} \ln (1+\xi h^2)$, or $h^2 \simeq (e^{\sqrt{2/3}\widetilde{h}}-1)/\xi$. The potential behaves as
	\begin{align}
	V(\widetilde{h})\simeq  \frac{\lambda}{4\xi^2}\left( 1- e^{-\sqrt{2/3}\widetilde{h}} \right)^2.
	\end{align}
	
	Since we want to embed the Higgs inflation with non-minimal coupling to gravity into supergravity, we consider non-minimal coupling of Higgs superfields to supergravity.
	It can be written as an $F$-term invariant or a $D$-term invariant,
	\begin{align}
	\int \text{d}^2 \Theta 2\ms{E} \mc{R} H_u H_d +\text{H.c.}=&- \frac{1}{8}\int \text{d}^2 \Theta 2\ms{E} (\bar{\ms{D}}\bar{\ms{D}}-8\mc{R})H_u H_d  + \text{H.c.} \nonumber \\
	=& - \frac{1}{16}\int \text{d}^2 \Theta 2\ms{E} (\bar{\ms{D}}\bar{\ms{D}}-8\mc{R})(H_u H_d  + \overline{H_u}\overline{H_d}) + \text{H.c.}
	\end{align}
	The operand of the chiral projection operator in the last line can be regarded as the ``frame function'' or the exponential of a K\"{a}hler potential.
	Motivated by this fact, we consider the following K\"{a}hler potential.
	\begin{align}
	K=-3 \ln \left( 1 - \frac{1}{3} \left( |H_u|^2+|H_d|^2 \right) +\frac{x}{2}\left( H_u H_d +\overline{H_u}\overline{H_d}  \right) \right),
	\end{align}
	where $x$ is the constant parameter characterizing the non-minimal coupling.
	
	Einhorn and Jones tried to embed the Higgs inflation with non-minimal coupling into MSSM, but they fount that the potential becomes negative~\cite{Einhorn2010}.  In the same paper, they proposed to embed it into NMSSM.
	Although the tachyonic mass problem of the singlet was subsequently pointed out~\cite{Lee2010, Ferrara2010}, it is easily solved by adding a quartic term of the singlet~\cite{Lee2010, Ferrara2011}.
	Variants of Higgs inflation have been discussed \textit{e.g.}~in Refs.~\cite{Ben-Dayan2010, Nakayama2011, Nakayama2014}.
	
	Let us briefly discuss the MSSM case.  
	We do not repeat the original discussion, but provide a somewhat heuristic explanation.
	The reason why the model fails to realize large field inflation is essentially similar to the previous discussion in Section~\ref{sec:SSSS}, but it is not directly compared since we have now several superfields.  It becomes clear by the following argument.
	In the strong coupling limit, $x\gg 1$, the K\"{a}hler potential is dominated by the non-minimal coupling terms.  We define a composite chiral superfield $\Phi=\sqrt{3}x H_u H_d/2$.
	Neglecting the canonical terms, the K\"{a}hler potential and the MSSM superpotential become
	\begin{align}
	K\simeq& -3 \ln \left( 1+ \frac{1}{\sqrt{3}}(\Phi+\bar{\Phi}) \right), \\
	W\simeq& \frac{2}{\sqrt{3}x}\mu \Phi + W_0.
	\end{align}
	The scalar potential following from these potentials is
	\begin{align}
	V=& \frac{4 \mu^2 }{x^2 (\sqrt{3} + (\Phi+\bar{\Phi}) )^2)} \left( 1-\frac{2}{\sqrt{3}}(\Phi+\bar{\Phi} ) -3 \text{Re} \frac{ x W_0}{\mu}\right) \nonumber \\
	\simeq & - \frac{8\mu^2}{\sqrt{3}x^2 (\Phi+\bar{\Phi})},
	\end{align}	
	where the second equality holds in the large field limit, and the asymptotic expression reproduces the result in Ref.~\cite{Einhorn2010}.\footnote{
	It seems that eq.~(3.10) in Ref.~\cite{Einhorn2010} misses a factor $2$.
	}
	 The potential becomes negative in the large field region.
	 
	 In the NMSSM, we consider the following K\"{a}hler potential and superpotential,
	 \begin{align}
	K=& -3 \ln \left( 1 - \frac{1}{3} \left( |H_u|^2+|H_d|^2 +|S|^2\right) +\frac{x}{2}\left( H_u H_d +\overline{H_u}\overline{H_d}  \right) + \frac{\zeta}{9} |S|^4 \right), \\
	W=& \lambda' S H_u H_d + \frac{\rho}{3}S^3+W_0.
	 \end{align}
	The quartic term of $S$ is introduced to cure the tachyonic mass of the singlet $S$.
	Note that the singlet $S$ of NMSSM can be naturally identified with the stabilizer superfield $S$ for inflation.
	We neglect the constant term $W_0$ in the superpotential to simplify the analysis.
	This is valid when SUSY breaking scale is lower than the inflation scale.
	Since $S$ is stabilized at the origin, the $\rho S^3$ has no effect on the inflaton potential.
	Similarly to the previous exercise, we focus on the large $x$ limit where only $S$ and $\Phi=\sqrt{3}x H_u H_d/2$ are relevant.   Although $S$ does not have factors of $x$, we should keep it since inflation is driven by its $F$-term.
	After these simplification, the K\"{a}hler potential and the superpotential become
 	\begin{align}
	K\simeq & -3 \ln \left( 1 +\frac{1}{\sqrt{3}}\left( \Phi+\bar{\Phi} \right)  - \frac{1}{3}|S|^2 + \frac{\zeta}{9} |S|^4 \right), \\
	W\simeq & \frac{2\lambda'}{\sqrt{3}x} S \Phi .
	\end{align}
	The resultant scalar potential is
	\begin{align}
	V\simeq& \frac{4\lambda'^2|\Phi|^2}{3x^2\left( 1+\frac{1}{\sqrt{3}}\left(\Phi+\bar{\Phi}\right)\right)^2} , \nonumber \\
	=& \frac{\lambda'^2}{x^2}\left( 1 - e^{-\sqrt{2/3}\widetilde{\phi}}\right)^2,
	\end{align}
	where we have set $\text{Im}\Phi=0$ and canonically normalized the real part, $2\text{Re}\Phi= \sqrt{3} \left( e^{\sqrt{2/3}\widetilde{\phi}}-1\right)$ in the second equality.
	The potential of Higgs inflation or Starobinsky inflation has been successfully reproduced.
	Detailed analyses including masses of other fields we neglected here are found in Refs.~\cite{Einhorn2010, Ferrara2010, Ferrara2011}. 
	We will use the method of replacing $H_u H_d$ with $\Phi$ also in Section~\ref{sec:generalization}, where we will discuss some more justification of the method.
	
\section{Inflation with a linear or real superfield}\label{sec:vector}
	In the previous Section, we see the framework utilizing a chiral stabilizer superfield.
	There is another  method to embed inflation into supergravity without tuning.
	We review a framework utilizing a real linear or real superfield in this Section.
	
	Farakos, Kehagias, and Riotto revisited in Ref.~\cite{Farakos2013} a supersymmetric version of $R+\alpha R^2$ (Starobinsky) model in the new-minimal supergravity~\cite{Cecotti1988} in the inflationary context.  They consider the higher derivative new-minimal supergravity action and discuss its relation to a standard supergravity with a real (vector) superfields.
	The chiral multiplet and the massless vector multiplet merges into a massive vector multiplet.
	This is a SUSY version of the Higgs mechanism.  The imaginary part of the complex scalar in the chiral superfield is eaten by the gauge field in the real superfield.
	Thus, after Higgsing, there is only one real scalar field in the mass spectrum.
	This situation is desirable for inflationary applications because we do not have to worry about the stability along other scalars than the inflaton.
	 Although we have to eventually check it when we couple the inflaton sector to other sectors, but it is surely an advantage that the real and imaginary part in the same multiplet is separated (and the latter is eaten).
	The resultant potential of the scalar is that of the Starobinsky model, $V=3m^2\left(1-e^{-\sqrt{2/3}\phi}\right)^2/4$ where the inflaton mass is given by $m=\sqrt{6}gM_{\text{G}}$ with the gauge coupling constant $g$ and the reduced Planck mass $M_{\text{G}}$.
	They also considered the effects of higher derivative corrections corresponding to the $R^4$ term.
	The coefficients of such terms have to be suppressed enough so that the sufficient $e$-foldings are obtained.
	
	Seemingly inspired by their work, Ferrara, Kallosh, Linde, and Porrati proposed four types of theories~\cite{Ferrara2013}, two of which is old-minimal and the other new-minimal, and each two includes a theory of vector and a theory of tensor.  These theories are related by a web of dualities and equivalent.
	To see applications to inflation, we first focus on one of the theories (old-minimal with vector multiplet), and later discuss other theories and their relations.
	
	\subsection{Model with a massive vector multiplet}
	
	Let us consider the following theory~\cite{Ferrara2013},
	\begin{align}
	\mc{L}=-\frac{3}{2}\left [ \bar{S}_0 S_0 \Omega (gV+\Lambda+\bar{\Lambda}) \right ]_D + \frac{1}{4}\left [ \mc{W}(V)\mc{W}(V)\right]_F , \label{vectorOld}
	\end{align}
	where $S_0$ is the chiral compensator, $g$ is a gauge coupling constant, $\mc{W}(V)$ is the field strength of a real supermultiplet $V$, and $\Lambda$ is a chiral supermultiplet.
	This has a gauge symmetry, $V\rightarrow V -\Sigma -\bar{\Sigma}$ and $\Lambda \rightarrow \Lambda + g \Sigma$.
	Gauge fixing $\Lambda=0$, the theory contains only the real supermultiplet as well as the compensator,
	\begin{align}
	\mc{L}=-\frac{3}{2}\left [ \bar{S}_0 S_0 \Omega (gV) \right ]_D + \frac{1}{4} \left [ \mc{W}\mc{W}\right]_F , 
	\end{align}
	Setting $\bar{S}_0S_0 \Omega=1$, the bosonic part becomes Van Proeyen's  massive vector multiplet Lagrangian density~\cite{VanProeyen1980a}.
	\begin{align}
	\mc{L}=& -\frac{1}{2}R-\frac{1}{4}F_{\mu\nu}F^{\mu\nu}+\frac{g^2}{2}J''(C) B_{\mu}B^{\mu}+\frac{1}{2}J'' (C) \partial_{\mu}C\partial^{\mu}C - \frac{g^2}{2}J'^2, \label{LvectorOld}
	\end{align}
	where $F_{\mu\nu}$ is the field strength of the vector boson $B_{\mu}$, $C$ is a real scalar, and $J=\frac{3}{2}\ln \Omega +\text{const.}$
	 For the scalar kinetic term to have the physical sign, the second derivative of the function $J$ must be negative, $J''<0$.
	Here, the gauge $D$-term is equal to $J'$ (primes denote differentiation with respect to the argument), and the scalar potential is its square.
	
	To discuss inflation, let us neglect (integrate out) the massive vector with mass $g\sqrt{-J''}$.
	The scalar and gravity part of the Lagrangian density is
	\begin{align}
	\mc{L}=-\frac{1}{2}R+\frac{1}{2}J''(C) (\partial_{\mu} C)^2 - V(C),
	\end{align}
	with $J''<0$.
	The potential and its derivatives are 
	\begin{align}
	V=& \frac{g^2}{2}(J')^2=\frac{g^2}{2}D^2,\\
	V'=&g^2 J'J''=g^2DD' ,\\
	V''=&g^2(J'')^2+g^2J'J'''=g^2(D')^2+g^2DD'',
	\end{align}
	so the stationary condition $V'=0$ imply $J'=0$ at the vacuum, which in turn tells us that the potential vanishes at the vacuum $V=0$.  It is remarkable that the cosmological constant in this sector vanishes without fine tuning.
	Moreover, the mass squared of the scalar at the vacuum $V'=0$ is given by $V''=(gJ'')^2>0$, so it is not tachyonic.
	The potential is proportional to the gauge coupling squared, so it must be small $g\ll 1$ to reproduce the amplitude of the fluctuation of CMB.
	Then, the quantum corrections are suppressed by the same small factor $g^2$.
	This is another advantage of this class of theories.
	
	The potential increases from zero when we go away form the vacuum.
	$J'$ decreases from zero because $J''$ is negative.
	The first derivative of the potential thus keeps being positive.
	The potential is a monotonously increasing function provided that $J''<0$ continues to be valid.
	If the first derivative vanishes at some point, it indicates $J''=0$ at the point because $J'$ has been decreasing until that point from zero at the vacuum.
	At the point, the kinetic term of the scalar vanishes, and it becomes the unphysical sign when we proceed further.  Therefore, the point where $V'=0$ other than the vacuum is where the description of the theory breaks down.  Turning it around, in the field range of validity of the theory, the potential is monotonously increasing away from the minimum.
	
	The scalar $C$ is not canonically normalized.
	The canonically normalized scalar $\varphi$ is determined by $\left(\text{d}\varphi/\text{d}C\right)^2=-\text{d}^2J/\text{d}C^2$.  It can be rewritten as 
	\begin{align}
	\frac{\text{d}\varphi}{\text{d}C}=-\frac{\text{d}D}{\text{d}\varphi}. \label{DcanonicalR}
	\end{align}
	One can begin with the function $J(C)$ and derive the canonical field $\varphi=\varphi(C)$ and the $D$-term potential $D(C(\varphi))$.
	Also, one may begin with a desired $D$-term $D(\varphi)$ and obtain $C=C(\varphi)$ and the function $J(\varphi(C))$.
	Let us see examples in Ref.~\cite{Ferrara2013}.
	\paragraph{Example 1: quadratic potential}
	Consider a quadratic potential,
	\begin{align}
	V=\frac{m^2}{2}\varphi^2,
	\end{align}
	 in terms of the canonically normalized field $\varphi$ with $D$-term $D=\frac{m}{g}\varphi$.
	Using eq.~\eqref{DcanonicalR}, the relation between $\varphi$ and $C$ is obtained,
	\begin{align}
	\varphi = -\frac{m}{g}(C-C_0),
	\end{align}
	where $C_0$ is an integration constant corresponding to a choice of the origin of $C$.
	The function $J$ is obtained by integration of $D$ since $J'=D=\frac{m}{g}\varphi=-\frac{m^2}{g^2}C$.
	\begin{align}
	J=-\frac{m^2}{2g^2}(C-C_0)^2 +J_0,
	\end{align}
	where $J_0$ is another integration constant corresponding to a conformal frame.
	One can easily confirm that the sign of the kinetic term is canonical, $J''=-\frac{m^2}{g^2}<0$.
	
	\paragraph{Example 2: Starobinsky-like potential}
	Consider a Starobinsky-like potential,
	\begin{align}
	V=\frac{g^2}{2}\left( c \left( 1-a e^{-b\varphi}\right)\right)^2.
	\end{align}
	The $D$-term is given by $D=c(1-ae^{-b\varphi})$.
	Using eq.~\eqref{DcanonicalR}, the relation between $\varphi$ and $C$ is obtained,
	\begin{align}
	C-C_0&=-\frac{e^{b\varphi}}{ab^2c}, &  \left( \Leftrightarrow \varphi =  \frac{1}{b} \ln (-ab^2 c (C-C_0)) \right),
	\end{align}
	where $C_0$ is an integration constant.
	The function $J$ is
	\begin{align}
	J=c(C-C_0)+\frac{1}{b^2}\ln (C-C_0)+J_0,
	\end{align}
	where $J_0$ is another integration constant.
	It is easily checked that $J''=-\frac{1}{b^2C^2}<0$, so the kinetic term has the physical sign.
	
	Higher order corrections to the theories were studied in Ref.~\cite{Ferrara2013a} by the same authors.
	They considered the following generic correction terms,
	\begin{align}
	\left [ a_{k,l,p}g^n \frac{W^2\bar{W}^2}{\left( \Omega (gV)\bar{S}_0S_0 \right)^2} \left( \Sigma \frac{\bar{W}^2}{\left( \Omega (gV)\bar{S}_0S_0 \right)^2} \right)^k \left( \bar{\Sigma} \frac{W^2}{\left( \Omega (gV)\bar{S}_0S_0 \right)^2} \right)^l \left(  \frac{DW}{ \Omega (gV)\bar{S}_0S_0 } \right)^p \Psi_n (gV) \right ]_D,
	\end{align}
	where $a_{k,l,p}$ is a real coefficient, $\Sigma$ is the chiral projection operator, $\Psi_n (gV)$ is a Hermitian function, and indices satisfy $n=4+2K+2l+p$ with $k,l,p\geq0$ $(n\geq4)$.
	The chiral compensator is inserted to make the total Weyl weight $2$.
	This superconformal expression leads to the bosonic component Lagrangian density of the following form,
	\begin{align}
	\mc{L}^{k,l,p}= a_{k,l,p}g^n (F^{+2}-D^2)^{1+k}(F^{-2}-D^2)^{1+l}D^p \Psi_n (gC),
	\end{align}
	where $F^{+}$ and $F^{-}$ are self-dual and anti-self dual part of the field strength.
	The part of the bosonic Lagrangian density depending on the auxiliary $D$ field is of the form
	\begin{align}
	\mc{L}=\frac{1}{2}D^2 +gJ'(C)D+\sum_{n}\xi_n g^n \Psi_n (gC) D^n,
	\end{align}
	and the equation of motion of $D$ gives the potential 
	\begin{align}
	V=\frac{g^2}{2}(J'(C))^2-\sum_n \xi_n g^n \Psi_n (gC) (-gJ'(C))^n+\dots.
	\end{align}
	If we assume $J(C)$ and $J'(C)$ is of order one, the gauge coupling $g$ is the scale of the inflaton mass (in units of the reduced Planck mass).  To match the normalization of CMB fluctuation, it should be of order $10^{-5}$.
	Unless $\xi_n \Psi_n (gC) \gtrsim 10^{10(n-1)}$, the correction terms are subdominant, and the tree-level analysis can be believed.
	
	\subsection{Web of dualities}
	Now, let us move on other theories related to the previous theory.
	What was called a superconformal master model in Ref.~\cite{Ferrara2013} is
	\begin{align}
	\mc{L}=-\frac{3}{2}\left [ \bar{S}_0S_0\Omega (U) \right ]_D +\frac{3}{2}\left [ L(U-gV)\right ]_D + \frac{1}{4}\left [ \mc{W}(V) \mc{W}(V) \right ]_F, \label{masterOld}
	\end{align}
	where $S_0$ and $L$ are the chiral compensator and a linear supermultiplet, while $U$ and $V$ are real supermultiplets.
	Variation of this Lagrangian density with respect to the linear multiplet $L$ yields an equation $U=gV+\Lambda+\bar{\Lambda}$ with $\Lambda$ chiral.
	That is, the above theory~\eqref{masterOld} reduces to the previous theory~\eqref{vectorOld} after elimination of $L$ and $U$.
	They are classically equivalent theories.
	On the other hand, if we variate it with respect to the real multiplet $U$, we obtain an equation 
	\begin{align}
	L=\bar{S}_0S_0 \Omega'(U). \label{LandU}
	\end{align}
	Substituting this into eq.~\eqref{masterOld}, it becomes
	\begin{align}
	\mc{L}=\frac{3}{2}\left [ \bar{S}_0 S_0 F\left ( \frac{L}{\bar{S}_0S_0} \right)-gLV\right ]_D  + \frac{1}{4}\left [ \mc{W}(V) \mc{W}(V) \right ]_F, 
	\end{align}
	where $F$ is the Legendre transform of $\Omega$,
	\begin{align}
	F\left ( \frac{L}{\bar{S}_0S_0} \right)=\Omega'(U)U-\Omega(U),
	\end{align}
	evaluated with eq.~\eqref{LandU}.
	In contrast to a real linear supermultiplet whose vector component is constrained, a chiral supermultiplet can be variated freely.
	The real linear supermultiplet $L$ can be expressed in terms of a spinor chiral supermultiplet $L_{\alpha}$ as
	\begin{align}
	L=D^{\alpha}L_{\alpha}+\bar{D}_{\dot{\alpha}}\bar{L}^{\dot{\alpha}}.
	\end{align}
	In terms of these, $-g[LV]_D$ is expressed as $2g[L\mc{W}]_F$.
	Defining a new chiral spinor supermultiplet as $X_{\alpha}=\mc{W}+6gL_{\alpha}$, $-\frac{3}{2}g[LV]_D+\frac{1}{4}[\mc{W}\mc{W}]_F$ can be rewritten as $-9g^2 [L^{\alpha}L_{\alpha}]+\frac{1}{4}[X^{\alpha}X_{\alpha}]_F$.
	The latter term $[XX]_F$ is decoupled from other terms and has no kinetic terms.
	Therefore we neglect it.
	Finally, the theory becomes
	\begin{align}
	\mc{L}=\frac{3}{2}\left [ \bar{S}_0 S_0 F\left( \frac{D^{\alpha}L_{\alpha}+\bar{D}_{\dot{\alpha}}\bar{L}^{\dot{\alpha}}}{\bar{S}_0 S_0} \right) \right ]_D -9g^2 \left [ L^{\alpha}L_{\alpha} \right ]_F.
	\end{align}
	Gauge fixing the compensator, the bosonic part becomes
	\begin{align}
	\mc{L}=-\frac{1}{2}R-\frac{1}{2}(J'')^{-1}\left( \partial_{[\mu}B_{\rho\sigma]}\right)^2 -\frac{g^2}{4}B_{\rho\sigma}{}^2 +\frac{1}{2}J''\partial_{\mu}C\partial^{\mu}C-\frac{g^2}{2}(J')^2, \label{linearOld}
	\end{align}
	where $[\dots ]$ denotes antisymmetrization.
	This is a theory of a real scalar $C$ and a massive tensor $B_{\mu\nu}$ as well as gravity.
	This is equivalent to the Lagrangian density in eq.~\eqref{LvectorOld}~\cite{Cecotti1987, Ferrara2013}.
	
	As we have seen above, the superconformal master model~\eqref{masterOld} is classically equivalent to the theories~\eqref{vectorOld} and \eqref{linearOld}.  It generates the former when we variate it with respect to $L$ and the latter when we variate it with respect to $U$.
	Interestingly, the superconformal master model~\eqref{masterOld} is equivalent to another superconformal master model with a linear compensator.
	The master Lagrangian density is given by
	\begin{align}
	\mc{L}=\frac{3}{2}\left [ L_0 \ln \left( \frac{L_0}{\bar{S}_0 S_0} \right) +\frac{1}{3}L_0 \mc{J}(U)+L(U-gV) \right ]_D +\frac{1}{4} \left [ \mc{W}(V)\mc{W}(V) \right]_F, \label{masterNew}
	\end{align}
	where $L_0$ is the real linear compensator, $L$ is another real linear supermultiplet, $S_0$ is a chiral supermultiplet, and $U$ and $V$ are real supermultiplets.
	To show the equivalence between the master models~\eqref{masterOld} and \eqref{masterNew}, let us rewrite the first two terms in eq.~\eqref{masterNew} since the latter two terms are also present in the old minimal version~\eqref{masterOld}.
	\begin{align}
	\mc{L}=\frac{3}{2}\left [ L_0 \ln \left( \frac{L_0}{\bar{S}_0 S_0} \right) +\frac{1}{3}L_0 \mc{J}(U) +i L_0 (\varphi -\bar{\varphi}) \right ]_D, \label{masterNew2}
	\end{align}
	where $L_0$ is now an unconstrained real superfield.
	Variation of this with respect to $\varphi$ ensures $L_0$ is linear.
	On the other hand, variation of the Lagrangian density with respect to $L_0$ gives
	\begin{align}
	L_0 = \bar{S}_0 S_0 e^{-1-\frac{1}{3}\mc{J}(U)-i(\varphi - \bar{\varphi})}.
	\end{align}
	Substituting this into eq.~\eqref{masterNew2}, and redefining the chiral supermultiplets as $S_0 \rightarrow e^{i\varphi +\frac{1}{2}}S_0$ and $\bar{S}_0 \rightarrow e^{-i\varphi +\frac{1}{2}}\bar{S}_0$, it becomes
	\begin{align}
	-\frac{3}{2}\left [ \bar{S}_0 S_0 e^{-\frac{1}{3}\mc{J}(U)} \right ]_D.
	\end{align}
	Relating $\mc{J}(U)$ and $\Omega (U)$ by an equation $\Omega=e^{-\mc{J}/3}$ or $\mc{J}=-3\ln\Omega$, equivalence of the two master actions are proven.
	
	The new-minimal master model, when variated with respect to $L$, yields $U=gV+\Lambda+\bar{\Lambda}$, and reduces to a vector theory,
	\begin{align}
	\mc{L}=\frac{3}{2}\left [ L_0 \ln \left( \frac{L_0}{\bar{S}_0 S_0} \right) + \frac{1}{3}L_0 \mc{J}(gV+\Lambda+\bar{\Lambda}) \right ]_D +\frac{1}{4} \left [ \mc{W}(V) \mc{W}(V) \right ]_F. \label{vectorNew}
	\end{align}
	If we instead variate it with respect to $U$, the relation between $U$, $L_0$ and $L$ is determined,
	\begin{align}
	L=-\frac{1}{3}L_0 \mc{J}'(U). \label{LandU2}
	\end{align}
	Substituting this into eq.~\eqref{masterNew}, it becomes a theory of the linear multiplet,
	\begin{align}
	\mc{L}=\frac{3}{2}\left[ L_0 \ln \left( \frac{L_0}{\bar{S}_0 S_0} \right) + L_0 M \left( \frac{L}{L_0} \right) \right ]_D -9g^2 \left [ L^{\alpha}L_{\alpha} \right ]_F, \label{linearNew}
	\end{align}
	where the function $M$ is the Legendre transform of $\mc{J}$,
	\begin{align}
	M\left(\frac{L}{L_0}\right)= \frac{1}{3} (\mc{J}(U)-U\mc{J}'(U)),
	\end{align}
	with $U$ evaluated with eq.~\eqref{LandU2}.
	The theories~\eqref{vectorNew} and \eqref{linearNew} are equivalent because these are derived from one master theory~\eqref{masterNew}.
	
	To summarize this Section, we reviewed minimal inflationary models~\eqref{vectorOld} and \eqref{linearOld}, both derivable from the master model~\eqref{masterOld} in the old-minimal supergravity, and models~\eqref{vectorNew} and \eqref{linearNew}, both derivable from the master model~\eqref{masterNew} in the new-minimal supergravity, which are all equivalent at least classically~\cite{Ferrara2013}.
	In particular, we studied the scalar sector of the model~\eqref{vectorOld}, and discussed the characteristics of the potential, canonical normalization, and relations between the $D$-term and the function $J$ present in the action.
	A distinguishing feature of these models is that there is only one scalar in the theory, so once we construct a suitable inflationary potential, we do not have to worry about stability of other scalars in the inflaton sector.
	Moreover, we have seen that the inflaton potential is monotonously increasing function away from the minimum, where the cosmological constant vanishes without tuning, as far as the sign of the kinetic term of the inflaton is canonical.
	The higher order corrections to the action~\cite{Ferrara2013} are typically subdominant because of the tiny gauge coupling constant.
	See also related models~\cite{Farakos2014} discussed after BICEP2~\cite{Ade2014b}.	
	
\section{Purely supergravitational theories}
\label{sec:pureSUGRA}

Our proposal in Chapter~\ref{ch:main} is to remove the stabilizer superfield to decrease the necessary degrees of freedom for inflation in supergravity without giving up possibilities to describe various kinds of inflationary models.
By decreasing the degrees of freedom to one supermultiplet, we can have a \emph{minimal theory of inflation in supergravity} as in Section~\ref{sec:sGoldstino}, but we will see that we can describe large field inflation as well as small field inflation in Chapter~\ref{ch:main}.
We here discuss other inflationary models from a bit different perspective on minimality.

	When we put something `by hand' into a theory, it is often not natural or minimal.
	It is also the case for inflation.
	People put a scalar field, `inflaton', by hand to explain accelerated expansion of the universe.
	It is elegant, however, if the theory of spacetime, (super)gravity and/or universe automatically explain the phenomenon.
	In this Section, we review such a kind of minimalistic approaches in gravity and supergravity though the number of degrees of freedom in the theory is not necessarily minimal.

	\subsection{Starobinsky model and its SUSY extension}
	\label{subsec:Starobinsky}
	
	An important example is the Starobinsky model~\cite{Starobinsky1980, Starobinsky1981a, Starobinsky1983a, Mukhanov1981}.
	Quantum corrections by matter loop generates higher order gravitational terms in the action.
	The second order curvature invariants are $R^2$, $R^{\mu\nu}R_{\mu\nu}$, and $R^{\mu\nu\rho\sigma}R_{\mu\nu\rho\sigma}$, but one combination can be eliminated by the Gauss-Bonnet theorem.
	Also, Weyl tensor vanishes in FLRW spacetime, so the higher curvature terms up to quadratic order are expressed solely by $R^2$ term~\cite{Mijic1986}.
	Then, the Starobinsky model is expressed as
	\begin{align}
	e^{-1}\mc{L}= -\frac{1}{2}R+ \frac{1}{12m^2} R^2 , \label{Starobinsky}
	\end{align}
	where $m$ is a mass dimensional parameter of order $m\sim 10^{-5}$.
	A propagating degree of freedom has been added to the General Relativity, in which two helicity states of graviton propagate.  This higher derivative purely gravitational theory can be recast into the form of the Einstein gravity plus a real scalar field~\cite{Whitt1984, Maeda1988},
	\begin{align}
	e^{-1}\mc{L}= -\frac{1}{2}R -\frac{1}{2}\partial^{\mu} \phi \partial_{\mu}\phi - \frac{3m^2}{4}\left( 1- e^{-\sqrt{2/3}\phi}\right)^2. \label{StarobinskyST}
	\end{align}
	  The degrees of freedom match between the original and the new form of the theory.
	The scalar is sometimes called the scalaron.
	
	The equivalence between the Starobinsky model~\eqref{Starobinsky} and the model~\eqref{StarobinskyST} is a particular case of equivalence between $f(R)$ gravity and scalar-tensor theories~\cite{Sotiriou2010, DeFelice2010e}.
	We will discuss generalization of the equivalence in supergravity below, so let us review the bosonic case now.
	Consider an $f(R)$ theory,
	\begin{align}
	\mc{L}=-ef(R). \label{f(R)}
	\end{align}
	As in the previous Section, we consider a `master action',
	\begin{align}
	\mc{L}=-e \left(  f(Q)+\Lambda \left( R-Q \right) \right), \label{f(R)master}
	\end{align}
	where $Q$ is a scalar field, and $\Lambda$ is a Lagrange multiplier scalar field.
	The equation of motion of the Lagrange multiplier is $Q=R$, and the Lagrangian density~\eqref{f(R)master} reduces to that of the original $f(R)$ model~\eqref{f(R)}.
	The point is that $R$ is now linear in eq.~\eqref{f(R)master}.
	We may use the equation of motion of $Q$, \textit{i.e.}~$\Lambda=f'(Q)$.
	The Lagrangian density is now
	\begin{align}
	\mc{L}=-e\left( f'(Q)R+f(Q)-f'(Q)Q \right).
	\end{align}
	Next, we rescale the metric (or vierbein) to move to the Einstein frame.
	Under the Weyl transformation, fields transform as in eqs.~\eqref{WeylT_metric}, \eqref{WeylT_det}, and \eqref{WeylT_R}.
	Choosing it as $e^{2\sigma}=(2 f '(Q))^{-1}$, we can go to the Einstein frame with an emergent dynamical scalar.
	\begin{align}
	\mc{L}=&e \left( -\frac{1}{2}R- \frac{3}{4(f'(Q))^2}\partial^{\mu}f'(Q)\partial_{\mu}f'(Q)- \frac{1}{4(f'(Q))^2}\left( f'(Q)Q-f(Q) \right)  \right) \nonumber \\
	=&e\left( -\frac{1}{2}R- \frac{1}{2}\partial^{\mu}\phi \partial_{\mu}\phi- e^{-2\sqrt{2/3}(\phi-\phi_0)}\left(\frac{1}{2}e^{\sqrt{2/3}(\phi-\phi_0)}Q(\phi)-f(Q(\phi)) \right)  \right) . \label{f(R)ST}
	\end{align}
	The canonically normalized scalaron $\phi$ is determined by the equation $f'(Q)=\frac{1}{2}e^{\sqrt{2/3}(\phi-\phi_0)}$, where $\phi_0$ is an integration constant, and it corresponds to the freedom of the choice of the origin of $\phi$.
	  In the case of the Starobinsky model, $f(R)=\frac{1}{2}R-\frac{1}{12m^2}R^2$, the scalar-tensor Lagrangian density~\eqref{f(R)ST} becomes eq.~\eqref{StarobinskyST} with the choice $\phi_0=0$.
	In this way, equivalence between $f(R)$ theories~\eqref{f(R)} and scalar-tensor theories~\eqref{f(R)ST} is established.
	Although $f(R)$ gravity is a higher derivative theory, it is clear that it does not have a ghost state (a state with unphysical sign of its kinetic term) in the scalar-tensor side.
	
	We will consider a supersymmetric version of the Starobinsky model from now on as one of the minimalistic way of describing inflation in supersymmetric theories.
	Since the Starobinsky model is a higher curvature (hence higher derivative) modified theory of gravity, its supersymmtrized version is necessarily a supergravity model, a local (gauged) version of SUSY.
	The Starobinsky was embedded in the old-minimal supergravity by Cecotti~\cite{Cecotti1987f}, and in the new-minimal supergravity by Cecotti \textit{et al.}~\cite{Cecotti1988}.
	Equivalence between these higher curvature (higher derivative) supergravity theories and standard supergravity theories (with additional matter) was also established in each reference.
	These SUSY Starobinsky model was revisited after the Planck 2013 result~\cite{Ade2014a, Ade2014} in Ref.~\cite{Kallosh2013a} for the old-minimal case and in Ref.~\cite{Farakos2013} for the new-minimal case.
	We will discuss these models later in this subsection.
	
	Cecotti developed rather general higher curvature action of the old-minimal supergravity, depending on supercurvature $\mc{R}$, its conjugate $\bar{\mc{R}}$, and superderivatives $\Sigma \bar{R}$ \textit{etc.}, already in Ref.~\cite{Cecotti1987f}, and vacuum structure of generic higher curvature models, depending on $\mc{R}$ and $\bar{\mc{R}}$, was studied by Hindawi \textit{et al.}~\cite{Hindawi1996b}.
	They also studied SUSY breaking in this extended supergravity sector and its mediation to the visible sector (like the MSSM sector) in Ref.~\cite{Hindawi1996d}.
	Generally, interactions among various fields take place at least with gravitational strength in supergravity even if super- and K\"{a}hler potentials are decoupled.
	In the modified supergravity case studied by Hindawi~\textit{et al.}, however, soft SUSY breaking in the visible sector can be suppressed if one can ensure decoupling between matter sector and higher derivative supergravity sector in the $D$-term action ($\sim$ K\"{a}hler potential).
	This possibility, SUSY breaking in the SUSY (supergravity) sector, is conceptually elegant and minimalistic, and was recently revisited in the inflationary context in Ref.~\cite{Dalianis2014}.
	
	Years later, a reduced class of modified supergravity theories depending on chiral supercurvature $\mc{R}$ was rediscovered by Gates and Ketov~\cite{Gates2009}.
	The Lagrangian density is~\cite{Hindawi1996b, Gates2009}
	\begin{align}
	\mc{L}= \int \text{d}^2 \Theta 2 \ms{E} F(\mc{R}) + \text{H.c.}, \label{F(R)SUGRA}
	\end{align}
	 and it was dubbed as $F(\mc{R})$ supergravity (similarly to $f(R)$ gravity).
	Developments of $F(\mc{R})$ supergravity is reviewed in Ref.~\cite{Ketov2013b}.
	The present author with Ketov considered a generalization to include a spinor chiral superfield $\mc{W}$ containing Weyl tensor $C$, and a bosonic $f(R, C)$ modified gravity~\cite{Ketov2013}.
	The former is the general holomorphic pure supergravity theory in the old-minimal supergravity up to (super)derivatives.
	
	Analogously to the bosonic case, it is shown that this $F(\mc{R})$ supergravity is equivalent to the standard supergravity with an additional chiral superfield.  First, we extend the theory with a Lagrange multiplier $T$,
	\begin{align}
	\mc{L}=\int \text{d}^2 \Theta 2 \ms{E} \left( F(Q) + T(Q-\mc{R}) \right)+ \text{H.c.},
	\end{align}
	where $Q$ and $T$ are chiral superfields.
	The equation of motion of $T$ let it back to the original theory.
	On the other hand, the equation of motion of $Q$  tells us $T=-F'(Q)$.
	We implicitly solve the inverse $Q=Q(T)$, and the theory becomes
	\begin{align}
	\mc{L}=&\int \text{d}^2 \Theta 2 \ms{E} \left( \frac{1}{8} \left( \bar{\ms{D}}\bar{\ms{D}}-8\mc{R}\right)T + W(T) \right) + \text{H.c.} \nonumber \\
	=&\int \text{d}^2 \Theta 2 \ms{E} \left( \frac{3}{8} \left( \bar{\ms{D}}\bar{\ms{D}}-8\mc{R}\right)e^{-K/3} + W(T) \right) + \text{H.c.} ,
	\end{align}
	where we have used the fact that $\int \text{d}^2 2 \ms{E}  \left( \bar{\ms{D}}\bar{\ms{D}}-8\mc{R}\right) (S-\bar{S}) + \text{H.c.}$ is a total derivative for any superfield $S$~\cite{Hindawi1996b, Cremmer1979}.
	The superpotential of the `scalaron superfield' $T$ is given by the Legendre transform of $F(Q)$, $W(T)=F(Q(T))+Q(T)T$, and the K\"{a}hler potential of $T$ is
	\begin{align}
	K=-3 \ln \left( \frac{T+\bar{T}}{6} \right).
	\end{align}
	This is the standard supergravity with a chiral superfield $T$ of the no-scale type K\"{a}hler potential.
	Therefore, the number of bosonic and fermionic degrees of freedom of the standard supergravity, $16+16$ off-shell and $2+2$ on-shell, has been increased to $20+20$ off-shell and $4+4$ on-shell.
	Also, there are no ghost states in the theory.
	
	Unfortunately, large-field inflation (of the Starobinsky-model type) in this theory is found to be difficult.
	Discussions~\cite{Ellis2013, Ferrara2013b, Ketov2013a, Ferrara2013c} are compelling, though not exact proofs.
	Simply summarizing, the reason is as follows: to obtain an asymptotically flat (scale invariant) potential, the power of the superpotential (assumed to be dominated by a monomial term) must be $3/2$ in the ``scalar-tensor supergravity'' side (or equivalently $F(\mc{R})\sim \mc{R}^3$), but the resultant scalar potential is negative, or a negative norm state (the unphysical kinetic sign) appears.
	Extensions including superderivatives of supercurvature $\mc{R}$ involve ghost states~\cite{Cecotti1987f, Ketov2013a}, so not promising.
	
	A successful supergravity embedding of the Starobinsky model is found in generic higher order pure supergravity depending on $\mc{R}$ and $\bar{\mc{R}}$~\cite{Cecotti1987f, Hindawi1996b, Ketov2013a},
	\begin{align}
	\mc{L}=\int \text{d}^2\Theta 2\ms{E} \left(- \frac{1}{8} \left( \bar{\ms{D}}\bar{\ms{D}}-8\mc{R}\right) N(\mc{R},\bar{\mc{R}}) + F(\mc{R}) \right) +\text{H.c.} \label{N+F}
	\end{align}
	The bosonic part of the action was studied in Refs.~\cite{Hindawi1996b, Ketov2013a}.
	See also Ref.~\cite{Ferrara1978a} where the simple case $N\propto \bar{\mc{R}}\mc{R}+\text{const.}$ was studied at linearized level, and propagating degrees of freedom were identified.
	The above higher derivative supergravity theory~\eqref{N+F} is equivalent to the standard supergravity with two additional superfields.
	The procedures to show the equivalence are similar to the previous examples, so we skip them and refer the readers to \textit{e.g.}~Refs.~\cite{Cecotti1987f, Ketov2013e}.
	The number of the bosonic and fermionic degrees of freedom is now $24+24$ off-shell and $6+6$ on-shell.
	In spite of the higher derivative action, it is a  ghost-free theory, and the absence of ghosts are clearly shown by the fact that it is equivalent to the standard supergravity with the standard chiral matter.
	The latter Lagrangian density is given by 
	\begin{align}
	\mc{L}= &\int \text{d}^2 \Theta 2 \ms{E} \left( \frac{3}{8} \left( \bar{\ms{D}}\bar{\ms{D}}-8\mc{R}\right)e^{-K/3} + W(T, S) \right) + \text{H.c.} , 
	\end{align}	
	where $T$ and $S$ are the two chiral superfields.
	Their super- and K\"{a}hler potentials are given by
	\begin{align}
	K=& -3 \ln \left( \frac{T+\bar{T}-2N(S,\bar{S})}{6} \right) , \label{KN+F}\\
	W=& ST + F(S). \label{WN+F}
	\end{align}
	The origin of $T$ is the Lagrange multiplier.
	In the application to the Starobinsky model of inflation, $S$ plays the role of the sGoldstino (stabilizer) in Section~\ref{sec:SSSS}.
	It is remarkable that the dependence on $T$ is completely fixed by the structure of the theory.
	For example, it appears linearly both in the super- and K\"{a}hler potentials.
	This feature is preserved even when we couple this higher supergravity sector to other matter sectors~\cite{Hindawi1996d, Terada2015}.
	On the other hand, $S$ enters in the super- and K\"{a}hler potentials with generic functional forms.
	The functions $N$ and $F$ in eqs.~\eqref{KN+F} and \eqref{WN+F} are same as these in eq.~\eqref{N+F}.
	
	One may na\"{i}vely expect that generic functions $N(\mc{R},\bar{\mc{R}})$ and $F(\mc{R})$ result various powers of Ricci scalar $R$.
	This is not true, and the schematic form of the Lagrangian is $\mc{L}\sim C_0 (X)+C_1 (X)R + C_2 (X) R^2$, where $C_i (X)$ denotes field dependent coefficients, and $X$ collectively denotes other fields in the theory.
	This is because Ricci scalar $R$ is contained not in the leading component but in the $\Theta \Theta$ component of the superfield $\mc{R}$.  Therefore there are at most second powers of $R$ unless one introduces superderivatives of $\mc{R}$ and $\bar{\mc{R}}$.
	Here is an interesting and important point.
	The generic higher (super)curvature supergravity theories~\eqref{N+F} without superderivatives on $\mc{R}$ and $\bar{\mc{R}}$ lead to the quadratic curvature gravity, which can be a suitable arena to realize the Starobinsky model.
	
	But why do the superderivative terms vanish?
	This is the SUSY version of the question in the original (bosonic) Starobinsky model: why do higher order curvature terms $R^n$ with $n\geq 3$ vanish?
	This is a weak point in the Starobinsky model.
	In the context of asymptotic safety scenario of quantum gravity~\cite{Weinberg1979x}, smallness of the inverse of the coefficient of the second order term $R^2$ can be understood as a consequence of existence of asymptotically free UV fixed point~\cite{Copeland2013}.
	Or perhaps the smallness is just by accident.
	But there are in general no known reasons or mechanisms for an expansion series continued to some order (second order in our case) with an expansion parameter (the large $m^{-2}\sim 10^{10}$ in our case) stops there\footnote{
	These points were stressed to the author by T. Kawano and K. Kohri.
		}
	 except in the cases of quantum anomaly and renormalization of the gauge kinetic function in SUSY theories.
	Instead of being absent, the higher order terms may be present but suppressed.
	The potential of Higgs inflation~\cite{Bezrukov2008}, which asymptotes to that of the Starobinsky model, is protected by the scale symmetry~\cite{Bezrukov2011}.  A similar argument is possible for the Starobinsky model.  See \textit{e.g.} Refs.~\cite{Gorbunov2014, Salvio2014a, Kannike2015, Bamba2014, Kounnas2014, Rinaldi2014, Einhorn2014} for this kind of approach.  Another interpretation of the peculiar form of the Starobinsky model based on extra dimensions is found in Ref.~\cite{Asaka2015}. 
Anyway, it can be said that data are suggesting and favoring the Starobinsky-like flat inflationary potential, and it is interpreted,  in the purely (super)gravitational approach, that the higher order terms are somehow suppressed to the negligible level.
This approach is phenomenological, and the mechanism, if exists at all, should be understood eventually.

	A particular choice of the functions $N$ and $F$ suitable for realization of the Starobinsky model was studied by Kallosh and Linde~\cite{Kallosh2013a}.
	It is specified by
	\begin{align}
	N= & -3 + \frac{12}{m^2}\bar{\mc{R}}\mc{R} - \frac{\zeta}{m^4} (\bar{\mc{R}}\mc{R})^2 , \label{KLKahler} \\
	F=& 0 ,
	\end{align}
	where $m$ is a mass dimensional parameter of order $10^{-5}$ corresponding to the inflaton mass.
	The constant term is the same as the Einstein supergravity whereas the quadratic term generates the $R^2$ term.
	We will explain the role of the quartic term below.
	The role of the inflaton is played by the real part of $T$.
	This is essentially the scalaron in the original Starobinsky model.
	In the absence of the quartic term in eq.~\eqref{KLKahler}, $S$ becomes tachyonic in the large inflaton region.  It is stabilized by the SUSY breaking mass term $\sim \zeta G_{\bar{S}}G_{S} \bar{S}S$ since the sGoldstino $S$ breaks SUSY during inflation.
	This is the reason to introduce the quartic term.
	Then, $S$ can be integrated out, and the mass of the imaginary part of $T$ is also larger than the Hubble parameter during inflation.
	The inflationary trajectory is effectively of the single-field type, and the Starobinsky model is reproduced.

	In the above example, we have not exploited degrees of freedom in the theory.
	There are possibilities to use other fields, $\text{Im}T$ and the real and imaginary parts of $S$, to describe different inflationary scenarios or the present accelerating expansion of the universe (dark energy).
	Supersymmetrization of some bosonic inflationary models or modified theories of gravity not just adds
fermionic superpartners, but also provides scalar superpartners and sometimes additional multiplets too.
	The SUSY Starobinsky model is an example of this.
	Inspired by the announcement of detection of $B$-mode polarization of CMB of the primordial origin by BICEP2~\cite{Ade2014b}, several people including the author examined the possibility to realize quadratic inflation in the SUSY Starobinsky model~\cite{Ferrara2014, Kallosh2014, Ellis2014, Hamaguchi2014}.
	We noticed that the potential in the $\text{Im}T$ direction is quadratic, which looks suitable for chaotic inflation generating the tensor-to-scalar ratio of order 0.1.
	The potential of $S$ is unstable without the stabilization, but it becomes so steep in the large $\text{Re}T$ domain after the stabilization without tuning that $S$ quickly settles to the origin.  In the smaller region of $\text{Re}T$, modification of the super- and K\"{a}hler potentials of $S$ leads to curved inflationary trajectories and/or local minima with cosmological constants.
	Note in passing that this feature is positively accepted in Ref.~\cite{Dalianis2014} in which the Starobinsky-type inflationary model leading to a SUSY breaking vacuum with the vanishing cosmological constant with tuning.
	We found it not so easy to produce the quadratic potential in the $S$ sector, so let us focus on the complex $T$ sector.
	For simplicity, we stabilize $S$ strongly enough so that it vanishes and decouple from the inflation sector.  Alternatively, the same effective theory can be described by Akulov-Volkov supergravity~\cite{Antoniadis2014}, where the sGoldstino $S$ obeys the nilpotency condition, $S^2=0$~\cite{Komargodski2009}.
	The two-field analyses of the model were performed in Refs.~\cite{Kallosh2014, Ellis2014, Hamaguchi2014}.
	In contrast to the apparent quadratic potential, inflationary dynamics is not the quadratic type, but the Starobinsky type.
	This is because the real and imaginary parts are mixed to and interact with each other.
	The steep exponential potential of $\text{Re}T$ in the finite $\text{Im}T$ region strongly drives the field to too large $\text{Re}T$ direction for canonically normalized $\text{Im}T$ to drives quadratic inflation.
	For quadratic inflation to take place, the potential should be modified so that the real part is strongly stabilized at a constant value~\cite{Kallosh2014, Ellis2014, Ferrara2014}.  With such a modification, however, the theory is no longer related to the pure higher derivative supergravity~\cite{Kallosh2014} (but see Ref.~\cite{Ferrara2014} too).  Finally we mention that the stabilization mechanism is also employed in our proposal in Chapter~\ref{ch:main}.
		
	As mentioned at the beginning of Section~\ref{sec:vector}, the counterpart on the SUSY Starobinsky model in the new-minimal supergravity was studied by Farakos \textit{et al.}~\cite{Farakos2013}.
	They considered the following pure supergravity.
	\begin{align}
	\mc{L}= \frac{3}{2}\left [ L_0 V_{R} \right ]_D + \frac{\lambda_2 }{4}\left[ \mc{W}(V_{R})\mc{W}(V_{R}) \right ]_F , \label{FKR}
	\end{align}
	where the real supermultiplet $V_{R}$ is defined in terms of the linear compensator $L_0$ as follows (\textit{cf.}~eq.~\eqref{pureSUGRAnew}),
	\begin{align}
	V_{R}=\ln \left( \frac{L_0}{\bar{S}_0 S_0} \right). \label{realNMmultiplet}
	\end{align}
	This multiplet contains graviton, gravitino, and ``would-be-auxiliary fields'' in the new-minimal supergravity.
	Fixing the superconformal gauge freedom, the Lagrangian density in terms of curved superspace is~\cite{Farakos2013}
	\begin{align}
	\mc{L}=\int \text{d}^2 \Theta 2 \mc{E} \left(  -\frac{3}{8}\bar{\ms{D}}\bar{\ms{D}} V_{R} + \frac{\lambda_2}{4} \mc{W}(V_R)\mc{W}(V_R) \right) + \text{H.c.}
	\end{align}
	The former term is the FI term of $V_{R}$, which is the Einstein supergravity term containing $R$, and the latter term is the kinetic term of $V_{R}$, which produces the $R^2$ correction.
	
	Similarly to the previous examples we have seen many times, it can be rewritten as the standard supergravity with additional matter.
	In the intermediate (or ``master'') theory, $\mc{W}(V_R)$ is replaced by $\mc{W}(V)$, and the term $\left [ L' (V-V_R ) \right ]_D$ is added, where $V$ is real and $L'$ is real linear.
	The equation of motion of $L'$ equates $V$ and $V_{R}$ up to shift (gauge transformation) by the real part of a chiral supermultiplet, $V=V_R - \ln \Phi - \ln \bar{\Phi}$ with $\Phi$ chiral.
	Eliminating $V$ reproduces eq.~\eqref{FKR}, while eliminating $L_0$ via eq.~\eqref{realNMmultiplet} gives us an alternative expression,
	\begin{align}
	\mc{L}=-\frac{3}{2}\left [ \bar{S}_0 S_0 \bar{\Phi}e^V \Phi \ln \left( \bar{\Phi}e^V \Phi \right) \right]_D + \frac{\lambda_2}{4} \left[ \mc{W}(V_{R})\mc{W}(V_{R}) \right ]_F .
	\end{align}
	This is the standard matter-coupled gauged supergravity with the K\"{a}hler potential
	\begin{align}
	K=-3 \ln \left( -\frac{1}{3} \bar{\Phi}e^V \Phi \ln \left( \bar{\Phi}e^V \Phi \right)  \right).
	\end{align}

	\subsection{Inflation driven by gravitino condensation}\label{subsec:gravitino}
	
	All of the above examples in this Section are about Starobinsky model and its SUSY extension, and the role of the inflaton is played by the scalar mode (scalaron) in metric.
	In supergravity, the bound state of gravitinos can be an inflaton~\cite{Ellis2013c, Alexandre2013, Alexandre2014} (see also a review~\cite{Alexandre2014a}).
	In this novel scenario, the gravitino meson field $\sigma$ rolls down on a one-loop effective potential.
	During the slow-roll, the meson develops a finite value, and it represents gravitino condensation.
	Inflation is a process of phase transition to the vacuum with nonzero gravitino condensation.
	The bound state of a pair of gravitinos as an inflaton is thus absent before the phase transition, and inflation is actually the process of emergence of the inflaton.
	
	The torsion term in supergravity is the four-gravitino interaction term,~\cite{Alexandre2014a}
	\begin{align}
	\mc{L}_{\text{torsion}}=&-\frac{1}{8}\left( \left( \bar{\psi}^{\rho}\gamma^{\mu}\psi^{\nu} \right)  \left( \bar{\psi}_{\rho}\gamma_{\mu}\psi_{\nu}+2\bar{\psi}_{\rho}\gamma_{\nu}\psi_{\mu} \right) \right) \nonumber \\
	=& \lambda_{\text{S}} \left( \bar{\psi}^{\rho}\psi_{\rho} \right)^2 +  \lambda_{\text{PS}} \left( \bar{\psi}^{\rho}\gamma_{5}\psi_{\rho} \right)^2 +  \lambda_{\text{AV}} \left( \bar{\psi}^{\rho}\gamma_{5}\gamma_{\mu}\psi_{\rho} \right)^2,
	\end{align}
	where we have used Fierz	 identities in the second equality, and the coefficients are constrained so that $\lambda_{\text{S}}-\lambda_{\text{PS}}+\lambda_{\text{AV}}=-3/4$.
	There are two remaining ambiguous degrees of freedom in the parametrization above, and this issue was discussed in Ref.~\cite{Alexandre2014a}.
	The above terms are dimension six operators, and they are implicitly multiplied by the inverse square of the reduced Planck mass.
	This coupling constant $\tilde{\kappa}$ can be made larger than $\kappa=M_{\text{G}}^{-1}$ if we introduce a dilaton $\phi$ while keeping the coefficient of the Einstein term unchanged.  Actually, this is required for the Hubble scale during inflation to match the observational bound, $H\lesssim \mathcal{O}(10^{16})$ GeV.  A typical value in Refs.~\cite{Ellis2013c, Alexandre2013, Alexandre2014, Alexandre2014a} is $\tilde{\kappa}=e^{-\phi}\kappa \simeq (10^{3}\sim 10^4 ) \kappa$, and the dilaton is supposed to be stabilized by an unspecified mechanism.
	Let us focus on the scalar bilinear part since it is relevant for cosmological discussion.
	To linearize the gravitino bilinear, we introduce an auxiliary field $\sigma$, which will eventually becomes a propagating degree of freedom -- inflaton, like
	\begin{align}
	\frac{1}{4} \left( \bar{\psi}^{\mu}\psi_{\mu} \right)^2 \sim \sigma \left( \bar{\psi}^{\mu}\psi_{\mu} \right) -\sigma^2.
	\end{align}
	Substituting the equation of motion of $\sigma$ in the right-hand side reproduces the left-hand side.
	Nonzero value of $\sigma$ is interpreted as a dynamically generated mass contribution.
	
	The effective Lagrangian density describing the meson $\sigma$ is given by~\cite{Alexandre2014a}
	\begin{align}
	\mc{L}_{\text{eff}}=-\frac{Z(\sigma)}{2}\partial^{\mu}\sigma\partial_{\mu}\sigma -V_{\text{eff}}(\sigma),
	\end{align}
	where $Z(\sigma)$ is the so-called wave-function renormalization factor, schematically represented as 
	\begin{align}
	Z\sim \frac{1}{2\pi^2} \ln \left( \frac{\sigma^2}{M_{\text{G}}^2} \right),
	\end{align}
	where we have identified the cut-off scale as the reduced Planck mass $M_{\text{G}}$, and the effective potential $V_{\text{eff}}$ integrating out graviton and gravitino at one-loop is of the form of the standard ``wine bottle'' or ``Mexican hat''  symmetry-breaking potential consisting of positive quartic term and negative quadratic term with some logarithmic terms.
	The exact expression involves many numerical factors, and we refer readers to the original papers~\cite{Alexandre2013, Alexandre2014a}.
	The wave-function renormalization factor is of order one at the minimum, but it is tremendously large near the origin.  It helps flattening of the potential in terms of the canonically normalized inflaton, $\tilde{\sigma} \simeq \sqrt{Z} \sigma$. Expanding the potential around the origin of $\sigma$, it is approximated as the tree-level one,
	\begin{align}
	V_{\text{eff}}= f^2 - \sigma^2+\dots,
	\end{align}
	where $f$ is the SUSY breaking order parameter, which is the vestige of Volkov-Akulov Lagrangian~\cite{Volkov1972, Volkov1973} after the super-Higgs effect~\cite{DeserZumino1977, Cremmer1979}, and dots represent loop-induced higher order terms.
	SUSY breaking mechanism is not specified here.  It may be triggered by $F$- or $D$-term of some matter superfields.
	Inflation is driven by the SUSY breaking energy $f^2$, and it is classified as hill-top inflation.
	
To summarize this Chapter, the score sheet of the methods for inflation in supergravity discussed in this Chapter is presented as Table~\ref{tab:scoresheet}.

 \begin{table}[htb]
  \begin{center}
    \caption{Score sheet of various ways of inflation in supergravity.
    A check mark indicates that the method can describe a small or large field inflationary model, or an arbitrary potential depending on the column.  The column of $F$/$D$ shows that inflation is driven by the $F$ or $D$-term.  Our new method discussed in Chapter~\ref{ch:main} as well as the standard $D$-term inflationary models are also included for comparison.  MD, broken(*), and preserved(*) denote model-dependent, generically broken, and preserved by tuning, respectively.}
        \scalebox{0.945}{	
    \begin{tabular}{|c|l||c|c|c|c|c|c|} \hline
      section & method & small & large & arbitrary & number of & $F$/$D$ & SUSY after \\
      &  & field & field & potential & scalar d.o.f. & & inflation \\ \hline \hline
      \ref{subsec:sGoldstino} & sGoldstino inflation & \checkmark &   &   & 2 & $F$ & broken \\
      \ref{subsec:inflection}  & (inflection point inflation) & \checkmark & &  &  2 & $F$ & MD \\ \hline
      \ref{subsec:discreteR} & discrete $R$ symmetry & \checkmark &   &   & 2 & $F$ & preserved \\ \hline
       & shift symmetry and &  & & & & & \\
      \ref{subsec:stabilizer} & \quad stabilizer superfield & \checkmark & \checkmark & \checkmark & 4 &$F$ & preserved \\ 
      \ref{subsec:nilpotent} & \quad stabilizer (nilpotent) & \checkmark & \checkmark & \checkmark & 2 &$F$ & broken \\
\hline
      \ref{sec:vector} & vector or linear multiplet & \checkmark & \checkmark & \checkmark & 1 &$D$ & preserved \\ \hline
      \ref{subsec:Starobinsky} & SUSY Starobinsky &   & \checkmark &   & 4 &$F$ & preserved \\ \hline
      \ref{subsec:gravitino} & gravitino condensation & \checkmark &   &   & 1 & MD & broken \\ \hline \hline
       -- & $D$-term inflation & \checkmark & \checkmark &  & MD & $D$ &  preserved \\ \hline
        & quartic term & & & & &   & \\
      \ref{sec:strategy} & \quad in generic $K$ & \checkmark & \checkmark & \checkmark & 2 &$F$ & broken(*) \\
      \ref{sec:arbitrary} & \quad in no-scale-like $K$ & \checkmark & \checkmark & \checkmark & 2 &$F$ & preserved(*) \\ \hline
    \end{tabular}
    	}
    \label{tab:scoresheet}
  \end{center}
\end{table}

\chapter{Inflation in Supergravity with a Single Chiral Superfield}
\label{ch:main}

In this Chapter, we discuss our new alternative framework to embed inflation (including large field models) into supergravity without using the stabilizer superfield.
Thus, the proposed class of models requires only a single chiral matter superfield, \textit{i.e.}~one including the inflaton in its lowest bosonic component.  In contrast to many existing attempts for inflation in supergravity with a single chiral superfield, we utilize the quartic term in the K\"{a}hler potential to stabilize the non-inflaton part of the complex scalar field which includes the inflaton.  The stabilization mechanism is similar to the case of stabilization of the stabilizer superfield in the double superfield approach.  Our method invokes renewed interests to the single superfield models of inflation.  Because of the stabilization mechanism, it effectively reduces to single field inflationary models.  When tuning is allowed, it is possible to approximately embed arbitrary positive semi-definite scalar potentials of a real scalar field into supergravity.

\section{Basic strategy and implementations}\label{sec:strategy}
As we saw in the previous Chapter, two difficulties of implementation of positive flat potential in supergravity are the exponential factor and the negative definite term.  We impose shift symmetry in the K\"{a}hler potential  as in the usual approach in the literature.  Once imposing the shift symmetry, the potential tends to become negative in the large field region of the inflaton.  In the standard approach, one introduces a vanishing field to suppress (actually eliminate) the negative definite term proportional to the superpotential squared in the scalar potential.  It is also possible to enhance the positive definite term ($F$-term SUSY breaking part) in the scalar potential so that it becomes larger than the absolute value of the negative term.

From na\"{i}ve dimensional consideration, the superpotential term $|W|^2$ dominates its derivatives $|W_{\Phi}|^2$ in the large field region.  For example, polynomial super potentials $W=\sum_{i=0}^n c_i \Phi^i$ generically have this property.  Therefore, we require that the coefficient of the former term be positive,
\begin{align}
K^{\bar{\Phi}\Phi}K_{\Phi}K_{\bar{\Phi}} \geq 3. \label{KphiCondition}
\end{align}
This is a crucial condition, and may be the most important equation in this thesis.
We have to satisfy this inequality by somehow adjusting the expectation value of the non-inflaton field $\phi=(\Phi+\bar{\Phi})/\sqrt{2}$ to have a positive potential in the large field region.
In the following, we consider two representative forms of K\"{a}hler potential, namely the minimal one and logarithmic one, and their deformations.

	\subsection{Minimal K\"{a}hler potential}
	Consider the following K\"{a}hler potential, 
	\begin{align}
	K=\frac{1}{2}\left( \Phi+\bar{\Phi}\right)^2 . \label{minimalKshift}
	\end{align}
	Its derivative is $K_{\Phi}=\Phi+\bar{\Phi}$.
	If we can ensure that the expectation value of the non-inflaton $\phi=(\Phi+\bar{\Phi})/\sqrt{2}$ is large enough to satisfy \eqref{KphiCondition}, the potential becomes positive.
	Suppose that $\phi$'s value is somehow fixed to $\sqrt{2}\Phi_0$ during inflation as we provide such a mechanism later, the scalar potential becomes
	\begin{align}
	V=& e^{2\Phi_0^2}\left( \left|W_{\Phi}+2\Phi_0 W\right|^2 -3 \left|W\right|^2 \right) \nonumber \\
	=& e^{2\Phi_0^2}\left( \left|W_{\Phi}\right|^2+2\Phi_0 \left( \bar{W}W_{\Phi}+W\bar{W}_{\bar{\Phi}} \right) +\left(4\Phi_0^2-3\right)\left|W\right|^2 \right) .
	\end{align}	
	Now, the superpotential is the function of $\Phi=\Phi_0 +i \chi /\sqrt{2}$, where $\chi=(\Phi-\bar{\Phi})/2i$ is the imaginary part (inflaton).
	The condition \eqref{KphiCondition} reduces to $\Phi_0 > \sqrt{3}/2$.
	
	For illustration, let us take a monomial superpotential,
	\begin{align}
	W=c_{n}\Phi^{n}. \label{Wmonomial}
	\end{align}
	Then, the scalar potential becomes
	\begin{align}
	V=e^{2\Phi_{0}^{2}} |c_n|^{2} \left(\Phi_0 ^2+ \frac{1}{2}\chi^2 \right)^{n-1} \left(  n^2+ \Phi_0^2 +4\Phi_0^4 +\frac{1}{2}\left( 4\Phi_0^2 -3 \right)\chi^2 \right) .
	\end{align}
	In the case of a generic polynomial,
	\begin{align}
	W=\sum_n c_n \Phi^n ,
	\end{align}
	the potential becomes
	\begin{align}
	V=e^{2\Phi_0^2} \left( \left| \sum_n n \Phi^{n-1} \right|^2 +2\Phi_0 \sum_{n,m}\left( n c_n \bar{c_m} \Phi^{n-1}\bar{\Phi}^{m} +\text{H.c.} \right) +\left( 4\Phi_0^2 -3 \right) \left| \sum_n c_n \Phi^n \right|^2 \right).
	\end{align}
	
	For example, if we take a linear (Polonyi-like) superpotential,
	\begin{align}
	W=W_0+m \Phi, \label{WPolonyi}
	\end{align}
	the potential is
	\begin{align}
V=e^{2\Phi_{0}^{2}}|m|^2 \left( \frac{1}{2} \left( 4\Phi_0^2 -3 \right)\left(\chi-\chi_0 \right)^2  + \left( (4\Phi_0^2 -3)\widetilde{W}_0^2+(8\Phi_0^3-2\Phi_0 )\widetilde{W}_0 + 4\Phi_0^4+\Phi_0^2+1 \right)  \right). \label{VquadMinimal}
\end{align}
	where $\chi_0=-\sqrt{2}\text{Im}(W_0/m)$ is the VEV of $\chi$ and $\widetilde{W}_0=\text{Re}(W_0/m)$.
	When one first fix the phase of $\Phi$ so that the K\"{a}hler potential becomes the function of the real part $\phi=(\Phi+\bar{\Phi})/\sqrt{2}$, one can not rotate away the relative phase of $m$ and $W_0$, so the $\chi_0$ may be a nonzero value.
	The cosmological constant can be eliminated by tuning the constant term $W_0$ in the superpotential as follows,
	\begin{align}
	\widetilde{W}_0=\frac{-4\Phi_0^3+\Phi_0\pm \sqrt{3}}{4\Phi_0^2-3}.
	\end{align}
	Redefining the origin of the inflaton as $\chi-\chi_0 = \widetilde{\chi}$, the potential finally becomes
	\begin{align}
	V=\frac{1}{2}\widetilde{m}^2 \widetilde{\chi}^2,
	\end{align}
	where $\widetilde{m}^2=e^{2\Phi_0^2}(4\Phi_0^2-3)m^2$ is the mass of the (canonically normalized) inflaton $\tilde{\chi}$.  This quadratic potential can be used in the chaotic inflation scenario.
	After inflation, SUSY is broken by inflaton's $F$-term.
	Therefore, SUSY breaking scale coincides with the inflationary scale.
	The inflationary scale of quadratic inflation (as well as Starobinsky inflation) is about $10^{13}$ GeV, and such a high scale SUSY breaking gives large quantum correction to Higgs mass~\cite{Giudice2012}.  It may be marginally allowed assuming anomaly mediation of gravitino mass into gaugino mass and gaugino mediation of gaugino mass into squark and slepton mass taking into account large uncertainty in the strong coupling constant and top quark mass.
	Alternatively, particles beyond the MSSM may partially cancel the Higgs self energy correction.
	In addition to the Higgs mass concern, high energy SUSY breaking typically leads to too much energy density of dark matter because the neutralinos are heavy.
	They may be diluted by thermal inflation~\cite{LythStewart1996} or decay through $R$-parity violating interactions.
	Although these problems are present, it is interesting possibility that the inflaton breaks supersymmetry at the inflationary scale. 
	
\begin{figure}
  \begin{center}
    \includegraphics[clip, width=8cm]{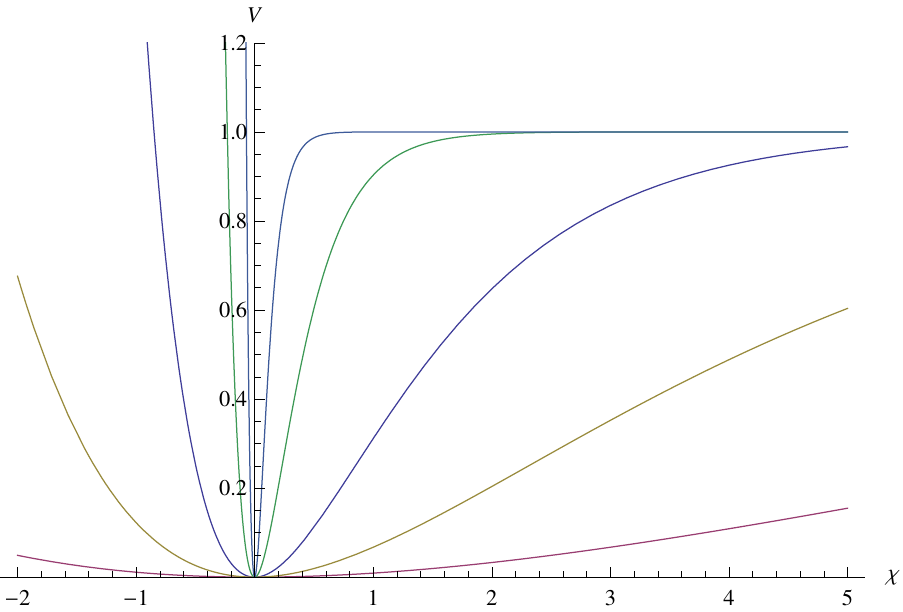}
    \caption{The deformed Starobinsky potential~\eqref{defStaro1}.  The parameter $a$ is set to $0.1, 0.3, \sqrt{2/3}$ (Starobinsky), $ 3$, and $10$ from bottom to top. The height of the potential is normalized to one.  This Figure is from our paper~\cite{Ketov2014a}.}
    \label{fig:defStaro1}
  \end{center}
\end{figure}

	It is possible to embed other inflationary scalar potentials.
	To show the wide capacity of the mechanism, let us consider the potential of Starobinsky model, $V(\chi)=\frac{3}{4}m^2 \left( 1-e^{-\sqrt{2/3}\chi} \right)^2$.
	Its prediction of the small tensor-to-scalar ratio is in sharp contrast to the quadratic potential we have just discussed above.  The potential has a very flat plateau and its embedding into supergravity is a priori non-trivial.  Let us consider the superpotential
	\begin{align}
	W= m \left( b - e^{\sqrt{2}ai \left( \Phi -\Phi_0 \right)} \right). \label{WStarobinsky}
	\end{align}
	Provided the real part is stabilized at $\phi=\sqrt{2}\Phi_0$, the scalar potential of the inflaton $\chi$ is
	\begin{align}
	V=|m|^2 e^{2\Phi_0^2}\left( \left(4\Phi_0^2-3\right) \left( \text{Re}b-e^{-a\chi}\right)^2+\left( 2\Phi_0 \text{Im}b-\sqrt{2}ae^{-a\chi} \right)^2 -3\left(\text{Im}b\right)^2 \right).
	\end{align}
	When $\Phi_0^2 \geq 3/4$ or $\Phi_0^2 \leq (3-a^2)/4$, there are two solutions of $b$ for the potential to coincide with the form $V\propto \left( 1-e^{-a\chi}\right)^2$, where the proportionality coefficient is a constant.  The latter choice leads to the negative potential, and hence we disregard it.  The exact forms of the solutions, $b=b(\Phi_0, a)$, are lengthy,  and it is not worth reproducing it here.
	When we satisfy the condition, the potential reads
	\begin{align}
	V=|m|^2 e^{2\Phi_0^2} \left( 4\Phi_0^2-3+a^2 \right) \left(1- e^{-a\chi} \right)^2. \label{defStaro1}
	\end{align}
	This potential is plotted in Fig.~\ref{fig:defStaro1}.
	The Starobinsky potential corresponds to the case $a=\sqrt{2/3}$.  Other values of $a$ are also viable.
	SUSY is broken at the vacuum, and the SUSY breaking scale is the inflationary scale.
	
	Now, we specify the mechanism of the stabilization of the non-inflaton part $\phi$.
	We use the quartic term of the non-inflaton part in the K\"{a}hler potential.
	\begin{align}
	K=\frac{1}{2}\left( \Phi+\bar{\Phi}\right)^2 - \frac{\zeta}{4} \left( \Phi+\bar{\Phi}-2\Phi_0 \right)^4. \label{minKstab}
	\end{align}
	The second term gives $\phi=(\Phi+\bar{\Phi})/\sqrt{2}$ the SUSY breaking mass during inflation, when the inflaton $\Phi$ breaks SUSY by its $F$-term, $m^2 \sim \zeta |F_{\Phi}|^2$, but does not give $\chi=(\Phi-\bar{\Phi})/\sqrt{2}i$ mass.
	Note that the quartic term preserves the shift symmetry.
	A similar stabilization term was first proposed in Ref.~\cite{Ellis1984b} in a no-scale model.
	The quartic term (without shift symmetry) of the stabilizer superfield is often used in the literature.
	See \textit{e.g.} Refs.~\cite{Lee2010, Ferrara2011, Kallosh2013a} and also Ref.~\cite{Kitano2006}.
	It is instructive to write derivatives of the K\"{a}hler potential,
	\begin{align}
	K_{\Phi}=& \left( \Phi+\bar{\Phi}\right)-\zeta \left(\Phi+\bar{\Phi}-2\Phi_0 \right)^3 , \\
	K_{\Phi\bar{\Phi}}=& 1 - 3 \zeta \left(\Phi+\bar{\Phi}-2\Phi_0 \right)^2 . \label{KmetricStabilized}
	\end{align}
	The sign of the K\"{a}hler metric, which is the coefficient of the kinetic term, changes when $\Phi+\bar{\Phi}$ goes far from $2\Phi_0$.
	Thus, the theory described by the K\"{a}hler potential \eqref{minKstab} should be understood as an effective theory valid only in the limited field range, $|\Phi+\bar{\Phi}-2\Phi_0| < 1/\sqrt{3\zeta}$, otherwise the kinetic term has the unphysical sign.
	It is possible that higher order terms modify this property near the boundary of the field space, but the truncated theory is a good starting point with the characteristic feature. 
	We have argued that $K_{\Phi}$ tend to be suppressed.  The argument assumed the strongest dependence of the scalar potential on the non-inflaton $\phi$ is in the exponential factor $e^K$.
	The trick in the current situation is that the inverse of the K\"{a}hler metric $K^{\bar{\Phi}\Phi}$ diverges in the boundary of the theory $|\Phi+\bar{\Phi}-2\Phi_0| \rightarrow 1/\sqrt{3\zeta}$ where the kinetic term vanishes.  Thus, $K^{\bar{\Phi}\Phi}$ also strongly depends on the value of $\phi$.
	The general formula of scalar mass term in supergravity contains the curvature of the K\"{a}hler manifold, $V_{i\bar{j}}=-e^{G}R_{i\bar{j}k\bar{l}}G^{k}G^{\bar{l}} +\cdots$ (even without using the conditions $V=0$ or $V_{i}=0$), where $G=K+\ln |W|^2$ is the K\"{a}hler-Weyl invariant potential.    This is nothing but the above mentioned SUSY breaking mass term.
	In our case, $R_{\Phi\bar{\Phi}\Phi\bar{\Phi}}\simeq G_{\Phi\bar{\Phi}\Phi\bar{\Phi}} \simeq -6\zeta$.
	In reality, the holomorphic mass term $V_{ij}$ is also relevant because of the shift symmetric K\"{a}hler potential, \textit{i.e.}~$\partial/\partial \Phi=\partial/\partial \bar{\Phi}$ on $K(\Phi+\bar{\Phi})$.  It has the similar contribution, $V_{\Phi \Phi}=-e^{G}G_{ijk\bar{l}}G^{k}G^{\bar{l}}+\cdots$.  In our case, $V_{\Phi\Phi}\simeq V_{\bar{\Phi}\bar{\Phi}} \simeq V_{\Phi \bar{\Phi}} \simeq 6\zeta e^{G}G^{\Phi}G^{\bar{\Phi}}$.  The masses of the real and imaginary parts are proportional to $V_{\Phi \Phi}+V_{\bar{\Phi}\bar{\Phi}} \pm 2 V_{\Phi\bar{\Phi}}$ respectively, and the terms we are talking about (proportional to $\zeta$) cancel for the imaginary part.  In this way, we give SUSY breaking mass to stabilize the real part, but not the imaginary part.	 Detailed analysis and numerical confirmation of the stabilization mechanism are presented in Section \ref{sec:stabilization}.

	It might be easier to understand the mechanism if we redefine the origin of the inflaton so that its VEV vanishes, $\tilde{\Phi}=\Phi-\Phi_0$.  Then, the nonzero value of $K_{\Phi}=c$ is provided by the linear term in the K\"{a}hler potential.
	\begin{align}
	K= c \left( \Phi+\bar{\Phi}\right)+\frac{1}{2} \left( \Phi+\bar{\Phi}\right)^2  -\frac{\zeta}{4} \left(\Phi+\bar{\Phi}\right)^4 \label{minKstabilizedShifted}
	\end{align}
	where $c=2\Phi_0$ and we have omitted tildes on the new field.
	The constant term in the K\"{a}hler potential, $2\Phi_0^2$, has been absorbed by the superpotential.
	But the linear and quadratic terms are not enough, and the quartic term plays the crucial role.
	If there is no quartic term, there is no guarantee that $K_{\Phi}=c+ (\Phi+\bar{\Phi})$ is equal to $c$, and the field value tends to cancel the constant $c$.

	\subsection{Logarithmic K\"{a}hler potential}
		
	The condition \eqref{KphiCondition}, $K^{\bar{\Phi}\Phi}K_{\Phi}K_{\bar{\Phi}} \geq 3$, can be satisfied by using other special forms of K\"{a}hler potentials.
	The no-scale-like K\"{a}hler potential,
	\begin{align}
	K=-a\ln \left( 1+\frac{1}{\sqrt{a}}\left( \Phi+\bar{\Phi} \right) \right) , \label{a-no-scale}
	\end{align}
	is such an example, where $a$ is a constant parameter.
	Here, the square root factor has been inserted to obtain the canonically normalized kinetic term under a condition (see below).
	The case of $a=3$ corresponds to the K\"{a}hler potential of the no-scale model.
	Possible values of $a$ are 1, 2, and 3 in the string models~\cite{Roest2013}, but we consider values larger than 3 too to explore more general possibilities in the general field theory context.
	Because derivatives of the K\"{a}hler potential are $K_{\Phi}=\frac{-\sqrt{a}}{1+\left( \Phi+\bar{\Phi}\right)/\sqrt{a}}$ and $K_{\bar{\Phi}\Phi}=\frac{1}{\left(1+\left(\Phi+\bar{\Phi}\right)/\sqrt{a}\right)^2}$, the combination satisfies
	\begin{align}
	K^{\bar{\Phi}\Phi}K_{\Phi}K_{\bar{\Phi}}=a,
	\end{align}
	independently of the value of $\Phi$.
	This implies the coefficient of the $|W|^2$ term is positive if we take $a>3$ without introducing the stabilization mechanism such as the quartic term in the K\"{a}hler potential.
	However, we assume again the real part $(\Phi+\bar{\Phi})$ is somehow fixed so that we can separate out the dynamics of the imaginary part (inflaton).
	In fact, if the real part is not stabilized at some value, the kinetic mixing effect makes the dynamics complicated~\cite{Kallosh2014, Ellis2014, Hamaguchi2014}.
	The kinetic term and the scalar potential resulting from the K\"{a}hler potential~\eqref{a-no-scale} and a superpotential $W(\Phi)$ are 
	\begin{align}
	&\mathcal{L}_{\text{kin}}= - \frac{1}{\left(1+\left(\Phi+\bar{\Phi}\right)/\sqrt{a}\right)^2} \partial_{\mu}\bar{\Phi}\partial^{\mu}\Phi, \label{kin_a-no-scale} \\
	&V= \left( 1+\left( \Phi +\bar{\Phi} \right)/\sqrt{a} \right)^{-a} \left( \left( 1+\left( \Phi +\bar{\Phi} \right)/\sqrt{a} \right)^{2} \left| W_{\Phi} - \frac{\sqrt{a}}{ 1+\left( \Phi +\bar{\Phi} \right)/\sqrt{a}} W \right|^2 -3 \left| W \right|^2  \right) . \label{V_a-no-scale}
	\end{align}
	Suppose that $(\Phi+\bar{\Phi})$ is stabilized at some value.
	After the shift of the field $\Phi$ and renormalization of the Newton constant, it is possible to keep the form of eq.~\eqref{a-no-scale} with vanishing $(\Phi+\bar{\Phi})$ in terms of the redefined variables.
	With this in mind, we assume that $(\Phi+\bar{\Phi})$ is fixed at the origin.
	For example, it can be accomplished by the following quartic term.
	\begin{align}
	K=-a\ln \left( 1+\frac{1}{\sqrt{a}}\left( \Phi+\bar{\Phi} \right) + \frac{\zeta}{a^2} \left( \Phi+\bar{\Phi} \right)^4 \right) . \label{K-a-no-scale}
	\end{align}
	This kind of modification of the no-scale model was proposed in Ref.~\cite{Ellis1984b}, and recently used in Refs.~\cite{Ellis2013, Ferrara2014, Ellis2014}.
	The key term in the fourth derivative of the K\"{a}hler potential is $K_{\Phi\bar{\Phi}\Phi\bar{\Phi}}\sim -24(\zeta /a)$, so the mass of the stabilized part is roughly $m^2 \sim 24(\zeta /a)e^{G}G^{\Phi}G^{\bar{\Phi}}$.
	Once the real part is fixed to the origin, 
	 the kinetic term of the imaginary part (as well as the real part) is canonically normalized, and the scalar potential is
	\begin{align}
	V=|W_{\Phi}|^2-\sqrt{a}\left( \bar{W}W_{\Phi}+W\bar{W}_{\bar{\Phi}}\right)+(a-3)|W|^2 . \label{V-a-no-scale}
	\end{align}
	As discussed above, $a>3$ generically allows the potential to be positive in the large field region $(|\Phi|\gg 1)$, whereas $a<3$ gives rise to a potential unbounded below.
	Another interesting possibility is $a=3$.
	In this case, the cross term generically has phase directions some of which becomes positive and other negative in the large field region. We will study a special case in which the cross term as well as the $|W|^2$ term vanishes in the next Section.

	One may guess that the shift symmetry is not needed for a logarithmic K\"{a}hler potential, because the exponential of the logarithm does not change drastically with respect to the field variables.
	However, it is not true unless the superpotential takes a special form.
	Let us take the following K\"{a}hler potential,\footnote{
	Actually, this is equivalent to the no-scale type K\"{a}hler potential~\eqref{a-no-scale} via a field redefinition up to a K\"{a}hler transformation~\cite{Cecotti2014}.  The K\"{a}hler transformation is equivalent to a particular modification of the superpotential.  Also, the original shift symmetry becomes obscured in the transformed K\"{a}hler potential~\eqref{logK}, so we discuss it separately.
	}
	\begin{align}
	K=-3\ln \left( 1-\frac{1}{3}\bar{\Phi}\Phi \right). \label{logK}
	\end{align}
	This K\"{a}hler potential leads to the minimal kinetic term in the original (Jordan) frame, in contrast to the case of the so-called minimal K\"{a}hler potential, which leads to the minimal kinetic term in the Einstein frame.
	The K\"{a}hler metric is $K_{\Phi\bar{\Phi}}=\frac{1}{\left( 1- |\Phi|^2/3 \right)^2}$, and diverges in the limit $|\Phi|\rightarrow \sqrt{3}$.
	To simplify the analysis, let us assume that the real or imaginary part is fixed at the origin.  For concreteness, we fix the imaginary part and neglect its dynamics just to see the shape of the potential of the real part.
	The real part is stretched exponentially upon canonical normalization, and the canonically normalized field is $\tilde{\Phi}= \sqrt{\frac{3}{2}}\ln \frac{\Phi+\sqrt{3}}{\Phi-\sqrt{3}}$ (implicitly neglecting the imaginary part).
	The scalar potential is
	\begin{align}
	V=& \left( 1- \frac{1}{3}|\Phi|^2 \right)^{-1}\left| W_{\Phi} \right|^2+  \left( 1- \frac{1}{3}|\Phi|^2 \right)^{-2}\left( \bar{W}W_{\Phi}\Phi+W\bar{W}_{\bar{\Phi}}\bar{\Phi} -3 |W|^2 \right)  \nonumber \\
	\rightarrow & \left( 1- \frac{1}{3}|\Phi|^2 \right)^{-2}\left( \bar{W}W_{\Phi}\Phi+W\bar{W}_{\bar{\Phi}}\bar{\Phi} -3 |W|^2 \right) , \qquad (\text{as } |\Phi|^2\rightarrow 3 ) \label{logKexponentialV}
	\end{align}
	where the last line expresses the large field behavior.
	In terms of the canonically normalized field, 
	the factor $\left( 1- \frac{1}{3}|\Phi|^2 \right)^{-2}=\cosh ^2 \tilde{\Phi}$ grows exponentially.

	Now, let us impose shift symmetry on the previous K\"{a}hler potential~\eqref{logK},
	\begin{align}
	K=-3\ln \left(1 - \frac{1}{6}\left( \Phi+\bar{\Phi}\right)^2 \right). \label{logKshift}
	\end{align}
	The scalar potential following from the above K\"{a}hler potential \eqref{logKshift} is
	\begin{align}
	V=\frac{1}{A_{+}A_{-}^{2}}\left( A_{-} |W_{\Phi}|^2+\left( \Phi+\bar{\Phi} \right) \left( \bar{W}W_{\Phi}+W\bar{W}_{\bar{\Phi}} \right) -3 |W|^2\right) , 
	\end{align}
	where
	\begin{align}
	A_{\pm} \equiv \left(1 \pm \frac{1}{6}\left( \Phi+\bar{\Phi}\right)^2 \right). 
	\end{align}
	The combination $K^{\bar{\Phi}\Phi}K_{\Phi}K_{\bar{\Phi}}=\frac{\left( \Phi+\bar{\Phi}\right)^2}{1+\left( \Phi+\bar{\Phi}\right)^2/6}$ can not exceed $3$ as long as the sign of the kinetic term of gravity is canonical.  The latter condition is equivalent to the positivity of the argument in the logarithm in the K\"{a}hler potential \eqref{logKshift}, \textit{i.e.} $(\Phi+\bar{\Phi})^2<6$.
	Thus, it is impossible to obtain a positive coefficient of $|W|^2$ with a single chiral superfield and its K\"{a}hler potential \eqref{logKshift}.
	
	Let us go back to the no-scale-like case~\eqref{a-no-scale} in which the inflaton superfield enters linearly in the logarithm, and see some example models.
	First, let us set $a=4$, and the effective single field potential~\eqref{V-a-no-scale} becomes
	\begin{align}
	V=|W_{\Phi}|^2 -2 \left( \bar{W}W_{\Phi}+W\bar{W}_{\bar{\Phi}} \right) + |W|^2 .
	\end{align}
	The potential is generically dominated  by the last term, $|W|^2$.
	Let us take a simple superpotential, $W=m\Phi+W_0$ where $m$ is taken as a real mass parameter.
	Then, the potential becomes
	\begin{align}
	V=\frac{m^2}{2}\left( \chi - \chi_0 \right)^2 +m^2 -4m \text{Re}W_0 +\left( \text{Re}W_0 \right)^2, \label{V4quadratic}
	\end{align}
	where $\chi_0 = -\sqrt{2}(\text{Im}W_0)/m$.
	Tuning the parameters so that they satisfy $\text{Re}W_0=(2\pm \sqrt{3})m$, the cosmological constant vanishes.  The quadratic potential is suitable for chaotic inflation.
	Again, SUSY is broken at the vacuum, and its scale is that of inflation.
	
	Next, we set $a=3$.  This time, the effective single field potential becomes
	\begin{align}
	V=|W_{\Phi}|^2 -\sqrt{3} \left( \bar{W}W_{\Phi}+W\bar{W}_{\bar{\Phi}} \right).  \label{V-no-scale}
	\end{align}
	The $|W|^2$ terms are cancelled by the no-scale K\"{a}hler potential.
	To see the basic features, let us consider a monomial superpotential~\eqref{Wmonomial}, $W=c_n \Phi^n$. The potential is very simple, because the second and third terms cancel due to the vanishing VEV of the real part.
	\begin{align}
	V=n^2 |c_n|^2 \left( \frac{\chi^2}{2} \right)^{n-1}.
	\end{align}
	The case $n=2$ leads to the quadratic potential. 
	In this case, SUSY is preserved at the vacuum.
	 When we add terms in the superpotential, the cancellation does not work generically. 
	As we have seen above, there are many possible embedding of a potential into supergravity with a single chiral superfield.
	
	It is worth noting that the real part can also be used as an inflaton in the case of the no-scale K\"{a}hler potential, if we take a suitable superpotential, because of exponential stretching of the field due to canonical normalization.
	Consider the following K\"{a}hler potential.
	\begin{align}
	K=-3\ln \left( \left(\Phi+\bar{\Phi}\right) +\zeta \left(i\left(\bar{\Phi}-\Phi\right) - 2\Phi_0 \right)^4 \right). \label{K-imaginary-fixed}
	\end{align}
	The effect of the quartic term is to stabilize the imaginary part around $\Phi_0$.
	The kinetic term is given by
	\begin{align}
	\mathcal{L}_{\text{kin}}= - \frac{3}{\left(\Phi+\bar{\Phi}\right)^2} \partial_{\mu}\bar{\Phi}\partial^{\mu}\Phi.
	\end{align}
	The canonically normalized inflaton field is $\tilde{\phi}=\sqrt{3/2} \ln \phi$ $(\Longleftrightarrow \phi = e^{\sqrt{2/3}\tilde{\phi}})$, with $\phi=\sqrt{2}\text{Re}\Phi$.
	In this case, not the real part but the imaginary part is fixed, so the effective single field potential is given by
	\begin{align}
	V=&\frac{1}{\left(\Phi+\bar{\Phi}\right)^2} \left( \frac{\Phi+\bar{\Phi}}{3}|W_{\Phi}|^2 -\left( \bar{W}W_{\Phi}+W\bar{W}_{\bar{\Phi}}  \right) \right). \label{V-imaginary-fixed}
	\end{align}
	Let us first try with a monomial superpotential~\eqref{Wmonomial}, $W=c_n \Phi^n$.  The potential is now
	\begin{align}
	V=&\frac{n(n-3)|c_n|^2 }{3\left(\Phi+\bar{\Phi}\right)}|\Phi|^{2n-2} \nonumber \\
	=& \frac{n(n-3)|c_n|^2 }{6}e^{-\sqrt{2/3}\tilde{\phi}} \left( \Phi_0^2 + \frac{1}{2}e^{2\sqrt{2/3}\tilde{\phi}} \right)^{n-1}.
	\end{align}
	Note that the potential vanishes when $n=0$ or $3$.  This is the property of no-scale models.
	Moreover, the potential becomes positive when $n>3$ or $n<0$, and becomes negative when $0<n<3$.
	Also, the potential of the canonically normalized inflaton has an exponential form.
	To obtain a flat potential we have to take a fractional value $n=3/2$ so that the exponential factors cancel in the large field region, but then the potential becomes negative.
	
	Now, we have to choose more general superpotentials to obtain flat potentials.
	The key idea is to take a polynomial superpotential and use cross terms. 
	To explain this point, we take the following binomial superpotential,
	\begin{align}
	W=c_m \Phi^m + c_n \Phi^n ,
	\end{align}
	with $m<n$ without loss of generality.
	The scalar potential for this superpotential is
	\begin{align}
	V=& \frac{1}{\left( \Phi+\bar{\Phi} \right)^2} \left( \frac{\Phi+\bar{\Phi}}{3} \left( n(n-3)|c_n|^2 |\Phi|^{2n-2} +m(m-3)|c_m|^2 |\Phi|^{2m-2} \right. \right. \nonumber \\
	& \left.\phantom{\frac{\Phi+\bar{\Phi}}{3}}  \left. \hfill +2 nm \text{Re} \left( c_m \bar{c_n} \Phi^{m-1}\bar{\Phi}^{n-1} \right) \right) -2n \text{Re} \left( c_n \bar{c_m} \Phi^{n-1}\bar{\Phi}^{m} \right) -2m \text{Re}\left( c_m \bar{c_n}\Phi^{m-1}\bar{\Phi}^n \right)  \right).
	\end{align}
	The first lines are just the sum of the monomial potentials.
	To have a flat (asymptotically constant) potential, the powers of the cross terms $m+n-3$ must vanish.
	We can satisfy this constraint keeping the first line vanishing with the choice $n=3$ and $m=0$,
	\begin{align}
	W=c_0 + c_3 \Phi^3. \label{W03}
	\end{align}
	Then, the potential reduces to
	\begin{align}
	V=\frac{3}{2 (\text{Re}\Phi )^2} \left( -\text{Re}\left( c_3 \bar{c_0} \right) \left( \text{Re}\Phi - \frac{\text{Im}\left( c_3 \bar{c_0} \right)}{\text{Re}\left( c_3 \bar{c_0} \right)} \Phi_0 \right)^2  +\left( \text{Re}\left( c_3 \bar{c_0} \right)-  \frac{\left(\text{Im}\left( c_3 \bar{c_0} \right)\right)^2}{\text{Re}\left( c_3 \bar{c_0} \right)} \right)  \Phi_0^2 \right), \label{Vlognonshift1}
	\end{align}
	where we have set $\text{Im}\Phi=\Phi_0$.  Note that the masses of the real and imaginary parts have different signs, so either one is tachyonic.  The imaginary part has the negative mass contribution but we assume that it is stabilized by the quartic term. The cosmological constant vanishes with the choice $\text{Re}(c_3 \bar{c_0})=\text{Im}(c_3\bar{c_0})$.
	The potential finally becomes that of the Starobinsky model,
	\begin{align}
	V=\frac{3}{4}m^2 \left( 1 - \sqrt{2}\Phi_0 e^{-\sqrt{2/3}\tilde{\phi}} \right)^2 , \label{Vlognonshift2}
	\end{align}
	with
	\begin{align}
	m^2 \equiv  - 2 \text{Re}\left( c_3 \bar{c_0} \right) . \label{m2}
	\end{align}
	The coefficient of the second term, $\sqrt{2}\Phi_0$, can be absorbed by the redefinition of the origin of $\tilde{\phi}$ unless $\Phi_0=0$, so it does not affect physical observables. 
	This model also breaks SUSY at the vacuum, and the SUSY breaking scale coincides  with the inflationary scale.
	
\begin{figure}
  \begin{center}
    \includegraphics[clip, width=8cm]{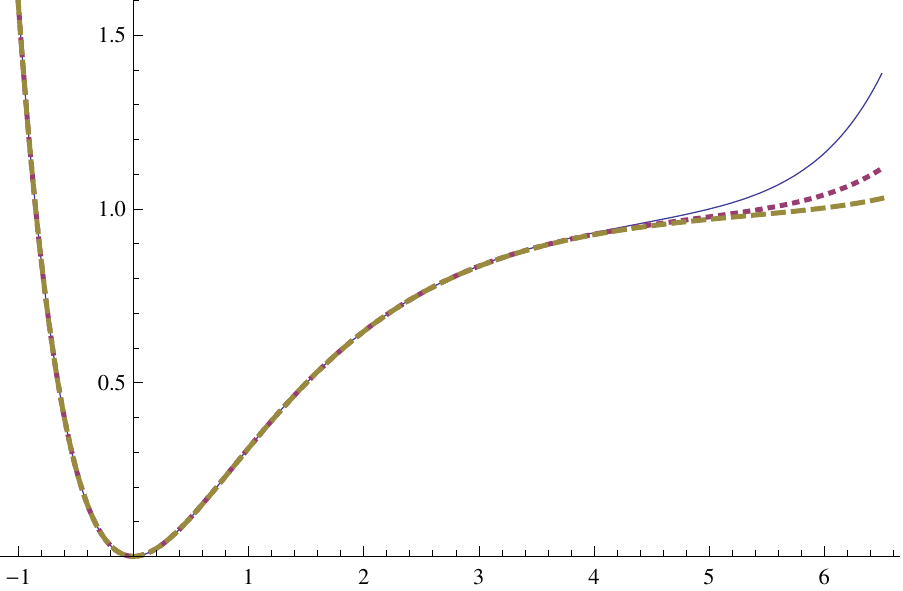}
    \caption{The deformed Starobinsky potential~\eqref{V03shifted}.  The relative magnitude of the correction $2\Phi_0^3|c_3|^2/m^2$ is set to $10^{-5}, 10^{-5.5}$, and $10^{-6}$ for the blue solid, red dotted, and yellow dashed lines, respectively.}
    \label{fig:defStaro2}
  \end{center}
\end{figure}

	The case of vanishing $\Phi_0$ was studied in our paper~\cite{Ketov2014}.
	Substituting $\Phi_0=0$ into the above scalar potential~\eqref{Vlognonshift1} or \eqref{Vlognonshift2}, the potential becomes just a constant.  It is flat and can be chosen as positive, but useless for slow-roll inflation.  It is a starting point for further modification.
	Emulating the above success due to the nonzero VEV of the imaginary part, we shift the field $\Phi$ in the superpotential to make the cancellation of $e^{\sqrt{2/3}\tilde{\phi}}$ factor incomplete.
	\begin{align}
	W= c_0 + c_3 \left( \Phi - \Phi_0 \right)^3. \label{W03shifted}
	\end{align}
	This $\Phi_0$ has nothing to do with the previous $\Phi_0$ in \eqref{K-imaginary-fixed} which is the expectation value of the imaginary part.  (The latter is taken as $0$ currently.)
	The scalar potential is
	\begin{align}
	V=&\frac{3}{4} m^2 \left(1- e^{-\sqrt{2/3}\tilde{\phi}} \right)^{2}+\frac{3}{2}\Phi_0^3 |c_3|^{2} \left(1- e^{-\sqrt{2/3}\tilde{\phi}} \right)^{4} e^{2\sqrt{2/3}\tilde{\phi}}, \label{V03shifted}
	\end{align}
	where $m^2$ is given by eq.~\eqref{m2}, and we have redefined the origin of the field as $\sqrt{2}\Phi_0 e^{-\sqrt{2/3}\phi} \rightarrow e^{-\sqrt{2/3}\phi}$.
	The first term is the Starobinsky potential, and the second term is a correction to it.
	The cross terms of the cubic and quadratic terms in the superpotential generate $\Phi^2 =e^{2\sqrt{2/3}\tilde{\phi}}$ behavior in the large field region.
	The potential is shown in Fig.~\ref{fig:defStaro2}.
	We started from the prototypical model~\eqref{W03} with the flat feature, and modified it as eq.~\eqref{W03shifted}, eventually obtaining an exponential correction.  But we need flatness only in the field range which corresponds to the observable sky of $50$ or $60$ e-foldings.
	For that purpose, tuning of $\Phi_0 |c_3|^2$ to be small enough is needed to maintain the flatness of the potential while keeping the value of $m^2=-2\text{Re}(c_3\bar{c_0})$ to be consistent with the amplitude of the fluctuation of CMB.
	This model also breaks SUSY at the vacuum.  The SUSY breaking scale is determined by $|c_0|^2$ and it is larger than the inflation scale $\text{Re}(c_3 \bar{c_0})$.
	
\begin{figure}
  \begin{center}
    \includegraphics[clip, width=8cm]{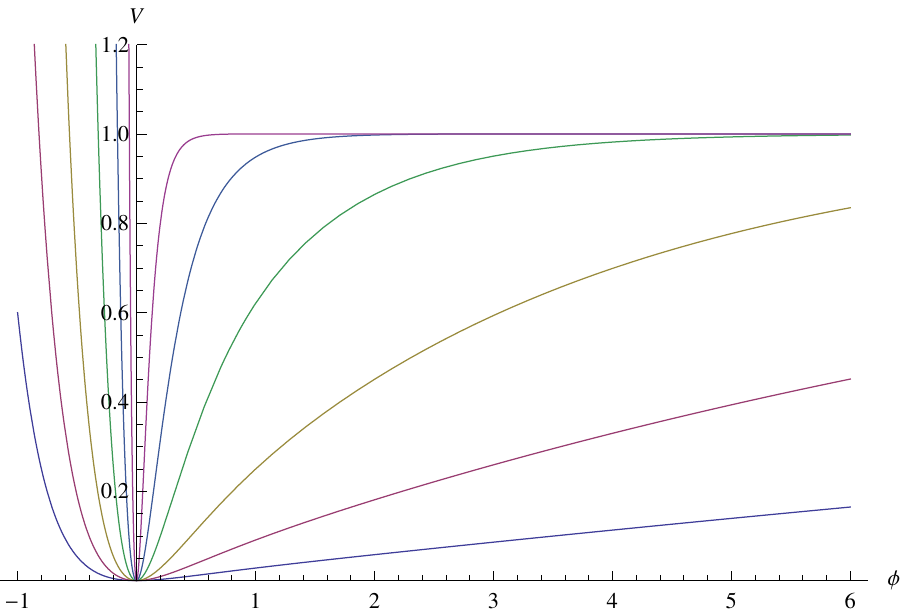}
    \caption{The deformed Starobinsky potential~\eqref{defStaro3}. The power $n$ is $-0.03, -0.1, -0.3, -1, -3$, and $-10$ from bottom to top. The parameter values are taken as $a=-b=-c=d=1$.  This Figure is from our paper~\cite{Ketov2014a}.}
    \label{fig:defStaro3}
  \end{center}
\end{figure}

	Let us consider one more modification of the model specified by the K\"{a}hler potential~\eqref{K-imaginary-fixed} and the superpotential~\eqref{W03}.   Let us consider the following superpotential,
	\begin{align}
	W=c_{-n}\Phi^{-n}+c_0 + c_3 \Phi^3. \label{W-n03}
	\end{align}
	We have introduced a negative power in the superpotential because it does not affect the large field behavior, \textit{i.e.} it keeps flatness.
	The effective single field scalar potential for $\text{Im}\Phi=\Phi_0=0$ is the following form
	\begin{align}
	V=& a + b e^{-n \sqrt{2/3}\tilde{\phi}}+ce^{-(n+3)\sqrt{2/3}\tilde{\phi}}+de^{-(2n+3)\sqrt{2/3}\tilde{\phi}}, \label{defStaro3}
	\end{align}
	where 
	\begin{align}
	a=-\frac{81}{2}\text{Re}(c_0 \bar{c_3}), && b=\frac{-27(3+n)}{2}\text{Re}(c_{-n}\bar{c_3}), && c=\frac{27n}{2}\text{Re}(c_{-n}\bar{c_0}), && d=\frac{9n(3+n)}{2}|c_{-n}|^2. 
	\end{align}
	If we take $a=-b=-c=d>0$, the potential satisfies both $V=0$ and $V_{\tilde{\phi}}=0$ at the vacuum $\tilde{\phi}=0$ for an arbitrary $n$.  It is satisfied by the choice, $c_0=-(n+3/n)c_3$ and $c_{-n}=(3/n)c_3$.  
	The potential in this case is plotted in Fig.~\ref{fig:defStaro3}.
	The mass squared at the vacuum is given by $m^2 = bn^2+c(n+3)^2+d(2n+3)^2$.  Although we do not present the origin of the negative power, it is remarkable that all of the models in this class~\eqref{W-n03} lead to very flat potentials. Moreover, these models preserve SUSY at the vacuum.
	
\begin{figure}[htbp]
  \begin{center}
    \includegraphics[clip,width=10.70cm]{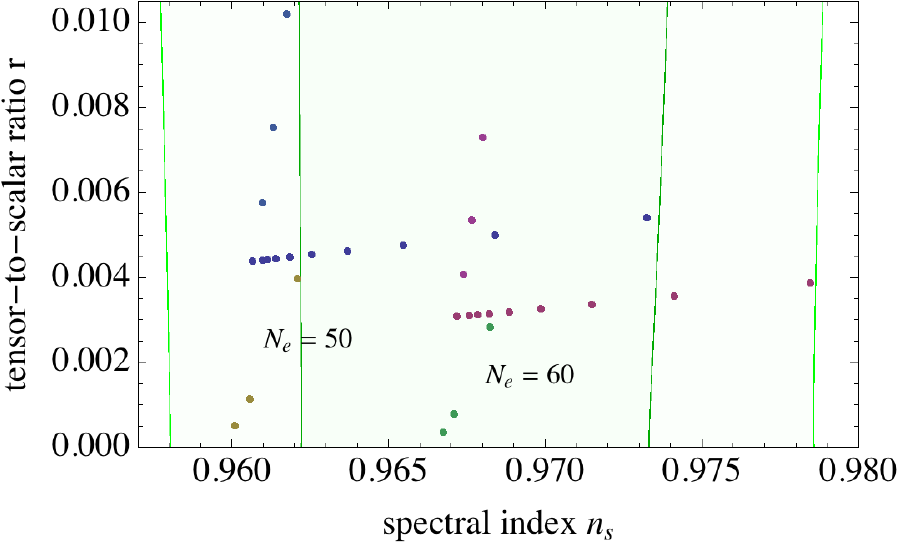}
    \caption{The spectral index and the tensor-to-scalar ratio of some deformations of the Starobinsky model, \eqref{defStaro1}, \eqref{V03shifted} and \eqref{defStaro3}.  The prediction of \eqref{defStaro1} corresponds to blue points ($N_e=50$) and red points ($N_e=60$) lined vertically.  The lowest point in each case is the prediction of the Starobinsky model ($a=\sqrt{2/3}$), and the above points correspond to $a=0.7, 0.6$, and 0.5 from bottom to top.  The prediction of \eqref{V03shifted} corresponds to blue points ($N_e =50$) and red points ($N_e =60$) lined horizontally.  The most left point in each case is the prediction of the Starobinsky model.  The relative coefficient $-c_3 / 3 c_0$ of the correction term is set to $10^{-8}, 10^{-7.8}, 10^{-7.6}, \dots$ from left to right.  We take real parameters and $\Phi_0=1$.   The prediction of \eqref{defStaro3} corresponds to the yellow ($N_e =50$) and green ($N_e =60$) points with the same parameters as in Fig.~\ref{fig:defStaro3}. The power $-n$ is taken as $-1, -2,$ and $-3$ from top to bottom.  The 1$\sigma$ and 2$\sigma$ contours of the Planck TT+lowP+BKP+lensing+BAO+JLA+$H_0$ constraints (traced from Fig. 21 in Ref.~\cite{Ade2015a}) are also shown by light green shading.}
    \label{fig:ns_r}
  \end{center}
\end{figure}

	The differences of the predictions of these deformed Starobinsky potentials on the inflationary observables, the spectral index and the tensor-to-scalar ratio, are shown in Fig.~\ref{fig:ns_r}.
	This Figure is just to compare different modification of the Starobinsky potential, and do not imply our method is restricted to the small $r$ region indicated in the Figure.
	
	We have seen that various K\"{a}hler potential and superpotential leads to potentials suitable for large field inflation assuming the non-inflaton field is fixed some value due to the quartic stabilization term in the K\"{a}hler potential.  In the next Section, we focus on a special K\"{a}hler potential which is useful to embed almost arbitrary positive semidefinite scalar potential into supergravity with a single chiral superfield.

\section{Embedding arbitrary scalar potentials}\label{sec:arbitrary}

Many models in the previous Section have complicated expression for the scalar potential, and breaks SUSY even after inflation.
Moreover, we had to search superpotential that leads to the desired scalar potential by trials and errors.
In this Section, we concentrate on a special subset of the proposed class of single superfield inflationary models.
The models in the subset allow us to easily specify the superpotential, and also to easily make the vacuum supersymmetric.
Consider the following K\"{a}hler potential,
\begin{align}
K=-3 \ln \left( 1+\frac{1}{\sqrt{3}} \left( \Phi+\bar{\Phi}\right)  \right)
\end{align}
This is eq.~\eqref{a-no-scale} with $a=3$.
The kinetic term and the scalar potential are given by eqs.~\eqref{kin_a-no-scale} and \eqref{V_a-no-scale} with $a=3$, respectively.
We assume that the real part is fixed to its origin by some mechanism.
Any such mechanism could be used, but to be specific, we again call for the quartic term in the K\"{a}hler potential.
\begin{align}
K=-3 \ln \left( 1+\frac{1}{\sqrt{3}} \left( \Phi+\bar{\Phi}\right) + \frac{\zeta}{9}\left( \Phi+\bar{\Phi}\right)^4 \right) . \label{K-arbitrary-stabilized}
\end{align}
Assuming we can set $\Phi+\bar{\Phi}=0$ (we examine it in Section~\ref{sec:stabilization}), the kinetic term becomes canonical, and the potential becomes (\textit{cf.} \eqref{V-no-scale})
\begin{align}
V= |W_{\Phi}|^2 - \sqrt{3} \left( \bar{W}W_{\Phi}+W\bar{W}_{\bar{\Phi}} \right). \label{V-arbitrary-0}
\end{align}
Note that the $|W|^2$ terms are cancelled by the no-scale structure.

At this stage, it is possible to examine model by model whether inflation occurs or not by specifying the superpotential.
Here, we impose a constraint on the superpotential: Every coefficients of Taylor series expansion of the superpotential has the same phase, so that they are all real up to an overall phase, which is unphysical.
We do not mean by this that the superpotential is not a holomorphic function.  It is just a constraint on coefficients.
Now we parametrize the superpotential as
\begin{align}
W(\Phi)=\frac{1}{\sqrt{2}}\widetilde{W}(-\sqrt{2}i\Phi).
\end{align}
The potential reduces to the very simple form like the global SUSY $F$-term,
\begin{align}
V=\left( \widetilde{W}'(\chi) \right)^2, \label{V-arbitrary}
\end{align}
where $\chi=\sqrt{2}\text{Im}\Phi$, and the prime denotes differentiation with respect to the argument.
This potential is very convenient to embed one's favorite scalar potential.
Substituting the desired scalar potential into the left-hand side~\eqref{V-arbitrary}, taking the square root, and integrating with respect to the inflaton, one can obtain the superpotential that leads to the desired scalar potential.
It may be difficult or impossible to obtain an exact analytic formula of the superpotential, but approximating the function by a polynomial function, it is always possible to solve the superpotential.
In this sense, it is possible to realize almost arbitrary positive semidefinite scalar potential in supergravity with a single chiral superfield.
Our method is as powerful as the existing methods to realize arbitrary scalar potentials~\cite{Kallosh2010, Kallosh2011, Ferrara2013}, but is minimal and does not rely on the stabilizer superfield.

It is not difficult to choose the superpotential so that the cosmological constant vanishes at the vacuum.
It is always possible to tune the constant term in the superpotential so that the $F$-term of the inflaton,
$D_{\Phi}W=W_{\Phi} + K_{\Phi}W$,
vanishes.  The derivative $W_{\Phi}$ vanishes at the vacuum because its square is the potential, and $K_{\Phi}=-\sqrt{3}$ so $W$ must vanish at the vacuum to preserve SUSY. 
This is possible because the potential~\eqref{V-arbitrary} depends only on the derivative of the superpotential, and does not depend on the constant term in the superpotential.  Therefore we can always tune the constant so that $W$ vanishes at the vacuum.
In this scenario, SUSY should be broken by some other sector.
It is consistent with the low-energy SUSY scenario.

By the way, the above `reality condition' may seem a strong condition, but a similar condition is used in the two-superfield case~\cite{Kallosh2010, Kallosh2011}.
The condition is automatically satisfied in the case of the monomial superpotential.
Also, once the condition is satisfied, it continues to be satisfied because of the non-renormalization theorem of the superpotential~\cite{Grisaru1979, Seiberg1993, Weinberg1998}.

\subsection{Examples of the (super)potential} \label{subsec:examples}
Let us see some examples below.

\begin{figure}[htbp]
 \begin{center}
  \subfigure[Example 1: quadratic]{	
   \includegraphics[width=.45\columnwidth]{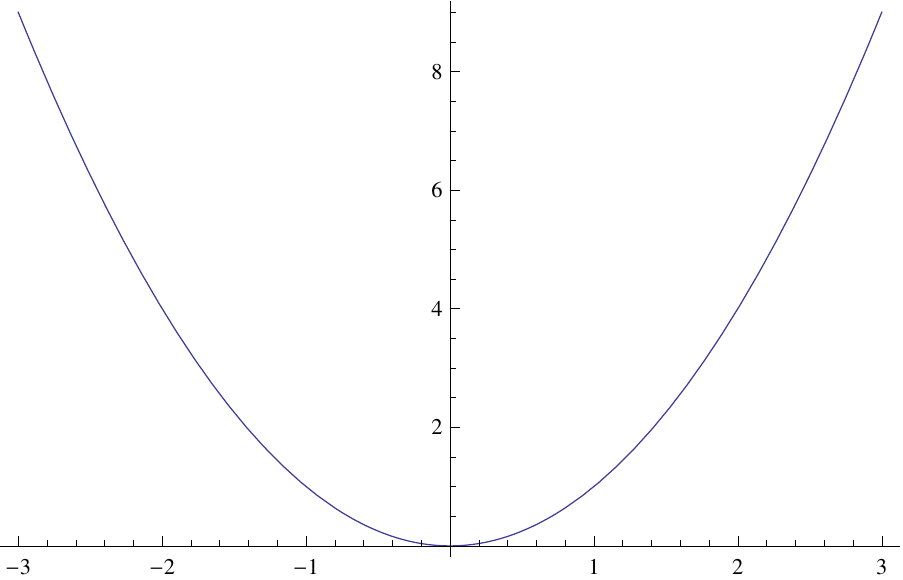}
  }~
    \subfigure[Example 2: Starobinsky-like]{	
   \includegraphics[width=.45\columnwidth]{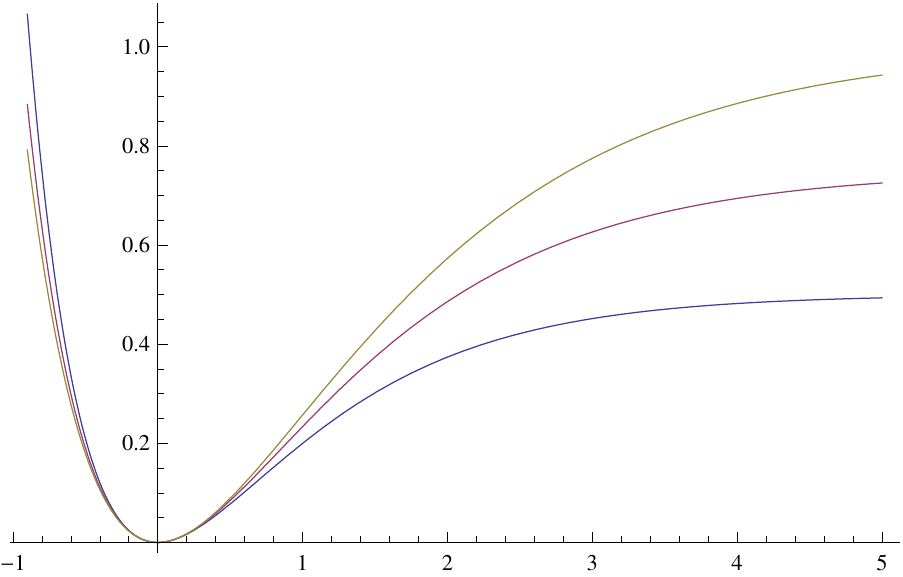}
  } \\
    \subfigure[Example 3: ``symmetry breaking'']{	
   \includegraphics[width=.45\columnwidth]{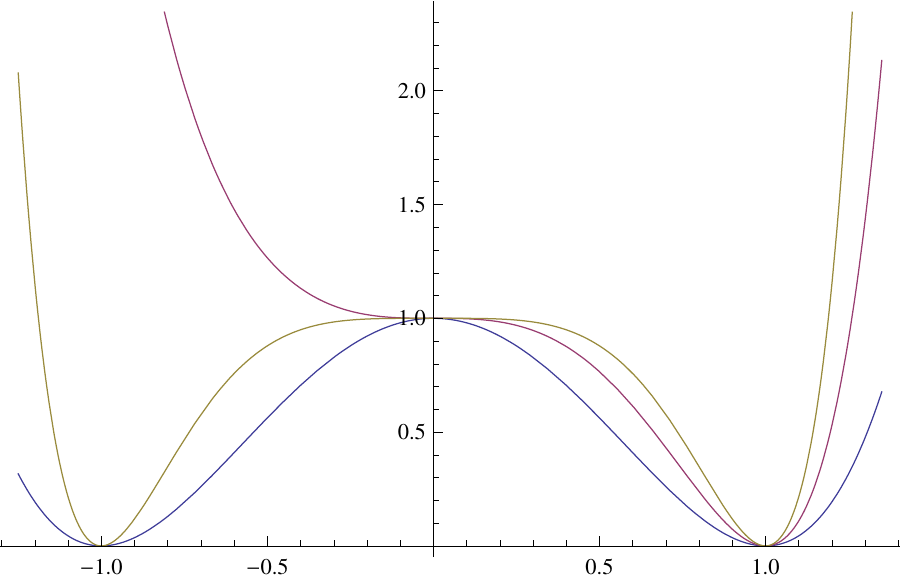}
  }~
    \subfigure[Example 4: sinusoidal]{	
   \includegraphics[width=.45\columnwidth]{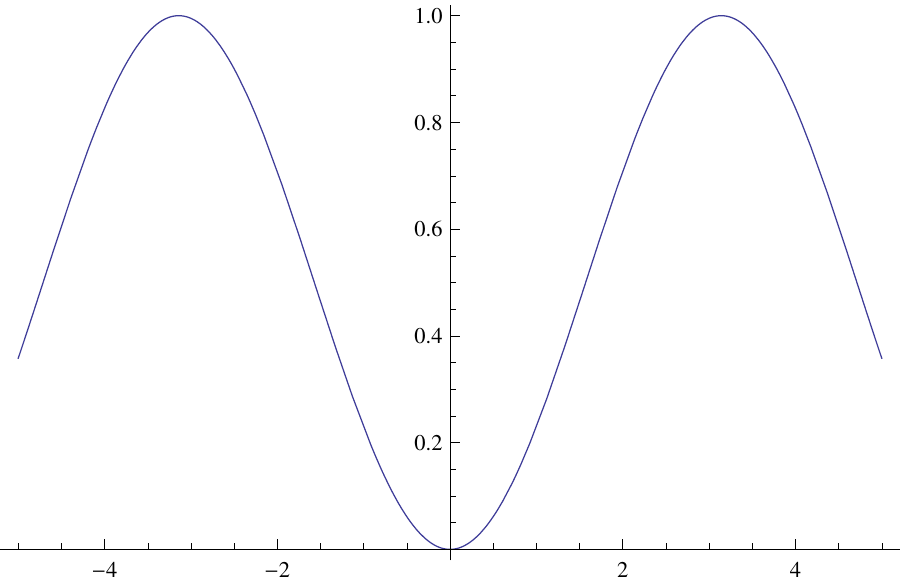}
  }\\
    \subfigure[Example 5: polynomial]{	
   \includegraphics[width=.45\columnwidth]{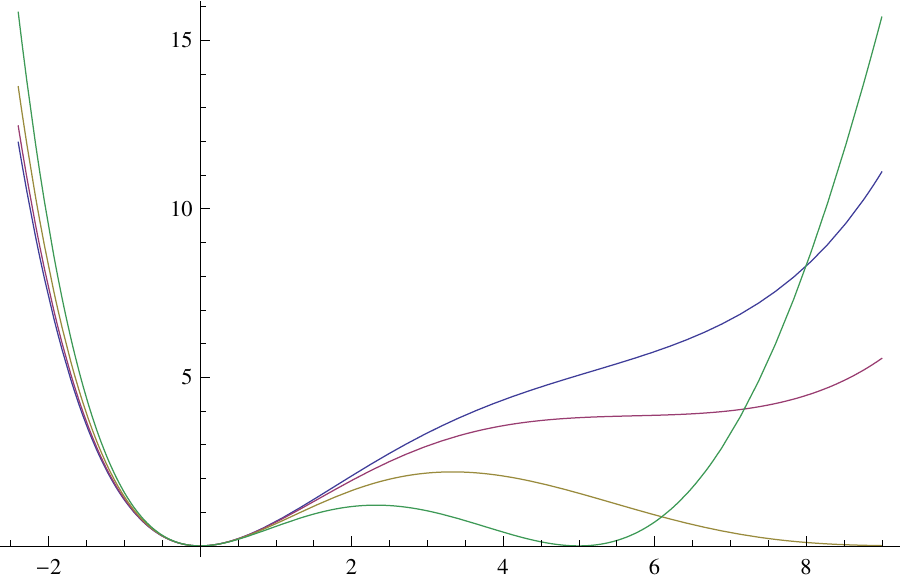}
  }~
    \subfigure[Example 6: T-model]{	
   \includegraphics[width=.45\columnwidth]{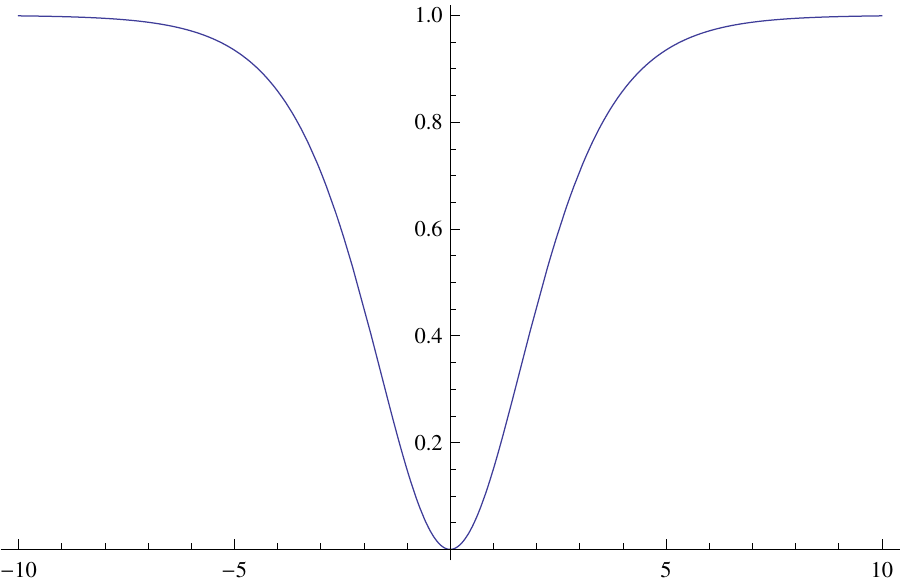}
  }
  \caption{Examples of the scalar potentials.  For (b), $\alpha=2/3, 1,$ and $4/3$ for the blue, red, and yellow line, respectively.  For (c), $n=2, 3$, and $4$ for the blue, red, and yellow line, respectively.  For (e), $b=0.01$ for all the line, and $a=0.16, 0.172, 0.2$, and $0.25$ for the blue, red, yellow, and green line, respectively.} 
  \label{fig:EXpotentials}
 \end{center}
\end{figure}

\paragraph{Example 1: monomial potential}
The monomial scalar potential $V= n^2 |c_n|^2 \chi^{2n-2}$ follows from the superpotential function
\begin{align}
\widetilde{W}(\chi )= c_n \chi^n.
\end{align}
In particular, the quadratic potential, $V=\frac{1}{2}m^2 \chi^2$, follows from the quadratic superpotential, $\widetilde{W}(\chi)=\frac{1}{2\sqrt{2}}m\chi^2$.

\paragraph{Example 2: Starobinsky-like potential}
A set of modified Starobinsky potential,
\begin{align}
V= \frac{3 \alpha}{4}m^2 \left( 1 - e^{-\sqrt{2/3\alpha}\chi} \right)^2 , 
\end{align}
parametrized by $\alpha$, is derived from the following superpotential,
\begin{align}
\widetilde{W}(\chi)=\frac{\sqrt{3\alpha}}{2}m \left( \chi + \sqrt{\frac{3\alpha}{2}}\left( e^{-\sqrt{2/3\alpha}\chi}-1 \right)  \right).
\end{align}
The original Starobinsky potential corresponds to $\alpha=1$.

\paragraph{Example 3: ``symmetry breaking''-type potential}
The potential often used for a new~\cite{Linde1982, Albrecht1982}, chaotic~\cite{Linde1983}, or topological~\cite{Linde1994, Linde1994a, Vilenkin1994} inflationary model (see a review~\cite{Yamaguchi2011}),
\begin{align}
V=\lambda \left( \chi^2 - v^2 \right)^2 ,
\end{align}
follows from a binomial superpotential,
\begin{align}
\widetilde{W}(\chi)=\sqrt{\lambda}\left( \frac{1}{3}\chi^3 - v^2 \chi \right).
\end{align}
Its generalization, $V=\lambda \left( \chi^{n}-v^{n}\right)^2$, was studied in Ref.~\cite{Kallosh2010}.
This potential is generated from the superpotential
\begin{align}
\widetilde{W}(\chi)=\sqrt{\lambda} \left(  \frac{1}{n+1}\chi^{n+1}-v^n \chi \right). \label{W-SB}
\end{align}

\paragraph{Example 4: sinusoidal potential}
The sinusoidal scalar potential, $V=V_0 \left( 1- \cos n\chi \right)/2$, which appears in natural inflation~\cite{Freese1990, Adams1993} follows from the following superpotential,
\begin{align}
\widetilde{W}(\chi)= \frac{2\sqrt{V_0}}{n} \sqrt{1-\cos n \chi}\cot \frac{n\chi}{2}.
\end{align}

\paragraph{Example 5: polynomial chaotic model}
The following polynomial potential, 
\begin{align}
V=c^2 \chi^2 \left( 1- a\chi + b\chi^2 \right)^2 ,
\end{align} studied in Ref.~\cite{Ferrara2013} follows from the quite simple polynomial superpotential,
\begin{align}
\widetilde{W}(\chi)=c \chi^2 \left(\frac{1}{2}-\frac{1}{3}a\chi + \frac{1}{4}b\chi^2 \right).
\end{align}

\paragraph{Example 6: T-model}
A Starobinsky-like potential, $V=V_{0}\tanh ^2 \left( \frac{\chi}{\sqrt{6}}\right)$, looks like a letter `T', and is called T-model~\cite{Kallosh2013}.
It follows from the superpotential,
\begin{align}
\widetilde{W}(\chi)=\sqrt{6V_0}\ln \left( \cosh \frac{\chi}{\sqrt{6}} \right).
\end{align}

These potentials are depicted in Figure~\ref{fig:EXpotentials}.

\section{Effects of the finite stabilization term}\label{sec:stabilization}

In the previous Sections, we studied effective single field models with an implicit assumption of the ideal stabilization.
We study deviation from the ideal case and how large the corrections are.
Below we examine the models considered above one by one taking the effect of the stabilization term into account.
We will see that the previous treatments are approximately true with a sufficiently large coefficient of the stabilization term.

The kinetic term and the scalar potential following from the minimal shift-symmetric K\"{a}hler potential augmented with the quartic term~\eqref{minKstab},
\begin{align*}
	K=\frac{1}{2}\left( \Phi+\bar{\Phi}\right)^2 - \frac{\zeta}{4} \left( \Phi+\bar{\Phi}-2\Phi_0 \right)^4,
	\end{align*} 
 are 
\begin{align}
\mathcal{L}_{\text{kin}}=& -\frac{1}{1-3 \zeta \left( 2 \text{Re}\Phi - 2\Phi_0 \right)^2} \partial_{\mu}\bar{\Phi}\partial^{\mu}\Phi, \\
V=& \frac{e^{2\Phi_0^2-\frac{\zeta}{4} \left( 2 \text{Re}\Phi - 2\Phi_0 \right)^4}}{1-3\zeta \left( 2 \text{Re}\Phi - 2\Phi_0 \right)^2 } \left( |W_{\Phi}|^2 + 2\left( \text{Re}\Phi-\frac{1}{2}\zeta \left( 2 \text{Re}\Phi - 2\Phi_0 \right)^3  \right) \left( W_{\Phi}\bar{W}+\bar{W}_{\bar{\Phi}}W\right)  \right. \nonumber \\
& \qquad  \left.  +4 \left( \text{Re}\Phi-\frac{1}{2}\zeta \left( 2 \text{Re}\Phi - 2\Phi_0 \right)^3  \right)^2 |W|^2 \right) -3e^{2\Phi_0^2-\frac{\zeta}{4} \left( 2 \text{Re}\Phi - 2\Phi_0 \right)^4}|W|^2~.
\end{align}
There are no cancellations in the potential unlike the case of no-scale models, so the small deviation of the real part from $\Phi_0$ merely perturbs the potential, which will be confirmed below.
We solve the VEV of the real part $\phi=\sqrt{2}\text{Re}\Phi$ by Taylor expanding the potential around $\phi=\sqrt{2}\Phi_0$ up to the order $\mathcal{O}(\phi^2)$.
Since it is hard to deal with general superpotentials, we neglect derivatives of the superpotential for dimensional reasons, $\partial W/ \partial \Phi \sim W/ \Phi$ and $\partial^2 W / \partial \Phi^2 \sim W/ \Phi^2$ etc.  This approximation is somewhat crude but is valid for generic polynomial superpotential during large field inflation.  After all, it is generally impossible to solve the field value analytically and exactly.
The field deviation is
\begin{align}
\vev{\text{Re}\Phi}-\Phi_0\simeq - \frac{\Phi_0 \left( 4\Phi_0^2-1 \right) }{24\zeta \Phi_0^2+16\Phi_0^4+8\Phi_0^2-1}\simeq \mc{O}\left(10^{-1}\zeta^{-1}\right) .   \label{dev_minK}
\end{align}
The mass squared of the real (stabilized) part is
\begin{align}
V_{\phi \phi}\simeq& \frac{6\left(16 \Phi_0^4+\left( 24\zeta +8 \right)\Phi_0^2-1  \right)}{4\Phi_0^2-3}H^2,
\end{align}
where $H$ is the Hubble parameter, and we have used the Friedmann equation $3H^2 \simeq V\simeq  e^{2\Phi_0^2}\left(4\Phi_0^2-3 \right) |W|^2$.
This has been evaluated at $\phi=\sqrt{2}\Phi_0$ to avoid the complicated expression: we just want to know the leading order value of the mass to see how strong it is stabilized in contrast to the cases of field value (eq.~\eqref{dev_minK}) and potential (see eq.~\eqref{dev_V_minK} below) in which we want to examine the corrections themselves.
The mass easily becomes larger than the Hubble scale, so the real part is indeed stabilized at the value of eq.~\eqref{dev_minK}.
Then, the kinetic term becomes a constant, and approximately canonically normalized.
Compared to the approximate potential $V_0=e^{2\Phi_0^2}\left(4\Phi_0^2-3 \right) |W|^2$ up to the leading order $\mc{O}(\phi^0)$, the potential up to $\mc{O}(\phi^2)$ evaluated with eq.~\eqref{dev_minK} is modified as
\begin{align}
V=V_0 \left( 1 + \frac{ 32\Phi_0^6-48\Phi_0^4 -6\Phi_0^2+4 }{\zeta \left( 96\Phi_0^4 -72\Phi_0^2  \right)+48\Phi_0^4-24\Phi_0^2-1} \right). \label{dev_V_minK}
\end{align}
The correction term goes to $1$ in the limit $\zeta \rightarrow \infty$.
In this sense, the model of the minimal shift-symmetric K\"{a}hler potential~\eqref{minimalKshift} with $\text{Re}\Phi$ set to $\Phi_0$ is not drastically modified when including the effects of $\text{Re}\Phi \neq \Phi_0$.   Numerically, the second term in the parenthesis is $-0.38, -0.19$, and $-0.068$ for $ \zeta =1, 3$, and 10, setting $\Phi_0=1$.  Thus, the stabilization parameter $\zeta$ of $\mc{O}(10)$ is required to sustain the original prediction of the model.

\begin{figure}
  \begin{center}
    \includegraphics[clip, width=10cm]{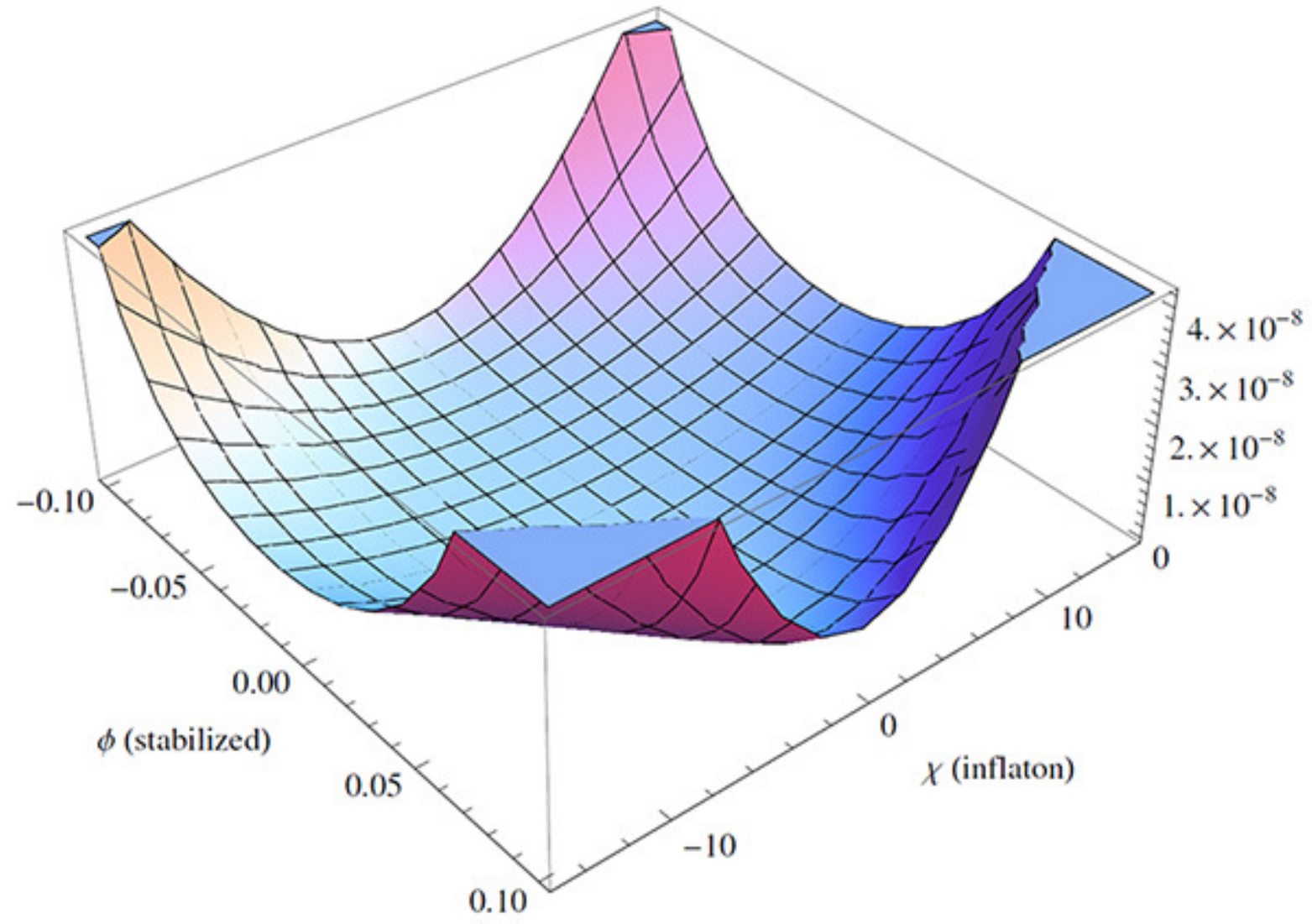}
    \caption{The scalar potential of the stabilized quadratic model. The mass scale and the coefficients of the linear and quartic terms are set to $m=10^{-5}$, $c=2$, and $\zeta=10$, respectively.}
    \label{fig:quadratic_p_simpleK}
  \end{center}
\end{figure}

\begin{figure}
  \begin{center}
    \includegraphics[clip, width=8cm]{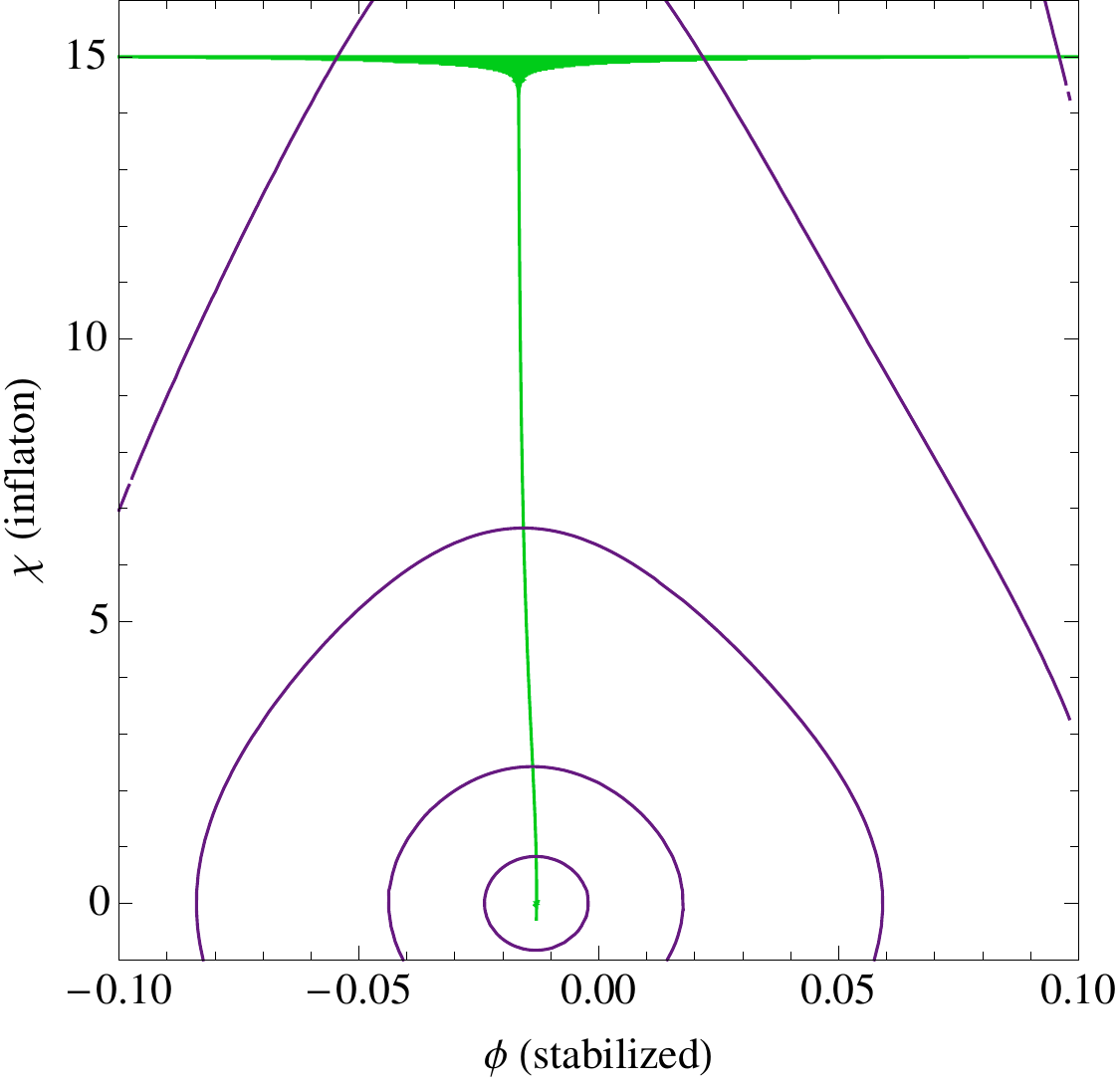}
    \caption{The inflaton trajectory (green) in the stabilized quadratic model. The initial conditions are $\phi=0.1$, $\chi=15$, $\dot{\phi}=0$, and $\dot{\chi}=0$.  The model parameters are same as those in Figure~\ref{fig:quadratic_p_simpleK}.  The contour plot of logarithm of the potential is shown in purple.}
    \label{fig:quadratic_t_simpleK}
  \end{center}
\end{figure}

\begin{figure}
  \begin{center}
    \includegraphics[clip, width=10cm]{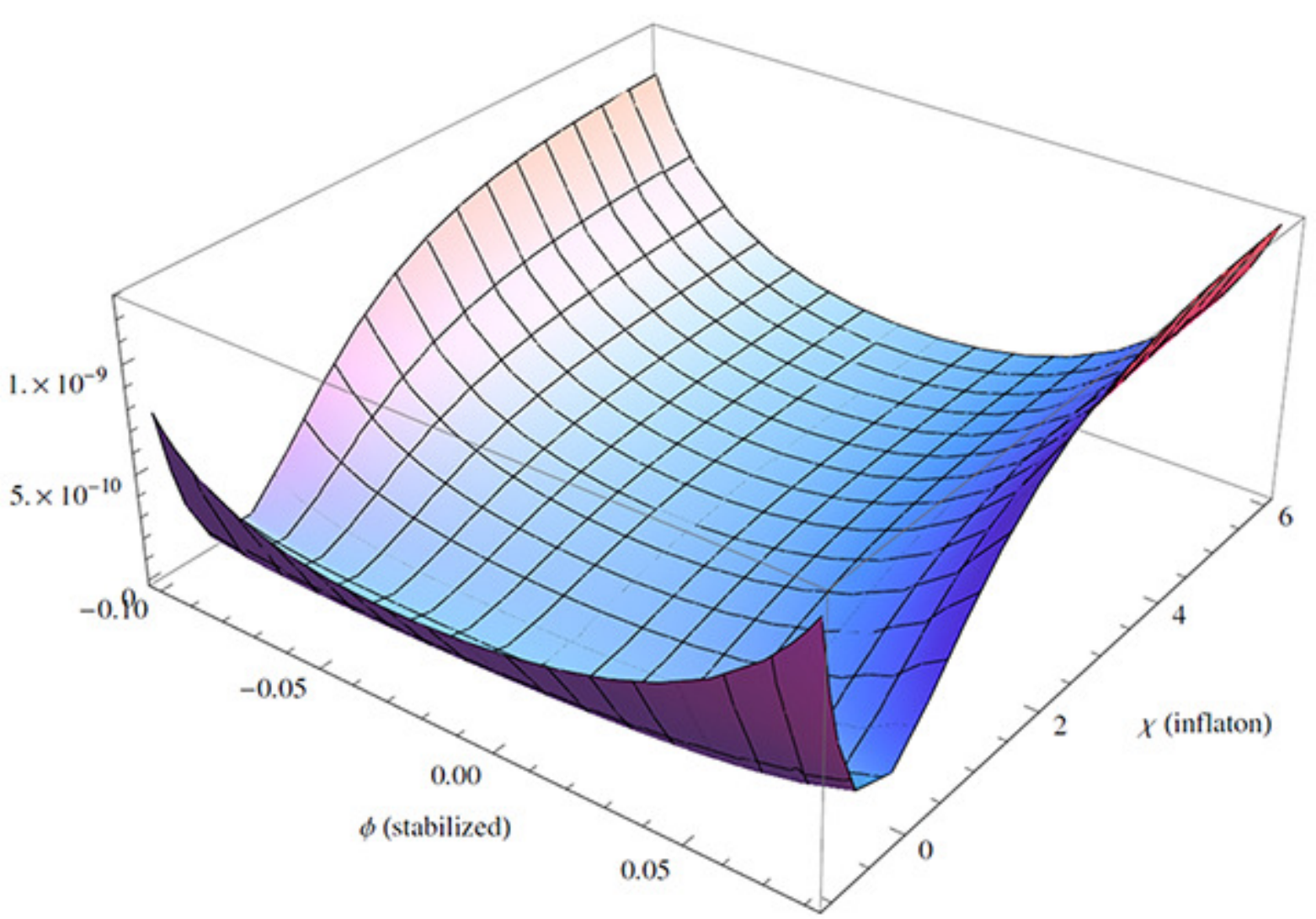}
    \caption{The scalar potential of the stabilized Starobinsky model. The mass scale, the coefficients of the linear and quartic terms, and the parameters in the superpotential are set to $m=10^{-5}$, $c=2$, $\zeta=10$, $a=\sqrt{2/3}$, and $b=1.53414+0.346062 i$, respectively.}
    \label{fig:Starobinsky_p_simpleK}
  \end{center}
\end{figure}

\begin{figure}
  \begin{center}
    \includegraphics[clip, width=8cm]{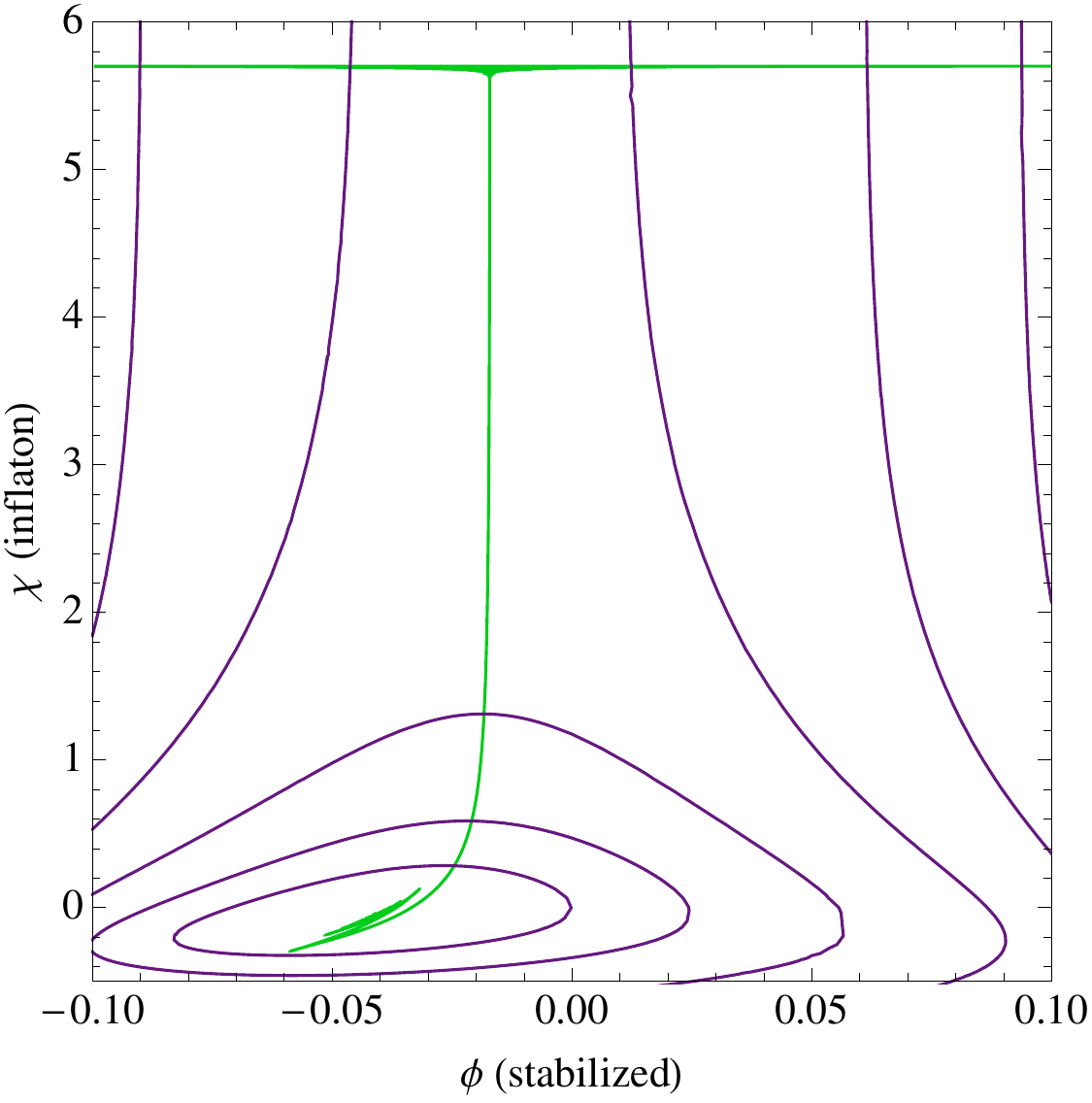}
    \caption{The inflaton trajectory (green) in the stabilized Starobinsky model. The initial conditions are $\phi=0.1$, $\chi=5.7$, $\dot{\phi}=0$, and $\dot{\chi}=0$.  The model parameters are same as those in Figure~\ref{fig:Starobinsky_p_simpleK}.  The contour plot of logarithm of the potential is shown in purple.}
    \label{fig:Starobinsky_t_simpleK}
  \end{center}
\end{figure}

\allowbreak
Let us take some examples of superpotential to discuss it quantitatively without relying on the dimensional estimation like $\partial W/ \partial \Phi \sim W/ \Phi$.
We take the models of quadratic potential~\eqref{WPolonyi} and the Starobinsky potential~\eqref{WStarobinsky} as benchmark models~\footnote{
As can be seen from Figs.~\ref{fig:quadratic_t_simpleK} and \ref{fig:Starobinsky_t_simpleK}, the fields deviate from the origin at the end of inflation.
If we adopt the same parameters for these models as we discussed in Section~\ref{sec:strategy}, the potentials at the vacuum have a small negative value.
To uplift the potential, we added a small constant to the superpotentials.
For the quadratic case, we take
\begin{align}
W=m\left(\Phi - \frac{1}{c-\sqrt{3}-0.01}\right),
\end{align}
where the $0.01$ is inserted to make the cosmological constant a tiny positive value.
Similarly we tune $b$ in eq.~\eqref{WStarobinsky} to obtain a small positive cosmological constant at the vacuum for the Starobinsky case.
In the above argument, ``small'' means that it is smaller by order of magnitudes than the typical inflationary scale, $V\sim m^2 M_{\text{G}}^2$.
}.
The potentials on the complex $\Phi$ plane are shown in Figs.~\ref{fig:quadratic_p_simpleK} and \ref{fig:Starobinsky_p_simpleK} for the quadratic and Starobinsky models, and examples of inflaton trajectories obtained by numerical calculation are shown in Figs.~\ref{fig:quadratic_t_simpleK} and \ref{fig:Starobinsky_t_simpleK} for the quadratic and Starobinsky models, respectively.

Note that the scales of axes are different for $\phi$ (non-inflaton) and $\chi$ (inflaton).
It can be seen from Figs.~\ref{fig:quadratic_p_simpleK} and \ref{fig:Starobinsky_p_simpleK} of the potentials that $\phi$ is stabilized near the origin.
From Figs.~\ref{fig:quadratic_t_simpleK} and \ref{fig:Starobinsky_t_simpleK}, large parts of the trajectories can be seen as those of single field inflation.
As we set the initial conditions away from the instantaneous minimum, $\phi$ rapidly oscillates around the origin at the first stage.
After the rapid oscillation damps, the value of $\phi$ is approximately a constant of order $\mc{O}(10^{-2})$.
Here, the stabilization parameter $\zeta$ is taken as $\zeta=10$, and it is consistent with eq.~\eqref{dev_minK}.
At the final stage when inflation ends, the trajectories get curved, and the fields settle to a point away from the origin.

Let us move on to the no-scale type K\"{a}hler potential~\eqref{K-imaginary-fixed},
\begin{align*}
	K=-3\ln \left( \left(\Phi+\bar{\Phi}\right) +\zeta \left(i\left(\bar{\Phi}-\Phi\right) - 2\Phi_0 \right)^4 \right). 
	\end{align*}
 where the imaginary part is fixed and the real part becomes the inflaton.
The kinetic term and the scalar potential are
\begin{align}
\mc{L}_{\text{kin}}=& - \frac{3\left(1-12\zeta \left( \Phi+\bar{\Phi} \right) \left( -i (\Phi-\bar{\Phi})-2\Phi_0 \right)^2 +4 \zeta^2 \left( -i (\Phi-\bar{\Phi})-2\Phi_0 \right)^6 \right)}{\left( \Phi+\bar{\Phi}+ \zeta \left( -i (\Phi-\bar{\Phi})-2\Phi_0 \right)^4  \right)^2} \partial_{\mu}\bar{\Phi}\partial^{\mu}\Phi, 
\end{align}
\begin{align}
V  =& \frac{1}{3\left[\Phi+\bar{\Phi}+\zeta (-i\Phi+i\bar{\Phi}-2\Phi_0)^4 \right]^2} \nonumber \\
& \times \frac{1}{  1-12\zeta \left( \Phi+\bar{\Phi} \right)(-i\Phi+i\bar{\Phi}-2\Phi_0)^2 +4 \zeta^2  \left(-i \Phi+i\bar{\Phi}-2\Phi_{0} \right)^{6} } \nonumber \\
&\times  \left[ (\Phi+\bar{\Phi}+\zeta (-i\Phi+i\bar{\Phi}-2\Phi_0)^4 ) \left|W_{\Phi}\right|^{2}- 3 \left( \bar{W}W_{\Phi}+W\bar{W}_{\bar{\Phi}}\right)   \right.\nonumber \\
& \left. +12i\zeta (-i\Phi+i\bar{\Phi}-2\Phi_0)^3 \left( W\bar{W}_{\bar{\Phi}}-\bar{W}W_{\Phi} \right)  +108\zeta (-i\Phi+i\bar{\Phi}-2\Phi_0 )^{2}|W|^{2} \right].
\end{align}
The deviation of $\text{Im}\Phi$ from $\Phi_0$ is evaluated as
\begin{align}
\vev{\text{Im}\Phi} -\Phi_0 \simeq & i \frac{-3 \left( \bar{W}W_{\Phi\Phi}-W\bar{W}_{\bar{\Phi}\bar{\Phi}} \right) +2\text{Re}\Phi \left( \bar{W}_{\bar{\Phi}}W_{\Phi\Phi}-W_{\Phi}\bar{W}_{\bar{\Phi}\bar{\Phi}} \right) }{864\zeta |W|^2 -576\zeta \text{Re}\Phi \left( \bar{W}W_{\Phi}+W\bar{W}_{\bar{\Phi}} \right) +384 (\text{Re}\Phi)^2\zeta |W_{\Phi}|^2 } \nonumber \\
\sim& i \mc{O}(10^{-2}\zeta^{-1}(\text{Re}\Phi)^{-2}).
\end{align}
It is somewhat suppressed more than the previous case.
For example, it is also suppressed by the inflaton value, and becomes small during large field inflation.
The mass squared of the imaginary (stabilized) part $\chi=\sqrt{3/2}\text{Im}\Phi/\vev{\text{Re}\Phi}$ (evaluated at $\text{Im}\Phi=\Phi_0$ and the instantaneous expectation value $\text{Re}\Phi$) is 
\begin{align}
V_{\chi\chi}\simeq \zeta \left( 48|W|^2-32\text{Re}\Phi \left( \bar{W}W_{\Phi}+W\bar{W}_{\bar{\Phi}} \right)+\frac{64}{3} \left(\text{Re}\Phi\right)^2|W_{\Phi}|^2  \right).
\end{align}
This is expected to be $\mc{O}(10^2 \zeta (\text{Re}\Phi)^3)$ times larger than $H^2$.
The potential $V_0$ at the leading approximation~\eqref{V-imaginary-fixed} is modified as
\begin{align}
V\simeq V_0 \left(1 + \frac{A^2}{192\zeta BC}  \right)
\end{align}
where we have introduced shorthand notations
\begin{align}
A=&  2\text{Re}\Phi \left( \bar{W}''W' - W''\bar{W}' \right) +3 \left( \bar{W}W''-W\bar{W}'' \right) ,\nonumber \\
B=& 2\text{Re}\Phi |W'|^2 -3 \left( \bar{W}W'+W\bar{W}' \right) ,   \nonumber \\
C=&  8(\text{Re}\Phi)^2 |W'|^2 -6\text{Re}\Phi \left( \bar{W}W'+W\bar{W}'\right)+9|W|^2  . \nonumber
\end{align}
Here, primes denote differentiation with respect to the argument, \textit{e.g.} $W'=\partial W/ \partial \Phi$ and $\bar{W}'=\partial \bar{W}/\partial \bar{\Phi}$ etc.
The correction factor in the potential vanishes in the large $\zeta$ limit.
In addition, the dimensional counting like $W'\sim W/\Phi$ tells us that the correction factor behaves as $\Phi^{-3}$ in the large field region $\Phi \gg 1$.  Thus, the correction is well suppressed during large field inflation.  The actual value depends on parameters of the model, but it is about of order $10^{-4}$ to $10^{-6}$ using values like $1 \lesssim \zeta \lesssim 10$ and $3\lesssim \text{Re}\Phi \lesssim 10$.

Next, we consider the K\"{a}hler potential~\eqref{K-a-no-scale},
\begin{align*}
K=-a\ln \left( 1+\frac{1}{\sqrt{a}}\left( \Phi+\bar{\Phi} \right) + \frac{\zeta}{a^2} \left( \Phi+\bar{\Phi} \right)^4 \right) . 
\end{align*}
The constant $a$ is inserted in the logarithm (just as a convention).
The first $a$ in the logarithm is for approximate canonical normalization, and the second is due to the expectation that higher terms appear with the same expansion coefficient.
The kinetic and potential terms are
\begin{align}
\mc{L}_{\text{kin}}=& - \frac{1-\frac{12\zeta}{a}\left( \Phi+\bar{\Phi} \right)^2 -\frac{4\zeta}{a\sqrt{a}}\left( \Phi+\bar{\Phi} \right)^3 +\frac{4\zeta^2}{a^3}\left( \Phi+\bar{\Phi} \right)^6  }{\left( 1+\frac{1}{\sqrt{a}}\left( \Phi+\bar{\Phi} \right)+\frac{\zeta}{a^2}\left( \Phi+\bar{\Phi} \right)^4  \right)^2} \partial_{\mu} \bar{\Phi}\partial^{\mu}\Phi , \\
V=& \frac{1}{AB^{a}}\left( B^2 |W_{\Phi}|^2 -B\left( \sqrt{a}+\frac{4\zeta}{a}\left(\Phi+\bar{\Phi}\right)^3  \right) \left(\bar{W}W_{\Phi}+W\bar{W}_{\bar{\Phi}}\right) +C|W|^2 \right) , \label{full_V_a}
\end{align}
with
\begin{align}
A=&1-\frac{12\zeta}{a}\left(\Phi+\bar{\Phi}\right)^2-\frac{4\zeta}{a\sqrt{a}}\left(\Phi+\bar{\Phi}\right)^3 +\frac{4\zeta^2}{a^3}\left(\Phi+\bar{\Phi}\right)^6 , \\
B=& 1+\frac{1}{\sqrt{a}}\left( \Phi+\bar{\Phi} \right) + \frac{\zeta}{a^2} \left( \Phi+\bar{\Phi} \right)^4 , \\
C=& (a-3)+\frac{36\zeta}{a}\left(\Phi+\bar{\Phi}\right)^2 +\left(\frac{8}{\sqrt{a}}+\frac{12}{a\sqrt{a}}\right)\zeta \left(\Phi+\bar{\Phi}\right)^3+\left( \frac{16}{a^2}-\frac{12}{a^3}\right)\zeta^2 \left(\Phi+\bar{\Phi}\right)^6 .
\end{align}
In the generic case of $a>3$, we neglect the derivatives of superpotential by the dimensional argument.
The shift of the real part from its origin is evaluated as
\begin{align}
\phi\simeq \frac{\sqrt{a}(a-3)}{\sqrt{2} \left( 24\zeta + a^2 -2a -3 \right)}.
\end{align}
It is suppressed by the stabilization parameter $\zeta$.
The mass squared evaluated at the origin is
\begin{align}
V''\simeq \frac{6\left( 24 \zeta +a^2 -2a -3 \right)}{a-3}H^2,
\end{align}
where we have used $3H^2=V\simeq (a-3)|W|^2 $.
It is easy to increase the mass so that the real part becomes heavier than the Hubble scale.
The correction to the potential is as follows,
\begin{align}
V\simeq V_0 \left( 1 - \frac{a(a-3)}{2\left( 24\zeta + a^2 -2a -3 \right)}  \right),
\end{align}
where $V_0=(a-3)|W|^2$ is the leading order potential.
Setting $a=4$, the numerical value of the second term in the parenthesis is $0.069, 0.026$, and $0.0082$ for $\zeta= 1, 3$, and 10.

In the case of $a=3$ and a generic superpotential (for which we use na\"{i}ve dimensional analysis like $\partial W/\partial \Phi \sim W/ \Phi$), the expectation value of $\phi$ is
\begin{align}
\phi \simeq \frac{1}{12\sqrt{2}\zeta} \left( \frac{W'}{W} +\frac{\bar{W}'}{\bar{W}} \right).
\end{align}
This is suppressed by the parameter $\zeta$ and the linear power of the field $\Phi$, but a care should be taken because even a small value destroys the cancellation working in the no-scale type model.  That is, $\phi=0$ is a special point in the moduli space.
Before looking at the corrections to the potential, let us check the stabilization strength, \textit{i.e.} mass of the real part.
It is given by
\begin{align}
V_{\phi\phi} \simeq 48 \zeta |W|^2.
\end{align}
It can be larger than the Hubble scale.
The corrected potential is
\begin{align}
V\simeq -\sqrt{3}\left( \bar{W}W_{\Phi}+W\bar{W}_{\bar{\Phi}} \right) + |W_{\Phi}|^2 + \frac{\left( \bar{W}W_{\Phi}+W\bar{W}_{\bar{\Phi}} \right)^2}{4\zeta |W|^2}.
\end{align}
The second term, which is the subleading term in the uncorrected potential, is the same order (up to the $\zeta$ factor) to the third term, which is the leading correction term.
Once again, the relative value of the correction becomes small in the limit of the large $\zeta$ and the large inflaton field.

We are also interested in the special case discussed in the previous Section: the reality condition on the coefficients of the superpotential.
When we assume it, various expressions are simplified, and we do not need to use the approximation like $\partial W/\partial \Phi \sim W/\Phi$.
The expectation value of $\phi$ is derived as
\begin{align}
\phi = & \frac{\sqrt{a/2}\left( a(a-3)\widetilde{W}^2 +4(a-1)\widetilde{W}'^2 -2a \widetilde{W}\widetilde{W}'' \right)}{  a(24\zeta +a^2 -2a-3)\widetilde{W}^2+(48\zeta +7a^2-13a+4)\widetilde{W}'^2-a(5a-7)\widetilde{W}\widetilde{W}''+2a\widetilde{W}''^2-2a\widetilde{W}'\widetilde{W}''' } \nonumber \\
\simeq & \frac{\sqrt{2a}\left( a(a-3)E+4(a-1)-a\sqrt{2 \epsilon E} \right)}{  2a(24\zeta +a^2 -2a-3)E+2(48\zeta +7a^2-13a+4)-a(5a-7)\sqrt{2\epsilon E} }  \label{dev_specialK}
\end{align}
where  the slow-roll parameters are given by
\begin{align}
\epsilon=& \frac{1}{2}\left( \frac{V'}{V} \right)= 2\left( \frac{\widetilde{W}''}{\widetilde{W}'}\right)^2 , & \eta=& \frac{V''}{V}= 2\left( \left( \frac{\widetilde{W}''}{\widetilde{W}'}\right)^2+\frac{\widetilde{W}'''}{\widetilde{W}'} \right),
\end{align}
so that $(\widetilde{W}''/\widetilde{W}')=\sqrt{\epsilon/2}$ and $(\widetilde{W}'''/\widetilde{W}')=(\eta-\epsilon)/2$.
We have introduced an enhancement factor
\begin{align}
E(\chi )\equiv \left( \frac{\widetilde{W}}{\widetilde{W}'} \right)^2. \label{Efactor}
\end{align}
In the case of a monomial potential, $E=\left( \frac{\chi}{n} \right)^2$ and it is large $(E>1)$ during large field inflation $(\chi>1)$ for $n$ not too large.  In the case of the Starobinsky potential, $E=\left( \frac{\chi}{1-e^{-\sqrt{2/3}\chi}}-\frac{\sqrt{3}}{2} \right)^2$ and it is also large during large field inflation.
Typically, $E$ is of the order $\chi^2$, so the combination $\sqrt{\epsilon E}$ is roughly of the order one.
The first equality in eq.~\eqref{dev_specialK} holds both during and after inflation, whereas the second equality is valid during inflation since slow roll parameters not accompanied by the enhancement factor $E$ have been neglected.
The mass squared is
\fontsize{11pt}{20pt}
\begin{align}
 V_{\phi\phi}\simeq & \left(24\zeta +a^2-2a-3\right)\widetilde{W}^2+\frac{1}{a}\left( 48\zeta +7a^2-13a+4\right)\widetilde{W}'^2 -(5a-7)\widetilde{W}\widetilde{W}''+2\widetilde{W}''^2-2\widetilde{W}'\widetilde{W}''' \nonumber \\
\simeq & 3H^2 \left( \left(24\zeta +a^2-2a-3\right)E+\frac{1}{a}\left( 48\zeta +7a^2-13a+4\right) -(5a-7)\sqrt{\epsilon E/2} \right), 
\end{align}
\large
where the same approximation has been used in the second equality as above.
The mass easily becomes larger than the Hubble scale.
The corrected potential is of the form
\fontsize{11pt}{20pt}
\begin{align}
V=& \widetilde{W}'^2+\frac{a-3}{2}\widetilde{W}^2 \nonumber \\
 - & \frac{\left( a(a-3)\widetilde{W}^2+4(a-1)\widetilde{W}'^2-2a\widetilde{W}\widetilde{W}'' \right)^2 /4}{ a(24\zeta+a^2-2a-3)\widetilde{W}^2+(48\zeta+7a^2-13a+4)\widetilde{W}'^2-a(5a-7)\widetilde{W}\widetilde{W}''+2a\widetilde{W}''^2-2a\widetilde{W}'\widetilde{W}'''  }.
\end{align}
\large
The correction (second line) is only suppressed by a numerical factor including $\zeta$ for $a>3$.
When we set $a=3$, the field expectation value, the mass squared, and the corrected potential are
\begin{align}
\phi =&  \frac{\sqrt{6}\left( 4\widetilde{W}'^2-3 \widetilde{W}\widetilde{W}'' \right)}{  72\zeta \widetilde{W}^2+4(12\zeta +7)\widetilde{W}'^2-24\widetilde{W}\widetilde{W}''+6\widetilde{W}''^2-6\widetilde{W}'\widetilde{W}''' } \nonumber \\
\simeq &  \frac{\sqrt{6}\left( 4-3 \sqrt{\epsilon E/2} \right)}{  72\zeta E+4(12\zeta +7)-24\sqrt{\epsilon E/2}}, \label{devphi_arbitrary}\\
V_{\phi\phi}\simeq & 24\zeta \widetilde{W}^2+\frac{4}{3}\left( 12\zeta +7\right)\widetilde{W}'^2 -8\widetilde{W}\widetilde{W}''+2\widetilde{W}''^2-2\widetilde{W}'\widetilde{W}''' \nonumber \\
\simeq & H^2 \left( 72\zeta E+4\left( 12\zeta +7\right) -24\sqrt{\epsilon E/2}\right) , \label{stabmass_arbitrary} \\
V=& \widetilde{W}'{}^2 \left( 1 -\frac{\left( 4\widetilde{W}'{}^2-3\widetilde{W}\widetilde{W}'' \right)^2}{2\widetilde{W}'{}^2 \left( 36\zeta \widetilde{W}^2+2\left( 12\zeta+7\right)\widetilde{W}'{}^2-12\widetilde{W}\widetilde{W}''+3\widetilde{W}''{}^2-3\widetilde{W}'\widetilde{W}''' \right)}  \right) \nonumber \\
\simeq & \widetilde{W}'{}^2 \left( 1 -\frac{\left( 4-3\sqrt{\epsilon E/2} \right)^2}{2 \left( 36\zeta E+2\left( 12\zeta+7\right)-12\sqrt{\epsilon E/2} \right)}  \right). \label{Vcor_arbitrary}
\end{align}
In the second lines of these expressions, we have neglected slow-roll parameters unless it appears with the enhancement factor $E$ (eq.~\eqref{Efactor}), so the second lines are valid only during (large field) inflation.
The value of the real field is well suppressed, its mass can be easily larger than the Hubble scale, and the correction to the potential is subdominant.  Numerically, the second term in the parenthesis of the scalar potential is $-0.046, -0.0081$, and -0.00046 for $(E, \epsilon, \zeta)=(3, 0.01, 1), (5, 0.03, 3)$, and (10, 0.1, 10).

\begin{figure}
  \begin{center}
    \includegraphics[clip, width=10cm]{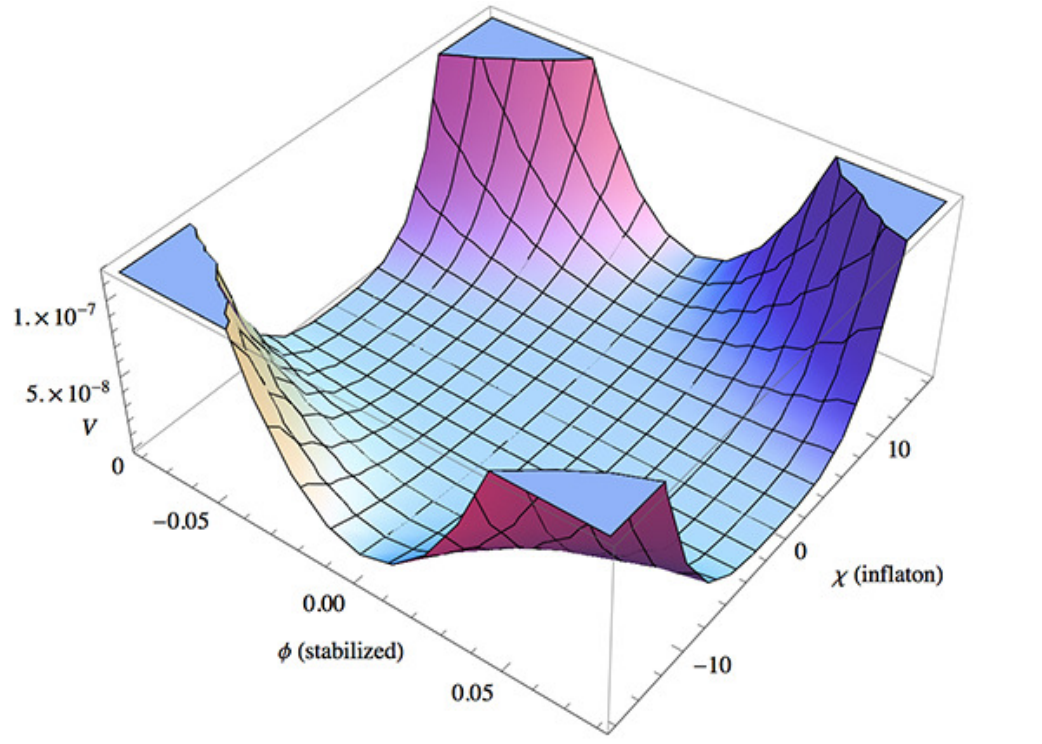}
    \caption{The scalar potential of the stabilized quadratic model. The mass scale and the stabilization strength are set to $m=10^{-5}$ and $\zeta=3\sqrt{3}$, respectively.  This Figure is from our paper~\cite{Ketov2014a}.}
    \label{fig:quadratic_potential}
  \end{center}
\end{figure}

\begin{figure}
  \begin{center}
    \includegraphics[clip, width=8cm]{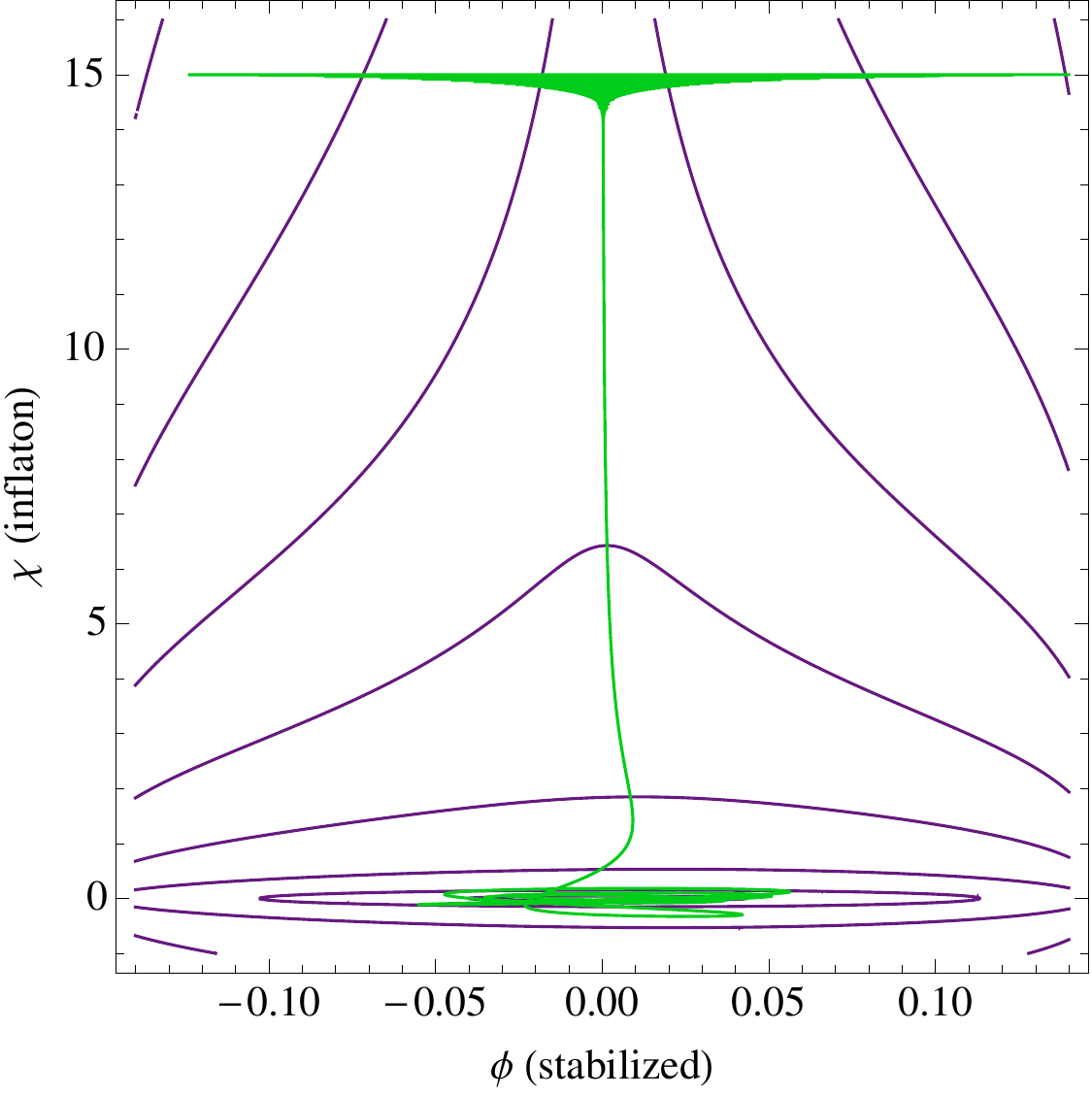}
    \caption{The inflaton trajectory (green) in the stabilized quadratic model. The initial conditions are $\phi=0.14$, $\chi=15$, $\dot{\phi}=0$, and $\dot{\chi}=0$.  The mass scale and the stabilization strength are set to $m=10^{-5}$ and $\zeta=3\sqrt{3}$,
    respectively. The contour plot of logarithm of the potential is shown in purple.   This Figure is from our paper~\cite{Ketov2014a}.}
    \label{fig:quadratic_trajectory}
  \end{center}
\end{figure}

\begin{figure}
  \begin{center}
    \includegraphics[clip, width=10cm]{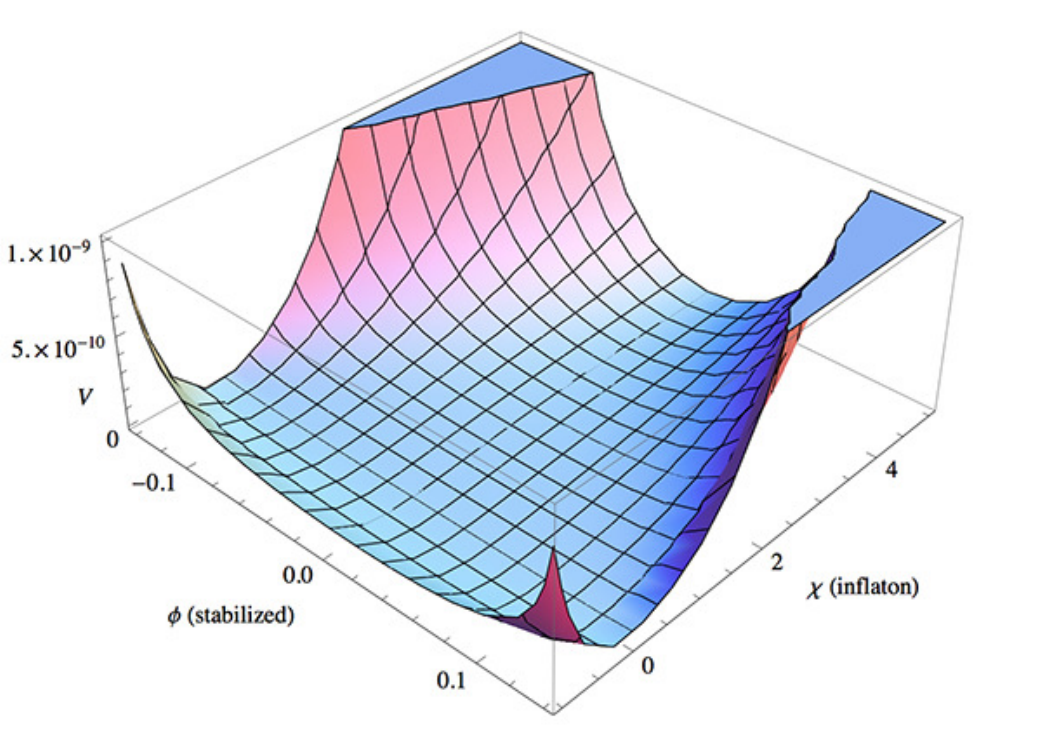}
    \caption{The scalar potential of the stabilized Starobinsky model. The mass scale and stabilization strength are set to $m=10^{-5}$ and $\zeta=3\sqrt{3}$, respectively.   This Figure is from our paper~\cite{Ketov2014a}.}
    \label{fig:Starobinsky_potential}
  \end{center}
\end{figure}

\begin{figure}
  \begin{center}
    \includegraphics[clip, width=8cm]{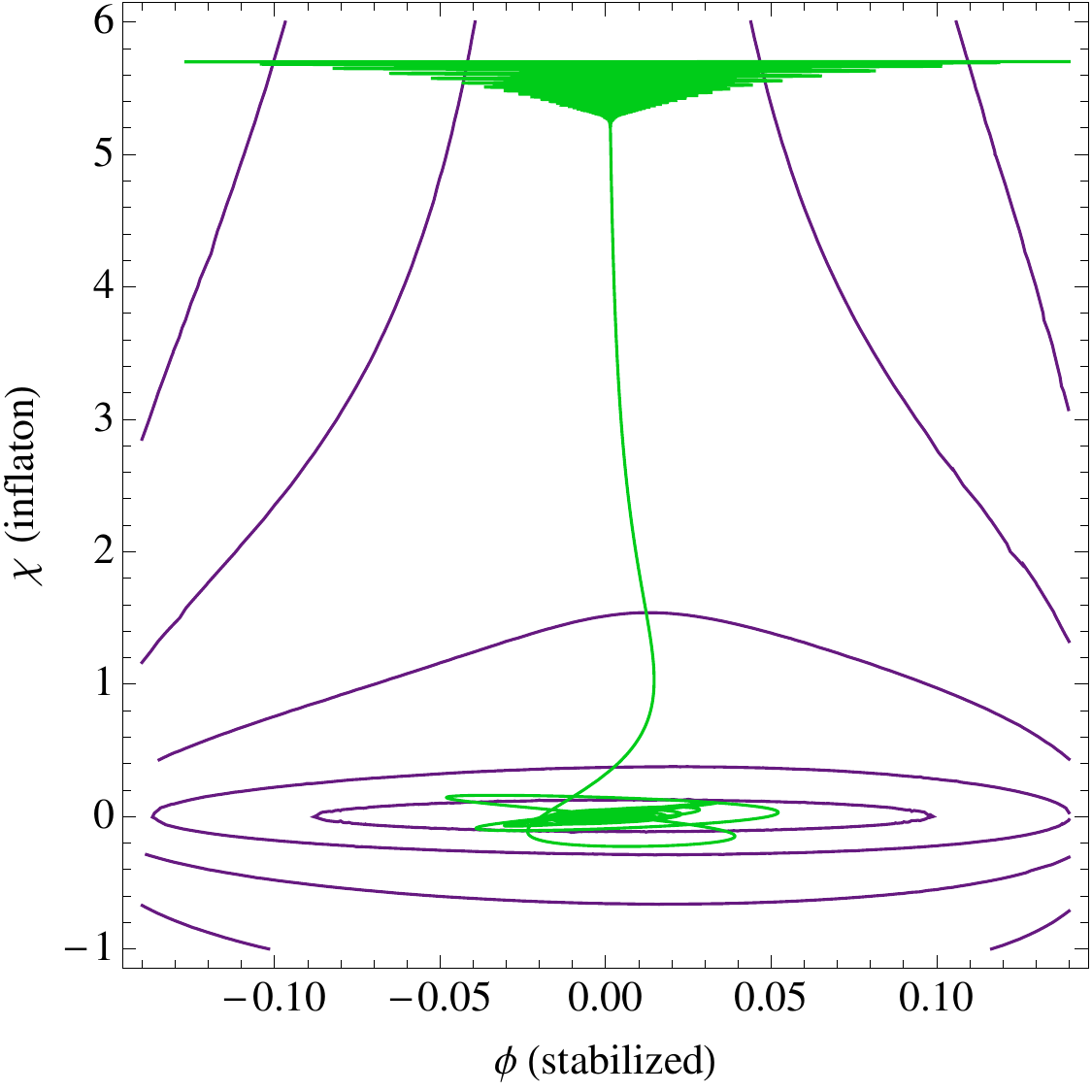}
    \caption{The inflaton trajectory (green) in the stabilized Starobinsky model. The initial conditions are $\phi=0.14$, $\chi=5.7$, $\dot{\phi}=0$, and $\dot{\chi}=0$.  The mass scale and the stabilization strength are set to $m=10^{-5}$ and $\zeta=3\sqrt{3}$, respectively. The contour plot of logarithm of the potential is shown in purple.   This Figure is from our paper~\cite{Ketov2014a}.}
    \label{fig:Starobinsky_trajectory}
  \end{center}
\end{figure}

In the following, we examine stabilization quality numerically.
Take the special K\"{a}hler potential~\eqref{K-arbitrary-stabilized} that can lead to arbitrary positive semidefinite scalar potentials.  We take two benchmarks as inflationary models: chaotic inflation with the quadratic potential and chaotic inflation with the Starobinsky potential.
The potentials in the complex plane are shown in Figs.~\ref{fig:quadratic_potential} and \ref{fig:Starobinsky_potential}.
The scales of the axes are different for the illustration purpose.
As we discussed it above, the stabilization becomes stronger when the inflaton takes a larger value.
Examples of the inflaton trajectory are shown in Figs.~\ref{fig:quadratic_trajectory} and \ref{fig:Starobinsky_trajectory} for the quadratic and Starobinsky potential cases, respectively.
The green trajectories are superimposed on purple logarithmic contours of the potentials.
We take initial conditions such that the inflaton is displaced from the instantaneous minimum (stabilization valley) with vanishing velocity.
As can be seen from the Figures, the inflaton rapidly oscillates and is damped to the bottom of the stabilization valley.
It seems that the inflaton is damped more efficiently in the quadratic potential (Fig.~\ref{fig:quadratic_trajectory}) than in the Starobinsky potential (Fig.~\ref{fig:Starobinsky_trajectory}).
This is because the inflaton takes  larger values in the former model  than in the latter model, and the large inflaton value, implying large enhancement factor $E$ in eq.~\eqref{Efactor}, suppress the value of the real field via the enhanced mass term~\eqref{stabmass_arbitrary}.
Afterward, the trajectories are essentially that of single field inflationary models.
Interestingly, the trajectories slightly deviate from the imaginary axis at the end of inflation, leading to final oscillation phase where the decay of inflaton reheats the universe.
The fractional difference between the usual potential (ideally stabilized case) and the actual potential value in our setup along the inflaton trajectory is within only 1.4\% for the quadratic case, and within 2.2\% for the Starobinsky case.  These values are taken near the end of inflation corresponding to the maximum of the real part $\phi$ before the final oscillation.  In most of the time, the fractional differences are much suppressed.

\subsection{Robustness of SUSY preservation} \label{subsec:robustSUSY}

The models of the no-scale-like ($a=3$) K\"{a}hler potential have two favorable features as we discussed in Section~\ref{sec:arbitrary}.
We can approximately embed arbitrary positive semidefinite scalar potentials into supergravity using the K\"{a}hler potential, and we can also tune the constant term in the superpotential to preserve SUSY after inflation.
These features are clear when the stabilization is ideal \textit{i.e.} in the $\zeta\rightarrow \infty$ limit.
In the above, we have analyzed how the former statement is modified quantitatively (eqs.~\eqref{devphi_arbitrary}, \eqref{stabmass_arbitrary}, and \eqref{Vcor_arbitrary}).
Now, let us examine whether the latter feature is preserved or not considering the effect of the finite stabilization term.
In terms of the non-inflaton $\phi$ and the inflaton $\chi$, the potential~\eqref{full_V_a} is written as
\begin{align}
V=& A^{-1}B^{-1} \left[  \left| \widetilde{W}'(-i\phi + \chi) \right|^2  + 12\zeta \phi^2 \left( 1+\sqrt{\frac{2}{3}}\phi+\frac{4}{9}\zeta \phi^4 \right) B^{-2} \left| \widetilde{W}(-i\phi +\chi) \right|^2\right. \nonumber \\
& \left. \qquad  - \sqrt{\frac{3}{2}}\left(1+\frac{8\sqrt{6}}{9}\zeta \phi^3 \right) B^{-1} i \left( \widetilde{W}'(i\phi+\chi)\widetilde{W}(-i\phi+\chi)-\widetilde{W}'(-i\phi+\chi)\widetilde{W}(i\phi+\chi) \right)  \right] ,
\end{align}
where 
\begin{align}
A=& 1-8 \zeta \phi^2 -\frac{8\sqrt{6}}{9}\zeta\phi^3+\frac{32}{27}\zeta^2\phi^6, \\
B=&1+\sqrt{\frac{2}{3}}\phi+\frac{4}{9}\zeta \phi^4.
\end{align}

The first derivatives of the potential have long expressions.
Those evaluated at $\phi=0$ as an ansatz are
\begin{align}
V_\phi |_{\phi=0}=& \sqrt{\frac{2}{3}} \left( -4 \widetilde{W}'(\chi)^2 +3 \widetilde{W}(\chi)\widetilde{W}''(\chi) \right) ,  \\
V_\chi |_{\phi=0}=& 2 \widetilde{W}'(\chi)\widetilde{W}''(\chi).
\end{align}
Note that these equations are independent of $\zeta$ since we substituted $\phi=0$.
Now we search a solution of $V_{\phi}|_{\phi=0}=V_{\chi}|_{\phi=0}=0$ for $\chi$.
If it exists, say at $\chi=\chi_0$, the point $(\phi=0, \, \chi=\chi_0 )$ is a stationary point.

Remember that the potential becomes $V(\chi)= \widetilde{W}'(\chi)^2$ in the ideal ($\zeta \rightarrow \infty$) case.
Suppose we construct an inflationary model such that this potential $V(\chi)$ vanishes at the vacuum $\chi=\chi_0$ where we tune the constant term in the superpotential such that SUSY is preserved, \textit{i.e.}
\begin{align}
\widetilde{W}'(\chi_0)=&0, & &\text{and}  & \widetilde{W}(\chi_0)=&0.
\end{align}
Note that this is automatically a solution of $V_{\phi}|_{\phi=0}=V_{\chi}|_{\phi=0}=0$ with an arbitrary $\zeta$.
Therefore, the point $(\phi=0, \, \chi=\chi_0 )$ is a stationary point.
At the point, the potential coincides with the ideal one, $V=\widetilde{W}'(\chi_0)^2$.

Is the stationary point a maximum, minimum, or saddle point?
In the inflaton ($\chi$) direction, it is a minimum if the inflationary model is meaningful.
Since SUSY is preserved ($V=\widetilde{W}=\widetilde{W}'=0$) at the point, the masses of $\phi$ and $\chi$ are same.
It follows that the stationary point is actually a minimum (vacuum) of the potential.
Thus, the consistency of the ansatz $\phi=0$ has been shown.
In conclusion, the cosmological constant and SUSY breaking can be both eliminated at the vacuum by tuning with an arbitrary stabilization parameter $\zeta$.

\subsection{Comments on small field inflation} \label{subsec:small}

We are primarily interested in large field inflation since our method enables us to realize it without the stabilizer superfield for the first time.
It is also interesting to study whether our method is also applicable to small field inflation or not.
In the above, we have seen that stabilization is strong enough at the large field region but it becomes loose near the end of inflation where inflaton has a smaller value.
One might have an impression that it is not suitable for small field inflation.
However, the essence of the stabilization mechanism is that the effective mass for the scalar superpartner of the inflaton is given by the inflaton $F$-term, which drives inflation.
Therefore, it is generally expected that the stabilization works during inflation.
If the SUSY is broken also at the vacuum, the stabilization still works.
Quantitative details depends on the value of the coefficient $\zeta$.

\begin{figure}[htbp]
  \begin{center}
    \includegraphics[clip, width=10cm]{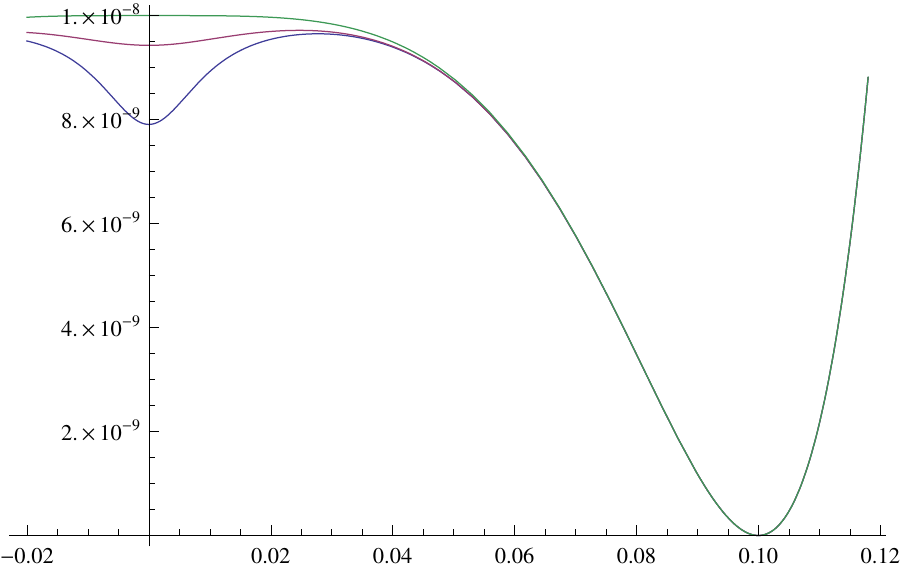}
    \caption{The corrected potential $V(\chi)$~\eqref{Vcor_arbitrary} for the small field model~\eqref{W-SB} with $n=4$.  The blue (bottom), red (middle), and green (top) lines correspond to $\zeta=1, 3\sqrt{3}$, and $\zeta \rightarrow \infty$.  Other parameters are set to $\lambda=1$ and $v=0.1$.}
    \label{fig:VSmallField}
  \end{center}
\end{figure}

We take the model of \textbf{Example 3} in Subsection~\ref{subsec:examples} (the K\"{a}hler potential~\eqref{K-arbitrary-stabilized} and the superpotential~\eqref{W-SB}) as an explicit example,
\begin{align}
K=&-3 \ln \left( 1+\frac{1}{\sqrt{3}} \left( \Phi+\bar{\Phi}\right) + \frac{\zeta}{9}\left( \Phi+\bar{\Phi}\right)^4 \right) , \\
\widetilde{W}=& \sqrt{\lambda} \left(  \frac{1}{n+1}\chi^{n+1} - v^n \chi  + c\frac{n}{n+1}v^{n+1} \right),
\end{align}
where we have added a constant term to the superpotential, and $c$ is the coefficient.
Since we are interested in small field inflation, we take $v\ll 1$.
When we take $c$=1, SUSY is preserved at the vacuum at $\chi=v$.
The relevant derivatives are
\begin{align}
\widetilde{W}'=& \sqrt{\lambda} \left( \chi^n - v^n \right), &
\widetilde{W}''=& n \sqrt{\lambda} \chi^{n-1} , &
\widetilde{W}'''=& n(n-1) \sqrt{\lambda} \chi^{n-2}.
\end{align}
Inflation begins with a small inflaton value $\chi \ll v$, so let us take the limit $\chi \rightarrow 0$.
Then, the constant terms in the superpotential and its first derivative remains, $\widetilde{W}(0)=c\sqrt{\lambda}(n/n+1)v^{n+1}$ and $\widetilde{W}'(0)=-\sqrt{\lambda}v^n$, while the second and third derivatives vanish, $\widetilde{W}''(0)=\widetilde{W}'''(0)=0$.
If we take the coefficient $c$ as an order one parameter, the superpotential is smaller than its first derivative because of the higher power of $v$.
So, we also neglect $\widetilde{W}(0)$ to obtain simple formulae.
The eqs.~\eqref{devphi_arbitrary}, \eqref{stabmass_arbitrary}, and \eqref{Vcor_arbitrary} become
\begin{align}
\phi|_{\chi=0}=&\frac{\sqrt{6}}{12\zeta + 7}, &
V_{\phi \phi}|_{\chi=0}=& 4( 12\zeta + 7) H^2 , & 
V(0)=& \widetilde{W}'(0)^2 \left( 1 - \frac{4}{12\zeta +7} \right).
\end{align}
As we have argued above, the mass of the non-inflaton $\phi$ is larger than the Hubble scale.
So, it is a good approximation to describe the model as a single field inflationary model.
The value of $\phi$ is also less than one, so the perturbation in $\phi$ used to obtain eqs.~\eqref{devphi_arbitrary}, \eqref{stabmass_arbitrary}, and \eqref{Vcor_arbitrary} is consistent.
However, the correction to the effective single field potential $V$ is not so small for a small value of $\zeta$, \textit{e.g.}~$\frac{4}{12\zeta+7} \simeq 0.2$ for $\zeta=1$.
This level of change of the potential can not be neglected especially for small field inflation.
The first and second derivatives of the potential are affected and inflationary observables change accordingly.
But the effect decreases when we increase the strength of stabilization $\zeta$.
We show the corrected potential in Fig.~\ref{fig:VSmallField}.

Although we have examined only one example, we have reasons to think other models share the same qualitative features.
During inflation, regardless of small field or large field, the slow-roll parameters should be small, so $\widetilde{W}''$ and $\widetilde{W}'''$ can be neglected.
The value of the superpotential depends on the functional form including the constant term.
Unless it is larger than its first derivative ($E>1$), its effect is subdominant (see eq.~\eqref{Vcor_arbitrary}).
Thus, our framework is also applicable for small field inflationary models although a larger stabilization term is required.

\subsection{Comments on the unitarity bound} \label{subsec:unitarity}

Our method utilizes the quartic stabilization term instead of the stabilizer superfield.
We have checked that the coefficient $\zeta$ of order one is sufficient for the large-field models of the no-scale-like K\"{a}hler potential~\eqref{K-arbitrary-stabilized}, but $\zeta$ of order ten or larger is needed for small-field cases  and  generic models of the minimal (polynomial) K\"{a}hler potential~\eqref{minKstab}.

Generally, a large coupling constant implies a low cut off scale.
In the context of the Starobinsky model and Higgs inflation with non-minimal coupling to gravity, the unitarity was questioned~\cite{Burgess2009, Burgess2010, Barbon2009, Hertzberg2010} since these models introduce a coupling constant of order $10^4$.  Due to the standard effective field theory analyses around the vacuum,  the cut off is apparently $10^{-4}M_{\text{G}}$, and it is too low for these inflationary models to be reliably described.
This is an illusion caused by the fact that only the physics around the vacuum was considered~\cite{Lerner2010a}, and the unitarity bound is not violated during inflation taking the full form of the potential~\cite{Ferrara2011} or background scalar field value~\cite{Bezrukov2011}.
The cut off is inflaton field-dependent, and it is higher than the typical energy scale of inflation.

In contrast to these models, a somewhat large parameter $\zeta$ in our case appears in front of the self-coupling term of not the inflaton $\chi$ but its scalar superpartner $\phi$.
The latter is fixed to the vacuum value, so the simple estimate of the cut off suffices.
The cut off is inversely proportional to the square root of the stabilization parameter, $\Lambda \sim 1/\sqrt{\zeta}$.
For example, the field displacement $\delta \phi$ from the VEV at which the kinetic term vanishes and the scalar potential diverges is $\delta \phi = 1/ \sqrt{6\zeta}$ for the polynomial K\"{a}hler potential~\eqref{minKstab} (or \eqref{minKstabilizedShifted}) and $\delta \phi \simeq \sqrt{a/24\zeta}$ for the logarithmic K\"{a}hler potential~\eqref{K-a-no-scale}.
Conservatively, we take $\Lambda \equiv 1/\sqrt{10\zeta}$ as the typical cut off scale.

The Hubble scale for the typical chaotic inflation models such as  the quadratic and Starobinsky models are of order $10^{-5}$ in the reduced Planck unit, and the energy density scale $E_{\text{inf}}\equiv=V^{1/4}$ is $10^{-2.5}$.
Even if we take $\zeta\simeq 10^2$, the cut off scale $\Lambda \simeq 10^{-1.5}$ is one order of magnitude larger than the energy scale of inflation $E_{\text{inf}}$.
When we take $\zeta \simeq 10^4$, these scales become comparable $\Lambda \simeq E_{\text{inf}}$, and the theory enters a strong coupling phase.
Then, the inflaton potential is significantly affected by couplings between the inflaton $\chi$ and the strongly coupled $\phi$.
The condition $\Lambda > E_{\text{inf}}$ is rewritten in terms of $\zeta$ as 
\begin{align}
\zeta < \frac{M_{\text{G}}^2}{10 E_{\text{inf}}^2}\simeq \frac{M_{\text{G}}}{10 m_{\chi}},
\end{align}
where $m_{\chi}$ is the inflaton mass, and we have used $E_{\text{inf}}=V^{1/4}\simeq \sqrt{m_{\chi}M_{\text{G}}}$.

\section{Effects of other terms in the K\"{a}hler potential}\label{sec:OtherTerms}
We are using higher order terms in the K\"{a}hler potential for the sake of stabilization of the scalar superpartner of the inflaton.
So, the following questions arise: ``Can we neglect other (lower or even higher) terms in the K\"{a}hler potential?'', or ``What are the effects of such terms, if they can not be simply neglected?''
In this Section, we consider the effects of terms that we have not considered until now.

We only consider shift-symmetric terms in the K\"{a}hler potential since the effects of symmetry breaking terms are higher orders in symmetry breaking parameter, which is proportional to inflation scale.
For simplicity of notation, we define $X\equiv \Phi+\bar{\Phi}$.
Then, the shift-symmetric K\"{a}hler potential or its exponential (so-called frame function) is expanded as a series of $X$.
To consistently neglect higher than quartic terms, we consider the parametrization in which $X$ is stabilized near the origin.  This is understood as follows.
Suppose that $X$ is anyway stabilized by effects of K\"{a}hler potential at some point in the field space $X=X_0$.
Then we shift $X$ by $X_0$ by redefining the origin of $X$: $X \rightarrow \widetilde{X}=X-X_0$.
In terms of $\widetilde{X}$, it is stabilized at the origin and higher order terms are well controlled by higher powers of (approximately) vanishing value of $\widetilde{X}$.
With this understanding, we will omit the tilde in the following.

The mechanism of the stabilization we considered is based on the quartic term.
We proposed it as just an example, and it may be replaced with some alternative, perturbative or non-perturbative mechanism.
At the same time, it is also true that the simplest and effective way of stabilization is due to the quartic term if we can expand the K\"{a}hler potential with respect to $X$ around the origin, which we always assume.
Actually the quartic term is the unique term which generates mass of (the scalar component of) $X$ using its SUSY breaking $F$-term like $|F^{\Phi}|^2 X^2$.\footnote{
If there are SUSY breaking sources other than the inflaton during inflation, $X$ may be stabilized by coupling them.  The term like $\zeta X^2 |z|^2$ works similarly where $z$ denotes the SUSY breaking field.
Also, effects of higher order terms such as $X^6$ and $X^8$ become important in the field region away from the origin.
}  See the discussion around eq.~\eqref{KmetricStabilized}. 

In summary, we expand the K\"{a}hler potential or the frame function with respect to $X=\Phi+\bar{\Phi}$ around the origin, and utilize the quartic term for stabilization as we studied until now.
We will study effects of lower order terms.
If $X$ is still stabilized at or near the origin, effects from higher order terms can be consistently neglected.
In the following, we first study the ``minimal'' K\"{a}hler potential, and then move on to the logarithmic K\"{a}hler potential, in which case we expand the argument of the logarithm.

\subsection{Minimal K\"{a}hler potential}\label{subsec:EffMinimalK}

Consider the following generic K\"{a}hler potential,
\begin{align}
K= \sum_{n=0} c_n X^n =c_0 + c_1 X + c_2 X^2 + c_3 X^3 + c_4 X^4 + \dots \label{minKparamet}
\end{align}
where $X\equiv \Phi+\bar{\Phi}$ with $\Phi$ chiral, and dots represent higher order terms.
The $c_0$ can be absorbed by the superpotential by K\"{a}hler transformation, and $c_1$ is primarily related to the value of the first derivative $K_{\Phi}$ at $X=0$.
The $c_2$ changes the normalization of the fields, and the $c_4$ is the strength of stabilization.
To understand the role of $c_3$, it is useful to rewrite eq.~\eqref{minKparamet} as follows,
\begin{align}
K=&\left( c_0 - \frac{{c_3}^4}{256{c_4}^3} \right) + \left( c_1 - \frac{{c_3}^3}{16 {c_4}^2} \right)X +\left( c_2 -\frac{3c_3^2}{8c_4} \right)X^2 + c_4 \left( X+ \frac{c_3}{4c_4} \right)^4 +\dots  \nonumber \\
\equiv& \widetilde{c_0} + \widetilde{c_1}X+\widetilde{c_2}X^2 + \widetilde{c_4} (X- X_0)^4 + \dots .
\end{align}
In this form, it is clear from our studies in the previous Sections that $-c_3 / 4c_4$ represents the VEV of $X$.
We interpret $c_3$ as the parameter that controls deviation of the VEV of $X$ from its origin.
This should be so because $X^3$ is odd under the sign flip of $X$ and it makes gradient at the origin.
From the discussion at the beginning of this Section, the VEV of $X$ has to be small for consistent truncation of higher order terms such as $c_5 X^5, c_6 X^6, \dots$.
This immediately places a condition $|X_0| \ll 1$, \textit{i.e.}~$|c_3| \ll 4|c_4|$.

This may not be so restrictive condition analogously with the following argument for a scalar potential.
Consider an arbitrary smooth scalar potential.
Its shape may be concave or convex.  It may increase and sometimes decrease.
In a certain domain of field range, one may find a minimum or some minima.
Wherever they are, the gradient of the potential vanishes there, and the mass (curvature of the potential) is positive.
Our situation is similar.
If the field is stabilized at all by the K\"{a}hler potential and the $F$-term SUSY breaking of $\Phi$, and wherever the point is, the third derivative of the K\"{a}hler potential vanishes there, and the mass is provided by the fourth derivative and $F$-term SUSY breaking of $\Phi$.

Now we approximately canonically normalize $\Phi$ by $\Phi=\widetilde{\Phi}/\sqrt{2\widetilde{c_2}}$, which leads to
\begin{align}
K=&\widetilde{c_0}+\frac{\widetilde{c_1}}{\sqrt{2 \widetilde{c_2}}}\widetilde{X}+ \frac{1}{2}\widetilde{X}^2 +\frac{\widetilde{c_4}}{4 \widetilde{c_2}^2}\left( \widetilde{X} - \sqrt{2 \widetilde{c_2}}X_0 \right)^4 + \dots \nonumber \\
\equiv & \widehat{c_0}+ \widehat{c_1}\widetilde{X}+ \frac{1}{2} \widetilde{X}^2 + \widehat{c_4} \left( \widetilde{X} - \widetilde{X}_0 \right)^4,
\end{align}
where $\widetilde{X}=\widetilde{\Phi}+\overline{\widetilde{\Phi}}$.
In terms of the original parameters, these new parameters are
\begin{align}
\widehat{c_0}=& c_0 - \frac{{c_3}^4}{256{c_4}^3}, &
\widehat{c_1}=& \frac{ c_1 - \frac{{c_3}^3}{16 {c_4}^2} }{\sqrt{2\left( c_2 -\frac{3c_3^2}{8c_4} \right)}}, &
\widehat{c_4}=& \frac{c_4}{4 \left(c_2 -\frac{3c_3^2}{8c_4} \right)^2}, &
 \widetilde{X}_0=& - \frac{c_3}{4c_4} \sqrt{2 \left( c_2 -\frac{3c_3^2}{8c_4}  \right)}. \label{parameterCorresp}
\end{align}
It is more appropriate to impose the consistent truncation condition on the canonically normalized field.  The condition $|X_0| \ll 1$ is replaced with 
\begin{align}
\left| \widetilde{X}_0 \right| \ll 1. \label{ConsistentTruncation}
\end{align}
Basically, this tells us that $|c_3|$ should be much smaller than $|c_4|$.
If it is the case, differences between $\widetilde{c_n}$ and $c_n$ for lower $n$ are smaller because it involves higher order of $X_0$.
One might hope that the VEV $X_0$ can be suppressed accidentally by cancellation between $c_2$ and $3{c_3}^2/8c_4$ (see eq.~\eqref{parameterCorresp}).
But this is not the case because $c_4$ is negative for positive stabilizing mass, while $c_2$ is positive for the physically viable kinetic term.
The constant term $\widehat{c_0}$ (and also the linear term $\widehat{c_1}\widetilde{X}$ if one wishes) can be transferred to the superpotential, $W\rightarrow e^{\widehat{c_0}/2}W$.
Thus, the generic K\"{a}hler potential with the only requirement~\eqref{ConsistentTruncation} is now almost the same form as eq.~\eqref{minKstabilizedShifted} with identification $c=\widehat{c_1}$ and $\zeta=-4\widehat{c_4}$.
This means that in the parametrization of eq.~\eqref{minKstab}, higher order terms are understood to be expanded around $\widetilde{X}_0$ like $\widehat{c_5} (\widetilde{X}-\widetilde{X}_0)^5+ \widehat{c_6} (\widetilde{X}-\widetilde{X}_0)^6+ \dots$.
We summarize the meaning of parameters in Table~\ref{tab:ParamInterp}.

\begin{table}[htbp]
\begin{center}
\caption{Interpretation of parameters in the K\"{a}hler potential~\eqref{minKparamet}. }
  \begin{tabular}{| l | l |}
 \hline
 $c_0$ & normalization of the superpotential \\ \hline
 $c_1$ & magnitude of $K_{\Phi}$ \\ \hline
 $c_2$ & normalization of the field \\ \hline
 $c_3$ & deviation of the VEV from the origin  \\ \hline
 $c_4$ & strength of stabilization \\ \hline
   \end{tabular}\label{tab:ParamInterp}
  \end{center}
\end{table} 

\subsection{Logarithmic K\"{a}hler potential}\label{subsec:EffLogK}

Consider the following K\"{a}hler potential,
\begin{align}
K=& -a \ln \left( d_0 + d_1 X + d_2 X^2 + d_3 X^3 + d_4 X^4+ \dots   \right), \label{logKparamet}
\end{align}
where dots represent higher order terms.
Similarly to the previous case, we transform the $d_3 X^3$ term into the shift of the center of stabilization.
Then we separate out the overall constant in the logarithm.
After that we normalize $X$ such that the linear term has the coefficient $1/\sqrt{a}$.
It becomes of the form
\begin{align}
K=-a\ln \left( 1+\frac{1}{\sqrt{a}}\widehat{X}+ \widehat{d_2} \widehat{X}^2+\frac{\widehat{d_4}}{a^2} \left( \widehat{X}-\widehat{X}_0 \right)^4 \right) -a \ln \widehat{d_0}, \label{logKparameT}
\end{align}
with
\begin{align}
\widehat{X}=&\frac{\sqrt{a}\left(d_1 - \frac{{d_3}^3}{16{d_4}^2}  \right)}{d_0 - \frac{{d_3}^4}{256 {d_4}^3} }X ,&
\widehat{X}_0=& -\frac{\sqrt{a} d_3 \left(d_1 - \frac{{d_3}^3}{16{d_4}^2}  \right)}{4 d_4 \left( d_0 - \frac{{d_3}^4}{256 {d_4}^3}\right)}, & & & \\
\widehat{d_0}=& d_0 -\frac{{d_3}^4}{256 {d_4}^3}, &
\widehat{d_2}=& \frac{\left( d_2 - \frac{3 {d_3}^2}{8d_4} \right) \left( d_0 - \frac{{d_3}^4}{256 {d_4}^3} \right)}{a\left(d_1 - \frac{{d_3}^3}{16{d_4}^2}  \right)^2}, &
\widehat{d_4}=& \frac{d_4 \left( d_0 - \frac{{d_3}^4}{256 {d_4}^3} \right)^3}{ \left( d_1 - \frac{{d_3}^3}{16{d_4}^2} \right)^4 }. \label{ParameterCorresp2}
\end{align}
Consistent truncation of higher order terms requires
\begin{align}
\left|\widehat{X}_0 \right| \ll 1.
\end{align}

The K\"{a}hler potential~\eqref{logKparameT} looks similar to eq.~\eqref{K-a-no-scale} after we transfer the constant term $-a \ln \widehat{d_0}$ to the superpotential by K\"{a}hler transformation as $W \rightarrow \widehat{d_0}^{-a/2} W$.
However, there is a notable difference.  The $\widehat{d_2}X^2$ term is a new element.
In the case of $a=3$ in particular, this term drastically changes the structure of the potential breaking the no-scale cancellation $K^{\bar{\Phi}\Phi}K_{\Phi}K_{\bar{\Phi}}-3=0$.
Also, $\widehat{X}_0$, even small, can not be neglected in contrast to the minimal K\"{a}hler case (Subsection~\ref{subsec:EffMinimalK}).  
So, instead of quantifying bearable amount of these parameters to keep the results of undeformed model~\eqref{K-arbitrary-stabilized}, we establish necessary conditions for viable large field inflationary model.

\begin{figure}[htbp]
  \begin{center}
    \includegraphics[clip, width=8cm]{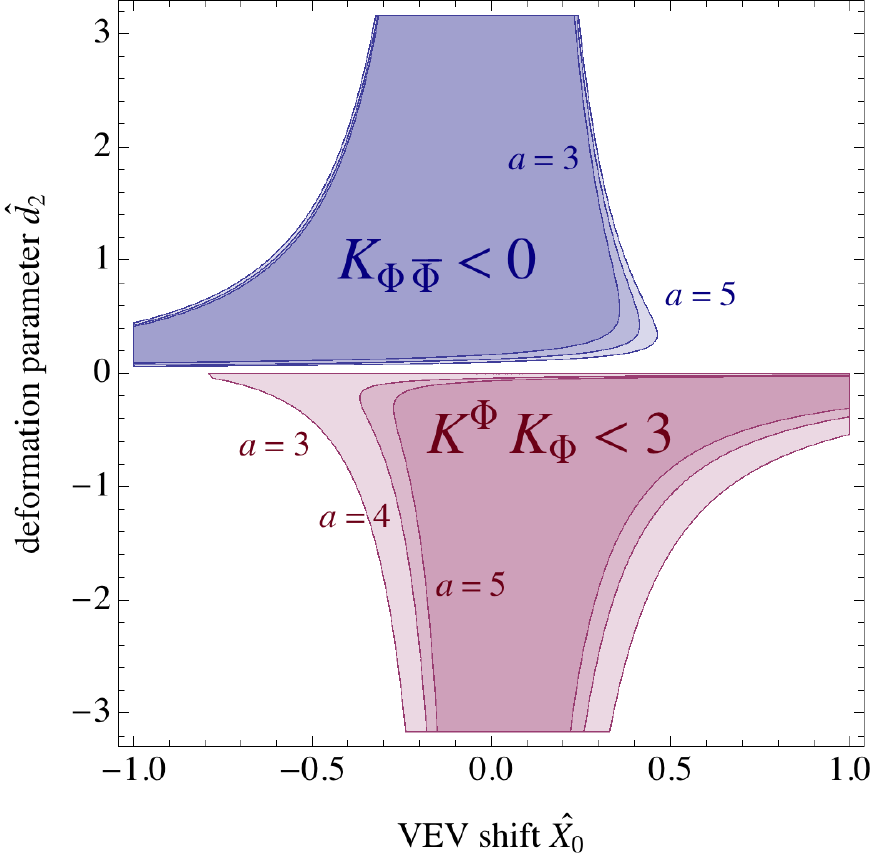}
    \caption{Constraints on the deformation parameters $\widehat{d_2}$ and $\widehat{X}_0$ in the K\"{a}hler potential~\eqref{logKparameT}.  The horizontal axis is $\widehat{X}_0$, and the vertical axis is $\widehat{d_2}$.  The theory is plagued by the ghost state ($K_{\widehat{\Phi}{\widehat{\bar{\Phi}}}}< 0$) and hence excluded in the blue region.  Large field inflation is improbable ($K^{\widehat{\bar{\Phi}}\widehat{\Phi}} K_{\widehat{\Phi}}K_{{\widehat{\bar{\Phi}}}}-3 < 0$) in the red region.  In the blue region, both constraints are not satisfied.  For the blue region, the three contours correspond to $a=3, 4$, and $5$ from inside to outside, while for the red region, the three contours correspond to $a=3, 4$, and $5$ from outside to inside.}
    \label{fig:EffOtherTerms}
  \end{center}
\end{figure}

The K\"{a}hler metric and the key combination $K^{{\widehat{\Phi}}\widehat{\bar{\Phi}}}K_{\widehat{\Phi}}K_{{\widehat{\bar{\Phi}}}}-3$ evaluated at $\widehat{X}=\widehat{X}_0$ are
\begin{align}
K_{\widehat{\Phi}{\widehat{\bar{\Phi}}}}=& \frac{1-2a \widehat{d_2}+2 \sqrt{a} \widehat{d_2}\widehat{X}_0 + 2a \widehat{d_2}^2\widehat{X}_0^2}{\left( 1+\frac{1}{\sqrt{a}}\widehat{X}_0 + \widehat{d_2} \widehat{X}_0^2   \right)^2} , \\
K^{\widehat{\bar{\Phi}}\widehat{\Phi}} K_{\widehat{\Phi}}K_{{\widehat{\bar{\Phi}}}}-3
=&\frac{-3+a +6a \widehat{d_2} + (4 a -6 ) \sqrt{a}\widehat{d_2}\widehat{X}_0+(4a^2 - 6a) \widehat{d_2}^2 \widehat{X}_0^2 }{1-2a \widehat{d_2}+2 \sqrt{a} \widehat{d_2}\widehat{X}_0 + 2a \widehat{d_2}^2\widehat{X}_0^2}.
\end{align}
For the case of $a=3$, these expressions reduce to
\begin{align}
K_{\widehat{\Phi}{\widehat{\bar{\Phi}}}}=& \frac{1-6 \widehat{d_2}+2 \sqrt{3} \widehat{d_2}\widehat{X}_0 + 6 \widehat{d_2}^2\widehat{X}_0^2}{\left( 1+\frac{1}{\sqrt{3}}\widehat{X}_0 + \widehat{d_2} \widehat{X}_0^2   \right)^2} , \\
K^{\widehat{\bar{\Phi}}\widehat{\Phi}} K_{\widehat{\Phi}}K_{{\widehat{\bar{\Phi}}}}-3=&\frac{6 \widehat{d_2}\left( 3+\sqrt{3}\widehat{X}_0+3 \widehat{d_2} \widehat{X}_0^2 \right)}{1-6 \widehat{d_2}+2 \sqrt{3} \widehat{d_2}\widehat{X}_0 + 6 \widehat{d_2}^2\widehat{X}_0^2}.
\end{align}
The K\"{a}hler metric must be positive to avoid the negative norm states, while the latter combination is the coefficient of the $|W|^2$ term and should be positive for large field inflation.
The contour plot of these constraints $K_{\widehat{\Phi}{\widehat{\bar{\Phi}}}}> 0$ and $K^{\widehat{\bar{\Phi}}\widehat{\Phi}} K_{\widehat{\Phi}}K_{{\widehat{\bar{\Phi}}}}-3 > 0$ on the parameters $\widehat{d_2}$ and $\widehat{X}_0$ is shown in Fig.~\ref{fig:EffOtherTerms}.
The colored regions are excluded.
In the blue region, the sign of the kinetic term is unphysical, and the three contours correspond to $a=3, 4$, and $5$ from inside to outside.
In the red region, the coefficient of the superpotential squared term is negative, and the three contours correspond to $a=3, 4$, and $5$ from outside to inside.

Our undeformed model in the previous Sections sits at the origin ($\widehat{d_2}=\widehat{X}_0=0$) in Fig.~\ref{fig:EffOtherTerms}.
Small deviation away from the origin destroys the structure of the K\"{a}hler potential (the no-scale structure for $a=3$) and changes the theory significantly entering the blue or red region in the Figure.
For example, $-\frac{1}{6}+\frac{1}{2a}\leq \widehat{d_2}< \frac{1}{2a}$ is required to be in the white region at $\widehat{X}_0=0$. 
Thus, the logarithmic models with quartic stabilization term~\eqref{K-arbitrary-stabilized} should be regarded as tuned models.
Note that an infinitesimal negative $\widehat{d_2}$ leads to the negative $|W|^2$ term in the large field region for generic superpotentials in the case of the no-scale type models ($a=3$).
This does not immediately lead to the conclusion that such cases are excluded for inflationary application.
This is because it is enough that the potential has a suitable form only in the field region corresponding to $50$ to $60$ e-foldings.
Another comment is that there is a narrow allowed band near $\widehat{d_2}\simeq 0$.
This can be understood as follows.
When we set $\widehat{d_2}=0$ in the K\"{a}hler potential~\eqref{logKparameT}, the shift of $\widehat{X}$ by $\widehat{X}_0$ does not generate quadratic or cubic term in $\widehat{X}$ and does not change the coefficient of the linear and quartic term.
It changes the constant term, but it can be again absorbed by the superpotential by K\"{a}hler transformation.
Thus, the K\"{a}hler potential keeps the same form but now without $\widehat{X}_0$.
One should keep track of the field redefinition since it also change the functional form of the superpotential.
However, we can not reliably shift the $X$ when the shift is large as we have discussed at the beginning of this Section.
That is why we restrict the domain of $\widehat{X}_0$ to $\left|\widehat{X}_0\right|\leq 1$ in Figure~\ref{fig:EffOtherTerms}.
We summarize the meanings of parameters in the K\"{a}hler potential~\eqref{logKparamet} in Table~\ref{tab:ParamInterp2}.
\begin{table}[htbp]
\begin{center}
\caption{Interpretation of parameters in the K\"{a}hler potential~\eqref{logKparamet}. }
  \begin{tabular}{| l | l |}
 \hline
 $d_0$ & normalization of the superpotential \\ \hline
 $d_1$ & normalization of the field \\ \hline
 $d_2$ & deformation of the K\"{a}hler structure \\ \hline
 $d_3$ & deviation of the VEV from the origin  \\ \hline
 $d_4$ & strength of stabilization \\ \hline
   \end{tabular}\label{tab:ParamInterp2}
  \end{center}
\end{table} 

\section{Matter coupling and inflaton decay}\label{sec:coupling}
We have established that inflation can occur with a single chiral superfield in supergravity, without the aid of other superfields.
But we have to consider other matter superfields possibly present in the theory at the end of the day.
After all, there should be Standard Model particle contents, and some SUSY breaking fields in our supergravity framework.
These matter are not introduced to realize inflation itself, but to describe our real world and a realistic cosmological history.
After inflation, the inflaton has to decay into particles to reheat the universe.
The energy stored in the inflaton field is converted to kinetic and potential energy of lighter particles and finally to thermal energy of radiation.
It is the beginning of the hot Big-Bang universe.
So, let us consider coupling of the inflaton sector to other matter sector in the theory.
If they are decoupled at the global SUSY level, coupling among various fields with gravitational strength are generically induced in supergravity.
We consider some examples of coupling, and study potential effects of inflaton on other sectors and vice versa.

For simplicity, we consider the minimal coupling between the inflaton and other sectors in the superpotential,
\begin{align}
W(\Phi, \phi^i )= W^{(\text{inf})}(\Phi) + W^{(\text{oth})}(\phi^i), \label{Wcoupling}
\end{align}
where $W^{\text{inf}}(\Phi)$ and $W^{\text{oth}}(\phi^i)$ are the superpotentials of the inflaton $\Phi$ and other fields collectively denoted by $\phi^i$.
This separation is preserved in the renormalization group running by the non-renormalization theorem~\cite{Grisaru1979, Seiberg1993, Weinberg1998}.

First, consider the ``minimal'' inflaton K\"{a}hler potential~\eqref{minKstab} and its minimal coupling to other superfields,
\begin{align}
K(\Phi, \phi^i, \bar{\Phi},\bar{\phi}^{\bar{j}})= \frac{1}{2} \left( \Phi + \bar{\Phi} \right) ^2 - \frac{\zeta}{4}\left( \Phi + \bar{\Phi} -2\Phi_0 \right)^4 + K^{(\text{oth})}(\phi ^i, \bar{\phi}^{\bar{j}} ) , \label{minKcoupling}
\end{align}
where $K^{\text{oth}}(\phi ^i, \bar{\phi}^{\bar{j}} )$ is the K\"{a}hler potential of other superfields.
This separation in the K\"{a}hler potential is not ensured by any symmetrical reasons.
For example, there may be shift symmetric terms like $\xi (\Phi+\bar{\Phi})|\phi^i |^2$, which affects the following results.
The choice of the minimal K\"{a}hler potential is based on simplicity, and we take it just as an example.
The situation may be realized accidentally at some renormalization scale (hopefully at the energy scale of inflation or reheating) during renormalization group running.
An advantage of the minimal K\"{a}hler potential is that there is no kinetic mixing between $\Phi$ and $\phi^i$.

The scalar potential following from eqs.~\eqref{Wcoupling} and \eqref{minKcoupling} is
\begin{align}
V= e^{2\Phi_0^2 + K ^{(\text{oth})}} \left( |W_{\Phi}|^2 + 2\Phi_0 \left( W \overline{W}_{\bar{\Phi}}+\overline{W}W_{\Phi} \right) +K^{(\text{oth})\bar{j}i}D_i W \bar{D}_{\bar{j}}\overline{W} + \left(4 \Phi_0^2-3 \right) |W|^2  \right). \label{VcouplingMinimal}
\end{align}
Derivatives of $\Phi$ or $\phi^i$ imply they are the quantity depending solely on each sector, \textit{e.g.}~$W_{\Phi}=W^{(\text{inf})}_{\Phi}$ and $K_{i}=K^{(\text{oth})}_i$ because of the assumption of minimal coupling.
Beware that  $D_iW=W_i + K_i W$ contains $W^{(\text{inf})}$.
We have substituted $\Phi+\bar{\Phi}=2\Phi_0$, but it should be kept in mind that the stabilization becomes loose when the SUSY breaking due to inflaton, which drives inflationary expansion of spacetime, becomes small.
If the minimum where the inflaton reaches  at the end of inflation is supersymmetric, both the real and imaginary components of the inflaton have the same mass.
In other cases, it has to be checked that the minimum is stable in both directions.
We consider the effective theory after integrating out heavier SUSY preserving fields than inflaton.
If lighter fields than the inflaton are present, its displacement from the low-energy values may cause moduli problem.
But they acquire Hubble induced masses.
The Hubble induced mass squared $m^2$ of $\phi^i$ is, assuming the minimal K\"{a}hler potential for $\phi^i$,
\begin{align}
m^2\simeq \frac{3H^2}{4\Phi_0^2-3},
\end{align}
where we have used the Friedmann equation $3H^2=V\simeq e^{2\Phi_0^2}(4\Phi_0^2-3)|W|^2$.
We take $4\Phi_0^2-3$ positive to have the positive $|W|^2$ term, but if we increase it so that $4\Phi_0^2-3>3$, the Hubble induced mass becomes smaller than the Hubble scale.
When the Hubble-induced mass is larger than the Hubble scale, the otherwise light fields are frozen near the origin.

These arguments are somewhat rough, so let us look at eq.~\eqref{VcouplingMinimal} more carefully.
A term proportional to $|W|^2$ is present in the $K^{\text{(oth)}}D_iWD_{\bar{j}}\overline{W}$ term in the form $K_i K^i |W|^2$.  This kind of term is in quite contrast to the framework based on the stabilizer superfield $S$ (Section~\ref{sec:SSSS}) in which $W^{\text{(inf)}}\propto S$ vanishes during inflation.
If there are fields whose VEVs are of order the Planck scale, this term affects the inflationary dynamics since it is the same order of the dominant term in the inflaton potential, $(4\Phi_0^2-3)|W|^2$.
But the VEVs of fields are expected to be suppressed by the Hubble induced masses, so the inflaton potential is not affected by the term.
If it is a good approximation to regard the last term in eq.~\eqref{VcouplingMinimal} as the dominant contribution to the inflaton potential, $|W_{\Phi}|$ should be small compared to $|W|$ during inflation because otherwise the slow-roll condition is not satisfied.  So, the second term in eq.~\eqref{VcouplingMinimal} does not have a large effect.
These remarks are less applicable when the scale of the coupled sector are not much below the inflation scale.
In particular, SUSY breaking sector may have not small VEVs and $F$-term, $K_{z}\neq 0$ and/or $W_{z}\neq 0$.
Moreover, we can not integrate out SUSY breaking sector and keep SUSY description for the resultant effective theory.
When the SUSY breaking scale is larger than the inflation scale, it can significantly affect the inflationary potential \textit{e.g.}~through the terms $2\Phi_0 \text{Re} (W^{\text{(oth)}}\overline{W}_{\bar{\Phi}})$ and $\text{Re}(D_{z}W K^{z}W^{\text{(inf)}})$.  
This feature that SUSY breaking scale higher than the inflation scale spoils inflationary dynamics is similar to the case of the framework using the stabilizer superfield~\cite{Buchmuller2014a, Abe2014}.
Some aspects of effects of SUSY breaking sector on the inflaton sector are easily understood from the superconformal approach~\cite{Abe2014}.
The scalar potential is scaled when we move from Jordan frame to Einstein frame by a conformal factor.
The conformal factor is a power of the frame function $(\Omega/3)=e^{-K/3}$.
When this factor contains inflaton, the inflaton potential can be drastically (exponentially) changed via the rescaling of the SUSY breaking term.
In our case, the K\"{a}hler potential is shift symmetric and does not depend on the inflaton, and there are no effects of the conformal rescaling in such cases~\cite{Abe2014}.
One still have to examine direct couplings already present in the Jordan frame, as we have done above.

One of the distinguishing features of our framework is that the first derivative of the K\"{a}hler potential with respect to the inflaton $K_{\Phi}$ is not suppressed.
This implies that the particle production from inflaton decay is not suppressed~\cite{AsakaNakamuraYamaguchi2006, EndoTakahashiYanagida2007a}.
The reheating is efficient, but production of unwanted long-lived particle, gravitino, is also not suppressed.
The two-body decay rate into scalars or spinors, and the three-body decay rate into three scalars or a scalar plus two spinors are~\cite{EndoTakahashiYanagida2007a}
\begin{align}
\Gamma^{\text{(2-body)}}\simeq & \frac{\Phi_0^2 m^2 m_{\chi}}{2\pi M_{\text{G}}^2} , &
\Gamma^{\text{(3-body)}}\simeq & \frac{\Phi_0^2 m_{\chi}^3 |y|^2}{64\pi^3 M_{\text{G}}^2},
\end{align}
where $m$ and $m_{\chi}$ are the mass of the final state particles and the inflaton, and $y$ is the relevant trilinear coupling constant (Yukawa coupling).
When there are heavy fields such as SUSY breaking field or right-handed neutrinos, decay into these particles are enhanced.  Concerning three-body decay, the top Yukawa coupling is the largest in the visible sector and it is of order one.
The three-body decay rate can be regarded as the lower bound of the decay rate.

In passing, let us introduce a direct coupling between inflaton and matter fields like
\begin{align}
W=c (\Phi-\Phi_0) A B,
\end{align}
where $c$ is the coupling constant, and $A$ and $B$ are chiral superfields such as Higgses $H_u$ and $H_d$ or right-handed neutrinos $N$.
The coupling constant $c$ is bounded as eq.~\eqref{coupling_constraint}, so the decay rate $\Gamma^{(\text{direct})} \simeq \frac{c^2 m_{\chi}}{4\pi}$ is also bounded from above.
Combining with the lower bound in the previous paragraph,
we can constrain the reheating temperature $T_{\text{R}}=\left( \frac{90}{\pi^2 g_{*}(T_{\text{R}})} \right)^{1/4}\sqrt{M_{\text{G}}\Gamma }$ as
\begin{align}
2\times 10^8 \phantom{ } \text{GeV}\left( \frac{m_{\chi}}{10^{13}\phantom{ }\text{GeV}} \right)^{\frac{3}{2}} \lesssim T_{\text{R}} \ll 7 \times 10^{13}\phantom{ } \text{GeV} \left( \frac{m_{\chi}}{10^{13}\phantom{ }\text{GeV}} \right)^{\frac{1}{2}},
\end{align}
where we took $g_{*}=240$ and $\Phi_0=1$.
This range of the reheating temperature is consistent with thermal leptogenesis~\cite{Fukugita1986}.

Corresponding to the anomaly of the shift symmetry, the following coupling may be present~\cite{Iwamoto2014}
\begin{align}
\frac{1}{4} \int \text{d}^2 \Theta c \Phi \mc{W}^{A}\mc{W}^{A}+ \text{H.c.},
\end{align}
where $c$ is the real coupling constant, and $\mc{W}^{A}$ is a gauge field strength chiral superfield.
The shift in the imaginary direction generates a total derivative, and the symmetry is broken non-perturbatively.
The decay rate into gauge bosons and gauginos through this coupling is
\begin{align}
\Gamma^{\text{(gauge)}}\simeq  \frac{N_{\text{g}}|c|^2 m_{\chi}^3}{128 \pi M_{\text{G}}^2},
\end{align}
where $N_{\text{g}}$ is the number of generators of the gauge algebra.
If the imaginary part continues to be a mass eigenstate at the time of decay, or mixing with the real part is ineffective, the inflaton can not decay through super-Weyl-K\"{a}hler and sigma model anomaly-induced channel~\cite{BaggerMoroiPoppitz2000, EndoTakahashiYanagida2008} into gauge bosons and gauginos.
Whether it continues to be a mass eigenstate depends on phases in the superpotential.
When it becomes no longer a mass eigenstate and mix with the real part, oscillation time scale $m_{3/2}^{-1}$ should be compared to the decay time scale $\Gamma^{-1}$.
Here, oscillation time scale is determined by mass splitting and it is due to SUSY breaking, hence the gravitino mass $m_{3/2}=e^{K/2}|W|$ is relevant.
For the inflaton whose decay is due to Planck-suppressed operators and inflaton mass of $\mc{O}(10^{13})$~GeV, the typical threshold for gravitino mass above which mixing is effective is $\mc{O}(1)$~GeV~\cite{Nakayama2014d, Terada2015}.
The super-Weyl-K\"{a}hler and sigma model anomaly-induced decay rate depends on the first derivative of the K\"{a}hler potential with respect to the parent particle~\cite{EndoTakahashiYanagida2008, EndoTakahashiYanagida2007a}, so the rate from the imaginary part (inflaton) vanishes.  This feature is same for all shift symmetric models~\cite{Ketov2014a, Terada2015}.

Let us see whether inflaton decay into gravitinos are kinematically possible or not for models in Section~\ref{sec:strategy} which break SUSY after inflation.
In both of the quadratic models~\eqref{VquadMinimal} and \eqref{V4quadratic} for the polynomial K\"{a}hler potential~\eqref{minKstab} and logarithmic one~\eqref{K-a-no-scale} with $a=4$, gravitino mass $m_{3/2}$ is related to the inflaton mass $m_{\chi}$ as $m_{3/2}=|2\pm \sqrt{3}|m_{\chi}$.
The origin of the sign options is two solutions of the condition $V=0$ with respect to the constant term in the superpotential.
Choosing the upper sign results in the heavier gravitino into which inflaton cannot decay.
On the other hand, the lower sign results in the lighter gravitino into which inflaton can decay.
Similarly, in the case of the model of the Starobinsky potential~\eqref{WStarobinsky} with polynomial K\"{a}hler potential, inflaton decays into gravitinos depending on the choice of the parameter $b$.
In the case of the K\"{a}hler potential~\eqref{K-imaginary-fixed} and the superpotential~\eqref{W03}, which is another realization of the Starobinsky potential, the inflaton mass is given by eq.~\eqref{m2} while the gravitino mass is given by $(|c_0|+2\sqrt{2}|c_3|\Phi_0 )^2/8\Phi_0^3$ where we have used $\text{Re}(c_3 \overline{c_0})=\text{Im}(c_3 \overline{c_0})<0$ (see the texts after eq.~\eqref{W03}).
In this case, the decay is forbidden.
For the Starobinsky-like model of the K\"{a}hler potential~\eqref{K-imaginary-fixed} with $\Phi_0=0$ and the superpotential~\eqref{W03shifted}, the inflaton mass scale squared is given by $\text{Re}(c_3 \overline{c_0})$ while that for gravitino is given by $|c_0|^2$.  Since $|c_3|$ has to be much smaller than $|c_0|$ in this model to obtain sufficiently flat Starobinsky-like potential, the gravitino mass is larger than the inflaton mass.
Again the decay is impossible.
Thus, inflaton can decay in some models but cannot in other models.

Let us consider the case in which the inflaton breaks SUSY also after inflation, $G_{\Phi}G^{\Phi}=3$.
In such a case, gravitino is copiously pair-produced by inflaton decay, if kinematically possible, with the rate~\cite{EndoHamaguchiTakahashi2006a, NakamuraYamaguchi2006}
\begin{align}
\Gamma ^{\text{(gravitino)}} \simeq \frac{ m_{\chi}^5}{96 \pi m_{3/2}^2 M_{\text{G}}^2} \sim \frac{m_{\chi}^3}{6\pi M_{\text{G}}^2}, \label{PairGravitinoRate0}
\end{align}
where we have substituted $m_{3/2}\sim m_{\chi}/4$ in the last equality since gravitino mass $m_{3/2}$ is of the order the inflation scale when the inflaton breaks SUSY after inflation, but it is less than the half of the inflaton mass when the decay is kinematically possible.
Actually, the above rate~\eqref{PairGravitinoRate0} is obtained using the known result with approximation $m_{\chi}\gg m_{3/2}$, but the order of magnitude is expected to be same in the case $m_{\chi}\sim m_{3/2}$.  It is just the typical Planck-suppressed decay rate.
Thus, gravitino becomes a main component of the universe.
Since the decay rate of gravitino is comparable to that of inflaton, gravitino decays shortly.
The lightest supersymmetric particle (LSP) are copiously produced by the gravitino decay, and its abundance is way too large to be consistent with the observed dark matter abundance.
$R$-parity breaking or thermal inflation~\cite{LythStewart1996} is needed to decrease the LSP abundance.

On the other hand, if inflaton preserves SUSY at the vacuum, gravitino production is less significant, but they are still produced by mixing effects with the SUSY breaking field~\cite{DineKitanoMorisseShirman2006, EndoHamaguchiTakahashi2006}.
The rate in such a case is given by
\begin{align}
\Gamma ^{\text{(gravitino)}} \simeq \frac{\left| \mc{G}_{\chi}^{\text{(eff)}}\right|^2 m_{\chi}^5}{288 \pi m_{3/2}^2 M_{\text{G}}^2} ,  \label{PairGravitinoRate}
\end{align}
where the effective coupling constant $\mc{G}_{\chi}^{\text{(eff)}}$ \cite{EndoHamaguchiTakahashi2006, Endo2012} is evaluated in our minimal coupling case as
\begin{align}
\left| \mc{G}_{\chi}^{\text{(eff)}}\right|^2 \simeq 36 \Phi_0 ^2 \left( \frac{m_{3/2} m_z ^2}{m_{\chi} \left( m_{\chi}^2 -m_z^2 \right)^2} \right)^2, \label{Geff_minK}
\end{align}
where $m_z$ is the mass of the SUSY breaking field $z$.
When we introduce non-minimal coupling like $\xi (\Phi+\bar{\Phi})|\phi^i|^2$, these decay rates change, 
but the above rates are thought to be the lower bound of decay rates since unless cancellation occurs the order of magnitude does not decrease.
The gravitino yield, $Y_{3/2}^{(\text{direct})}\equiv n_{3/2}/s$ where $n_{3/2}$ and $s$ are the gravitino number density and the entropy density, following from the above decay rate into gravitinos is
\begin{align}
Y_{3/2}^{(\text{direct})} \simeq 2\times 10^{-36} \left( \frac{240}{g_{*}(T_{\text{R}})}  \right)^{\frac{1}{4}} \left( \frac{1}{c_{\text{tot}}} \right) \left( \frac{\Phi_0}{1} \right)^2 \left( \frac{10^{13}\phantom{ }\text{GeV}}{m_{\chi}}\right)^{\frac{7}{2}} \left( \frac{m_{z}}{10^5 \phantom{ }\text{GeV}}\right)^4 ,
\end{align}
where we have defined the total decay rate of inflaton as $\Gamma_{\text{tot}}=\frac{c_{\text{tot}}^2 m_{\chi}}{4\pi}$.
The thermal gravitino yield is known to be~\cite{BolzBrandenburgBuchmuller2001, Pradler2007, Pradler2007a, Rychkov2007, KohriMoroiYotsuyanagi2006}
\begin{align}
Y_{3/2}^{(\text{thermal})}\simeq 2 \times 10^{-13}\left( 1+\frac{m_{\tilde{g}}}{3 m_{3/2}^2} \right)\left( \frac{T_{\text{R}}}{10^{9}\phantom{ }\text{GeV}} \right), \label{Y3/2thermal}
\end{align}
where $m_{\tilde{g}}$ is the gluino mass at zero temperature.
Thus, the thermal gravitino is dominant over the direct decay products from inflaton in our current setup.
From big-bang nucleosynthesis (BBN), the reheating temperature (gravitino yield) is constrained as $T_{\text{R}}\lesssim 10^{5}\phantom{ }\text{GeV} \sim 10^{10}\phantom{ }\text{GeV}$ ($Y_{3/2}\lesssim 10^{-17}\sim 10^{-12}$) depending on the gravitino mass~\cite{KohriMoroiYotsuyanagi2006}.
For complete analysis, gravitino production from the SUSY breaking sector should also be taken into account, but we do not go into these details.  See Refs.~\cite{Nakayama2012, Evans2014, Nakayama2014d, Terada2015}.
Also, when gravitino is heavier than about 30 TeV, constraints on gravitino decay itself is absent, but the LSP abundance from gravitino decay is constrained not to exceed the observed dark matter abundance,  $Y_{\text{CDM}}^{(\text{obs})}=5 \times 10^{-13}\left( 10^3 \phantom{ }\text{GeV}/m_{\text{CDM}} \right)$, where CDM denotes cold dark matter.
The LSP dark matter can pair-annihilate to decrease its number, so $Y_{\text{LSP}}$ is at most of order of $Y_{3/2}$ if order one LSPs are produced from one gravitino.
The actual value of $Y_{\text{LSP}}$ is model-dependent, so we stop our general discussion here.
The bottom line is that the low-energy SUSY breaking scenario $m_{\chi}\gg m_{3/2}$ in our single superfield inflation is consistent with (or saying more conservatively, not immediately excluded by) these observational constraints.

We have also used the logarithmic K\"{a}hler potential for single superfield inflation, so let us next consider the following coupling,
\begin{align}
K(\Phi, \phi^i , \bar{\Phi}, \bar{\phi}^{\bar{j}})=-3 \ln \left(1 +\frac{1}{\sqrt{3}}\left (\Phi+\bar{\Phi}\right) +\frac{\zeta}{9} \left (\Phi+\bar{\Phi}\right)^4  \right) + K^{\text{(oth)}}(\phi^{i},\bar{\phi}^{\bar{j}}).
\end{align}
This again leads to no kinetic mixing, and we can easily write down the scalar potential,
\begin{align}
V=e^{K^{\text{(oth)}}}\left( |W_{\Phi}|^2 - \sqrt{3} \left( W \overline{W}_{\bar{\Phi}} + \overline{W} W_{\Phi}  \right) + K^{\text{(oth)}\bar{j}i}D_{i}W\bar{D}_{\bar{j}}\overline{W}  \right).
\end{align}
We have again substitute the expectation value of the real part $\Phi+\bar{\Phi}=0$.
At the end of inflation, the stabilization is loosened.
If we impose the phase alignment condition (Subsection~\ref{sec:arbitrary}) on the inflaton superpotential, $W$ and its conjugate in the second term reduces to $W^{\text{(oth)}}$ and its conjugate.
If we further impose it on the full superpotential, the second term $- \sqrt{3} \left( W \overline{W}_{\bar{\Phi}} + \overline{W} W_{\Phi}  \right)$ vanishes.
The Hubble induced mass for fields $\phi^i$ whose K\"{a}hler potential is minimal is
\begin{align}
m^2 \simeq 3 H^2 + |W^{\text{(inf)}}|^2,
\end{align}
where we have used Friedmann equation $3H^2=V\simeq |W_{\Phi}|^2$, and we have added the term $|W^{\text{(inf)}}|^2$ omitted in Ref.~\cite{Terada2015}.
The fields are fixed near the origin during inflation.
Various features are qualitatively similar to the minimal K\"{a}hler case above except the absence of the last $|W|^2$ term in eq.~\eqref{VcouplingMinimal}.
For example, $K_{\Phi}\simeq -\sqrt{3}$ is not suppressed unless its value changes much at the end of inflation. 
Various partial decay rates are the same order as the minimal K\"{a}hler case.
The effective coupling~\eqref{PairGravitinoRate} for the rate into a pair of gravitinos is
\begin{align}
\left| \mc{G}_{\chi}^{\text{(eff)}}\right|^2 \simeq 27 \left( \frac{m_{3/2}m_z^2}{m_{\chi}\left( m_{\chi}^2 -m_{z}^2\right)} \right)^2 .
\end{align}
This is the same order as eq.~\eqref{Geff_minK}, and the gravitino is mainly produced thermally.

Finally, consider the minimal coupling inside the logarithm,
\begin{align}
K(\Phi, \phi^i , \bar{\Phi}, \bar{\phi}^{\bar{j}})=-3 \ln \left(1 +\frac{1}{\sqrt{3}}\left (\Phi+\bar{\Phi}\right) +\frac{\zeta}{9} \left (\Phi+\bar{\Phi}\right)^4 - J^{\text{(oth)}}(\phi^{i},\bar{\phi}^{\bar{j}}) \right) ,
\end{align}
where $J^{\text{(oth)}}(\phi^{i},\bar{\phi}^{\bar{j}})$ is a Hermitian kinetic function.
This is the form of conformal or geometrical sequestering~\cite{Randall1999}.
The K\"{a}hler metric and its inverse are obtained as
\begin{align}
K_{I\bar{J}}=& \frac{1}{(\Omega/3)^2} 
\begin{pmatrix}
1 & -\frac{1}{\sqrt{3}}J_{\bar{j}} \\
-\frac{1}{\sqrt{3}} J_i & (\Omega/3) J_{i\bar{j}} + \frac{1}{3}J_i J_{\bar{j}} 
\end{pmatrix}
, & 
K^{\bar{J}I}=& (\Omega/3) 
\begin{pmatrix}
\Omega/3 + \frac{1}{3}J_i J^i  &  \frac{1}{\sqrt{3}}J^i \\
\frac{1}{\sqrt{3}} J^{\bar{j}} & J^{\bar{j}i}
\end{pmatrix},
\end{align}
where $(\Omega /3 )=e^{-K/3}$ is the frame function characterizing the original Jordan frame, indices $I, J, \dots$ run over both the inflaton $\Phi$ and other fields $i, j, \dots$.
Indices of the matter ``K\"{a}hler potential'' $J$ are lowered (raised) by the ``K\"{a}hler metric'' $J_{i\bar{j}}$ (its inverse $J^{\bar{j}i}$).
The scalar potential is 
\begin{align}
V=& \frac{1}{(\Omega/3)^2} \left( \left( \Omega/3 + \frac{1}{3}J_i J^i \right)|W_{\Phi}|^2 -\sqrt{3} \left( W \overline{W}_{\bar{\Phi}}+\overline{W} W_{\Phi} \right) \right. \nonumber \\
& \left. \qquad \qquad  +\frac{1}{\sqrt{3}} \left( J^i W_i \overline{W}_{\bar{\Phi}} + J^{\bar{i}}\overline{W}_{\bar{i}}W_{\Phi} \right)  +J^{\bar{j}i}W_i \overline{W}_{\bar{j}}  \right).
\end{align}
Many points are qualitatively similar to the previous cases, but there are no terms proportional to $|W|^2$.
In particular, no coupling between $|W^{\text{(inf)}}|^2$ and other fields exists.
The Hubble induced mass squared for fields $\phi^i$ is
\begin{align}
m^2 \simeq 2 H^2,
\end{align}
so the field rapidly settle to the origin and fixed there.
Then VEVs of fields vanish, and $J_i J^i$ term in the first line and the first term in the second line 
can be neglected.
Again, if the SUSY breaking scale is low enough, the inflationary dynamics is not affected.

Regarding partial decay rates, three body decay rates are suppressed by the structure of the K\"{a}hler potential~\cite{EndoTakahashiYanagida2007a, Terada2015}. 
The two-body decay rate into scalars and spinors is of the order $|J_{ij}|^2 m_{\chi}^3/M_{\text{G}}^2$.
The holomorphic bilinear term $J_{ij}\phi^i\phi^j$ is motivated by Giudice-Masiero mechanism to yield the effective $\mu$ term~\cite{Giudice1988}.
The effective coupling of gravitino pair production is 
\begin{align}
\left| \mc{G}_{\chi}^{\text{(eff)}}\right|^2 \simeq  \left( \left| J_z + \frac{2W_z}{m_{\chi}} \right| \frac{m_z^2}{m_{\chi}^2-m_z^2} \right) ^2 .
\end{align}
This leads to the gravitino yield of
\begin{align}
Y_{3/2}^{(\text{direct})}\simeq 6\times 10^{-25}\left( \frac{240}{g_{*}(T_{\text{R}})}  \right)^{\frac{1}{4}} \left( \frac{1}{c_{\text{tot}}} \right) \left( \frac{10^{13}\phantom{ }\text{GeV}}{m_{\chi}}\right)^{\frac{3}{2}} \left( \frac{m_{3/2}}{10^3\phantom{ } \text{GeV}}\right)^2 ,
\end{align}
and again negligible compared to the thermal yield~\eqref{Y3/2thermal}.

Before concluding this Section, we briefly discuss possible non-minimal couplings like $\xi (\Phi+\bar{\Phi}) |\phi^i|^2$ and $\lambda (\Phi+\bar{\Phi})^2 |\phi|^2$.
The former coupling enhance the partial decay rate into a pair of $\phi^i$ particles.
If it is a right-handed (s)neutrino, it may help leptogenesis~\cite{Fukugita1986} via its decay.
If it is the SUSY breaking field, it enhances gravitino production~\cite{EndoHamaguchiTakahashi2006a, NakamuraYamaguchi2006, DineKitanoMorisseShirman2006, EndoHamaguchiTakahashi2006}.
The latter coupling more strongly stabilize the $\phi^i$ field during inflation provided $\lambda <0$ in the case of the minimal K\"{a}hler potential and $\lambda>0$ in the case of the logarithmic K\"{a}hler potential.

We have seen in this Section that fields are stabilized near the origin by the Hubble induced masses.
They do not affect the inflaton potential unless SUSY breaking scale is comparable or higher than the inflation scale.  In such a case, SUSY breaking effects manifest itself as $W$ and $W_z$.
Actually, $W$ is not necessarily related to SUSY breaking scale in de Sitter spacetime, and other fields that makes $W$ large if present should be also taken into account.

\section{Generalization to charged superfields: MSSM Higgs inflation}\label{sec:generalization}
Now that we have studied basic features, stabilization mechanism, and inflationary dynamics of the  sGoldstino inflation in supergravity, let us explore some generalizations of our results.
All of the above examinations are based on the assumption that the inflaton is a gauge singlet.
We now consider an inflaton charged under some symmetries.
Examples of such inflationary models include Higgs inflation~\cite{Bezrukov2008}, inflection point inflation in an MSSM flat direction~\cite{Allahverdi2006}, and $D$-term chaotic inflation~\cite{Kadota2007}.
For Higgs inflation, the Higgs field is charged under SU(2).  For MSSM-flat-direction inflation, the flat direction is gauge invariant combination of charged fields.  For $D$-term chaotic inflation, inflaton is a combination of some charged fields.
Among these inflationary models, Higgs inflation is embedded into supergravity with the aid of a stabilizer superfield, which can be naturally identified with the singlet in NMSSM~\cite{Einhorn2010, Lee2010, Ferrara2010, Ferrara2011,Ben-Dayan2010}.\footnote{
There is also a small-field inflection-point inflation model with higher order terms in the superpotential~\cite{Chatterjee2011}, and a large-field model driven by soft masses with a monodromy structure and stringy corrections~\cite{Ibanez2014, Ibanez2014a, Bielleman2015}.
}
So, let us study the possibility to remove the singlet stabilizer.
We thus consider Higgs inflation in MSSM rather than in NMSSM.

We first remember some necessary conditions to reproduce appropriate low-energy electroweak theory after Higgs inflation.
The inflaton should settle to SUSY preserving vacuum after inflation since SUSY breaking is supposed to happen in a hidden sector.
The mass and VEV of the Higgs have to vanish approximately after inflation since these are electroweak scale and supposed to be much lower than the inflation scale.

As we have discussed in Subsection~\ref{subsec:HiggsReview}, all the known models of Higgs inflation in supergravity are based on some non-minimal K\"{a}hler potential.
We will also consider non-minimal K\"{a}hler potentials to try to embed Higgs inflation in MSSM, while we take the superpotential same as that in MSSM with a constant to cancel the cosmological constant,
\begin{align}
W=\mu H_{u}H_{d}+W_0. \label{WMSSM}
\end{align}

The possibility of Higgs inflation in MSSM was discussed by Einhorn and Jones for logarithmic K\"{a}hler potential~\cite{Einhorn2010} and by Ben-Dayan and Einhorn for minimal K\"{a}hler potential~\cite{Ben-Dayan2010}.
It was found that Higgs inflation does not occur in both cases since the potential becomes negative.
We revisit MSSM Higgs inflation in supergravity, with a stabilization term in the K\"{a}hler potential.

The formulations and results in the previous Sections can not be straightforwardly reused in the case of charged fields.
For example, one will immediately face troubles in using the stabilization terms in eqs.~\eqref{minKstab} and \eqref{K-a-no-scale} with a charged field $C$.  The combination
\begin{align*}
\zeta (C+\bar{C} - C_0 )^4
\end{align*}
is not invariant with respect to the symmetry under which $C$ is charged unless the representation of $C$ is equivalent to its complex conjugate representation.  Moreover, the constant $C_0$ must vanish unless $C$ is a singlet.
The shift symmetry has been generalized in the literature~\cite{Takahashi2010, Ben-Dayan2010}, whose results are applicable to charged fields.
One way is as follows,~\cite{Ben-Dayan2010}
\begin{align}
H_{u}&\rightarrow H_{u}+c, & H_{d}&\rightarrow H_{d}-i\sigma_{2}\bar{c}, \label{shiftConj}
\end{align}
where $c$ is a constant SU(2) doublet, and $\sigma_{2}$ is the Pauli matrix.
A combination,
\begin{align}
H \equiv H_{u}-i\sigma_{2}\bar{H}_d, 
\end{align}
transforms coherently as a doublet $\textbf{2}$.  The gauge invariant can be formed as $\bar{H}H$.
This is possible because the fundamental representation \textbf{2} of SU(2) is equivalent to its complex conjugate representation $\bar{\textbf{2}}$.
Note that this $H$ is not a (left) chiral superfield (in the narrow sense; it is a sum of left chiral (chiral) and right chiral (anti-chiral) superfields).
Another way to implement a shift symmetry is~\cite{Takahashi2010, Nakayama2010a}~\footnote{
A solution of the transformation in terms of $H_u$ and $H_d$ is
\begin{align}
H_u \rightarrow H_u - \frac{\sigma_2 \overline{H_d}}{2n|H_d|^2}c+\dots, \\
H_d \rightarrow H_d+ \frac{\sigma_2 \overline{H_u}}{2n|H_u|^2}c+\dots,
\end{align}
where dots denote higher order terms in the transformation parameter $c$, and they can be solved order by order.
This highly non-linear transformation forbids the remaining building blocks, $|H_u|^2$ and $|H_d|^2$, of gauge invariant combinations in the K\"{a}hler potential.
We do not want to forbid these terms exactly, but we want the symmetry to appear approximately at a high energy scale as an ``accidental'' symmetry. 
}
\begin{align}
(H_u H_d )^n \rightarrow (H_u H_d )^n + i c, \label{shiftComp}
\end{align}
where $n$ is an integer, and  $c$ is a constant.
The doublets are contracted as usual: it is more precisely written as $H_u i \sigma_2 H_d = H_u^{+} H_d^{-}-H_u^{0}H_d^{0}$.
We can generalize this further by replacing $(H_u H_d)^n$ with an arbitrary holomorphic function $J(-H_u H_d)$ of $-H_u H_d$.
If we take $c$ real, the invariant combination is the real part, $\text{Re}J(-H_u H_d)$.
The two shift symmetries \eqref{shiftConj} and \eqref{shiftComp} are not compatible with each other.

Because we do not want Higgs fields to have larger VEVs than the electroweak scale, we are led to the stabilization at the origin.
We can not have gauge invariants linear in Higgs fields, so it is non-trivial to have nonzero first derivatives of the K\"{a}hler potential so that the condition \eqref{KphiCondition}, \textit{i.e.}~$K^{i\bar{j}}K_{i}K_{\bar{j}}>3$, is satisfied.
Notice the fact that the no-scale type K\"{a}hler potential~\eqref{K-a-no-scale} or \eqref{K-arbitrary-stabilized} does not require nonzero VEVs to satisfy the condition \eqref{KphiCondition}.
One may consider to use a no-scale modulus $T$ to realize $K^{\bar{T}T}K_{T}K_{\bar{T}}=3$ so that the superpotential squared terms cancel.
But if it preserves SUSY at the inflation scale, 
 the negative superpotential terms of the inflaton sector is regenerated after one introduce the inflaton sector.  That is, the the supersymmetrically stabilized K\"{a}hler modulus shifts so that its $F$-term vanishes, and it destroys the no-scale cancellation.  This fact was pointed out in Ref.~\cite{Buchmuller2014a}.
Also, if one introduce another superfield at all, the NMSSM model~\cite{Einhorn2010, Lee2010, Ferrara2010, Ferrara2011,Ben-Dayan2010} is more elegant.

Although we can not use linear Higgs terms since it breaks the gauge symmetry, we can obtain a nonzero first derivative of the K\"{a}hler potential using the inflaton value during inflation if we take a suitable K\"{a}hler potential.
Consider the following K\"{a}hler potential,
\begin{align}
K=&|H_u|^2 + |H_d|^2 + c \left( J(H_u H_d)+ \bar{J} ( \overline{H_u}\overline{H_d} ) \right)  \nonumber \\
& + \frac{1}{2} \left( J(H_u H_d)+ \bar{J} ( \overline{H_u}\overline{H_d} ) \right)^2 -\frac{\zeta}{4}\left( J(H_u H_d)+ \bar{J} ( \overline{H_u}\overline{H_d} ) \right)^4 , \label{KHiggsMinimal}
\end{align}
where $c$ and $\zeta$ are real constants, and $J(X)$ is an arbitrary holomorphic function of $X$.
We will consider a class of function $J(H_u H_d)$ that is much larger than the canonical terms $|H_u|^2$ and $|H_d|^2$ in the field domain corresponding to inflation.
Thus, this K\"{a}hler potential has an approximate shift symmetry of (the generalized version of) the second type~\eqref{shiftComp}.

Neglecting the canonical terms and defining a composite chiral superfield $\Phi=J(H_u H_d)$, the above K\"{a}hler potential becomes 
\begin{align}
K\simeq c ( \Phi + \bar{\Phi}) + \frac{1}{2} ( \Phi + \bar{\Phi}) ^2 - \frac{\zeta}{4} ( \Phi + \bar{\Phi}) ^4.
\end{align}
This is the same as eq.~\eqref{minKstabilizedShifted} for single superfield models of inflation!
In terms of $\Phi$, the superpotential is
\begin{align}
W= \mu J^{-1}(\Phi ) + W_0, \label{WHiggsGeneral}
\end{align}
where $J^{-1}$ is the inverse function of the arbitrary holomorphic function $J$ but not the reciprocal of the value of $J$.

Let us first consider a simple example, $\Phi=J(X)=\kappa X$.
The parameter $\kappa$ should be large, $\kappa \gg 10^3$, for the shift-symmetry breaking terms in the K\"{a}hler potential not to affect the inflationary potential.
The inverse function is $X=J^{-1}(\Phi)=\kappa^{-1}\Phi$.
Plugging this into the MSSM superpotential~\eqref{WMSSM}, it becomes 
\begin{align}
W= \mu \kappa^{-1} \Phi + W_0. \label{WHiggsMinimalQuadratic}
\end{align}
This is the superpotential of the Polonyi model~\cite{Polonyi}.
The scalar potential is
\begin{align}
V=& \left|\frac{\mu}{\kappa}\right|^2 \left( (c^2-3)\left( \frac{\chi}{\sqrt{2}} +\text{Im}\widetilde{W}_0 \right)^2 +(1+c \text{Re}\widetilde{W}_0)^2 - 3 ( \text{Re}\widetilde{W}_0 )^2 \right) \nonumber \\
=&   \left|\frac{\mu}{\kappa}\right|^2 \left( (c^2-3)\frac{\chi^2}{2} +(1+c \widetilde{W}_0)^2 -3\widetilde{W}_0^2 \right) ,
\label{VHiggsMinimalQuadratic}
\end{align}
where $\widetilde{W}_0=W_0 (\mu / \kappa)^{-1}$, and we have required that it is real in the second equality.
Otherwise, the Higgs inflation ends with the inflationary scale VEV, and the electroweak scale becomes too large.
Also, note that the inflation scale is determined by the combination $\mu/\kappa$, and it is greater than $\mu$ itself.
For typical large-field or chaotic inflation, the scale of the inflaton mass $(\sim \mu / \kappa )$ is $\mathcal{O}(10^{13})$~GeV.
  For the constant term not to dominate the potential, $W_0$ should be at most of the order of the inflationary scale, $|\mu /\kappa | \gtrsim |W_0|$.

At this stage, we comment on the relation between the electroweak scale and the inflation scale.
For the above K\"{a}hler potential~\eqref{KHiggsMinimal} with $J(X)=\kappa X$, the holomorphic term in the logarithm induces a contribution to the effective $\mu$ term through Giudice-Masiero mechanism~\cite{Giudice1988}.
The mechanism is tightly related to SUSY breaking.
The effective $\mu$ term contributions originates from the $F$-term of the SUSY breaking field $z$ and the $F$-term of the compensator (in the superconformal formalism) or the auxiliary field of the gravity supermultiplet (in the curved superspace formalism). 
Combining it with the original $\mu$ term, the effective $\mu$ term is,
\begin{align}
\mu_{\text{eff}}=  \mu +K_{ud} W_0 - K_{ud\bar{z}}K^{\bar{z}z}D_z W \label{effectiveMu}
\end{align}
where we have used $\vev{W}=W_0$. 
To reproduce the electroweak scale, we have to satisfy the following constraint among the $\mu_{\text{eff}}$, and soft SUSY breaking masses $m_{H_u}^2$, $m_{H_d}^2$, and $B\mu_{\text{eff}}$~\cite{Martin2011},
\begin{align}
m_{Z}^2= \frac{|m_{H_d}^2 - m_{H_u}^2|}{\sqrt{1-\sin^2 2\beta}} -m_{H_u}^2 -m_{H_d}^2-2|\mu_{\text{eff}}|^2, \label{EWcondition}
\end{align}
where 
\begin{align}
\sin^2 2\beta= \frac{2B\mu_{\text{eff}}}{m_{H_u}^2+m_{H_d}^2+2|\mu_{\text{eff}}|^2}.
\end{align}
That is, we reduce the $\mu$-problem by taking $\mu$ larger than the inflation scale, but enhance the little hierarchy problem to a hierarchy problem.
This fine-tuning is the necessary cost we have to pay to realize $F$-term Higgs inflation in MSSM.

Let us see more closely what hierarchies are required to satisfy the above equations.
First, look at eq.~\eqref{EWcondition}.
There is a possibility to drive inflation with soft SUSY breaking masses~\cite{Ibanez2014, Buchmuller2015}, but if we assume the soft masses are smaller than the inflation scale (see below), the effective $\mu$ term ($\mu_{\text{eff}}$) has to be also smaller than the inflation scale ($\mu/\kappa$), to reproduce much more small scale $m_{Z}$ by fine-tuning.
To suppress the effective $\mu$ term, another tuning is needed in eq.~\eqref{effectiveMu}.
Since $\mu$ is larger than $W_0$, $|\mu| \gg |\mu / \kappa | \gtrsim |W_0|$, $K_{ud}$ or $K_{ud\bar{z}}$ has to be of order $\kappa$ for the cancellation.  When we discuss other examples of $J(H_u H_d)$ like $J=(\kappa H_u H_d)^n$ below, the presence of the term like $\kappa H_u H_d $ in the function $J$ is therefore implicitly assumed.  It is subdominant during inflation, but important to reproduce the electroweak scale. 

We assume that couplings between the inflaton superfield and the SUSY breaking field respect the shift symmetry allowing terms like $z(\Phi+\bar{\Phi})$, $|z|^2 (\Phi+\bar{\Phi})$, and $|z|^2(\Phi+\bar{\Phi})^2$ \textit{etc}.
The first term induces the kinetic mixing between $z$ and $\Phi$, so it should be suppressed to some extent.
If we take parameters so that these terms do not change the value of $\Phi+\bar{\Phi}$ much during inflation, its supersymmetric effects (\textit{e.g.} change of field normalization) on the SUSY breaking field $z$ is negligible aside from the above mentioned kinetic mixing effect.
Similarly, the effects from the SUSY breaking field to the non-inflaton is negligible provided that the VEV of $z$ is suppressed.
On the other hand, these fields give masses to each other vie their SUSY breaking $F$-terms helping  stabilization of each field.
Because of the shift symmetry, the inflaton potential is not affected as far as the value of $\Phi+\bar{\Phi}$ is suppressed enough.
In contrast, these couplings give us some freedom to satisfy the phenomenological constraint~\eqref{EWcondition} at the vacuum.
So, we just assume, at the current level of investigation, that it is satisfied, and continue further to explore possibilities of MSSM Higgs inflation.

We need to cancel the cosmological constant at the vacuum, and it implies the SUSY breaking scale is the same as $W_0$.
When it is comparable to the inflation scale\footnote{
SUSY models with SUSY broken around the intermediate scale have been studied in the literature~\cite{Hall2009, Hall2013, Hall2014, Ibanez2012, Ibanez2013, Ibanez2014, Hebecker2012, Hebecker2013}.
It is motivated by the 125 GeV mass of the Higgs boson~\cite{ATLAS2015}, and by the fact that the Higgs quartic term crosses zero around the intermediate scale although the current uncertainty is sizable.
Connections of the scale to axion physics and right-handed neutrinos have also been discussed~\cite{Hall2014}.
}, we can not simply neglect the SUSY breaking sector.
A complete analysis involves the dynamics of the SUSY breaking field, but it depends on the SUSY breaking model.
One may want to take advantage of the nilpotent field to eliminate the dynamical scalar $z$ from the theory (see Section~\ref{subsec:nilpotent}).
But in our case, inflaton also breaks SUSY during inflation, and the consistency of the nilpotency condition only on $z$ is doubtful.
For simplicity, we add the following terms to the K\"{a}hler potential and the superpotential,
\begin{align}
\Delta K=& |z|^2 -\xi |z|^4, \label{DeltaK} \\
\Delta W=& lz,
\end{align}
where the $\xi$ term is introduced to freeze the dynamics of $z$ near the origin.
The mass of $z$ is proportional to $\sqrt{\xi}$.
The VEV of $z$ is inversely proportional to $\xi$, so we neglect it by taking a large $\xi(\gg 1)$.
Thus, we have dynamical $z$, but it is decoupled and the results are similar to the nilpotent case.
As discussed in the previous paragraph, there may be coupling terms between $z$ and $\Phi$, but we assume that their effects are insignificant during inflation because of their small VEVs ensured by the large stabilization parameters $\zeta$ and $\xi$.
At the vacuum ($K \simeq H_u\simeq H_d\simeq 0$), the condition $V=0$ implies $|l|^2\simeq 3W_0^2 =3\mu^2/c^2 \kappa^2$.
In our approximation, the $W_{z}\bar{W}$ term or its conjugate is absent since it is proportional to $K_{\bar{z}}$ or its conjugate and they in turn are proportional to $z$ or $\bar{z}$.
So we have to add 
\begin{align}
\Delta V=e^K K^{\bar{z}z}|W_{z}+K_{z}W|^2=e^K|l|^2 =3e^{K}|W_0|^2 . \label{DeltaV} 
\end{align}
With this uplift, the potential~\eqref{VHiggsMinimalQuadratic} becomes
\begin{align}
V=&   \left|\frac{\mu}{\kappa}\right|^2 \left( (c^2-3)\frac{\chi^2}{2} +(1+c \widetilde{W}_0)^2  \right) .
\end{align}
In this expression, the cosmological constant does not vanish at the origin, but the description in terms of $\Phi$ is not valid in the small field region.
In terms of $H_u$ and $H_d$, it is clear that it actually vanishes.

Although we must impose the phase alignment condition $\text{Im}\widetilde{W}_0=0$ for inflaton to go to the origin and the $Z$ mass condition~\eqref{EWcondition} around the origin to reproduce the electroweak scale, it is remarkable that the simple quadratic potential for the Higgs-inflaton emerges in MSSM.
It is implied that our strategy can be more easily applied to other charged superfields not necessarily related to the visible standard model sector.

Let us consider radiative corrections to the inflaton potential.\footnote{
Some of the following points became clear in discussion with T.~Kitahara, K.~Mukaida, and M.~Takimoto.
}
 During inflation, SUSY must be broken.
 Masses of bosons and fermions typically split by the Hubble scale $H$.
 If couplings between these supermultiplets and Higgses are not suppressed, self-energy corrections to the Higgses are of order the Hubble scale $H$.
 It means that the slow-roll parameter $\eta$ is affected significantly.
 This ruins the inflationary scenario of the model.
 In the standard case with the stabilizer superfield $S$, for example, SUSY breaking of $S$ is indeed the same scale as the Hubble scale, $3H^2=V=|W_{S}|^2$.
 Also in our case, SUSY breaking of the inflaton $\Phi$ is the same scale as the Hubble scale, $3H^2=V \sim |W|^2 \sim |D_{\Phi}W|^=|W_{\Phi}+K_{\Phi}W|^2$.
 Since the couplings between the inflaton and other fields break the shift symmetry, such couplings are supposed to be suppressed by the parameter governing the shift symmetry breaking, which is typically the energy scale of inflation or inflaton mass scale.
 However, in our case, the inflaton as the Higgses apparently has unsuppressed couplings to the MSSM sector.
 Actually, this is not the case if we take into canonical normalization into account.
 If we call the coupling constant of the Higgs $H_u$ or $H_d$ to some particle $y$ (\textit{e.g.}~Yukawa coupling), the coupling between the particle and the inflaton $\Phi = \kappa H_u H_d  $ is scaled as $y/\kappa$. 
 Since we assume $\kappa$ is large $\kappa \gg 1$, the quantum correction is suppressed by $\kappa$ and the loop factor such as $(16\pi)^{-1}$.

Let us revive the canonical terms in  the K\"{a}hler potential, and study the property of running kinetic term.
We first neglect the stabilization term and consider the remaining part.
Furthermore, we truncate the theory to the neutral sector along the lines of Refs.~\cite{Einhorn2010, Ferrara2010, Ferrara2011,Ben-Dayan2010}.
We will come back to these points later.
\begin{align}
K=&|H_u^0|^2 + |H_d^0|^2 + c \left( J(-H_u^0 H_d^0)+ \bar{J} ( -\overline{H_u^0}\overline{H_d^0} ) \right)  \nonumber \\
& + \frac{1}{2} \left( -J(H_u^0 H_d^0)+ \bar{J} (- \overline{H_u^0}\overline{H_d^0} ) \right)^2 -\frac{\zeta}{4}\left( J(-H_u^0 H_d^0)+ \bar{J} ( -\overline{H_u^0}\overline{H_d^0} ) \right)^4 , \label{KHiggsMinimalT}
\end{align}
The first derivatives in the $K$-flat direction~\cite{Ben-Dayan2010}  ($J+\bar{J}=0$) are $K_u=\overline{H_u^0}-c  H_d^0 J'$, and $H_d=\overline{H_d^0}-c H_u^0 J'$.
The K\"{a}hler metric and its inverse in the $K$-flat direction  are
\begin{align}
K_{i\bar{j}}=&
\begin{pmatrix}
1+ |H_d^0|^2 |J'|^2 &  H_d^0 \overline{H_u^0} |J'|^2 \\
H_u^0 \overline{H_d^0} |J'|^2 & 1+ |H_u^0|^2 |J'|^2
\end{pmatrix} , \\
K^{\bar{j}i}=& \frac{1}{1+ \left( |H_u^0|^2+|H_d^0|^2 \right)|J'|^2}
\begin{pmatrix}
1+ |H_u^0|^2 |J'|^2 &  - H_d^0 \overline{H_u^0} |J'|^2\\
-H_u^0 \overline{H_d^0} |J'|^2 & 1+ |H_d^0|^2 |J'|^2\end{pmatrix}.
\end{align}
The eigenvalues of the K\"{a}hler metric are 
\begin{align}
 1+ \left( |H_u^0|^2+|H_d^0|^2 \right)|J'|^2, & & \text{and} & & 1, \label{minKeigenvalues}
\end{align}
and the corresponding eigenvectors are
\begin{align}
\left( \vev{\frac{H_d^0}{H_u^0}}, 1 \right) , & & \text{and} & & \left( -\vev{\frac{\overline{H_u^0}}{\overline{H_d^0}}} , 1 \right), \label{minKeigenvectors}
\end{align}
respectively.
In the large $J$ limit, the latter eigenvalue becomes large, and this fact exactly match that the theory apparently depends only on $\Phi$ in the $J \rightarrow \infty$ limit.
So, we identify the latter mode with the inflaton.

Now, we parametrize the neutral Higgs fields as follows,
\begin{align}
H_u^0=&h \cos \beta , &  H_d^0=& h \sin \beta e^{i\alpha}, \label{HiggsParametrization}
\end{align}
where we have taken into account the fact that one phase degree of freedom is eaten by the gauge boson (Higgs mechanism), and we take $0\leq \beta \leq \pi/2$.
The mode corresponding to $\Phi$ is
\begin{align}
u \equiv \vev{\sin \beta e^{i \alpha}} h \cos \beta + \vev{\cos \beta} h \sin \beta e^{i \alpha} \sim h \sin 2\beta e^{i\alpha} , \label{u-field}
\end{align}
and the mode perpendicular to it is
\begin{align}
v \equiv \vev{\cos \beta} h \cos \beta -\vev{\sin \beta e^{-i \alpha}} h \sin \beta e^{i \alpha} \sim h \cos 2 \beta. \label{v-field}
\end{align}

Let us express the $K$- and $D$-flat directions in terms of the parametrization~\eqref{HiggsParametrization}.
For simplicity, we take $J=(-\kappa H_u^0 H_d^0)^n$, but other cases can be treated similarly.
The stabilization term  has the form $\zeta \kappa^{4n} ( (-H_u^0 H_d^0)^n + (-\overline{H_u^0}\overline{H_d^0})^n )^4 = \zeta  (2 \kappa^n h^{2n} \sin^n \beta \cos^n \beta \cos n\alpha)^4$.
Thus, the $K$-flat direction 
 in which field (approximately) does not experience the $e^{K}$ factor is 
\begin{align}
\sin^n \beta \cos^n \beta  \cos n \alpha =0 & & \text{($K$-flat direction)}. \label{K-flat}
\end{align}
The $D$-term potential
\begin{align}
V_{D}=&\frac{ (g^2+g'^2)}{8 }\left( |H_u^0|^2-|H_d^0|^2 \right)^2 \nonumber \\
=& \frac{ (g^2+g'^2)}{8 } h^4 \cos ^2 2\beta , \label{VHiggsD}
\end{align}
defines the $D$-flat direction,
\begin{align}
\cos 2 \beta = 0 & & \text{($D$-flat direction)}. \label{D-flat}
\end{align}
In the above, we have assumed that the gauge kinetic function is minimal, $H_{AB}=\delta_{AB}$.

The joint subspace of the $K$- and $D$-flat directions \eqref{K-flat} and \eqref{D-flat} is the inflaton candidate direction, $\beta = \pi/4$ and $\alpha=\pm \pi/2n$.
Substituting these into eqs.~\eqref{HiggsParametrization}, \eqref{u-field}, and \eqref{v-field}, we can extract the inflaton component,
\begin{align}
H_u^0=&\frac{h}{\sqrt{2}}= \frac{u}{\sqrt{2}}e^{\mp i \pi /2 n}  , &  H_d^0=& \frac{h}{\sqrt{2}}e^{\pm i \pi /2n}=\frac{u}{\sqrt{2}} , \label{InflatonInHiggses}
\end{align}
in the $K$- and $D$-flat direction.  Note that $v=0$ in the $D$-flat direction.

After the angular modes (and the charged fields) are stabilized, the inflaton is identified with the radial mode.
Substituting eq.~\eqref{InflatonInHiggses} and $J=(-\kappa H_u^0 H_d^0)^n$ into the eigenvalues of the K\"{a}hler metric, they become
$1$ and $1+2 n^2 \kappa^{2n} (h/2)^{4n-2}$.
The kinetic term of the inflaton is 
\begin{align}
-(1+4 n^2 \kappa^{2n} (h/2)^{4n-2}) \partial^{\mu}h\partial_{\mu}h \equiv -\frac{1}{2} \partial^{\mu}\widetilde{h}\partial_{\mu}\widetilde{h}, \label{HiggsKinetic}
\end{align}
where we have defined the canonically normalized inflaton $\widetilde{h}$.
When the field is sufficiently large, $4 n^2 \kappa^{2n} (h/2)^{4n-2} \gg 1$, the canonically normalized inflaton is $\widetilde{h}\simeq 2 \sqrt{2} \kappa^n (h/2)^{2n}$.
Taking $n=1$, this is identified with the canonically normalized inflaton $\chi= \sqrt{2} \text{Im}\Phi=\sqrt{2}\text{Im} J(-H_u^0 H_d^0)=\kappa h^2 /\sqrt{2}$ in eq.~\eqref{VHiggsMinimalQuadratic} justifying our previous treatment.

We now move on to the inflationary potentials originating from more generic $J(-H_u^0 H_d^0)$.
We can not differentiate the inverse function of an arbitrary function, and have to assume the functional form of $J$.
For simple choices such as a polynomial, its inverse function has poles and branch cuts in the field space of $\Phi$.
If the original function $J$ were such a function with poles and branch cuts, observables such as the effective gravitational coupling (or precisely the corresponding cross sections or rates) change when the phase of the Higgs rotate $2\pi$.
The Riemann surfaces of the Higgses would be physically inequivalent.
If it could be justified, one can obtain some holomorphic functions $J^{-1}$.
We do not pursue this further, and assume a legitimate holomorphic function $J$.
In contrast to the above discussion, poles and branch cuts are not necessarily a problem in terms of a composite superfield $\Phi$.
The inequivalent Riemann surfaces are artifacts of field redefinition $\Phi=J(-H_u^0 H_d^0)$.
This comment can be applied to the superpotentials in Section~\ref{sec:arbitrary}, if we assume the singlet inflaton $\Phi$ in the Section is also originated from some composite chiral superfields.

As the next example, we take $J= ( \kappa H_u H_d )^n$.
The qualitative feature is the same if we take an $n$-th order polynomial which also contains terms of powers less than $n$ if the largest power term is dominant in the large field region.
We identify the inflaton superfield $\Phi$ with $J= (-\kappa H_u^0 H_d^0 )^n$.
The inverse function is $H_u H_d=J^{-1}(\Phi )=\Phi^{1/n} / \kappa $.
That is, the superpotential~\eqref{WHiggsGeneral} is 
\begin{align}
W=  \frac{\mu}{\kappa}\Phi^{\frac{1}{n}} +W_0, \label{WHiggsFractional}
\end{align}
The scalar potential is
\begin{align}
V=& \left| \frac{\mu}{\kappa}\right|^2 (c^2-3) \left(  \left( \frac{\chi}{\sqrt{2}}\right)^{\frac{2}{n}} + 2 \left( \frac{\chi}{\sqrt{2}}\right)^{\frac{1}{n}} \left( \text{Re}\widetilde{W}_0 \cos\frac{\pi}{2n} + \text{Im}\widetilde{W}_0 \sin \frac{\pi}{2n} \right)  +|\widetilde{W}_0|^2  \right) \nonumber \\
& + \left| \frac{\mu}{\kappa}\right|^2 c^2 \left(  \frac{2}{n}\left( \frac{\chi}{\sqrt{2}}\right)^{\frac{1}{n}-1}\left( \text{Re}\widetilde{W}_0 \sin \frac{\pi}{2n} - \text{Im}\widetilde{W}_0 \cos \frac{\pi}{2n} \right) + \frac{1}{n^2} \left( \frac{\chi}{\sqrt{2}}\right)^{\frac{2}{n}-2}     \right).
\end{align}
It seems that the terms in the second line blow up at the origin, but it is an artifact of invalid extrapolation of description in terms of $\Phi$ into the small field region where that in terms of $H_u$ and $H_d$ is appropriate.  The description in terms of $\Phi$ is valid in the region satisfying $c\kappa \Phi^{(n-1)/n} \gg  1$.
The SUSY breaking vacuum energy $\Delta V=3 |\mu/ \kappa |^2 |\widetilde{W}_0|^2$ is to be added to the above potential to cancel the cosmological constant at the vacuum.
We have obtained a fractional power potential exploiting the mechanism of running kinetic term.

Next, we consider the following logarithmic K\"{a}hler potential,
\begin{align}
K=& -a \ln \left( 1+\frac{1}{\sqrt{a}}\left( J(H_uH_d)+\bar{J}(\overline{H_u}\overline{H_d} )\right) \right. \nonumber \\
& \left. \qquad \qquad -\frac{1}{a} \left( \left| H_u \right| ^2 + \left| H_d \right|^2 \right)+\frac{\zeta}{a^2} \left( J(H_uH_d)+\bar{J}(\overline{H_u}\overline{H_d}) \right)^4   \right), \label{KHiggsLog}
\end{align}
Defining a chiral superfield $\Phi$ as $\Phi \equiv J(H_u H_d)$, it becomes $K=-a \ln (1+(\Phi+\bar{\Phi})/\sqrt{a}+\dots)$.
Similarly to the previous, minimal K\"{a}hler case, we first neglect the canonical $|H_u|^2$ and $|H_d|^2$ terms.
In terms of the composite chiral superfield $\Phi$, the resulting K\"{a}hler potential is the same as eq.~\eqref{K-arbitrary-stabilized}.
So the superpotential squared term is cancelled $(a=3)$ or made positive $(a\geq 4)$ because of $K^{\bar{\Phi}\Phi}K_{\Phi}K_{\bar{\Phi}}=a$, see eq.~\eqref{KphiCondition}.
The MSSM superpotential~\eqref{WMSSM} morphs again into eq.~\eqref{WHiggsGeneral}.

Let us take a polynomial $\Phi=J(H_u H_d)=(\kappa H_u H_d)^n$, or $H_u H_d=J^{-1}(\Phi)=\Phi^{1/n} /\kappa$.
The superpotential is given by eq.~\eqref{WHiggsFractional}, $W=(\mu /\kappa )\Phi^{1/n} +W_0$. 
Plugging this into the potential obtained in the previous Section, eq.~\eqref{V-a-no-scale}, 
the potential becomes
\begin{align}
V=&\left| \frac{\mu}{\kappa} \right|^2 (a-3) \left( \left( \frac{\chi}{\sqrt{2}}\right)^{\frac{2}{n}} + 2 \left( \frac{\chi}{\sqrt{2}}\right)^{\frac{1}{n}} \left( \text{Re} \widetilde{W}_0 \cos \frac{\pi}{2n} +\text{Im}\widetilde{W}_0 \sin \frac{\pi}{2n}  \right)  +|\widetilde{W}_0|^2  \right) \nonumber \\
&+ \left| \frac{\mu}{\kappa} \right|^2 \left( \frac{2\sqrt{a}}{n}   \left( \frac{\chi}{\sqrt{2}}\right)^{\frac{1}{n}-1}   \left( \text{Im} \widetilde{W}_0 \cos \frac{\pi}{2n} -\text{Re}\widetilde{W}_0 \sin \frac{\pi}{2n}  \right) +\frac{1}{n^2} \left( \frac{\chi}{\sqrt{2}}\right)^{\frac{2}{n}-2}   \right), \label{VHiggsALog}
\end{align}
where $\widetilde{W}_0= W_0 (\mu / \kappa)^{-1}$.
For the no-scale like case $(a=3)$, the first line vanishes.
Let us take $n=1$ for example.  The above potential reduces to
\begin{align}
V=&\left| \frac{\mu}{\kappa} \right|^2 (a-3) \left( \left( \frac{\chi}{\sqrt{2}}\right)^{2} + 2 \text{Im}\widetilde{W}_0 \left( \frac{\chi}{\sqrt{2}}\right)  +|\widetilde{W}_0|^2  \right) + \left| \frac{\mu}{\kappa} \right|^2 \left( -2\sqrt{a} \text{Re}\widetilde{W}_0   +\frac{1}{n^2}   \right). \label{VHiggs3Log}
\end{align}
For the case of $a>3$, the potential becomes quadratic as in the previous $n=1$ example for the minimal K\"{a}hler potential.
For the no-scale like case $(a=3)$, the non-trivial potential vanishes and just a constant remains.
When we take $n>1$, the power of the leading term is $-2(n-1)/n$ (negative fractional; run-away) and $2/n$ (positive fractional) for $a=3$ and $a>3$ respectively.
That is, the qualitative behavior of the potential for the case of the logarithmic K\"{a}hler potential of $a>3$ is same as the counterpart for the minimal K\"{a}hler potential.
On the other hand, the situation in the no-scale like case is worse due to the absence of the superpotential squared term.  If we take some non-trivial function $J$, the no-scale case may also be able to realize positive power potential.

Actually, we should be careful about canonical normalization since exponential stretching occurs for the logarithmic K\"{a}hler potential.
Similarly to the previous exercise with the minimal K\"{a}hler potential, we truncate the theory to the neutral sector. 
\begin{align}
K=& -a \ln \left( 1 + \frac{1}{\sqrt{a}} \left(J(-H_u^{0} H_d^{0}) + \bar{J}(-\overline{H_u^{0}} \overline{H_d^{0}} ) \right) -\frac{1}{a} \left( \left| H_u^0 \right| ^2 + \left| H_d^0 \right|^2 \right)  \right). \label{KHiggsT}
\end{align}
The first derivatives in the $\text{Re}J(-H_u^0 H_d^0)$ direction (``$K$-flat direction''~\cite{Ben-Dayan2010}) are 
\begin{align}
K_u=& \frac{ \overline{H_u^0} -\sqrt{a} H_d^0 J'(H_u^0H_d^0)}{(\Omega /3 )}, &
K_d=& \frac{ \overline{H_d^0} -\sqrt{a} H_u^0 J'(H_u^0H_d^0)}{(\Omega /3 )},
\end{align}
where $\Omega/3$ denotes the argument of the logarithm, $\Omega/3=e^{-K/3}$.
The K\"{a}hler metric  in the $K$-flat direction is
\begin{align}
K_{u\bar{u}}=&  \frac{1}{(\Omega/3)^2}\left(  1 + \left( |J'|^2 -\frac{1}{a}\right) |H_d^0|^2 +\frac{1}{\sqrt{a}}\left( H_u^0 H_d^0 J'+\overline{H_u^0}\overline{H_d^0}\bar{J}'\right) \right) , \label{Kuub} \\
K_{u\bar{d}}=& \frac{1}{(\Omega/3)^2} \left(  \left( |J'|^2+\frac{1}{a} \right) H_d^0 \overline{H_u^0} + \frac{1}{\sqrt{a}}\left(\overline{H_u^0}^2\bar{J}'+{H_d^{0}}^{2}J'\right) \right). \label{Kudb}
\end{align}
$K_{d\bar{d}}$ and $K_{d\bar{u}}$ are obtained by exchanging $u$ and $d$ in these expressions.
After the angular modes (and the charged fields) are stabilized, the inflaton is identified with the radial mode.
Of the two eigenvalues of the K\"{a}hler metric, the coefficient of the kinetic term of the inflaton is the one that grows larger in the large $J$ limit. 
In the $K$-flat direction, the eigenvalue and the corresponding eigenvector are
\begin{align}
&\frac{\left(|H_d^0|^2+|H_u^0|^2 \right)  |J'|^2 + 1+\frac{2}{\sqrt{a}} \left( H_u^0 H_d^0 J'+\overline{H_u^0}\overline{H_d^0}\bar{J}' \right)  }{\left(1-\frac{1}{a} (|H_u^0|^2+|H_d^0|^2)\right)^2} , \label{inflatonKinetic} \\
&\left( \frac{\overline{H_u^0}+\sqrt{a}H_d^0 J'}{\overline{H_d^0}+\sqrt{a}H_u^0J'} , 1 \right). \label{logKmetU}
\end{align}
The condition for neglecting canonical terms is
\begin{align}
\frac{ 1}{|J'|^2  } \left( 1+\frac{2}{\sqrt{a}} \left( H_u^0 H_d^0 J'+\overline{H_u^0}\overline{H_d^0}\bar{J}' \right)\right)   \ll |H_u^0|^2+|H_d^0|^2 \ll a. \label{kappaNeglectC}
\end{align}
The left inequality comes from the numerator of eq.~\eqref{inflatonKinetic} and the right inequality comes from the denominator.
Eqs.~\eqref{VHiggsALog} and \eqref{VHiggs3Log} are valid in this range~\eqref{kappaNeglectC}.
We also record the other eigenvalue and the corresponding eigenvector of the K\"{a}hler metric,
\begin{align}
\frac{1}{1-\frac{1}{a} (|H_u^0|^2+|H_d^0|^2)}, & & \left(  -\frac{H_d^0 +\sqrt{a}\overline{H_u^0} \bar{J}'}{H_u^0 + \sqrt{a}\overline{H_d^0}\bar{J}'}  , 1 \right) \label{logKmetV}
\end{align}

When the function $J$ can be approximated as a monomial during inflation, 
the last term in the numerator of the eigenvalue~\eqref{inflatonKinetic} vanishes in the $K$-flat direction.
In the large field region satisfying the left inequality of eq.~\eqref{kappaNeglectC}, and using eqs.~\eqref{InflatonInHiggses},  the inflaton kinetic term is approximately
\begin{align}
-\left( \frac{ 2n \kappa^{n}(h/2)^{2n-1}}{1- \frac{1}{a}h^2} \right)^2 \partial^{\mu}h \partial_{\mu} h \simeq -\frac{1}{2}\partial^{\mu}\widetilde{h}\partial_{\mu}\widetilde{h}.
\end{align}
Integrating eq.~\eqref{HiggsKinetic}, we can express the canonical inflaton using a hypergeometric function.
Restricting ourselves to the case of $n=1$,  we have
\begin{align}
\widetilde{h}\simeq &-\frac{a\kappa}{\sqrt{2}} \ln \left( 1-\frac{1}{a}h^2 \right), & \text{or} & &  
h\simeq &  \sqrt{a  \left( 1-e^{-\sqrt{2}\widetilde{h}/a\kappa } \right) } . \label{HiggsCanonicalNormalization} \end{align}
Hence, the conformal factor $(\Omega/3)^{-1}=e^{K/3}=e^{\sqrt{2}\widetilde{h}/a \kappa}$ blows the potential up.
The description in terms of $\Phi$ is valid in the region $\frac{1}{2\kappa^2} \ll h^2 \ll a$.
If we take $\kappa$ large, the exponential deformation happens at the very large point $\widetilde{h}\sim a \kappa /\sqrt{2}$ for the canonical inflaton so that the relevant inflationary regime can be described by the composite superfield $\Phi$.
In fact, when the field is not too large, $\widetilde{h}\ll a \kappa /\sqrt{2}$, eq.~\eqref{HiggsCanonicalNormalization} reduces to the previous result, $\widetilde{h}\simeq \kappa h^2 /\sqrt{2}$.

Finally, let us check stability of non-inflaton scalar fields along the inflationary trajectory.
We first truncate it to the neutral sector, and deal with charged fields later.
The real part of the inflaton $\Phi=J=(-\kappa H_u^0 H_d^0)^n$  is stabilized by the $\zeta (\Phi+\bar{\Phi})^4$ term, while its imaginary part plays the role of inflaton.
The masses of these degrees of freedom during inflation is more easily understood in terms of $\Phi$ than in terms of $h$, $\alpha$ or $\beta$.
It is similar to the previous Sections, and the stabilized component has mass squared proportional to the stabilization coefficient $\zeta$.
It can be easily made heavier than the Hubble scale during inflation (Section~\ref{sec:stabilization}).

The eigenvectors~\eqref{logKmetU} and \eqref{logKmetV} for the logarithmic K\"{a}hler potential coincide with those for the minimal K\"{a}hler potential, eq.~\eqref{minKeigenvectors}, in the large filed (large $J$) limit.  So, we consider the mass of the non-inflaton field $v$ in eq.~\eqref{v-field}.
In the case of the logarithmic K\"{a}hler potentials, the K\"{a}hler metric and the $D$-term potential have the conformal factor $(\Omega /3)$.  During inflation, this factor should play a subdominant role, so we set them one.
Then, the mass term of the canonically normalized $v$, $\widetilde{v}=\sqrt{2}v$, in the $D$-term~\eqref{VHiggsD} is 
\begin{align}
V_D\simeq \frac{1}{2}\left( \frac{ (g^2+g'^2)}{8} h^2 \right)\widetilde{v}^2,
\end{align}
where $h^2=|u|^2+v^2$ is identified with the value of inflaton $\chi=h^2/2\sqrt{2}$ during inflation.
The mass of $\widetilde{v}$ can be readily read off from this expression.
The expression of the Hubble parameter depends on the model, but its scale is determined by $\mu \kappa^{-1}$.
Thus, the condition for neglecting $\widetilde{v}$ during inflation is roughly
\begin{align}
\sqrt{ \frac{g^2+g'^2}{\kappa}} \gg \frac{\mu }{\kappa}, \label{vNeglectC}
\end{align}
in the reduced Planck unit.  We have used $h\sim \kappa^{-1/2}$.
The right hand side of the inequality~\eqref{vNeglectC} is to be fixed to the inflationary scale, so the $\kappa$ can not become arbitrarily large.  This defines the upper bound on $\kappa$.
Combining with eq.~\eqref{kappaNeglectC}, the inflationary trajectory can not be arbitrarily extended.

We confirm stability of charged fields $H_u^{+}$ and $H_d^{-}$.
A half of degrees of freedom are eaten by gauge fields, so we keep only $H^{+}=H_u^{+}$ and set $H_d^{-}=0$.
We assume that the mass scale of $D$-term is larger than that of $F$-term.
In the $D$-flat direction, the $D$-term including the charged Higgs mass term is
\begin{align}
V_{D}=\frac{1}{2} \left( \frac{ g^2 h^2}{2} \right) \left| {H^{+}}\right|^2.
\end{align}
The condition that this mass is larger than the Hubble scale is also given by the inequality~\eqref{vNeglectC}.

We have checked the conditions that other field directions than the inflaton are stable against fluctuations perpendicular to the inflaton trajectory.
When these are satisfied, our analysis is consistent.

In conclusion, we have constructed the models of Higgs inflation in the MSSM for the first time at the cost of giving up naturalness of the electroweak scale.
The construction in this Section can be readily applied to building of composite or running-kinetic inflaton models not involving the MSSM Higgses where the issue on the electroweak naturalness is irrelevant.

\chapter{Conclusion}
\label{ch:conclusion}

In this thesis, we have constructed the new classes of inflationary models in supergravity without the stabilizer superfield, which does not contain the inflaton.
Moreover, the particular class of models discussed in Section~\ref{sec:arbitrary} can be used to embed approximately arbitrary positive semidefinite scalar potentials for inflation into supergravity.
The mechanism we employed for stabilization of the inflaton potential utilizes the inflaton quartic term in the K\"{a}hler potential.
The quartic stabilization term ensures the positivity of the inflaton potential in the large field region, and fixes the scalar superpartner of the inflaton to a certain value making its heavy enough to decouple from the theory during inflation.
We extensively studied the effects of the stabilization term in Section~\ref{sec:stabilization} such as the deviation of the non-inflaton scalar value, its mass, and the resultant correction to the inflaton scalar potential.
We have checked analytically and numerically that the stabilization mechanism works indeed.
In the numerical examples, we found that the stabilization parameter $\zeta$ of order one or ten is enough (we used $\zeta=3\sqrt{3}$).
The mechanism respects the approximate shift symmetry of the theory, and any terms preserving the symmetry generically appear in the effective field theory.
We discussed the naturalness and tuning issue in Section~\ref{sec:OtherTerms}, \textit{i.e.}~we studied effects of other terms allowed by the shift symmetry in the K\"{a}hler potential.
For the ``minimal'' K\"{a}hler potential, no fine tuning is required provided that if the non-inflaton scalar is stabilized at all.  That is, perturbation to the K\"{a}hler potential or superpotential just perturbs the kinetic terms and the scalar potential.
For the no-scale type K\"{a}hler potential, on the other hand, the theory is to be regarded as a fine tuned one, because a small perturbation to the K\"{a}hler potential breaks cancellation among terms and drastically change the structure of the theory.  It does not mean inflation is not possible in the perturbed theory, but it just means that one has to check again whether inflation occurs or not in the new theory.
We also briefly studied coupling of the inflaton sector and other sectors possibly present in the theory and inflaton decay into particles in these sectors in Section~\ref{sec:coupling}.
As a generalization and application of our technique, we constructed models of Higgs inflation in MSSM in Section~\ref{sec:generalization}.
Although it requires fine-tuning, potentials of Higgses that are suitable for large-field inflation have been constructed in MSSM (in the supergravity framework) for the first time.  Our approach is more easily applied for charged or composite superfields not directly related to the MSSM sector.

Related to the tuning issue of cubic term in the logarithmic K\"{a}hler potential, one would ask a question: ``What is the origin of the stabilization term if any?''
This is unanswered in this thesis and an open question.
The simplest answer is it requires no answer.
That is, the quartic term is present from the beginning at the tree-level action.
There are no practical problems, and the QFT with such a term is consistent at the same level with other QFT models.
A more ambitious attitude is to seek a possible origin in some UV-completed fundamental theory, but this is beyond the scope of this thesis.
An intermediate answer is that such a term can be generated integrating out heavy fields that couples to the inflaton superfield.
Although we can demonstrate that the effective quartic term as well as other terms like quadratic or cubic terms appears from such a procedure, but we have not found a way of coupling that improves naturalness or tuning status.  So, we will not go into these details.  Interested readers are referred to Appendix C of Ref.~\cite{Ketov2014a}.

How about predictions or falsifiability of our framework?
Generally speaking, it is difficult to pin down a particular model of inflation because of the number of inflationary observables we have is limited.
Even if we could pin down an inflationary model, it is again very difficult to prove it is embedded in supergravity theory because the fermionic sector does not play a major role in the inflationary context.
So, it is far more difficult to distinguish several embedding mechanisms of the inflationary model into supergravity.
If we would have infinite power to probe high-energy or microscopic physics, we could in principle exclude the model/framework or confirm its consistency.
Of course, we do not such power unfortunately.

However, \emph{if non-Gaussianity is discovered, our scenario is excluded} or at least disfavored or constrained\footnote{
Decreasing the stabilization paramter $\zeta$, multi-field dynamics may be possible.
}
 because the scalar superpartner $\phi$ of inflaton $\chi$ is stabilized and inflation is triggered effectively by a single field $\chi$.
Also, in \emph{some} (not all) of our models, inflaton breaks SUSY not only during but also after inflation.
\emph{Without tuning, the gravitino mass is as high as the inflaton mass scale.}
This has a huge impact on particle phenomenology.
The standard lore of low-energy SUSY suppose SUSY particles are at around TeV scale for the naturalness of Higgs mass.
The mass of the Higgs, $125$ GeV, implies SUSY scale may be as high as $10$ or $100$ TeV.
Such scales of SUSY breaking can explain the Higgs mass and keep gauge coupling unification.
The SUSY breaking scale in our scenario is much higher than this.
Because of such a high scale SUSY breaking, \emph{no SUSY particles will be found at the LHC} for our generic models in Section~\ref{sec:strategy}.
On the other hand, the special class of models in Section~\ref{sec:arbitrary} does not break SUSY after the end of inflation provided the constant in the superpotential is tuned, so the scale of SUSY breaking, which is supposed to take place in a hidden sector, can be much lower than the inflation scale.
As other examples of promising directions, relations between our single superfield models and baryon asymmetry of the universe, dark matter, or dark energy are interesting topics so to be studied in future.

Regarding dark energy, a work by Linde \textit{et al.}~\cite{Linde2014a} appeared recently in the finalizing process of this thesis.
They studied a possibility of realizing both the primordial (inflation) and current (dark energy) accelerating expansion of the universe in our single superfield framework.
 They found that the following na\"{i}ve guess is wrong: since some of our models, in particular the models accommodating arbitrary positive semidefinite potentials, do not break SUSY at the vacuum with the cosmological constant zero, it would be easy to obtain dark energy $V\sim 10^{-120}M_{\text{G}}^4$ by an infinitesimal deformation of parameters of the theory.
In fact, when one adds a small term to the superpotential, the vacuum energy goes negative.
Increasing the effect of the term, the vacuum energy decrease more, turn to increase at some point, and eventually cross zero to become positive.
But near the zero point, deformation of the original model is so large that SUSY breaking becomes of order of more or less the inflaton mass scale.
They pointed out that it is a manifestation of a theorem: 
\begin{quote}
\textit{
``a supersymmetric Minkowski vacuum
without flat directions cannot be continuously deformed into a non-supersymmetric vacuum.''}~\cite{Kallosh2014f}
\end{quote}
Although it would be more desirable to describe both inflation and dark energy by a single superfield, the fact that it is impossible is not really a problem at all.
We can just add a SUSY breaking sector to explain SUSY breaking, and assume usual fine-tuning of the cancellation between SUSY breaking $K^{\bar{z}z}|D_z W|^2$ and ``gravitino mass'' $3|W|^2$ to reproduce the tiny cosmological constant, as they also commented in their conclusions~\cite{Linde2014a}.
This evades the no-go theorem because the SUSY breaking field becomes a flat direction in the supersymmetric limit~\cite{Kallosh2014f}.
For more ambitious approach to describe both inflation and dark energy, one may use nilpotent superfields.
See Subsection~\ref{subsec:nilpotent}.

We point out some new possibilities.
First, our technique to embed arbitrary positive semidefinite scalar potentials in to supergravity without the stabilizer superfield is not necessarily restricted to inflation.
When one want to design a single field potential in supergravity, it becomes a useful tool.
Second, we have enlarged the viable domain of applicability of single superfield inflation (sGoldstino inflation) from small field inflation to small and large field inflation, so a new possibility naturally emerges.
 It might be possible to realize hybrid inflation entirely relying on two scalar components of the single superfield modifying its K\"{a}hler potential.
 This may deserve the name of ``self-hybrid inflation''.
 To construct it or exclude its possibility  is another possible future work.

In conclusion, we have pioneered an entirely new branch of inflation in supergravity, which enables us to realize inflation, in particular large field inflation, without the need for the stabilizer superfield.
We have reduced the number of degrees of freedom required for (large field) inflation by half.
Our framework is minimal in this sense, but at the same time powerful since various kinds of scalar potentials can be realized.
This is theoretically exciting and phenomenological or cosmological consequences are to be further explored in future.

\newgeometry{top=-50pt, left=70pt, right=50pt}

\chapter*{} \label{acknowledgement} 
\leavevmode \\
\vspace{-20pt}
\hypertarget{acknowledgement}{}
\section*{Acknowledgement}
\vspace{10pt}

First of all, I appreciate collaboration with S.~V.~Ketov.
Without him, this thesis would not exist.  I thank him also for kind advices on academic career \textit{etc.}
I would like to thank M.~Endo, K.~Hamaguchi, T.~Moroi, and K.~Nakayama for discussion in an informal seminar I did after our first paper~\cite{Ketov2014}.
Motivated by the discussion and questions I had received in the seminar, I decided to study the stabilization effect quantitatively in the second paper~\cite{Ketov2014a}.
It is my great pleasure to thank K.~Nakayama for both scientific discussion and kind advices on sociological things.
I have studied the contents of Section~\ref{sec:OtherTerms}, suggested by discussion with him.
I thank him for inviting me and other students often to puzzle and escape games for relaxation.
Also, I acknowledge H.~Murayama as the chief examiner of the thesis.
His comments were very useful for revision of the thesis, which was about the bound on the coupling constant of inflaton, investigation of SUSY preservation taking into the finite $\zeta$ effect into account, and so on.
Discussions with colleagues at my laboratory was very helpful.
I occasionally asked miscellaneous questions about SUSY phenomenology to T.~Kitahara, and he kindly answered them.  S.~Nakamura's simple questions triggered my conception of a new idea.  Discussions with K.~Mukaida and M.~Takimoto was also very useful.
It was suggested to me by K.~Nakayama, M.~Ibe, and K.~Harigaya that it may be easier to understand our method in terms of the shifted field such that its expectation value vanishes.
I am deeply indebted to K.~Hamaguchi, my advisor, for his kind support during my master and Ph.D. course.
  He suggested me to do numerical calculation done in Section~\ref{sec:stabilization}, which is based on the same techniques used in our previous collaboration~\cite{Hamaguchi2014}.
His introduction of me to S.~V.~Ketov let us start collaboration.  I am grateful to T.~Kawano and C.~Wieck for discussion from the symmetry and superstring perspectives.
I am also thankful to K.~Choi, K.~Kohri, K.~Schmitz, Y.~Sumitomo, and T.~T.~Yanagida for discussions.
Correspondence with G.~Dall'Agata and F.~Zwirner clarified similarities and differences between their work~\cite{DallAgata2014} and our work~\cite{Ketov2014a}.
Discussions with S.~Aoki and Y.~Yamada help me appreciate their work~\cite{Abe2014}.
I was benefited from M.~Ueda, my vice-advisor in the ALPS course.
I learned a lot from his general critical advicies as a scientist.
The anonymous referee of our second paper~\cite{Ketov2014a} made me to face seriously to possible origins of the stabilization term used in our approach.
I appreciate all comments and questions I received at conferences although I could not remember all of their faces and names.

I presented a talk on our works as an additional lecture, which is assigned for a work completed after the deadline but having significant importance, at Autumn Meeting of The Physical Society of Japan.
I would like to thank correspondence with K.~Ishikawa, N.~Maru, and Y.~Satoh.

I made a one minute trailer movie for our paper~\cite{Ketov2014} and uploaded it to YouTube.
The movie titled with \href{http://neko2.net/ssi}{``Single Superfield Inflation''} have been viewed over 11500 times as of Dec.~15, 2014.  I would like to express my gratitude for interests in our work (and the invention of somewhat cheating advertising method of an academic paper).
I thank K.~Oda for advising me to free access limitation of the movie.
I also thank those (re)tweeted about our work and the trailer.  The buzz is summarized in \href{http://togetter.com/li/708405}{Togetter}.
I wish to thank S.~Carroll for introducing the trailer as well as the contents of our paper in his \href{http://www.preposterousuniverse.com/blog/2014/08/17/single-superfield-inflation-the-trailer/}{blog article}.

Finally, I want to express my appreciation to my family for their continual hearty supports.

I had been supported since March 2012 until March 2014 by grants of Advanced Leading Graduate Course for Photon Science (ALPS) at the University of Tokyo, which is an example of the Program for Leading Graduate Schools, MEXT, Japan.
I have been supported since April 2014 by a Grant-in-Aid for JSPS Fellows, and a Grant-in-Aid of the JSPS under No.~26$\cdot$10619.

\restoregeometry

\appendix
\chapter{Bird's-eye Review of Supergravity} \label{ch:SUGRA}
Supergravity~\cite{FreedmanNieuwenhuizenFerrara1976, DeserZumino1976a} (see also textbooks and reviews~\cite{WB, Buchbinder1998d, Gates1983, Weinberg3, Freedman2012}) is the local or gauged version of supersymmetry (SUSY)~\cite{WessZumino1974b}.
It includes gravitation as a consequence of gauging of SUSY.
We are not intended to present a self-contained and detailed review of this large subject.
Instead, we give a bird's-eye view of some formulations of supergravity in preparation of discussion in the subsequent Chapters.

\subsection*{Conformal and Poincar\'{e} supergravity}
The standard supergravity is also called Poincar\'{e} supergravity, because the supergravity symmetry group contains Poincar\'{e} group as its subgroup.
The symmetry algebra is generated by the generators of translation $P_{\mu}$, Lorentz transformation $M_{\mu\nu}$, and supersymmetry $Q$.
Corresponding to these generators, there are gauge fields, vierbein (graviton) $e_{\mu}{}^{a}$, spin connection $\omega_{\mu}{}^{ab}$, and gravitino $\psi_{\mu}$.
The spin connection can be fixed by imposing a constraint.

One of the standard formalism is based on curved superspace.
As General Relativity can be formulated as a deffeomorphism invariant gauge theory in curved spacetime with a local Lorentz gauge group, Poincar\'{e} supergravity can be formulated as a deffeomorphism invariant gauge theory in curved superspace with a local Lorentz gauge group.
In short, supergravity is General Relativity on curved superspace.

There is a formulation called conformal supergravity or superconformal tensor calculus (see Refs.~\cite{Kugo1983, Kugo1985, VanNieuwenhuizen1981} and references therein), whose bosonic part of the symmetry group is conformal group, a larger group containing Poincar\'{e} group.
More concretely, the superconformal algebra is generated by the generators of dilatation $D$, conformal supersymmetry or `$S$-supersymmetry' $S$, $U(1)_{R}$ symmetry $T$, and conformal boost $K_{\mu}$ in addition to those of Poincar\'{e} supergravity.
Correspondingly, there are gauge fields, $b_{\mu}$, $\phi_{\mu}$, $A_{\mu}$, and $f_{\mu}{}^{a}$.
The gauge fields of $S$-supersymmetry $\phi_{\mu}$ and conformal boost $f_{\mu}{}^{a}$ as well as Lorentz symmetry $\omega_{\mu}{}^{ab}$ are fixed by constraints.
The purpose of the enhancement of the symmetry in conformal supergravity is to describe the same physics as Poincar\'{e} supergravity more simply and elegantly.
These additional symmetries are ``gauge fixed'' to reproduce the action of Poincar\'{e} supergravity.
When one constructs an action, compensator superfields are employed to make an original action superconformally invariant.
In contrast to the formalism in curved superspace, the superconformal approach is not based superspace, 
and the action is given in terms of invariant action formulae.

The closure of the supergravity algebra off-shell requires introduction of auxiliary fields in the supergravity multiplet.
The choice of the auxiliary fields are not unique, and there are two minimal formulations: the old-minimal~\cite{Ferrara1978, Stelle1978, Fradkin1978} and the new-minimal~\cite{Sohnius1981} formulations.

\subsection*{Old-minimal supergravity}
For the old-minimal choice of auxiliary field, the supergravity supermultiplet is composed of the graviton $e_{\mu}{}^{a}$, gravitino $\psi_{\mu}$, a real vector $A_{\mu}$, and a complex scalar $h$.
Taking the gauge fixing of local Lorentz symmetry into account, these have 10, 16, 4, and 2 degrees of freedom off-shell.  These reduce to 2, 2, 0, and 0 on-shell.  The bosonic and fermionic number of degrees of freedom match both off- and on-shell.

The latter two, auxiliary fields have the following origin in the curved superspace formalism.
The basic ingredients of the formalism are the vierbein superfield $E_{M}{}^{A}(x,\theta, \bar{\theta})$ and the superfield of the connection of the local Lorentz symmetry $\phi_{MA}{}^{B}(x,\theta,\bar{\theta})$.
The action is constructed in terms of (super)geometric objects like torsion $T^{A}=\ms{D}E^{A}$ and curvature $R_{A}{}^{B}=\text{d}\phi_{A}{}^{B}+\phi_{A}{}^{C}\phi_{C}{}^{B}$.
These satisfy the Bianchi identities: $\ms{D}T^{A}=E^{B}R_{B}{}^{A}$ and $\ms{D}R_{A}{}^{B}=0$.
These superfields have too many independent components, so we put constraints on the torsion to reduce the number of independent components.  The constraints are $T_{cb}{}^{a}=T_{c\underline{\beta}}{}^{a}=T_{\gamma \beta}{}^{a}=T_{\dot{\gamma}\dot{\beta}}{}^{a}=T_{\underline{\gamma} \underline{\beta}}{}^{\underline{\alpha}}=0$ and $T_{\gamma \dot{\beta}}{}^{a}=2i\sigma^{a}_{\gamma\dot{\beta}}$ where $\underline{\alpha}=\alpha, \dot{\alpha}$.
Consistency with these constraints and the Bianchi identities tells us that all the components of the torsion and curvature are written in terms of the three Lorentz-irreducible superfields, $\mc{R}$, $\mc{G}_{\mu}$, and $\mc{W}_{\alpha \beta \gamma}$.  The auxiliary field $A_{\mu}$ is the lowest component of $\mc{G}_{\mu}$, and $h$ is the lowest component of $\mc{R}$, up to a constant coefficient. 
The three superfields obey the following constraints as consequences of the Bianchi identities: $\bar{\ms{D}}\mc{R}=\bar{\ms{D}}\mc{W}_{\alpha\beta\gamma}=0$ ($\mc{R}$ and $\mc{W}$ are chiral), $\bar{\mc{G}}_{\mu}=\mc{G}_{\mu}$ ($\mc{G}_{\mu}$ is Hermitian), $\ms{D}^{\beta}\mc{G}_{\beta\dot{\alpha}}=\bar{D}_{\dot{\alpha}}\mc{R}$, and $\ms{D}^{\rho}\mc{W}_{\rho\alpha\beta}=i\ms{D}_{(\alpha}{}^{\dot{\rho}}\mc{G}_{\beta)\dot{\rho}}$.
$\mc{R}$, $\mc{G}_{\mu}$, and $\mc{W}$ contain Ricci scalar, Ricci tensor, and Weyl tensor, respectively.

The minimal supergravity Lagrangian density that leads to Einstein-Hilbert Lagrangian density as the bosonic part is (see \textit{e.g.}~Ref.~\cite{Hindawi1996b} and earlier works~\cite{Arnowitt1975, WessZumino1978})
\begin{align}
\mc{L}=-3 \int \text{d}^4\theta  E  \label{pureSUGRAoldFull},
\end{align}
where $E$ is the density of superspace~\cite{Arnowitt1975}, which is the analogue of the invariant measure $e=\det e_{\mu}{}^{a}=\sqrt{-g}$ in supergravity.
Remarkably, the action is just an integration of a constant (times the density) over full superspace.
It is convenient to rewrite this in chiral superspace.
\begin{align}
\mc{L}=&-\int \text{d}^2 \Theta 2 \ms{E} 3 \mc{R} + \text{H.c.} \nonumber \\
=&e\left( -\frac{1}{2}R + \frac{1}{2}\epsilon^{\mu\nu\rho\sigma} \left( \bar{\psi}_{\mu}\bar{\sigma}_{\nu}\ms{D}_{\rho}\psi_{\sigma}-\psi_{\mu}\sigma_{\nu}\ms{D}_{\rho}\bar{\psi}_{\sigma}\right) -\frac{1}{3}\bar{h}h+\frac{1}{3}A^{a}A_{a} \right), \label{pureSUGRAold}
\end{align}
where $\Theta$ is the so-called new theta variable~\cite{Wess1978, Ramirez1988}, and $\ms{E}$ is the chiral density~\cite{Wess1978}.
Variation of the auxiliary fields eliminates themselves, and the Einstein-Hilbert action and the gravitino kinetic term remains.
The equivalence between eqs.~\eqref{pureSUGRAoldFull} and \eqref{pureSUGRAold} can be seen as a generalization of the formula $\int \text{d}^4 x \text{d}^4 \theta  F =-\frac{1}{8} \int \text{d}^4 x \text{d}^2 \theta F + \text{H.c.}$ for any Hermitian $F$ in global SUSY~\cite{WB}.
When coupled to matter, the Lagrangian density is of the following form in superspace,
\begin{align}
\mc{L}=\int \text{d}^2\Theta 2 \ms{E} \left( \frac{3}{8}\left( \bar{\ms{D}}\bar{\ms{D}}-8R \right)e^{-K/3}+W+\frac{1}{4}H_{AB}\mc{W}^{A}\mc{W}^{B}  \right)+\text{H.c.}, \label{matterSUGRAoldSS}
\end{align}
where $K$ is the K\"{a}hler potential, $W$ is the superpotential, $H_{AB}$ is the gauge kinetic function, and $\mc{W}=\frac{1}{8}(\bar{\ms{D}}\bar{\ms{D}}-8\mc{R})e^{2V}\ms{D}e^{-2V}$ is the field strength of gauge superfields.
This Lagrangian density in terms of component fields and details of the above arguments are found in Ref.~\cite{WB}.

In the conformal supergravity, the old-minimal formulation utilize a chiral compensator, $S_0 = [z, \chi, h]$.
The gauge freedom of the conformal boost ($K$-gauge) is used to eliminate the gauge field of dilatation, $b_{\mu}=0$.
Similarly, one can eliminate the fermionic component of the compensator supermultiplet $\chi$ and set the complex scalar component $z$ to $1$ by the gauge choice of the conformal supersymmetry ($S$-gauge) and dilatation/$U(1)_R$ ($D$- and $T$-gauge), respectively.
The auxiliary field in the old-minimal supergravity, $A_{\mu}$ and $h$, can thus be interpreted as the gauge field of $U(1)_{R}$ symmetry and the auxiliary field of the chiral compensator multiplet.
After gauge fixing, the superconformal theory reduces to the Poincar\'{e} one.

For example, the pure supergravity~\eqref{pureSUGRAold} has a superconformal form
\begin{align}
\mc{L}=-\frac{3}{2}\left[\bar{S}_{0} S_0 \right]_{D}.
\end{align}
This is analogous to eq.~\eqref{pureSUGRAoldFull}, and the matter-coupled supergravity corresponding to eq.~\eqref{matterSUGRAoldSS} is described by
\begin{align}
\mc{L}=-\frac{3}{2}\left [  \bar{S}_0 S_0 e^{-K/3} \right ]_D + \left[ S_0^3 W \right]_F + \frac{1}{4}\left[ H_{AB} \mc{W}^{A}\mc{W}^{B}\right]_F. \label{matterSUGRAold}
\end{align}

\subsection*{New-minimal supergravity}
In the case of the new-minimal supergravity, the supergravity multiplet is composed of the graviton $e_{\mu}{}^{a}$, gravitino $\psi_{\mu}$, a real vector $A_{\mu}$, and a real two-form field $a_{\mu\nu}$.
The last one only appears in the combination $V_{\mu}\sim \epsilon_{\mu\nu\rho\sigma}\partial^{\nu}a^{\rho\sigma}$, so this real vector is constrained to be divergenceless, $\partial^{\mu}V_{\mu}=0$. It has three independent components.
The new-minimal supergravity has a gauged $R$-symmetry, and its gauge fixes a component of $A_{\mu}$.
The numbers of degrees of freedom are thus 10, 16, 3, and 3, respectively off-shell.
On-shell, these reduce to 2, 2, 0, and 0, and the spectrum of the theory coincides with that of the old-minimal supergravity.
In fact, the component Lagrangian is given by
\begin{align}
\mc{L}=&e\left( -\frac{1}{2}R + \frac{1}{2}\epsilon^{\mu\nu\rho\sigma} \left( \bar{\psi}_{\mu}\bar{\sigma}_{\nu}\ms{D}_{\rho}\psi_{\sigma}-\psi_{\mu}\sigma_{\nu}\ms{D}_{\rho}\bar{\psi}_{\sigma}\right) -(2A_{\mu}+V_{\mu})V^{\mu} \right)
\end{align}
up to surface terms. The gauge transformation of $A_{\mu}\rightarrow A_{\mu}+\partial_{\mu}\lambda$ produces only a surface term because of the constraint $\partial_{\mu}V^{\mu}$. The auxiliary fields $A_{\mu}$ and $V_{\mu}$ vanish on-shell, and then the Lagrangian becomes the same as that in the old-minimal one.

In curved superspace, basic superfields in the new-minimal supergravity are a chiral superfield $\mc{W}_{\alpha}$, which contains Ricci scalar, a real linear superfield\footnote{
Linear superfields do not appear as often as chiral and real superfields in the literature, so we here let the reader remember the definition.
These superfields are constrained ones.
For example, a (left-)chiral super field $\Phi$ is characterized by $\bar{D}_{\dot{\alpha}}\Phi=0$, and a real superfield $V$ is characterized by $\bar{V}=V$.
A (complex) linear superfield $\mc{L}$ is characterized by $(DD+\bar{D}\bar{D})\mc{L}=0$, and a real linear superfield $L$ is characterized by $DDL=\bar{D}\bar{D}L=0$.
The operator $\frac{1}{4}(DD+\bar{D}\bar{D})$ is the chiral projection operator such that a general superfield $S$ operated by the operator becomes chiral (in a broad sense; the sum of chiral and anti-chiral superfields in a narrow sense) $S'=\frac{1}{4}(DD+\bar{D}\bar{D})S$.
Thus, a linear superfield is such a superfield that its chiral projection vanishes.
A real linear superfield contains a real scalar $C$ and a divergenceless real scalar $V_{\mu}$ as independent bosonic fields, and a Weyl spinor $\omega$ and its conjugate $\bar{\omega}$ as independent fermionic fields.
These are four bosonic plus four fermionic degrees of freedom off-shell, which reduces to $2+2$ on-shell.
The expansion of a real linear superfield is as follows,
\begin{align}
L=C+i\theta \omega -i \bar{\theta}\bar{\omega}-\theta \sigma^{\mu}\bar{\theta}V_{\mu}+\dots.
\end{align}
} $\mc{E}_{\mu}$ (or chiral one $\bar{\ms{D}}\mc{E}$), which contains Ricci tensor, and a chiral superfield $\mc{W}_{\mu\nu\alpha}$ (or $\mc{W}_{\alpha\beta\gamma}=\frac{i}{4}(\sigma^{\mu\nu})_{\alpha\beta}\mc{W}_{\mu\nu\gamma}$), which contains Weyl tensor.
The Bianchi identities imply $\bar{\ms{D}}_{\dot{\alpha}}\mc{W}_{\alpha}=0$, $\mc{W}_{\alpha}=-\frac{1}{2}\bar{\ms{D}}^{\dot{\alpha}}\mc{E}_{\dot{\alpha}\alpha}$, $\bar{\ms{D}}_{\dot{\alpha}}\mc{W}_{\alpha\beta\gamma}=0$, and $\ms{D}_{\alpha}\mc{W}^{\alpha}{}_{\beta\gamma}=\frac{1}{2}\partial_{(\beta}{}^{\dot{\alpha}}\mc{E}_{\dot{\alpha}\gamma)}+\frac{i}{3}\ms{D}_{(\beta}\mc{W}_{\gamma)}$.

In the superconformal tensor calculus, the choice of the compensator for the new-minimal supergravity is a real linear supermultiplet, $L_0=[C, \omega, V_{\mu}]$.
We can eliminate the gauge field of dilatation $b_{\mu}$ and the spinor component of the linear compensator $\omega$ by $K$-gauge.
We are also able to fix the real scalar component of the compensator $C$ to 1 by $D$-gauge.
This reduces the superconformal theory to the Poincar\'{e} version.

The pure supergravity Lagrangian density is expressed by~\cite{Wit1982}
\begin{align}
\mc{L}=\frac{3}{2}\left [ L \ln \left( \frac{L}{\bar{S}_0 S_0 } \right) \right ]_D, \label{pureSUGRAnew}
\end{align}
where $S_0$ is an auxiliary chiral supermultiplet in the sense that the Lagrangian density does not depend on the value of $S_0$.  In fact, transforming it as $S_0\rightarrow S_0 e^{\Lambda}$ with $\Lambda$ chiral supermultiplet, the variation term $-[L(\Lambda+\bar{\Lambda})]_D=[\text{(linear)}\times \text{(chiral)}]_D$ vanishes.  This is called ``the additional gauge symmetry''~\cite{Ferrara1983}.  
Because the compensator is not chiral but linear in the new-minimal case, the superpotential term can be incorporated in the theory only if it is superconformal with itself.
The matter-coupled general Lagrangian density is 
\begin{align}
\mc{L}= \frac{3}{2}\left [ L \left( \ln \left( \frac{f}{\bar{S}_0 S_0} \right)  +g_{R}V_{R} \right) \right ]_{D} + \left[W\right]_F + \frac{1}{4}\left[ H_{AB} \mc{W}^{A}\mc{W}^{B}\right]_F + \frac{1}{4}\left[ H_{R} \mc{W}^{R}\mc{W}^{R}\right]_F, \label{matterSUGRAnew}
\end{align}
where $V_{R}$ is a real supermultiplet which has an FI term, and $f$ is a Hermitian function (like K\"{a}hler potential) of the linear compensator and matter.

It has been proven that the matter-coupled old-minimal supergravity~\eqref{matterSUGRAold} is equivalent to (can be rewritten in the language of) the matter-coupled new-minimal supergravity~\eqref{matterSUGRAnew} if and only if the superpotential of the old-minimal side respects $R$-symmetry~\cite{Ferrara1983}.
However, this equivalence no longer holds when one consider higher order terms in supergravity (like terms containing Ricci scalar squared $R^2$).

\chapter{Short Review of Inflation}
\label{ch:inflation}

We consider single field canonically normalized slow-roll inflation in the Einstein frame described by the Lagrangian density
\begin{align}
\mc{L}=\sqrt{-g} \left( -\frac{1}{2}R -\frac{1}{2}g^{\mu}\partial_{\mu}\phi \partial_{\nu}\phi - V(\phi) \right),
\end{align}
where $g$ is the determinant of the metric, $R$ is the Ricci scalar curvature, $\phi$ is the inflaton, and $V$ is its potential.
The equations of motion for the metric (Einstein equation) and the scalar are
\begin{align}
&-R_{\mu\nu}+\frac{1}{2} g_{\mu\nu}R=T_{\mu\nu} ,\\
&\frac{1}{\sqrt{-g}}\partial_{\mu} \left( \sqrt{-g}g^{\mu\nu}\partial_{\nu}\phi \right)- V'=0,
\end{align}
where $R_{\mu\nu}$ is the Ricci tensor, $T_{\mu\nu}$ is the energy momentum tensor, and the prime denotes differentiation with respect to the inflaton $\phi$.
The energy momentum tensor is given by
\begin{align}
T^{\mu\nu}=\partial^{\mu}\phi\partial^{nu}\phi +g^{\mu\nu} \left( -\frac{1}{2}g^{\rho\sigma}\partial_{\rho}\phi\partial_{\sigma}\phi -V(\phi) \right).
\end{align}
Neglecting spatial derivatives in the homogeneous and isotropic background (Friedmann-Lema\^itre-Robertson-Walker spacetime), the equation of motion for the scalar becomes
\begin{align}
\ddot{\phi}+3H\dot{\phi}+V'=0,
\end{align}
where $H$ is the Hubble variable.

In the slow-roll approximation, we neglect the first term leading to $3H\dot{\phi}+V'=0$.
The Einstein equation reduces to slow-roll Friedmann equations,
\begin{align}
3H^2 =\frac{3\ddot{a}}{a}=V.
\end{align}
Slow-roll parameters are defined as 
\begin{align}
\epsilon=&\frac{1}{2} \left( \frac{V'}{V} \right)^2, &
\eta=& \frac{V''}{V}, &
\xi^2=& \frac{V'V'''}{V^2}.
\end{align}
The slow-roll approximation is expressed as $\epsilon \ll 1$.
To keep inflationary expansion sufficiently long, another condition is usually required, $|\eta|\ll 1$.
The $e$-folding number $N_{*}$ for a mode of wavenumber $k_{*}$ from its horizon crossing to the end of inflation is 
\begin{align}
N_{*}=\int_{t_{*}}^{t_{\text{end}}} \text{d}t H = \int _{\phi_{\text{end}}}^{\phi_{*}}\text{d}\phi \, \frac{1}{\sqrt{2\epsilon}} ,
\end{align}
and its value corresponding to the mode we observe today should be in the range $50\lesssim N_{*}\lesssim 60$ depending on the details of reheating.

Quantum fluctuation of the inflaton is approximately Gaussian and scale invariant.
It is expanded by inflation and becomes the seed of density perturbation.
A part of inflaton fluctuation is converted to fluctuation of the scalar part of the metric called curvature perturbation.
Similarly, the transverse traceless part of the metric produces quantum fluctuation called tensor perturbation.
The power spectra of these perturbations are parameterized as
\begin{align}
\mc{P}_{\text{s}}(k)=& A_{\text{s}} \left( \frac{k}{k_{*}} \right)^{n_{\text{s}}-1 + \frac{1}{2} \frac{\text{d} n_{\text{s}}}{\text{d}\ln k}\ln \frac{k}{k_{*}}+ \dots },  & 
\mc{P}_{\text{t}}(k)=& A_{\text{t}} \left( \frac{k}{k_{*}} \right)^{n_{\text{t}} + \frac{1}{2} \frac{\text{d} n_{\text{t}}}{\text{d}\ln k} \ln \frac{k}{k_{*}}+ \dots} ,
\end{align}
where $A_{i}$ $(i=\text{s}, \text{t})$ are the amplitudes, $n_{i}$ are the spectral indices, and $\text{d}n_{i}/\text{d}\ln k$ are the runnings of the spectral indices.
They are given by
\begin{align}
A_{\text{s}}=&\frac{V}{24 \pi^2 \epsilon},  &
A_{\text{t}}=& \frac{2V}{3\pi^2}, \\
n_{\text{s}}-1=& -6 \epsilon +2 \eta , &
n_{\text{t}}=& -2 \epsilon , \\
\frac{\text{d}n_{\text{s}}}{\text{d}\ln k}=& 16 \epsilon \eta - 24 \epsilon^2 -2 \xi^2 , &
\frac{\text{d}n_{\text{t}}}{\text{d}\ln k}=& 4 \epsilon \eta - 8 \epsilon^2 .
\end{align}
The tensor-to-scalar ratio of the amplitudes of the power spectra is 
\begin{align}
r=16 \epsilon .
\end{align}
Detection of $r$ implies inflaton field excursion of order Planck mass during inflation because of the Lyth bound, $\Delta \phi \gtrsim 0.4 ( r / 0.05 )^{1/2}$~\cite{Lyth1997}.

Let us look at some inflationary model predictions.
For the monomial potential case,
\begin{align}
V =c_n \phi^n ,
\end{align}
where $c_n$ is a constant, the the first two slow-roll parameters are
\begin{align}
\epsilon=& \frac{n^2}{2\phi^2}, &
\eta=& \frac{n (n-1)}{\phi^2}.
\end{align}
The scalar amplitude, the scalar spectral index, and the tensor-to-scalar ratio are
\begin{align}
A_{\text{s}}=&\frac{c_n \phi^{n+2}}{12\pi^2 n^2}=\frac{c_n \left( 2n\left(N+\frac{n}{4}\right) \right)^{\frac{n+2}{2}}}{12\pi^2 n^2}, \\
1-n_{\text{s}}=& \frac{n(n+2)}{\phi^2}=\frac{n+2}{2\left(N+\frac{n}{4}\right)}, \\
r=& \frac{8n^2}{\phi^2}= \frac{4n}{N+\frac{n}{4}},
\end{align}
where we have defined the end of inflation as the time when $\epsilon$ becomes one, $\epsilon_{\text{end}}=(n^2 /2 \phi_{\text{end}}^2)=1$, so that the $e$-folding number is $N=\frac{\phi^2}{2n}-\frac{n}{4}$.

In the case of the Starobinsky-like model,
\begin{align}
V=V_0 \left( 1-e^{-a\phi}\right)^2,
\end{align}
where $V_0$ and $a$ are constants, the slow-roll parameters are
\begin{align}
\epsilon=& \frac{2a^2 e^{-2a\phi}}{(1-e^{-a\phi})^2} , & 
\eta=& \frac{-2a^2 e^{-a\phi}(1-2e^{-a\phi})}{(1-e^{-a\phi})^2}.
\end{align}
The amplitude, the spectral index and the tensor-to-scalar ratio are
\begin{align}
A_{\text{s}}=&\frac{V_0 e^{2a\phi}(1-e^{-a\phi})^4}{48\pi^2 a^2} \simeq \frac{a^2N^2 V_0\left(1-\frac{1}{2a^2 N} \right)^4 }{12\pi^2}, \\
1-n_{\text{s}}=& \frac{4a^2e^{-a\phi}(1+e^{-a\phi})}{(1-e^{-a\phi})^2}\simeq \frac{2 \left( 1+\frac{1}{2a^2N}\right)}{N\left(1- \frac{1}{2a^2N}\right)^2} , \\
r=& \frac{32a^2 e^{-2a\phi}}{(1-e^{-a\phi})^2}\simeq \frac{8}{a^2N^2 \left(1-\frac{1}{2a^2 N}\right)^2},
\end{align}
where the $e$-folding number is
\begin{align}
N=\frac{1}{2a^2} \left( e^{a\phi}-a\phi -(1+\sqrt{2}a) + \ln (1+\sqrt{2}a) \right) \simeq \frac{e^{a\phi}}{2a^2}.
\end{align}
Inflation ends at $\phi_{\text{end}}=a^{-1}\ln(1+\sqrt{2}a)$ where $\epsilon_{\text{end}}=1$.

Inflationary models must be confronted with data.
According to the Planck 2015 data~\cite{Ade2015a, Ade2015}, the amplitude of the curvature perturbation and the scalar spectral index are 
\begin{align}
A_{\text{s}}=& 2.198^{+0.076}_{-0.085} \times 10^{-9} &   (68\% \text{CL}, \text{Planck TT}+\text{lowP}), \\
n_{\text{s}}=&0.9655\pm0.0062 &   (68\% \text{CL}, \text{Planck TT}+\text{lowP}),
\end{align}
in the $\Lambda$CDM model with the pivot scale $k_0=0.05 \text{Mpc}^{-1}$. 
Allowing tensor perturbation, the spectral index and the tensor-to-scalar ratio are constrained with various data sets as 
\begin{align}
n_{\text{s}}=&0.9666\pm0.0062   &
r_{0.002}<&0.103 &  &(\text{Planck TT}+\text{lowP}), \\
n_{\text{s}}=&0.9688\pm0.0061,   &
r_{0.002}<&0.114 &  &(\text{Planck TT}+\text{lowP}+\text{lensing}), \\
n_{\text{s}}=&0.9680\pm0.0045,   &
r_{0.002}<&0.113 &  &(\text{Planck TT}+\text{lowP}+\text{BAO}), \\
n_{\text{s}}=&0.9652\pm0.0047,   &
r_{0.002}<&0.099 & & (\text{Planck TT, TE, EE}+\text{lowP}). 
\end{align}
where the confidence levels of the constraints are 68\% and 95\% for $n_{\text{s}}$ and $r$, respectively, and the tensor-to-scalar ratios are evaluated at the pivot scale $0.002 \text{Mpc}^{-1}$.

The tensor-to-scalar ratio is independently constrained by the joint analysis of BICEP2/Keck Array-Planck (BKP)~\cite{Ade2015b} as
\begin{align}
r_{0.05}<&0.12 &  (95\% \text{CL}, \text{BKP}),
\end{align}
where $r_{0.05}$ is the tensor-to-scale evaluated at the pivot scale $0.05 \text{Mpc}^{-1}$.
Combined with the above Planck data, it is more severely constrained as~\cite{Ade2015}
\begin{align}
r_{0.002}<&0.08 & (95\% \text{CL}, \text{Planck TT}+\text{lowP}+\text{BKP}).
\end{align}
Using other data sets or allowing more free parameters, these constraints become milder.  See Refs.~\cite{Ade2015b, Ade2015a, Ade2015} for more details.

\bibliographystyle{utphys}
\bibliography{ref}
\printindex
\end{document}